\documentstyle[twoside]{article} 

\input{amssym.def} \input{amssym.tex} 
\newtheorem{Definition}{Definition}[subsection]
\newtheorem{Lemma}[Definition]{Lemma}
\newtheorem{Theorem}[Definition]{Theorem}
\newtheorem{Proposition}[Definition]{Proposition}
\newtheorem{Example}[Definition]{Example}
\newtheorem{Corollary}[Definition]{Corollary}
\newtheorem{Construction}[Definition]{Construction}
\newcommand{\setc}{\setcounter}

\newcommand{\su}{\subsection}

\renewcommand{\ll}{\label}
\newcommand{\be}{\begin{equation}}
\newcommand{\ee}{\end{equation}}
\newcommand{\bea}{\begin{eqnarray}}
\newcommand{\eea}{\end{eqnarray}}
\newcommand{\nn}{\nonumber}

\newcommand{\qm}{quantum mechanics}

\newcommand{\ca}{$C^*$-algebra}

\newcommand{\rep}{representation}
\newcommand{\irrep}{irreducible representation}
\newcommand{\Hs}{Hilbert space}
\newcommand{\Bs}{Banach space}
\newcommand{\BA}{Banach algebra}

\newcommand{\sta}{$\mbox{}^*$-algebra}

\newcommand{\HCM}{Hilbert $C^*$-module}
\newcommand{\vN}{von Neumann}
\newcommand{\vNa}{von Neumann algebra}
\newcommand{\zb}{\overline{z}}
\newcommand{\ovl}{\overline}
\newcommand{\wt}{\widetilde}
\newcommand{\til}{\tilde}
\newcommand{\raw}{\rightarrow}

\newcommand{\Raw}{\Rightarrow}

\newcommand{\lraw}{\leftrightarrow}

\newcommand{\rlh}{\rightleftharpoons}
\newcommand{\n}{\parallel}
\newcommand{\ot}{\otimes}

\newcommand{\la}{\langle}
\newcommand{\ra}{\rangle}

\newcommand{\x}{\times}
\newcommand{\hb}{\hbar}

\newcommand{\ran}{{\rm ran}}
\newcommand{\Tr}{\mbox{\rm Tr}\,}

\newcommand{\KH}{{\frak B}_0({\cal H})}
\newcommand{\BH}{{\frak B}({\cal H})}

\newcommand{\cin}{C^{\infty}}
\newcommand{\cci}{C^{\infty}_c}
\newcommand{\half}{\mbox{\footnotesize $\frac{1}{2}$}}
\newcommand{\third}{\mbox{\footnotesize $\frac{1}{3}$}}
\newcommand{\quar}{\mbox{\footnotesize $\frac{1}{4}$}}

\newcommand{\eb}{\partial_e K}

\newcommand{\Ar}{{\frak A}_{\Bbb R}}

\newcommand{\Hlg}{{\cal H}_{\chi}}
\newcommand{\Hug}{{\cal H}^{\chi}}
\newcommand{\plg}{\pi_{\chi}}
\newcommand{\pug}{\pi^{\chi}}
\newcommand{\Ulg}{U_{\chi}}

\newcommand{\PA}{{\cal P}({\frak A})}
\newcommand{\SA}{{\cal S}({\frak A})}

\newcommand{\q}{{\cal Q}_{\hbar}}

\newcommand{\ddt}{\frac{d}{dt}}

\newcommand{\inv}{^{-1}}

\newcommand{\Exp}{{\rm Exp}}

\newcommand{\SHG}{{\sf H}^{\ch}}
\newcommand{\HG}{{\cal H}^{\ch}}
\newcommand{\Hg}{{\cal H}_{\ch}}

\newcommand{\bb}{\rangle_{\frak B}}
\newcommand{\ba}{\rangle_{\frak A}}

\newcommand{\Meq}{\stackrel{M}{\sim}}
\newcommand{\pir}{\pi_{\mbox{\tiny R}}}
\newcommand{\pil}{\pi_{\mbox{\tiny L}}}
\newcommand{\AI}{{\frak A}_{\Bbb I}}
\newcommand{\DA}{\Delta({\frak A})}
\newcommand{\al}{\alpha}
\newcommand{\bt}{\beta}
\newcommand{\gm}{\gamma}
\newcommand{\Gm}{\Gamma}
\newcommand{\dl}{\delta}
\newcommand{\Dl}{\Delta}
\newcommand{\ep}{\epsilon}

\newcommand{\th}{\theta}

\newcommand{\lm}{\lambda}
\newcommand{\Lm}{\Lambda}
\newcommand{\rh}{\rho}
\newcommand{\sg}{\sigma}
\newcommand{\Sg}{\Sigma}
\newcommand{\ta}{\tau}

\newcommand{\Ph}{\Phi}
\newcommand{\phv}{\varphi}
\newcommand{\ch}{\chi}
\newcommand{\ps}{\psi}
\newcommand{\Ps}{\Psi}
\newcommand{\om}{\omega}
\newcommand{\Om}{\Omega}
\newcommand{\A}{{\frak A}}
\newcommand{\B}{{\frak B}}
\newcommand{\GC}{{\frak C}}

\newcommand{\GI}{{\frak I}}

\newcommand{\M}{{\frak M}}
\newcommand{\GN}{{\frak N}}
\newcommand{\GU}{{\frak U}}

\newcommand{\g}{{\frak g}}

\newcommand{\CA}{{\cal A}}
\newcommand{\CB}{{\cal B}}

\newcommand{\CD}{{\cal D}}
\newcommand{\CE}{{\cal E}}

\renewcommand{\H}{{\cal H}}

\newcommand{\CI}{{\cal I}}
\newcommand{\CK}{{\cal K}}

\newcommand{\CN}{{\cal N}}
\newcommand{\CS}{{\cal S}}
\newcommand{\CO}{{\cal O}}

\newcommand{\CQ}{{\cal Q}}

\newcommand{\CT}{{\cal T}}
\newcommand{\CV}{{\cal V}}

\renewcommand{\P}{{\cal P}}

\newcommand{\C}{{\Bbb C}}

\newcommand{\I}{{\Bbb I}}
\newcommand{\N}{{\Bbb N}}
\newcommand{\R}{{\Bbb R}}

\newcommand{\SB}{{\sf B}}

\newcommand{\SP}{{\sf P}}

\newcommand{\SV}{{\sf V}}

\makeatletter
\newskip\tempskip
\def\endproof{{\parfillskip24\p@ plus\@ne fil\@@par}\tempskip\prevdepth
  \ifdim\lastskip=\z@\tempskip\z@\else\vskip-\lastskip
    \ifdim\tempskip>4\p@ \tempskip.5\tempskip \else \tempskip\z@\fi\fi
  \nobreak\vskip-\baselineskip\vskip-\tempskip\noindent\hbox 
to\hsize{\hfill
    $\blacksquare$}\par\vskip\tempskip\vskip\abovedisplayskip\@doendpe}
\makeatother
\makeatletter
\newskip\tempskip
\def\endiproof{{\parfillskip24\p@ plus\@ne fil\@@par}\tempskip\prevdepth
  \ifdim\lastskip=\z@\tempskip\z@\else\vskip-\lastskip
    \ifdim\tempskip>4\p@ \tempskip.5\tempskip \else \tempskip\z@\fi\fi
  \nobreak\vskip-\baselineskip\vskip-\tempskip\noindent\hbox 
to\hsize{\hfill
    $\Box$}\par\vskip\tempskip\vskip\abovedisplayskip\@doendpe}
\makeatother
\newcommand{\enp}{\endproof}
\newcommand{\eip}{\endiproof}
\topmargin = - 0.5 cm \textheight = 23 cm \textwidth = 15 cm
\begin{document} 
\pagenumbering{arabic} \setlength{\unitlength}{1cm}
\thispagestyle{empty}
\begin{center}
{\huge Lecture Notes on \\ \mbox{}\hfill \\
$C^*$-Algebras, Hilbert $C^*$-modules, \\ \mbox{}\hfill \\
and Quantum Mechanics}
\vspace{1cm}\\Draft: 8 April 1998 \vspace{2cm}\\
{\Large  N.P.~Landsman}
    \\ \mbox{} \hfill \\
 Korteweg-de Vries Institute for Mathematics, University of Amsterdam,\\
 Plantage Muidergracht 24,\\
1018 TV AMSTERDAM, THE NETHERLANDS \vspace{6mm} \\
{\em email:} npl@wins.uva.nl\\
{\em homepage:} http://turing.wins.uva.nl/$\mbox{}^{\sim}$npl/ \\
{\em telephone:} 020-5256282\\
{\em office:} Euclides 218a\vspace{1cm}
\end{center}
\newpage
\cleardoublepage
\tableofcontents
\cleardoublepage
\section{Historical notes}
\su{Origins in functional analysis and quantum mechanics} The
emergence of the theory of operator algebras may be traced back to (at
least) three developments.
\begin{itemize}
\item
The work of Hilbert and his pupils in G\"{o}ttingen on integral
equations, spectral theory, and infinite-dimensional quadratic forms
(1904-);
\item
The discovery of quantum mechanics by Heisenberg (1925) in
G\"{o}ttingen and (independently) by Schr\"{o}dinger in Z\"{u}rich
(1926);
\item
The arrival of John von Neumann in G\"{o}ttingen (1926) to become
Hilbert's assistant.
\end{itemize}
Hilbert's memoirs on integral equations appeared between 1904 and
1906.  In 1908 his student E. Schmidt defined the space $\ell^2$ in
the modern sense.  F. Riesz studied the space of all continuous linear
maps on $\ell^2$ (1912), and various examples of $L^2$-spaces emerged
around the same time.  However, the abstract concept of a \Hs\ was
still missing.

Heisenberg discovered a form of \qm, which at the time was called
`matrix mechanics'. Schr\"{o}dinger was led to a different formulation
of the theory, which he called `wave mechanics'. The relationship and
possible equivalence between these alternative formulations of \qm,
which at first sight looked completely different, was much discussed
at the time.  It was clear from either approach that the body of work
mentioned in the previous paragraph was relevant to \qm.

Heisenberg's paper initiating matrix mechanics was followed by the
`Drei\-m\"{a}n\-nerarbeit' of Born, Heisenberg, and Jordan (1926); all
three were in G\"{o}ttingen at that time. Born was one of the few
physicists of his time to be familiar with the concept of a matrix; in
previous research he had even used infinite matrices (Heisenberg's
fundamental equations could only be satisfied by infinite-dimensional
matrices).  Born turned to his former teacher Hilbert for mathematical
advice.  Hilbert had been interested in the mathematical structure of
physical theories for a long time; his Sixth Problem (1900) called for
the mathematical axiomatization of physics.  Aided by his assistants
Nordheim and von Neumann, Hilbert thus ran a seminar on the
mathematical structure of \qm, and the three wrote a joint paper on
the subject (now obsolete).

It was von Neumann alone who, at the age of 23, saw his way through
all structures and mathematical difficulties. In a series of papers
written between 1927-1932, culminating in his book {\em Mathematische
Grundlagen der Quantenmechanik} (1932), he formulated the abstract
concept of a \Hs, developed the spectral theory of bounded as well as
unbounded normal operators on a \Hs, and proved the mathematical
equivalence between matrix mechanics and wave mechanics.  Initiating
and largely completing the theory of self-adjoint operators on a \Hs,
and introducing notions such as density matrices and quantum entropy,
this book remains the definitive account of the mathematical structure
of elementary \qm. (von Neumann's book was preceded by Dirac's {\em
The Principles of Quantum Mechanics} (1930), which contains a
heuristic and mathematically unsatisfactory account of \qm\ in terms
of linear spaces and operators.)  \su{Rings of operators (von Neumann
algebras)} In one of his papers on \Hs\ theory (1929), von Neumann
defines a {\bf ring of operators} $\M$ (nowadays called a {\bf von
Neumann algebra}) as a $\mbox{}^*$-subalgebra of the algebra $\BH$ of
all bounded operators on a \Hs\ $\H$ (i.e, a subalgebra which is
closed under the involution $A\raw A^*$) that is closed (i.e.,
sequentially complete) in the weak operator topology.  The latter may
be defined by its notion of convergence: a sequence $\{A_n\}$ of
bounded operators weakly converges to $A$ when $(\Ps,A_n\Ps)\raw
(\Ps,A\Ps)$ for all $\Ps\in\H$.  This type of convergence is partly
motivated by \qm, in which $(\Ps,A\Ps)$ is the expectation value of
the observable $A$ in the state $\Ps$, provided that $A$ is
self-adjoint and $\Ps$ has unit norm.

For example, $\BH$ is itself a \vNa. (Since the weak topology is
weaker than the uniform (or norm) topology on $\BH$, a \vNa\ is
automatically norm-closed as well, so that, in terminology to be
introduced later on, a \vNa\ becomes a \ca\ when one changes the
topology from the weak to the uniform one. However, the natural
topology on a \vNa\ is neither the weak nor the uniform one.)

In the same paper, von Neumann proves what is still the basic theorem
of the subject: a $\mbox{}^*$-subalgebra $\M$ of $\BH$, containing the
unit operator $\I$, is weakly closed iff $\M''=\M$.  Here the {\bf
commutant} $\M'$ of a collection $\M$ of bounded operators consists of
all bounded operators which commute with all elements of $\M$, and the
bicommutant $\M''$ is simply $(\M')'$.  This theorem is remarkable, in
relating a topological condition to an algebraic one; one is reminded
of the much simpler fact that a linear subspace $\CK$ of $\H$ is
closed iff $\CK^{\perp\perp}$, where $\CK^{\perp}$ is the orthogonal
complement of $\CK$ in $\H$.

Von Neumann's motivation in studying rings of operators was
plurifold. His primary motivation probably came from \qm; unlike many
physicists then and even now, he knew that all \Hs s of a given
dimension are isomorphic, so that one cannot characterize a physical
system by saying that `its \Hs\ of (pure) states is
$L^2(\R^3)$'. Instead, \vN\ hoped to characterize quantum-mechanical
systems by algebraic conditions on the observables.  This programme
has, to some extent been realized in algebraic quantum field theory
(Haag and followers).

Among von Neumann's interest in \qm\ was the notion of entropy; he
wished to define states of minimal information. When $\H=\C^n$ for
$n<\infty$, such a state is given by the density matrix $\rh=\I/n$,
but for infinite-dimensional \Hs s this state may no longer be
defined.  Density matrices may be regarded as states on the \vNa $\BH$
(in the sense of positive linear functionals which map $\I$ to 1).  As
we shall see, there are \vNa s on infinite-dimensional \Hs s which do
admit states of minimal information that generalize $\I/n$, viz.\ the
factors of type II$\mbox{}_1$ (see below).

Furthermore, \vN\ hoped that the divergences in quantum field theory
might be removed by considering algebras of observables different from
$\BH$.  This hope has not materialized, although in algebraic quantum
field theory the basic algebras of local observables are, indeed, not
of the form $\BH$, but are all isomorphic to the unique hyperfinite
factor of type III$\mbox{}_1$ (see below).

Motivation from a different direction came from the structure theory
of algebras. In the present context, a theorem of Wedderburn says that
a \vNa\ on a finite-dimensional \Hs\ is (isomorphic to) a direct sum
of matrix algebras. Von Neumann wondered if this, or a similar result
in which direct sums are replaced by direct integrals (see below),
still holds when the dimension of $\H$ is infinite.  (As we shall see,
it does not.)

Finally, \vN's motivation came from group \rep s.  Von Neumann's
bicommutant theorem implies a useful alternative characterization of
von Neumann algebras; from now on we add to the definition of a von
Neumann algebra the condition that $\M$ contains $\I$.

The commutant of a group $\GU$ of unitary operators on a \Hs\ is a
\vNa, and, conversely, every \vNa\ arises in this way.  In one
direction, one trivially verifies that the commutant of any set of
bounded operators is weakly closed, whereas the commutant of a set of
bounded operators which is closed under the involution is a \sta. In
the opposite direction, given $\M$, one takes $\GU$ to be the set of
all unitaries in $\M'$.

This alternative characterization indicates why \vNa s are important
in physics: the set of bounded operators on $\H$ which are invariant
under a given group \rep\ $U(G)$ on $\H$ is automatically a
\vNa. (Note that a given group $\GU$ of unitaries on $\H$ may be
regarded as a \rep\ $U$ of $\GU$ itself, where $U$ is the identity
map.)  \su{Reduction of unitary group \rep s} The (possible) reduction
of $U(G)$ is determined by the \vNa s $U(G)''$ and $U(G)'$.  For
example, $U$ is irreducible iff $U(G)'=\C\I$ (Schur's lemma).  The
\rep\ $U$ is called {\bf primary} when $U(G)''$ has a trivial center,
that is, when $U(G)''\cap U(G)'=\C\I$. When $G$ is compact, so that
$U$ is discretely reducible, this implies that $U$ is a multiple of a
fixed irreducible \rep\ $U_{\gm}$ on a \Hs\ $\H_{\gm}$, so that
$\H\simeq \H_{\gm}\ot \CK$, and $U\simeq U_{\gm}\ot\I_{\CK}$.

When $G$ is not compact, but still assumed to be locally compact,
unitary \rep s may be reducible without containing any irreducible
subrepresentation. This occurs already in the simplest possible cases,
such as the regular \rep\ of $G=\R$ on $\H=L^2(\R)$; that is, one puts
$U(x)\Ps(y)=\Ps(y-x)$. The irreducible would-be subspaces of $\H$
would be spanned by the vectors $\Ps_p(y):=\exp(ipy)$, but these
functions do not lie in $L^2(\R)$. The solution to this problem was
given by \vN\ in a paper published in 1949, but written in the
thirties (the ideas in it must have guided \vN\ from at least 1936
on).

Instead of decomposing $\H$ as a direct sum, one should decompose it
as a {\bf direct integral}. (To do so, one needs to assume that $\H$
is separable.) This means that firstly one has a measure space
$(\Lm,\mu)$ and a family of \Hs s $\{\H_{\lm}\}_{\lm\in\Lm}$. A {\bf
section} of this family is a function $\Ps:\Lm\raw
\{\H_{\lm}\}_{\lm\in\Lm}$ for which $\Ps(\lm)\in \H_{\lm}$. To define
the direct integral of the $\H_{\lm}$ with respect to the measure
$\mu$, one needs a sequence of sections $\{\Ps_n\}$ satisfying the two
conditions that firstly the function $\lm\raw
(\Ps_n(\lm),\Ps_m(\lm))_{\lm}$ be measurable for all $n,m$, and
secondly that for each fixed $\lm$ the $\Ps_n$ span $\H_{\lm}$.  There
then exists a unique maximal linear subspace $\Gm_0$ of the space
$\Gm$ of all sections which contains all $\Ps_n$, and for which all
sections $\lm\raw (\Ps_{\lm},\Ph_{\lm})_{\lm}$ are measurable.
 
For $\Ps,\Ph\in\Gm_0$ it then makes sense to define
$$
(\Ps,\Ph):=\int_{\Lm}d\mu(\lm)\, (\Ps(\lm),\Ph(\lm))_{\lm}.
$$
The {\bf direct integral}
$$
\int_{\Lm}^{\oplus}d\mu(\lm)\, \H_{\lm}
$$
is then by definition the subset of $\Gm_0$ of functions $\Ps$ for
 which $(\Ps,\Ps)<\infty$. When $\Lm$ is discrete, the direct integral
 reduces to a direct sum.

An operator $A$ on this direct integral \Hs\ is said to be {\bf
diagonal} when
$$
A\Ps(\lm)=A_{\lm}\Ps(\lm)
$$
 for some (suitably measurable) family of operators $A_{\lm}$ on
$\H_{\lm}$. We then write
$$
A=\int_{\Lm}^{\oplus}d\mu(\lm)\, A_{\lm}.
$$

Thus a unitary group \rep\ $U(G)$ on $\H$ is diagonal when
$$
U(x)\Ps(\lm)=U_{\lm}(x)\Ps_{\lm}
$$
for all $x\in G$, in which case we, of course, write
$$
U=\int_{\Lm}^{\oplus}d\mu(\lm)\, U_{\lm}.
$$

Reducing a given \rep\ $U$ on some \Hs\ then amounts to finding a
unitary map $V$ between $\H$ and some direct integral \Hs, such that
each $\H_{\lm}$ carries a \rep\ $U_{\lm}$, and $VU(x)V^*$ is diagonal
in the above sense, with $A_{\lm}=U_{\lm}(x)$.  When $\H$ is
separable, one may always reduce a unitary \rep\ in such a way that
the $U_{\lm}$ occurring in the decomposition are primary, and this
{\bf central decomposition} of $U$ is essentially unique.

To completely reduce $U$, one needs the $U_{\lm}$ to be irreducible,
so that $\Lm$ is the space $\hat{G}$ of all equivalence classes of
irreducible unitary \rep s of $G$.  Complete reduction therefore calls
for a further direct integral decomposition of primary \rep s; this
will be discussed below.

For example, one may take $\Lm=\R$ with Lebesgue measure $\mu$, and
take the sequence $\{\Ps_n\}$ to consist of a single strictly positive
measurable function. This leads to the direct integral decomposition
$$
L^2(\R)=\int_{\R}^{\oplus}dp\, \H_p,
$$
in which each $\H_p$ is $\C$. To reduce the regular \rep\ of $\R$ on
$L^2(\R)$, one simply performs a Fourier transform $V:L^2(\R)\raw
L^2(\R)$, i.e.,
$$
V\Ps(p)=\int_{\R}dy\, e^{-ipy}\Ps(y).
$$
This leads to $VU(x)V^*\Ps(p)=\exp(ipx)\Ps(p)$, so that $U$ has been
diagonalized: the $U_{\lm}(x)$ above are now the one-dimensional
operators $U_p(x)=\exp(ipx)$ on $\H_p=\C$. We have therefore
completely reduced $U$.

As far as the reduction of unitary \rep s is concerned, there exist
two radically different classes of locally compact groups (the class
of all locally compact groups includes, for example, all
finite-dimensional Lie groups and all discrete groups).  A primary
\rep\ is said to be of {\bf type I} when it may be decomposed as the
direct sum of irreducible sub\rep s; these sub\rep s are necessarily
equivalent.  A locally compact group is said to be {\bf type I} or
{\bf tame} when every primary \rep\ is a multiple of a fixed
irreducible \rep; in other words, a group is type I when all its
primary \rep s are of type I.  If not, the group is called {\bf
non-type I} or {\bf wild}.  An example of a wild group, well known to
von Neumann, is the free group on two generators. Another example,
discovered at a later stage, is the group of matrices of the form
$$
\left(
\begin{array}{ccc}
e^{it} & 0 & z \\ 0 & e^{i\al t} & w \\ 0 & 0 & 1
\end{array}
\right) ,
$$
where $\al$ is an irrational real number, $t\in\R$, and $z,w\in\C$.

When $G$ is wild, curious phenomena may occur.  By definition, a wild
group has primary unitary \rep s which contain no irreducible sub\rep
s. More bizarrely, \rep s of the latter type may be decomposed in two
alternative ways
$$
U=\int_{\hat{G}}^{\oplus}d\mu_1(\gm)\, U_{\gm}=
\int_{\hat{G}}^{\oplus}d\mu_2(\gm)\, U_{\gm},
$$
where the measures $\mu_1$ and $\mu_2$ are disjoint (that is,
supported by disjoint subsets of $\hat{G}$).

A reducible primary \rep\ $U$ may always be decomposed as $U=U_h\oplus
U_h$.  In case that $U$ is not equivalent to $U_h$, and $U$ is not of
type I, it is said to be a \rep\ of {\bf type II}. When $U$ is neither
of type I nor of type II, it is of {\bf type III}. In that case $U$ is
equivalent to $U_h$; indeed, all (proper) sub\rep s of a primary type
III \rep\ are equivalent.  \su{The classification of factors} Between
1936 and 1953 \vN\ wrote 5 lengthy, difficult, and profound papers (3
of which were in collaboration with Murray) in which the study of his
`rings of operators' was initiated. (According to I.E. Segal, these
papers form `perhaps the most original major work in mathematics in
this century'.)

The analysis of Murray and \vN\ is based on the study of the
projections in a \vNa\ $\M$ (a projection is an operator $p$ for which
$p^2=p^*=p$); indeed, $\M$ is generated by its projections. They
noticed that one may define an equivalence relation $\sim$ on the set
of all projections in $\M$, in which $p\sim q$ iff there exists a
partial isometry $V$ in $\M$ such that $V^*V=p$ and $VV^*=q$.  When
$\M\subseteq \BH$, the operator $V$ is unitary from $p\H$ to $q\H$,
and annihilates $p\H^{\perp}$. Hence when $\M=\BH$ one has $p\sim q$
iff $p\H$ and $q\H$ have the same dimension, for in that case one may
take any $V$ with the above properties.

An equivalent characterization of $\sim$ arises when we write
$\M=U(G)'$ for some unitary \rep\ $U$ of a group $G$ (as we have seen,
this always applies); then $p\sim q$ iff the sub\rep s $pU$ and $qU$
(on $p\H$ and $q\H$, respectively), are unitarily equivalent.

Moreover, Murray and \vN\ define a partial orderering on the
collection of all projections in $\M$ by declaring that $p\leq q$ when
$pq=p$, that is, when $p\H\subseteq q\H$.  This induces a partial
orderering on the set of equivalence classes of projections by putting
$[p]\leq [q]$ when the equivalence classes $[p]$ and $[q]$ contain
representatives $\til{p}$ and $\til{q}$ such that $\til{p}\leq
\til{q}$.  For $\M=\BH$ this actually defines a total ordering on the
equivalence classes, in which $[p]\leq [q]$ when $p\H$ has the same
dimension as $q\H$; as we just saw, this is independent of the choice
of $p\in[p]$ and $q\in[q]$.

More generally, Murray and \vN\ showed that the set of equivalence
classes of projections in $\M$ is totally ordered by $\leq$ whenever
$\M$ is a {\bf factor}.  A \vNa\ $\M$ is a factor when
$\M\cap\M'=\C\I$; when $\M=U(G)'$ this means that $\M$ is a factor iff
the \rep\ $U$ is primary. The study of \vNa s acting on separable \Hs
s $\H$ reduces to the study of factors, for \vN\ proved that every
\vNa\ $\M\subseteq\BH$ may be uniquely decomposed, as in
\begin{eqnarray*}
\H & = & \int_{\Lm}^{\oplus}d\mu(\lm)\, \H_{\lm};\\ \M & = &
\int_{\Lm}^{\oplus}d\mu(\lm)\, \M_{\lm},
\end{eqnarray*}
where (almost) each $\M_{\lm}$ is a factor.  For $\M=U(G)'$ the
decomposition of $\H$ amounts to the central decomposition of $U(G)$.

As we have seen, for the factor $\M=\BH$ the dimension $d$ of a
projection is a complete invariant, distinguishing the equivalence
classes $[p]$. The dimension is a function from the set of all
projections in $\BH$ to $\R^+\cup\infty$, satisfying
\begin{enumerate}
\item
$d(p)>0$ when $p\neq 0$, and $d(0)=0$;
\item
$d(p)=d(q)$ iff $[p]\sim[q]$;
\item
$d(p+q)=d(p)+d(q)$ when $pq=0$ (i.e., when $p\H$ and $q\H$ are
orthogonal;
\item
$d(p)<\infty$ iff $p$ is finite.
\end{enumerate}
Here a projection in $\BH$ is called finite when $p\H$ is
 finite-dimensional.  Murray and \vN\ now proved that on any factor
 $\M$ (acting on a separable \Hs) there exists a function $d$ from the
 set of all projections in $\M$ to $\R^+\cup\infty$, satisfying the
 above properties.  Moreover, $d$ is unique up to finite rescaling.
 For this to be the possible, Murray and \vN\ define a projection to
 be {\bf finite} when it is not equivalent to any of its (proper)
 sub-projections; an {\bf infinite} projection is then a projection
 which has proper sub-projections to which it is equivalent. For
 $\M=\BH$ this generalized notion of finiteness coincides with the
 usual one, but in other factors all projections may be infinite in
 the usual sense, yet some are finite in the sense of Murray and
 \vN. One may say that, in order to distinguish infinite-dimensional
 but inequivalent projections, the dimension function $d$ is a
 `renormalized' version of the usual one.

A first classification of factors (on a separable \Hs) is now
performed by considering the possible finiteness of its projections
and the range of $d$.  A projection $p$ is called {\bf minimal} or
{\bf atomic} when there exists no $q<p$ (i.e., $q\leq p$ and $q\neq
p$). One then has the following possibilities for a factor $\M$.
\begin{itemize}
\item 
{\bf type I$\mbox{}_n$}, where $n<\infty$: $\M$ has minimal
projections, all projections are finite, and $d$ takes the values
$\{0,1,\ldots,n\}$. A factor of type I$\mbox{}_n$ is isomorphic to the
algebra of $n\x n$ matrices.
\item
{\bf type I$\mbox{}_{\infty}$}: $\M$ has minimal projections, and $d$
takes the values $\{0,1,\ldots,\infty\}$. Such a factor is isomorphic
to $\BH$ for separable infinite-dimensional $\H$.
\item
{\bf type II$\mbox{}_1$}: $\M$ has no minimal projections, all
projections are infinite-dimensional in the usual sense, and $\I$ is
finite. Normalizing $d$ such that $d(\I)=1$, the range of $d$ is the
interval $[0,1]$.
\item
{\bf type II$\mbox{}_{\infty}$}: $\M$ has no minimal projections, all
nonzero projections are infinite-dimensional in the usual sense, but
$\M$ has finite-dimensional projections in the sense of Murray and
\vN, and $\I$ is infinite. The range of $d$ is $[0,\infty]$.
\item
{\bf type III}: $\M$ has no minimal projections, all nonzero
projections are infinite-dimensional and equivalent in the usual sense
as well as in the sense of Murray and \vN, and $d$ assumes the values
$\{0,\infty\}$.
\end{itemize}

With $\M=U(G)'$, where, as we have seen, the \rep\ $U$ is primary iff
$\M$ is a factor, $U$ is of a given type iff $\M$ is of the same type.

One sometimes says that a factor is {\bf finite} when $\I$ is finite
(so that $d(\I)<\infty$); hence type $I_n$ and type II$\mbox{}_1$
factors are finite. Factors of type I$\mbox{}_{\infty}$ and
II$\mbox{}_{\infty}$ are then called {\bf semifinite}, and type III
factors are {\bf purely infinite}.

It is hard to construct an example of a II$\mbox{}_1$ factor, and even
harder to write down a type III factor. Murray and \vN\ managed to do
the former, and \vN\ did the latter by himself, but only 5 years after
he and Murray had recognized that the existence of type III factors
was a logical possibility. However, they were unable to provide a
further classification of all factors, and they admitted having no
tools to study type III factors.

Von Neumann was fascinated by II$\mbox{}_1$ factors. In view of the
range of $d$, he believed these defined some form of continuous
geometry. Moreover, the existence of a II$\mbox{}_1$ factor solved one
of the problems that worried him in \qm.  For he showed that on a
II$\mbox{}_1$ factor $\M$ the dimension function $d$, defined on the
projections in $\M$, may be extended to a positive linear functional
$tr$ on $\M$, with the property that $tr(UAU^*)=tr(A)$ for all
$A\in\M$ and all unitaries $U$ in $\M$.  This `trace' satisfies
$tr(\I)=d(\I)=1$, and gave \vN\ the state of minimal information he
had sought. Partly for this reason he believed that physics should be
described by II$\mbox{}_1$ factors.

At the time not many people were familiar with the difficult papers of
Murray and \vN, and until the sixties only a handful of mathematicians
worked on operator algebras (e.g., Segal, Kaplansky, Kadison, Dixmier,
Sakai, and others).  The precise connection between \vNa s and the
decomposition of unitary group \rep s envisaged by \vN\ was worked out
by Mackey, Mautner, Godement, and Adel'son-Vel'skii.

In the sixties, a group of physicists, led by Haag, realized that
operator algebras could be a useful tool in quantum field theory and
in the quantum statistical mechanics of infinite systems. This has led
to an extremely fruitful intercation between physics and mathematics,
which has helped both subjects.  In particular, in 1957 Haag observed
a formal similarity between the collection of all \vNa s on a \Hs\ and
the set of all causally closed subsets of Minkowksi space-time.  Here
a region $\CO$ in space-time is said to be {\bf causally closed} when
$\CO^{\perp\perp}=\CO$, where $\CO^{\perp}$ consists of all points
that are spacelike separated from $\CO$.  The operation
$\CO\raw\CO^{\perp}$ on causally closed regions in space-time is
somewhat analogous to the operation $\M\raw\M'$ on \vNa s. Thus Haag
proposed that a quantum field theory should be defined by a {\bf net
of local observables}; this is a map $\CO\raw\M(\CO)$ from the set of
all causally closed regions in space-time to the set of all \vNa s on
some \Hs, such that $\M(\CO_1)\subseteq \M(\CO_2)$ when
$\CO_1\subseteq \CO_2$, and $\M(\CO)'=\M(\CO^{\perp})$.

This idea initiated {\bf algebraic quantum field theory}, a subject
that really got off the ground with papers by Haag's pupil Araki in
1963 and by Haag and Kastler in 1964. From then till the present day,
algebraic quantum field theory has attracted a small but dedicated
group of mathematical physicists. One of the result has been that in
realistic quantum field theories the local algebras $\M(\CO)$ must all
be isomorphic to the unique hyperfinite factor of type III$\mbox{}_1$
discussed below. (Hence \vN's belief that physics should use
II$\mbox{}_1$ factors has not been vindicated.)

A few years later (1967), an extraordinary coincidence took place,
which was to play an essential role in the classification of factors
of type III. On the mathematics side, Tomita developed a technique in
the study of \vNa s, which nowadays is called {\bf modular theory} or
{\bf Tomita-Takesaki theory} (apart from clarifying Tomita's work,
Takesaki made essential contributions to this theory). Among other
things, this theory leads to a natural time-evolution on certain
factors.  On the physics side, Haag, Hugenholtz, and Winnink
characterized states of thermal equilibrium of infinite quantum
systems by an algebraic condition that had previously been introduced
in a heuristic setting by Kubo, Martin, and Schwinger, and is
therefore called the {\bf KMS condition}.  This condition leads to
type III factors equipped with a time-evolution which coincided with
the one of the Tomita-Takesaki theory.

In the hands of Connes, the Tomita-Takesaki theory and the examples of
type III factors provided by physicists (Araki, Woods, Powers, and
others) eventually led to the classification of all {\bf hyperfinite
factors} of type II and III (the complete classification of all
factors of type I is already given by the list presented
earlier). These are factors containing a sequence of
finite-dimensional subalgebras $\M_1\subset\M_2\ldots\subset\M$, such
that $\M$ is the weak closure of $\cup_n\M_n$. (Experience shows that
all factors playing a role in physics are hyperfinite, and many
natural examples of factors constructed by purely mathematical
techniques are hyperfinite as well.) The work of Connes, for which he
was awarded the Fields Medal in 1982, and others, led to the following
classification of hyperfinite factors of type II and III (up to
isomorphism):
\begin{itemize}
\item 
There is a unique hyperfinite factor of type II$\mbox{}_1$.  (In
 physics this factor occurs when one considers KMS-states at infinite
 temperature.)
\item
There is a unique hyperfinite factor of type II$\mbox{}_{\infty}$,
namely the tensor product of the hyperfinite
II$\mbox{}_{\infty}$-factor with $\B(\CK)$, for an
infinite-dimensional separable \Hs\ $\CK$.
\item
There is a family of type III factors, labeled by $\lm\in [0,1]$. For
$\lm\neq 0$ the factor of type III$\mbox{}_{\lm}$ is unique. There is
a family of type III$\mbox{}_0$ factors, which in turn is has been
classified in terms of concepts from ergodic theory.
\end{itemize}

As we have mentioned already, the unique hyperfinite III$\mbox{}_1$
factor plays a central role in algebraic quantum field theory. The
unique hyperfinite II$\mbox{}_1$ factor was crucial in a spectacular
development, in which the theory of inclusions of II$\mbox{}_1$
factors was related to knot theory, and even led to a new knot
invariant. In 1990 Jones was awarded a Fields medal for this work, the
second one to be given to the once obscure field of operator algebras.
\su{$C^*$-algebras} In the midst of the Murray-\vN\ series of papers,
Gel'fand initiated a separate development, combining operator algebras
with the theory of Banach spaces. In 1941 he defined the concept of a
Banach algebra, in which multiplication is (separately) continuous in
the norm-topology.  He proceeded to define an intrinsic spectral
theory, and proved most basic results in the theory of commutative
Banach algebras.

In 1943 Gel'fand and Neumark defined what is now called a \ca\ (some
of their axioms were later shown to be superfluous), and proved the
basic theorem that each \ca\ is isomorphic to the norm-closed \sta\ of
operators on a \Hs.  Their paper also contained the rudiments of what
is now called the GNS construction, connecting states to \rep s. In
its present form, this construction is due to Segal (1947), a great
admirer of \vN, who generalized \vN's idea of a state as a positive
normalized linear functional from $\BH$ to arbitrary \ca s. Moreover,
Segal returned to \vN's motivation of relating operator algebras to
\qm.

As with \vNa s, the sixties brought a fruitful interaction between \ca
s and quantum physics.  Moreover, the theory of \ca s turned out to be
interesting both for intrinsic reasons (structure and \rep\ theory of
\ca s), as well as because of its connections with a number of other
fields of mathematics.  Here the strategy is to take a given
mathematical structure, try and find a \ca\ which encodes this
structure in some way, and then obtain information about the structure
through proving theorems about the \ca\ of the structure.

The first instance where this led to a deep result which has not been
proved in any other way is the theorem of Gel'fand and Raikov (1943),
stating that the unitary \rep s of a locally compact group separate
the points of the group (that is, for each pair $x\neq y$ there exists
a unitary \rep\ $U$ for which $U(x)\neq U(y)$. This was proved by
constructing a \ca\ $C^*(G)$ of the group $G$, showing that \rep s of
$C^*(G)$ bijectively correspond to unitary \rep s of $G$, and finally
showing that the states of an arbitrary \ca\ $\A$ separate the
elements of $\A$.

Other examples of mathematical structures that may be analyzed through
an appropriate \ca\ are group actions, groupoids, foliations, and
complex domains.  The same idea lies at the basis of {\bf
non-commutative geometry} and {\bf non-commutative topology}. Here the
starting point is another theorem of Gel'fand, stating that any
commutative \ca\ (with unit) is isomorphic to $C(X)$, where $X$ is a
compact Hausdorff space. The strategy is now that the basic tools in
the topology of $X$, and, when appropriate, in its differential
geometry, should be translated into tools pertinent to the \ca\
$C(X)$, and that subsequently these tools should be generalized to
non-commutative \ca s.

This strategy has been successful in $K$-theory, whose non-commutative
version is even simpler than its usual incarnation, and in (de Rham)
cohomology theory, whose non-commutative version is called {\bf cyclic
cohomology}. Finally, homology, cohomology, $K$-theory, and index
theory haven been unified and made non-commutative in the {\bf
$KK$-theory} of Kasparov.  The basic tool in $KK$-theory is the
concept of a {\bf Hilbert $C^*$-module}, which we will study in detail
in these lectures.
\section{Elementary theory of $C^*$-algebras}
\su{Basic definitions} All vector spaces will be defined over $\C$,
and all functions will be $\C$-valued, unless we explicitly state
otherwise.  The abbreviation `iff' means `if and only if', which is
the same as the symbol $\Leftrightarrow$. An equation of the type
$a:=b$ means that $a$ is by definition equal to $b$.
\begin{Definition}\ll{defnorm} A {\bf norm} on a vector space $\CV$ is a map
$\n\:\n\,:\CV\raw\R$ such that
\begin{enumerate}
\item
$\n v\n\,\geq 0$ for all $v\in\CV$;
\item
$\n v\n=0$ iff $v=0$;
\item
$\n\lm v\n=|\lm|\; \n v\n$ for all $\lm\in\C$ and $v\in\CV$;
\item
$\n v+w\n\,\leq \,\n v\n+\n w\n$ (triangle inequality).
\end{enumerate}
A norm on $\CV$ defines a metric $d$ on $\CV$ by $d(v,w):=\n v-w\n$.
A vector space with a norm which is complete in the associated metric
(in the sense that every Cauchy sequence converges) is called a {\bf
Banach space}. We will denote a generic Banach space by the symbol
$\CB$.\end{Definition}

The two main examples of Banach spaces we will encounter are Hilbert
spaces and certain collections of operators on \Hs s.
\begin{Definition}\ll{defip}
 A {\bf pre-inner product} on a vector space $\CV$ is a map $(\,
,\,):\CV\x\CV\raw\C$ such that
\begin{enumerate}
\item
$(\lm_1 v_1+\lm_2 v_2,\mu_1 w_1+\mu_2 w_2)=\ovl{\lm_1}\mu_1(v_1,w_1)+
\ovl{\lm_1}\mu_2(v_1,w_2)+\ovl{\lm_2}\mu_1(v_2,w_1)+\ovl{\lm_2}\mu_2(v_2,w_2)$
for all $\lm_1,\lm_2,\mu_1,\mu_2\in\C$ and $v_1,v_2,w_1,w_2\in\CV$;
\item
$(v,v)\geq 0$ for all $v\in\CV$.
\end{enumerate}
An equivalent set of conditions is
\begin{enumerate}
\item
$\ovl{(v,w)}=(w,v)$ for all $v,w\in\CV$;
\item
$(v,\lm_1 w_1+\lm_2 w_2)=\lm_1(v,w_1)+\lm_2(v,w_2)$ for all
$\lm_1,\lm_2\in\C$ and $v,w_1,w_2\in\CV$;
\item
$(v,v)\geq 0$ for all $v\in\CV$.
\end{enumerate}
A pre-inner product for which $(v,v)=0$ iff $v=0$ is called an {\bf
inner product}.  \end{Definition}

The equivalence between the two definitions of a pre-inner product is
elementary; in fact, to derive the first axiom of the second
characterization from the first set of conditions, it is enough to
assume that $(v,v)\in\R$ for all $v$ (use this reality with $v\raw
v+iw$).  Either way, one derives the {\bf Cauchy-Schwarz inequality}
\be |(v,w)|^2\leq (v,v)(w,w), \ll{csb} \ee for all $v,w\in\CV$. Note
that this inequality is valid even when $(\, ,\,)$ is not an inner
product, but merely a pre-inner product.

It follows from these properties that an inner product on $\CV$
defines a norm on $\CV$ by $\n v\n:=\sqrt{(v,v)}$; the triangle
inequality is automatic.
\begin{Definition}\ll{defHs}
A {\bf Hilbert space} is a vector space with inner product which is
complete in the associated norm. We will usually denote \Hs s by the
symbol $\H$. \end{Definition}

A \Hs\ is completely characterized by its dimension (i.e., by the
cardinality of an arbitrary orthogonal basis). To obtain an
interesting theory, one therefore studies operators on a \Hs, rather
than the \Hs\ itself. To obtain a satisfactory mathematical theory, it
is wise to restrict oneself to bounded operators. We recall this
concept in the more general context of arbitrary Banach spaces.
\begin{Definition}\ll{defbobs} 
A {\bf bounded operator} on a Banach space $\CB$ is a linear map
$A:\CB\raw\CB$ for which \be \n A\n:= \sup\, \{ \n A v\n\,
|\,v\in\CB,\, \n v\n=1\}<\infty.  \ll{banalnorm} \ee
\end{Definition}
 
The number $\n A\n$ is the {\bf operator norm}, or simply the norm, of
$A$.  This terminology is justified, as it follows almost immediately
from its definition (and from the properties of the the norm on $\CB$)
that the operator norm is indeed a norm.

(It is easily shown that a linear map on a Banach space is continuous
iff it is a bounded operator, but we will never use this
result. Indeed, in arguments involving continuous operators on a
Banach space one almost always uses boundedness rather than
continuity.)

When $\cal B$ is a \Hs\ $\H$ the expression (\ref{banalnorm}) becomes
\be \n A\n:= \sup\, \{ ( A\Ps,A \Ps)^{\half}\, |\,\Ps\in\H,\,
(\Ps,\Ps)=1\}.  \ll{BHnorm} \ee

When $A$ is bounded, it follows that \be \n A v\n\,\leq\, \n A\n\; \n
v\n \ll{opprop} \ee for all $v\in\CB$.  Conversely, when for $A\neq 0$
there is a $C>0$ such that $\n A v\n\,\leq C \n v\n$ for all $v$, then
$A$ is bounded, with operator norm $\n A\n$ equal to the smallest
possible $C$ for which the above inequality holds.
\begin{Proposition}\ll{BCB}
The space $\B(\CB)$ of all bounded operators on a Banach space $\CB$
is itself a Banach space in the operator norm. \end{Proposition}

In view of the comments following (\ref{BHnorm}), it only remains to
be shown that $\B(\CB)$ is complete in the operator norm.  Let
$\{A_n\}$ be a Cauchy sequence in $\B(\CB)$. In other words, for any
$\ep>0$ there is a natural number $N(\ep)$ such that $\n
A_n-A_m\n\,<\ep$ when $n,m>N(\ep)$. For arbitrary $v\in\CB$, the
sequence $\{A_n v\}$ is a Cauchy sequence in $\CB$, because \be \n
A_nv-A_mv\n\,\leq\, \n A_n-A_m\n\, \n v\n\,\leq\, \ep \n v\n \ll{KRp}
\ee for $n,m>N(\ep)$. Since $\CB$ is complete by assumption, the
sequence $\{A_n v\}$ converges to some $w\in\CB$. Now define a map $A$
on $\CB$ by $Av:=w=\lim_n A_n v$. This map is obviously linear.
Taking $n\raw\infty$ in (\ref{KRp}), we obtain \be \n Av-A_mv\n\,\leq
\ep \n v\n \ll{AvAmv} \ee for all $m>N(\ep)$ and all $v\in\CB$. It now
follows from (\ref{banalnorm}) that $A-A_m$ is bounded. Since
$A=(A-A_m)+A_m$, and $\B(\CB)$ is a linear space, we infer that $A$ is
bounded. Moreover, (\ref{AvAmv}) and (\ref{banalnorm}) imply that $\n
A-A_m\n\,\leq\, \ep$ for all $m>N(\ep)$, so that $\{A_n\}$ converges
to $A$. Since we have just seen that $A\in\B(\CB)$, this proves that
$\B(\CB)$ is complete. \enp

We define a {\bf functional} on a Banach space $\CB$ as a linear map
$\rh:\CB\raw\C$ which is continuous in that $\rh(v)|\,\leq C\n v\n$
for some $C$, and all $v\in\CB$. The smallest such $C$ is the norm \be
\n\rh\n:= \sup\, \{ |\rh(v)|,\,v\in\CB,\, \n v\n=1\}.  \ee The {\bf
dual} $\CB^*$ of $\CB$ is the space of all functionals on $\CB$.
Similarly to the proof of \ref{BCB}, one shows that $\B^*$ is a Banach
space.  For later use, we quote, without proof, the fundamental {\bf
Hahn-Banach theorem}.  \begin{Theorem}\ll{HB} For a functional $\rh_0$
on a linear subspace $\CB_0$ of a Banach space $\CB$ there exists a
functional $\rh$ on $\CB$ such that $\rh=\rh_0$ on $\CB_0$ and
$\n\rh\n=\n\rh_0\n$. In other words, each functional defined on a
linear subspace of $\CB$ has an extension to $\CB$ with the same norm.
\end{Theorem}
\begin{Corollary}\ll{HB2} When $\rh(v)=0$ for all
$\rh\in\CB^*$ then $v=0$. \end{Corollary}

For $v\neq 0$ we may define a functional $\rh_0$ on $\C v$ by
$\rh_0(\lm v)=\lm$, and extend it to a functional $\rh$ on $\CB$ with
norm 1. \enp

Recall that an {\bf algebra} is a vector space with an associative
bilinear operation (`multiplication') $\cdot:\A\x\A\raw\A$; we usually
write $AB$ for $A\cdot B$.  It is clear that $\B(\CB)$ is an algebra
under operator multiplication. Moreover, using (\ref{opprop}) twice,
for each $v\in\CB$ one has
$$
\n ABv\n\,\leq\,\n A\n\,\n Bv\n\,\leq\, \n A\n\,\n B\n\,\n v\n.
$$
Hence from (\ref{banalnorm}) we obtain $\n AB\n\,\leq\, \n A\n\,\n
B\n$.
\begin{Definition}\ll{defbanachal}
A {\bf Banach algebra} is a Banach space $\A$ which is at the same
time an algebra, in which for all $A,B\in\A$ one has \be \n
AB\n\,\leq\, \n A\n\,\n B\n\ . \ll{Cstax1} \ee
\end{Definition}

It follows that multiplication in a Banach algebra is separately
continuous in each variable.

As we have just seen, for any \Bs\ $\CB$ the space $\B(\CB)$ of all
bounded operators on $\CB$ is a Banach algebra.  In what follows, we
will restrict ourselves to the case that $\CB$ is a \Hs\ $\H$; this
leads to the Banach algebra $\BH$.  This algebra has additional
structure.
\begin{Definition}\ll{definv}
An {\bf involution} on an algebra $\A$ is a real-linear map
$A\rightarrow A^*$ such that for all $A,B\in\A$ and $\lm\in\C$ one has
\bea A^{**}& = & A; \ll{AssisA}\\ (AB)^* & =& B^* A^*;\\ (\lm A)^* & =
& \ovl{\lm}A^*.  \eea
 
A {\bf $\mbox{}^*$-algebra} is an algebra with an
 involution. \end{Definition}

The operator adjoint $A\raw A^*$ on a \Hs, defined by the property
$(\Ps,A^*\Ph):= (A\Ps,\Ph)$, defines an involution on $\BH$. Hence
$\BH$ is a \sta.  As in this case, an element $A$ of a \ca\ $\A$ is
called {\bf self-adjoint} when $A^*=A$; we sometimes denote the
collection of all self-adjoint elements by \be \Ar:=\{A\in\A|\,
A^*=A\}. \ll{defAr} \ee Since one may write \be A=A'+i A'':=
\frac{A+A^*}{2}+i\frac{A-A^*}{2i}, \ll{aaa} \ee every element of $\A$
is a linear combination of two self-adjoint elements.

To see how the norm in $\BH$ is related to the involution, we pick
$\Ps\in\H$, and use the Cauchy-Schwarz inequality and (\ref{opprop})
to estimate
 $$
\n A\Ps\n^2=(A\Ps,A\Ps)=(\Ps,A^*A\Ps)\leq\,\n\Ps\n\,\n A^*A\Ps\n\,\leq
 \n A^*A\n\, \n\Ps\n^2.
$$
Using (\ref{BHnorm}) and (\ref{Cstax1}), we infer that \be \n
A\n^2\,\leq\,\n A^*A\n\, \leq\, \n A^*\n\,\n A\n.  \ll{hulp} \ee This
leads to $\n A\n\,\leq\,\n A^*\n$. Replacing $A$ by $A^*$ and using
(\ref{AssisA}) yields $\n A^*\n\,\leq\,\n A\n$, so that $\n A^*\n=\n
A\n$. Substituting this in (\ref{hulp}), we derive the crucial
property $\n A^* A\n = \n A\n^2$.

This motivates the following definition.
\begin{Definition}\ll{defcstaralgebra} 
  A {\bf $C^*$-algebra} is a complex Banach space $\A$ which is at the
same time a $\mbox{}^*$-algebra, such that for all $A,B\in\A$ one has
\bea \n A B\n \, & \leq & \, \n A\n\ \, \n B\n; \ll{cstax1}\\ \n A^*
A\n & = & \n A\n^2.\ll{cstax2} \eea In other words, a \ca\ is a Banach
\sta\ in which (\ref{cstax2}) holds.  \end{Definition}

Here a Banach \sta\ is, of course, a Banach algebra with
involution. Combining (\ref{cstax2}) and (\ref{cstax1}), one derives
$\n A\n\,\leq\,\n A^*\n$; as in the preceding paragraph, we infer that
for all elements $A$ of a \ca\ one has the equality \be \n A^*\n=\n
A\n. \ll{astisa} \ee The same argument proves the following.
\begin{Lemma}\ll{ineqcs}
A Banach \sta\ in which $\n A\n^2\,\leq\,\n A^*A\n$ is a
\ca. \end{Lemma}

We have just shown that $\BH$ is a \ca. Moreover, each (operator)
norm-closed \sta\ in $\BH$ is a \ca\ by the same argument. A much
deeper result, which we will formulate precisely and prove in due
course, states the converse of this: each \ca\ is isomorphic to a
norm-closed \sta\ in $\BH$, for some \Hs\ $\H$.  Hence the axioms in
\ref{defcstaralgebra} characterize norm-closed \sta s on \Hs s,
although the axioms make no reference to \Hs s at all.

For later use we state some self-evident definitions.
\begin{Definition}\ll{defmor}
A {\bf morphism} between \ca s $\A,\B$ is a (complex-) linear map
$\phv:\A\raw\B$ such that \bea \phv(AB) & = & \phv(A)\phv(B);
\ll{phvmult} \\ \phv(A^*) & = & \phv(A)^* \ll{phvstar} \eea for all
$A,B\in\A$.  An {\bf isomorphism} is a bijective morphism. Two \ca s
are {\bf isomorphic} when there exists an isomorphism between them.
\end{Definition}

One immediately checks that the inverse of a bijective morphism is a
morphism.  It is remarkable, however, that an injective morphism (and
hence an isomorphism) between \ca s is automatically isometric. For
this reason the condition that an isomorphism be isometric is not
included in the definition.  \su{Banach algebra basics\ll{Bab}} The
material in this section is not included for its own interest, but
because of its role in the theory of \ca s. Even in that special
context, it is enlightening to see concepts such as the spectrum in
their general and appropriate setting.

Recall Definition \ref{defbanachal}. A {\bf unit} in a Banach algebra
$\A$ is an element $\I$ satisfying $\I A=A\I=A$ for all $A\in\A$, and
\be \n\I\n=1. \ll{norm1} \ee A \BA\ with unit is called {\bf
unital}. We often write $z$ for $z\I$, where $z\in\C$. Note that in a
\ca\ the property $\I A=A\I=A$ already implies, (\ref{norm1}); take
$A=\I^*$, so that $\I^*\I=\I^*$; taking the adjoint, this implies
$\I^*=\I$, so that (\ref{norm1}) follows from (\ref{cstax2}).

When a \BA\ $\A$ does not contain a unit, we can always add one, as
follows.  Form the vector space \be \AI:=\A\oplus\C, \ll{defAI} \ee
and make this into an algebra by means of \be
(A+\lm\I)(B+\mu\I):=AB+\lm B+\mu A+\lm\mu\I, \ll{muAI} \ee where we
have written $A+\lm\I$ for $(A,\lm)$, etc.  In other words, the number
1 in $\C$ is identified with $\I$.  Furthermore, define a norm on
$\AI$ by \be \n A+\lm\I\n:=\n A\n +|\lm|. \ll{norminAI} \ee In
particular, $\n\I\n=1$.  Using (\ref{cstax1}) in $\A$, as well as
\ref{defnorm}.3, one sees from (\ref{muAI}) and (\ref{norminAI}) that
 $$
\n (A+\lm\I)(B+\mu\I)\n\,\leq\, \n A\n\,\n B\n +|\lm|\,\n B\n+
|\mu|\,\n A\n +|\lm|\, |\mu|=\n A+\lm\I\n\,\n B+\mu\I\n,
$$ 
so that $\AI$ is a Banach algebra with unit. Since by (\ref{norminAI})
the norm of $A\in\A$ in $\A$ coincides with the norm of $A+0\I$ in
$\AI$, we have shown the following.
\begin{Proposition}\ll{BAun}
For every \BA\ without unit there exists a unital \BA\ $\AI$ and an
isometric (hence injective) morphism $\A\raw\AI$, such that
$\AI/\A\simeq\C$.
\end{Proposition}

As we shall see at the end of section \ref{comCS}, the {\bf
unitization} $\AI$ with the given properties is not unique.
\begin{Definition}\ll{defspectrum}
Let $\A$ be a unital \BA. The {\bf resolvent} $\rh(A)$ of $A\in\A$ is
the set of all $z\in\C$ for which $A-z\I$ has a (two-sided) inverse in
$\A$.

The {\bf spectrum} $\sg(A)$ of $A\in\A$ is the complement of $\rh(A)$
in $\C$; in other words, $\sg(A)$ is the set of all $z\in\C$ for which
$A-z\I$ has no (two-sided) inverse in $\A$.

When $\A$ has no unit, the resolvent and the spectrum are defined
through the embedding of $\A$ in $\AI=\A\oplus\C$.  \end{Definition}

When $\A$ is the algebra of $n\x n$ matrices, the spectrum of $A$ is
just the set of eigenvalues. For $A=\BH$, Definition \ref{defspectrum}
reproduces the usual notion of the spectrum of an operator on a \Hs.

When $\A$ has no unit, the spectrum $\sg(A)$ of $A\in\A$ always
contains zero, since it follows from (\ref{muAI}) that $A$ never has
an inverse in $\AI$.
\begin{Theorem}\ll{thmsp}
The spectrum $\sg(A)$ of any element $A$ of a Banach algebra is
\begin{enumerate}
\item
 contained in the set $\{z\in\C|\, |z|\leq\,\n A\n\}$;
\item
compact;
\item
not empty.
\end{enumerate} \end{Theorem}

The proof uses two lemmas. We assume that $\A$ is unital.
\begin{Lemma}\ll{l1}
When $\n A\n\,< 1$ the sum $\sum_{k=0}^n A^k$ converges to
$(\I-A)\inv$.

Hence $(A-z\I)\inv$ always exists when $|z|\,>\,\n A\n$. \end{Lemma}

We first show that the sum is a Cauchy sequence. Indeed, for $n>m$ one
has
$$
\n \sum_{k=0}^n A^k-\sum_{k=0}^m A^k\n=\n\sum_{k=m+1}^n A^k\n\,\leq
 \sum_{k=m+1}^n \n A^k\n\,\leq \sum_{k=m+1}^n \n A\n^k.
$$ 
For $n,m\raw\infty$ this goes to 0 by the theory of the geometric
series.  Since $\A$ is complete, the Cauchy sequence $\sum_{k=0}^n
A^k$ converges for $n\raw\infty$. Now compute
$$
\sum_{k=0}^n A^k(\I-A)=\sum_{k=0}^n (A^k-A^{k+1})=\I-A^{n+1}.
$$
Hence
$$
\n\I-\sum_{k=0}^n A^k(\I-A)\n=\n A^{n+1}\n\,\leq\,\n A\n^{n+1},
$$
which $\raw 0$ for $n\raw\infty$, as $\n A\n\,< 1$ by assumption.
Thus $$ \lim_{n\raw\infty}\sum_{k=0}^n A^k(\I-A)=\I.$$ By a similar
argument,
 $$
\lim_{n\raw\infty}(\I-A)\sum_{k=0}^n A^k=\I.$$ so that, by continuity
of multiplication in a \BA, one finally has \be
\lim_{n\raw\infty}\sum_{k=0}^n A^k=(\I-A)\inv.  \ee

The second claim of the lemma follows because $(A-z)\inv
 =-z\inv(\I-A/z)\inv$, which exists because $\n A/z\n\,<1$ when
 $|z|\,>\,\n A\n$.  \enp

To prove that $\sg(A)$ is compact, it remains to be shown that it is
 closed.
\begin{Lemma}\ll{l2}
The set \be G(\A):=\{A\in\A|\, A\inv\: {\rm exists}\} \ll{GA} \ee of
invertible elements in $\A$ is open in $\A$. \end{Lemma}

Given $A\in G(\A)$, take a $B\in\A$ for which $\n B\n\,<\, \n
A\inv\n\inv$.  By (\ref{Cstax1}) this implies \be \n A\inv
B\n\,\leq\,\n A\inv\n\;\n B\n\,<\, 1. \ll{e1} \ee Hence
$A+B=A(\I+A\inv B)$ has an inverse, namely $(\I+A\inv B)\inv A\inv$,
which exists by (\ref{e1}) and Lemma \ref{l1}.  It follows that all
$C\in\A$ for which $\n A-C\n\,<\ep$ lie in $G(\A)$, for $\ep\leq \n
A\inv\n\inv$.\enp

To resume the proof of Theorem \ref{thmsp}, given $A\in\A$ we now
define a function $f:\C\raw\A$ by $f(z):=z-A$. Since $\n
f(z+\dl)-f(z)\n=\dl$, we see that $f$ is continuous (take $\dl=\ep$ in
the definition of continuity). Because $G(\A)$ is open in $\A$ by
Lemma \ref{l2}, it follows from the topological definition of a
continuous function that $f\inv(G(\A))$ is open in $\A$. But
$f\inv(G(\A))$ is the set of all $z\in\C$ where $z-A$ has an inverse,
so that $f\inv(G(\A))=\rh(A)$. This set being open, its complement
$\sg(A)$ is closed.

Finally, define $g:\rh(A)\raw\A$ by $g(z):=(z-A)\inv$. For fixed
$z_0\in\rh(A)$, choose $z\in\C$ such that $|z-z_0|\, <\,
\n(A-z_0)\inv\n\inv$.  From the proof of Lemma \ref{l2}, with $A\raw
A-z_0$ and $C\raw A-z$, we see that $z\in\rh(A)$, as $\n
A-z_0-(A-z)\n=|z-z_0|$. Moreover, the power series
$$
\frac{1}{z_0-A}\sum_{k=0}^{n}\left(\frac{z_0-z}{z_0-A}\right)
$$ 
converges for $n\raw\infty$ by Lemma \ref{l1}, because
$$
\n (z_0-z)(z_0-A)\inv\n=|z_0-z|\;\n(z_0-A)\inv\n\,<1.
$$
By Lemma \ref{l1}, the limit $n\raw\infty$ of this power series is
$$
\frac{1}{z_0-A}\sum_{k=0}^{\infty}\left(\frac{z_0-z}{z_0-A}\right)=
\frac{1}{z_0-A}\left( 1-\left(\frac{z_0-z}{z_0-A}\right)\inv\right)=
\frac{1}{z-A}=g(z).
$$
Hence \be g(z)=\sum_{k=0}^{\infty}(z_0-z)^k (z_0-A)^{k-1} \ll{gz} \ee
is a norm-convergent power series in $z$.  For $z\neq 0$ we write $\n
g(z)\n=|z|\inv\n(\I-A/z)\inv\n$ and observe that $\lim_{z\raw\infty}
\I-A/z=\I$, since $\lim_{z\raw\infty} \n A/z\n=0$ by
\ref{defnorm}.3. Hence $\lim_{z\raw\infty}(\I-A/z)\inv=\I$, and \be
\lim_{z\raw\infty} \n g(z)\n=0. \ll{limgz} \ee Let $\rh\in\A^*$ be a
functional on $\A$; since $\rh$ is bounded, (\ref{gz}) implies that
the function $g_{\rh}:z\raw\rh(g(z))$ is given by a convergent power
series, and (\ref{limgz}) implies that \be \lim_{z\raw\infty}
g_{\rh}(z)=0. \ll{limgz2} \ee Now suppose that $\sg(A)=\emptyset$, so
that $\rh(A)=\C$. The function $g$, and hence $g_{\rh}$, is then
defined on $\C$, where it is analytic and vanishes at infinity.  In
particular, $g_{\rh}$ is bounded, so that by Liouville's theorem it
must be constant.  By (\ref{limgz2}) this constant is zero, so that
$g=0$ by Corollary \ref{HB2}. This is absurd, so that $\rh(A)\neq\C$
hence $\sg(A)\neq\emptyset$.  \enp

The fact that the spectrum is never empty leads to the following {\bf
Gel'fand-Mazur theorem}, which will be essential in the
characterization of commutative \ca s.
\begin{Corollary}\ll{GM}
If every element (except 0) of a unital Banach algebra $\A$ is
invertible, then $\A\simeq\C$ as \BA s. \end{Corollary}

Since $\sg(A)\neq\emptyset$, for each $A\neq 0$ there is a $z_A\in\C$
for which $A-z_A\I$ is not invertible. Hence $A-z_A\I=0$ by
assumption, and the map $A\raw z_A$ is the desired algebra
isomorphism. Since $\n A\n=\n z\I\n=|z|$, this isomorphism is
isometric.\enp

Define the {\bf spectral radius} $r(A)$ of $A\in\A$ by \be
r(A):=\sup\{|z|, z\in\sg(A)\}. \ll{defrA} \ee From Theorem
\ref{thmsp}.1 one immediately infers \be r(A)\leq\,\n A\n. \ll{rAb}
\ee
\begin{Proposition}\ll{srt}
For each $A$ in a unital \BA\ one has \be r(A)=\lim_{n\raw\infty} \n
A^n\n^{1/n}. \ll{rAfor} \ee \end{Proposition}

 By Lemma \ref{l1}, for $|z|\,>\n A\n$ the function g in the proof of
Lemma \ref{l2} has the norm-convergent power series expansion \be
g(z)=\frac{1}{z}\sum_{k=0}^{\infty}\left(\frac{A}{z}\right)^k.\ll{gz2}
\ee On the other hand, we have seen that for any $z\in\rh(A)$ one may
find a $z_0\in\rh(A)$ such that the power series (\ref{gz})
converges. If $|z|\,>r(A)$ then $z\in\rh(A)$, so (\ref{gz}) converges
for $|z|\,>r(A)$.  At this point the proof relies on the theory of
analytic functions with values in a Banach space, which says that,
accordingly, (\ref{gz2}) is norm-convergent for $|z|\,>r(A)$,
uniformly in $z$.  Comparing with (\ref{rAb}), this sharpens what we
know from Lemma \ref{l1}. The same theory says that (\ref{gz2}) cannot
norm-converge uniformly in $z$ unless $\n A^n\n/|z|^n\,<1$ for large
enough $n$. This is true for all $z$ for which $|z|\,>r(A)$, so that
\be \lim\sup_{n\raw\infty} \n A\n^{1/n}\,\leq r(A). \ll{limsup} \ee To
derive a second inequality we use the following {\bf polynomial
spectral mapping property}.
\begin{Lemma}\ll{psmp} 
For a polynomial $p$ on $\C$, define $p(\sg(A))$ as $\{p(z)|\,
z\in\sg(A)\}$. Then \be p(\sg(A))=\sg(p(A)). \ll{psgAsgpA} \ee
\end{Lemma}

 To prove this equality, choose $z,\al\in\C$ and compare the
factorizations \bea p(z)-\al & = & c\prod_{i=1}^n(z-\bt_i(\al)); \nn
\\ p(A)-\al\I & = & c\prod_{i=1}^n(A-\bt_i(\al)\I).  \eea Here the
coefficients $c$ and $\bt_i(\al)$ are determined by $p$ and $\al$.
When $\al\in\rh(p(A))$ then $p(A)-\al\I$ is invertible, which implies
that all $A-\bt_i(\al)\I$ must be invertible. Hence $\al\in\sg(p(A))$
implies that at least one of the $A-\bt_i(\al)\I$ is not invertible,
so that $\bt_i(\al)\in\sg(A)$ for at least one $i$. Hence
$p(\bt_i(\al))-\al=0$, i.e., $\al\in p(\sg(A))$. This proves the
inclusion $\sg(p(A))\subseteq p(\sg(A))$.

Conversely, when $\al\in p(\sg(A))$ then $\al=p(z)$ for some
$z\in\sg(A)$, so that for some $i$ one must have $\bt_i(\al)=z$ for
this particular $z$. Hence $\bt_i(\al)\in\sg(A)$, so that
$A-\bt_i(\al)$ is not invertible, implying that $p(A)-\al\I$ is not
invertible, so that $\sl\in\sg(p(A))$. This shows that
$p(\sg(A))\subseteq \sg(p(A))$, and (\ref{psgAsgpA}) follows.  \enp

To conclude the proof of Proposition \ref{srt}, we note that since
$\sg(A)$ is closed there is an $\al\in\sg(A)$ for which
$|\al|=r(A)$. Since $\al^n\in\sg(A^n)$ by Lemma \ref{psmp}, one has
$|\al^n|\,\leq\,\n A^n\n$ by (\ref{rAb}). Hence $\n A^n\n^{1/n}\,\geq
|\al|=r(A)$.  Combining this with (\ref{limsup}) yields
$$
\lim\sup_{n\raw\infty} \n A\n^{1/n}\leq r(A)\leq \n A^n\n^{1/n}.
$$
Hence the limit must exist, and $$ \lim_{n\raw\infty} \n A\n^{1/n}\,
=\inf_n \n A^n\n^{1/n}\, = r(A).  $$
\enp
\begin{Definition}\ll{defideal} An {\bf ideal} in a Banach algebra
$\A$ is a closed linear subspace $\GI\subseteq\A$ such that $A\in\GI$
implies $AB\in\GI$ and $BA\in\GI$ for all $B\in\A$.

A {\bf left-ideal} of $\A$ is a closed linear subspace $\GI$ for which
 $A\in\GI$ implies $BA\in\GI$ for all $B\in\A$.

A {\bf right-ideal} of $\A$ is a closed linear subspace $\GI$ for
 which $A\in\GI$ implies $AB\in\GI$ for all $B\in\A$.

A {\bf maximal ideal} is an ideal $\GI\neq\A$ for which no ideal
$\til{\GI}\neq\A$, $\til{\GI}\neq \GI$, exists which contains
$\GI$. \end{Definition}

In particular, an ideal is itself a Banach algebra.  An ideal $\GI$
that contains an invertible element $A$ must coincide with $\A$, since
$A\inv A=\I$ must lie in $\GI$, so that all $B=B\I$ must lie in
$\GI$. This shows the need for considering \BA s with and without
unit; it is usually harmless to add a unit to a \BA\ $\A$, but a given
proper ideal $\GI\neq\A$ does not contain $\I$, and one cannot add
$\I$ to $\GI$ without ruining the property that it is a proper ideal.
\begin{Proposition}\ll{idBA}
If $\GI$ is an ideal in a \BA\ $\A$ then the quotient $\A/\GI$ is a
\BA\ in the norm \be \n \ta(A)\n:=\inf_{J\in\GI} {\n A+J\n} \ll{normq}
\ee and the multiplication \be \ta(A)\ta(B):=\ta(AB). \ll{mid} \ee
Here $\ta:\A\raw\A/\GI$ is the canonical projection.  If $\A$ is
unital then $\A/\GI$ is unital, with unit $\ta(\I)$.
\end{Proposition}

We omit the standard proof that $\A/\GI$ is a Banach space in the norm
(\ref{normq}). As far as the \BA\ structure is concerned, first note
that (\ref{mid}) is well defined: when $J_1,J_2\in\GI$ one has
$$
\ta(A+J_1)\ta(B+J_2)=\ta(AB+AJ_2+J_1B+J_1J_2)=\ta(AB)=\ta(A)\ta(B),
$$
since $AJ_2+J_1B+J_1J_2\in\GI$ by definition of an ideal, and
$\ta(J)=0$ for all $J\in\GI$.  To prove (\ref{Cstax1}), observe that,
by definition of the infimum, for given $A\in\A$, for each $\ep>0$
there exists a $J\in\GI$ such that \be \n\ta(A)\n +\ep\geq\,\n
A+J\n. \ll{idine} \ee For if such a $J$ would not exist, the norm in
$\A/\GI$ could not be given by (\ref{normq}).  On the other hand, for
any $J\in\GI$ it is clear from (\ref{normq}) that \be \n\ta(A)\n
=\n\ta(A+J)\n\,\leq\, \n A +J \n. \ll{nq2} \ee For $A,B\in\A$ choose
$\ep>0$ and $J_1,J_2\in\GI$ such that (\ref{idine}) holds for $A,B$,
and estimate \bea \n\ta(A)\ta(B)\n & = &
\n\ta(A+J_1)\ta(B+J_2)\n=\n\ta((A+J_1)(B+J_2))\n \nn\\ & \leq & \n
(A+J_1)(B+J_2)\n\,\leq\,\n A+J_1\n\:\n B+J_2\n \nn\\ & \leq &
(\n\ta(A)\n+\ep)(\n\ta(B)\n+\ep).  \eea Letting $\ep\raw 0$ yields
$\n\ta(A)\ta(B)\n\,\leq\,\n\ta(A)\n\: \n\ta(B)\n$.

When $\A$ has a unit, it is obvious from (\ref{mid}) that $\ta(\I)$ is
a unit in $\A/\GI$.  By (\ref{nq2}) with $A=\I$ one has
$\n\ta(\I)\n\,\leq\,\n\I\n=1$. On the other hand, from (\ref{Cstax1})
with $B=\I$ one derives $\n\ta(\I)\n\,\geq 1$. Hence $\n\ta(\I)\n= 1$.
\enp \su{Commutative Banach algebras} We now assume that the \BA\ $\A$
is {\bf commutative} (that is, $AB=BA$ for all $A,B\in\A$).
\begin{Definition}\ll{defDl}
The {\bf structure space} $\Dl(\A)$ of a commutative \BA\ $\A$ is the
set of all nonzero linear maps $\om:\A\raw\C$ for which \be
\om(AB)=\om(A)\om(B) \ll{pABAB} \ee for all $A,B\in\A$. We say that
such an $\om$ is {\bf multiplicative}.

In other words, $\Dl(\A)$ consists of all nonzero homomorphisms from
 $\A$ to $\C$.\end{Definition}
\begin{Proposition}\ll{elDl} Let $\A$ have a unit $\I$.
\begin{enumerate}
\item
Each $\om\in\Dl(\A)$ satisfies \be \om(\I)=1; \ll{phvI} \ee
\item 
each $\om\in\Dl(\A)$ is continuous, with norm \be \n\om\n=1; \ll{nph1}
\ee hence \be |\om(A)|\,\leq\,\n A\n \ll{phvbound} \ee for all
$A\in\A$.
\end{enumerate} \end{Proposition}

The first claim is obvious, since $\om(\I A)=\om(\I)\om(A)=\om(A)$,
and there is an $A$ for which $\om(A)\neq 0$ because $\om$ is not
identically zero.

For the second, we know from Lemma \ref{l1} that $A-z$ is invertible
when $|z|\,>\,\n A\n$, so that $\om(A-z)=\om(A)-z\neq 0$, since $\om$
is a homomorphism.  Hence $|\om(A)|\neq|z|$ for $|z|\,>\,\, A\n$, and
(\ref{phvbound}) follows.  \enp
\begin{Theorem}\ll{G1}
Let $\A$ be a unital commutative \BA.  There is a bijective
correspondence between $\Dl(\A)$ and the set of all maximal ideals in
$\A$, in that the kernel $\ker(\om)$ of each $\om\in\DA$ is a maximal
ideal $\GI_{\om}$, each maximal ideal is the kernel of some
$\om\in\DA$, and $\om_1=\om_2$ iff
$\GI_{\om_1}=\GI_{\om_2}$. \end{Theorem}

The kernel of each $\om\in\DA$ is closed, since $\om$ is continuous by
\ref{elDl}.2. Furthermore, $\ker(\om)$ is an ideal since $\om$
satisfies (\ref{pABAB}). The kernel of every linear map
$\om:\CV\raw\C$ on a vector space $\CV$ has codimension one (that is,
$\dim(\CV/\ker(\om))=1$), so that $\ker(\om)$ is a maximal ideal.
Again on any vector space, when $\ker(\om_1)=\ker(\om_2)$ then $\om_1$
is a multiple of $\om_2$. For $\om_i\in\DA$ this implies $\om_1=\om_2$
because of (\ref{phvI}).

We now show that every maximal ideal $\GI$ of $\A$ is the kernel of
some $\om\in\DA$. Since $\GI\neq\A$, there is a nonzero $B\in\A$ which
is not in $\GI$. Form
$$
\GI_B:=\{BA+J|, A\in\A,J\in\GI\}.
$$ 
This is clearly a left-ideal; since $\A$ is commutative, $\GI_B$ is
even an ideal. Taking $A=0$ we see $\GI\subseteq\GI_B$. Taking $A=\I$
and $J=0$ we see that $B\in\GI_B$, so that $\GI_B\neq\GI$. Hence
$\GI_B=\A$, as $\GI$ is maximal. In particular, $\I\in\GI_B$, hence
$\I=BA+J$ for suitable $A\in\A,J\in\GI$. Apply the canonical
projection $\ta:\A\raw\A/\GI$ to this equation, giving
$$
\ta(\I)=\I=\ta(BA)=\ta(B)\ta(A),$$ because of (\ref{mid}) and
$\ta(J)=0$. hence $\ta(A)=\ta(B)\inv$ in $\A/\GI$. Since $B$ was
arbitrary (though nonzero), this shows that every nonzero element of
$\A/\GI$ is invertible. By Corollary \ref{GM} this yields
$\A/\GI\simeq\C$, so that there is a homomorphism $\ps:\A/\GI\raw\C$.
Now define a map $\om:\A\raw\C$ by $\om(A):=\ps(\ta(A))$. This map is
clearly linear, since $\ta$ and $\ps$ are. Also,
$$
\om(A)\om(B)= \ps(\ta(A))\ps(\ta(B))=\ps(\ta(A)\ta(B))=\ps(\ta(AB))=
\om(AB),$$ because of (\ref{mid}) and the fact that $\ps$ is a
homomorphism.

Therefore, $\om$ is multiplicative; it is nonzero because $\om(B)\neq
0$, or because $\om(\I)=1$.  Hence $\om\in\DA$. Finally,
$\GI\subseteq\ker(\om)$ since $\GI=\ker(\ta)$; but if $B\notin\GI$ we
saw that $\om(B)\neq 0$, so that actually $\GI=\ker(\om)$.\enp

By \ref{elDl}.2 we have $\DA\subset\A^*$.  Recall that the {\bf
weak$\mbox{}^*$-topology}, also called {\bf $w^*$-topology}, on the
dual $\CB^*$ of a Banach space $\CB$ is defined by the convergence
$\om_n\raw\om$ iff $\om_n(v)\raw\om(v)$ for all $v\in\CB$.  The {\bf
Gel'fand topology} on $\DA$ is the relative $w^*$-topology.
\begin{Proposition}\ll{DAcpt}
The structure space $\DA$ of a unital commutative \BA\ $\A$ is compact
and Hausdorff in the Gel'fand topology. \end{Proposition}

 The convergence $\om_n\raw\om$ in the $w^*$-topology by definition
means that $\om_n(A)\raw\om(A)$ for all $A\in\A$.  When $\om_n\in\DA$
for all $n$, one has
$$
|\om(AB)-\om(A)\om(B)|=|\om(AB)-\om_n(AB)+\om_n(A)\om_n(B)-
\om(A)\om(B)|$$ $$ \leq |\om(AB)-\om_n(AB)|+|\om_n(A)\om_n(B)-
\om(A)\om(B)|.$$ In the second term we write
$$
\om_n(A)\om_n(B)- \om(A)\om(B)=(\om_n(A)-\om(A))\om_n(B)+
\om(A)(\om_n(B)-\om(B)).$$ By (\ref{phvbound}) and the triangle
inequality, the absolute value of the right-hand side is bounded by
$$
\n B\n\: |\om_n(A)-\om(A)|+\n A\n\: |\om_n(B)-\om(B)|.
$$
All in all, when $\om_n\raw\om$ in the $w^*$-topology we obtain
 $|\om(AB)-\om(A)\om(B)|=0$, so that the limit $\om\in\DA$.  Hence
 $\DA$ is $w^*$-closed.

From (\ref{nph1}) we have $\DA\in\A^*_1$ (the unit ball in $\A^*$,
consisting of all functionals with norm $\leq 1$). By the
Banach-Alaoglu theorem, the unit ball in $\A^*$ is
$w^*$-compact. Being a closed subset of this unit ball, $\DA$ is
$w^*$-compact. Since the $w^*$-topology is Hausdorff (as is immediate
from its definition), the claim follows.\enp

We embed $\A$ in $\A^{**}$ by $A\raw\hat{A}$, where \be
\hat{A}(\om):=\om(A). \ll{GT} \ee When $\om\in\DA$, this defines
$\hat{A}$ as a function on $\DA$.  By elementary functional analysis,
the $w^*$-topology on $\A^*$ is the weakest topology for which all
$\hat{A}$, $A\in\A$, are continuous.  This implies that the Gel'fand
topology on $\DA$ is the weakest topology for which all functions
$\hat{A}$ are continuous.  In particular, a basis for this topology is
formed by all open sets of the form \be
\hat{A}\inv(\CO)=\{\om\in\DA|\, \om(A)\in\CO\},\ll{Gtop} \ee where
$A\in\A$ and $\CO$ is an open set in $\C$.

Seen as a map from $\A$ to $C(\DA)$, the map $A\raw\hat{A}$ defined by
(\ref{GT}) is called the {\bf Gel'fand transform}.
 
For any compact Hausdorff space $X$, we regard the space $C(X)$ of all
continuous functions on $X$ as a Banach space in the {\bf sup-norm}
defined by \be \n f\n_{\infty}:=\sup_{x\in X} |f(x)|.  \ll{supnorm}
\ee A basic fact of topology and analysis is that $C(X)$ is complete
in this norm.  Convergence in the sup-norm is the same as uniform
convergence.  What's more, it is easily verified that $C(X)$ is even a
commutative \BA\ under pointwise addition and multiplication, that is,
\bea (\lm f+\mu g)(x) & := & \lm f(x)+\mu g(x); \nn \\ (fg)(x) & := &
f(x)g(x). \ll{baco} \eea Hence the function $1_X$ which is 1 for every
$x$ is the unit $\I$.  One checks that the spectrum of $f\in C(X)$ is
simply the set of values of $f$.

We regard $C(\DA)$ as a commutative \BA\ in the manner explained.
\begin{Theorem}\ll{GTGT} Let $\A$ be a unital commutative \BA.
\begin{enumerate}
\item
The Gel'fand transform is a homomorphism from $\A$ to $C(\DA)$.
\item
The image of $\A$ under the Gel'fand transform separates points in
$\DA$.
\item
The spectrum of $A\in\A$ is the set of values of $\hat{A}$ on $\DA$;
in other words, \be \sg(A)=\sg(\hat{A})=\{\hat{A}(\om)|\,
\om\in\DA\}.\ll{sgAGT} \ee
\item
The Gel'fand transform is a contraction, that is, \be
\n\hat{A}\n_{\infty}\,\leq\,\n A\n. \ll{GTnin} \ee
\end{enumerate} \end{Theorem}

The first property immediately follows from (\ref{GT}) and
(\ref{pABAB}).  When $\om_1\neq\om_2$ there is an $A\in\A$ for which
$\om_1(A)\neq \om_2(A)$, so that $\hat{A}(\om_1)\neq
\hat{A}(\om_2)$. This proves \ref{GTGT}.2.

If $A\in G(\A)$ (i.e., $A$ is invertibe), then $\om(A)\om(A\inv)=1$,
so that $\om(A)\neq 0$ for all $\om\in\DA$. When $A\notin G(\A)$ the
ideal $\GI_A:=\{AB|, B\in\A\}$ does not contain $\I$, so that it is
contained in a maximal ideal $\GI$ (this conclusion is actually
nontrivial, relying on the axiom of choice in the guise of Hausdorff's
maximality priciple).  Hence by Theorem \ref{G1} there is a
$\om\in\DA$ for which $\om(A)=0$.  All in all, we have showed that
$A\in\ G(\A)$ is equivalent to $\om(A)\neq 0$ for all
$\om\in\DA$. Hence $A-z\in G(\A)$ iff $\om(A)\neq z$ for all
$\om\in\DA$. Thus the resolvent is \be \om(A)=\{z\in\C|\, z\neq
\om(A)\,\forall\om\in\DA\}.  \ee Taking the complement, and using
(\ref{GT}), we obtain (\ref{sgAGT}).

Eq.\ (\ref{GTnin}) then follows from (\ref{defrA}), (\ref{rAb}),
(\ref{GT}), and (\ref{supnorm}).\enp

We now look at an example, which is included for three reasons:
firstly it provides a concrete illustration of the Gel'fand transform,
secondly it concerns a commutative \BA\ which is not a \ca, and
thirdly the \BA\ in question has no unit, so the example illustrates
what happens to the structure theory in the absence of a unit. In this
connection, let us note in general that each $\om\in\DA$ has a
suitable extension $\til{\om}$ to $\AI$, namely \be
\om(A+\lm\I):=\om(A)+\lm. \ll{omtil} \ee The point is that $\til{\om}$
remains multiplicative on $\AI$, as can be seen from (\ref{muAI}) and
the definition (\ref{pABAB}). This extension is clearly unique. Even
if one does not actually extend $\A$ to $\AI$, the existence of
$\til{\om}$ shows that $\om$ satisfies (\ref{phvbound}), since this
property (which was proved for the unital case) holds for $\til{\om}$,
and therefore certainly for the restriction $\om$ of $\til{\om}$ to
$\A$.

Consider $\A=L^1(\R)$, with the usual linear structure, and norm \be
\n f\n_1:=\int_{\R}dx\, |f(x)|.  \ee The associative product $*$
defining the Banach algebra structure is convolution, that is, \be
f*g(x):=\int_{\R}dy\, f(x-y)g(y). \ll{convR} \ee Strictly speaking,
this should first be defined on the dense subspace $C_c(\R)$, and
subsequently be extended by continuity to $L^1(\R)$, using the
inequality below. Indeed, using Fubini's theorem on product integrals,
we estimate
$$
\n f*g\n_1=\int_{\R}dx\,|\int_{\R}dy\, f(x-y)g(y)|\leq \int_{\R}dy\,
|g(y)| \int_{\R}dx\,|f(x-y)|$$
$$
=\int_{\R}dy\, |g(y)| \int_{\R}dx\,|f(x)|=\n f\n_1\,\n g\n_1,$$ which
is (\ref{Cstax1}).

 There is no unit in $L^1(\R)$, since from (\ref{convR}) one sees that
the unit should be Dirac's delta-function (i.e., the measure on $\R$
which assigns 1 to $x=0$ and 0 to all other $x$), which does not lie
in $L^1(\R)$.

We know from the discussion following (\ref{omtil}) that every
multiplicative functional $\om\in\Dl(L^1(\R))$ is continuous. Standard
Banach space theory says that the dual of $L^1(\R)$ is
$L^{\infty}(\R)$. Hence for each $\om\in\Dl(L^1(\R))$ there is a
function $\hat{\om}\in L^{\infty}(\R)$ such that \be
\om(f)=\int_{\R}dx\, f(x)\hat{\om}(x). \ll{omhatom} \ee

The multiplicativity condition (\ref{pABAB}) then implies that
$\hat{\om}(x+y)=\hat{\om}(x)\hat{\om}(y)$ for almost all $x,y\in\R$.
This implies \be \hat{\om}(x)=\exp(ipx) \ll{omp} \ee for some
$p\in\C$, and since $\hat{\om}$ is bounded (being in $L^{\infty}(\R)$)
it must be that $p\in\R$. The functional $\om$ corresponding to
(\ref{omp}) is simply called $p$. It is clear that different $p$'s
yield different functionals, so that $\Dl(L^1(\R))$ may be identified
with $\R$. With this notation, we see from (\ref{omhatom}) and
(\ref{omp}) that the Gel'fand transform (\ref{GT}) reads \be
\hat{f}(p)=\int_{\R}dx\, f(x)e^{ipx}. \ll{FT1} \ee Hence the Gel'fand
transform is nothing but the Fourier transform (more generally, many
of the integral transforms of classical analysis may be seen as
special cases of the Gel'fand transform).  The well-known fact that
the Fourier transform maps the convolution product (\ref{convR}) into
the pointwise product is then a restatement of Theorem
\ref{GTGT}.1. Moreover, we see from \ref{GTGT}.3 that the spectrum
$\sg(f)$ of $f$ in $L^1(\R)$ is just the set of values of its Fourier
transform.

Note that the Gel'fand transform is strictly a contraction, i.e.,
there is no equality in the bound (\ref{GTnin}). Finally, the
Riemann-Lebesgue lemma states that $f\in L^1(\R)$ implies $\hat{f}\in
C_0(\R)$, which is the space of continuous functions on $\R$ that go
to zero when $|x|\raw\infty$.  This is an important function space,
whose definition may be generalized as follows.
\begin{Definition}\ll{defc0}
Let $X$ be a Hausdorff space $X$ which is {\bf locally compact} (in
that each point has a compact neighbourhood).  The space $C_0(X)$
consists of all continuous functions on $X$ which {\bf vanish at
infinity} in the sense that for each $\ep>0$ there is a compact subset
$K\subset X$ such that $|f(x)|<\ep$ for all $x$ outside
$K$. \end{Definition}

So when $X$ is compact one trivially has $C_0(X)=C(X)$.  When $X$ is
not compact, the sup-norm (\ref{supnorm}) can still be defined, and
just as for $C(X)$ one easily checks that $C_0(X)$ is a \BA\ in this
norm.

We see that in the example $\A=L^1(\R)$ the Gel'fand transform takes
values in $C_0(\DA)$. This may be generalized to arbitrary commutative
non-unital \BA s. The non-unital version of Theorem \ref{GTGT} is
\begin{Theorem}\ll{GTGT0} Let $\A$ be a non-unital commutative \BA.
\begin{enumerate}
\item
The structure space $\DA$ is locally compact and Hausdorff in the
Gel'fand topology.
\item
The space $\Dl(\AI)$ is the one-point compactification of $\DA$.
\item
The Gel'fand transform is a homomorphism from $\A$ to $C_0(\DA)$.
\item
The spectrum of $A\in\A$ is the set of values of $\hat{A}$ on $\DA$,
with zero added (if 0 is not already contained in this set).
\item
The claims 2 and 4 in Theorem \ref{GTGT} hold.
\end{enumerate} \end{Theorem}

Recall that the {\bf one-point compactification} $\til{X}$ of a
non-compact topological space $X$ is the set $X\cup\infty$, whose open
sets are the open sets in $X$ plus those subsets of $X\cup\infty$
whose complement is compact in $X$.  If, on the other hand, $\til{X}$
is a compact Hausdorff space, the removal of some point `$\infty$'
yields a locally compact Hausdorff space
$X=\til{X}\backslash\{\infty\}$ in the relative topology (i.e., the
open sets in $X$ are the open sets in $\til{X}$ minus the point
$\infty$), whose one-point compactification is, in turn, $\til{X}$.

To prove \ref{GTGT0} we add a unit to $\A$, and note that \be
\Dl(\AI)=\Dl(\A)\cup\infty,\ll{DLAI} \ee where each $\om\in\DA$ is
seen as a functional $\til{\om}$ on $\AI$ by (\ref{omtil}), and the
functional $\infty$ is defined by \be \infty(A+\lm\I):=\lm . \ll{infy}
\ee There can be no other elements $\phv$ of $\Dl(\AI)$, because the
restriction of $\phv$ has a unique multiplicative extension
(\ref{omtil}) to $\AI$, unless it identically vanishes on $\DA$. In
the latter case (\ref{infy}) is clearly the only multiplicative
possibility.

By Proposition \ref{DAcpt} the space $\Dl(\AI)$ is compact and
Hausdorff; by (\ref{infy}) one has \be
\Dl(\A)=\Dl(\AI)\backslash\{\infty\} \ee as a set.  In view of the
paragraph following \ref{GTGT0}, in order to prove \ref{GTGT0}.1 and
2, we need to show that the Gel'fand topology of $\Dl(\AI)$ restricted
to $\DA$ coincides with the Gel'fand topology of $\DA$
itself. Firstly, it is clear from (\ref{Gtop}) that any open set in
$\DA$ (in its own Gel'fand topology) is the restriction of some open
set in $\Dl(\AI)$, because $\A\subset\AI$.  Secondly, for any
$A\in\A$, $\lm\in\C$, and open set $\CO\subset\C$, from (\ref{omtil})
we evidently have
$$
\{\phv\in\Dl(\AI)|\, \phv(A+\lm\I)\in\CO\}\backslash\{\infty\}=
\{\om\in\DA|\, \om(A)\in \CO-\lm\}.
$$
(When $\infty$ does not lie in the set $\{\ldots\}$ on the left-hand
side, one should here omit the ``$\backslash\{\infty\}$''.) With
(\ref{Gtop}), this shows that the restriction of any open set in
$\Dl(\AI)$ to $\DA$ is always open in the Gel'fand topology of
$\DA$. This establishes \ref{GTGT0}.1 and 2.

 It follows from (\ref{GTGT}) and (\ref{infy}) that \be
\hat{A}(\infty) =0 \ee for all $A\in\A$, which by continuity of
$\hat{A}$ leads to \ref{GTGT0}.3.

The comment preceding Theorem \ref{thmsp} implies \ref{GTGT0}.4.  The
final claim follows from the fact that it holds for $\AI$.\enp
\su{Commutative $C^*$-algebras\ll{comCS}} The \BA\ $C(X)$ considered
in the previous section is more than a \BA. Recall Definition
\ref{definv}. The map $f\raw f^*$, where \be f^*(x):=\ovl{f(x)},
\ll{stcx} \ee evidently defines an involution on $C(X)$, in which
$C(X)$ is a commutative \ca\ with unit. The main goal of this section
is to prove the converse statement; cf.\ Definition \ref{defmor}
\begin{Theorem}\ll{CCA} Let $\A$ be a commutative \ca\ with
unit. Then there is a compact Hausdorff space $X$ such that $\A$ is
(isometrically) isomorphic to $C(X)$. This space is unique up to
homeomorphism.  \end{Theorem}

The isomorphism in question is the Gel'fand transform, so that
$X=\DA$, equipped with the Gel'fand topology, and the isomorphism
$\phv:\A\raw C(X)$ is given by \be \phv(A):=\hat{A}. \ll{phvGT} \ee We
have already seen in \ref{GTGT}.1 that this transform is a
homomorphism, so that (\ref{phvmult}) is satisfied. To show that
(\ref{phvstar}) holds as well, it suffices to show that a self-adjoint
element of $\A$ is mapped into a real-valued function, because of
(\ref{aaa}), (\ref{stcx}), and the fact that the Gel'fand transform is
complex-linear.

We pick $A\in\Ar$ and $\om\in\DA$, and suppose that $\om(A)=\al+i\bt$,
where $\al,\bt\in\R$. By (\ref{phvI}) one has $\om(B)=i\bt$, where
$B:=A-\al\I$ is self-adjoint. Hence for $t\in\R$ one computes \be
|\om(B+it\I)|^2=\bt^2+2t\bt+t^2. \ll{btb} \ee On the other hand, using
(\ref{phvbound}) and (\ref{cstax2}) we estimate
$$
|\om(B+it\I)|^2\leq\,\n B+it\I\n^2=\n (B+it\I)^*(B+it\I)\n= \n
B^2+t^2\n\,\leq\,\n B\n^2+t^2.
$$
Using (\ref{btb}) then yields $\bt^2 +t\bt\leq\, \n B\n^2$ for all
$t\in\R$.  For $\bt>0$ this is impossible. For $\bt<0$ we repeat the
argument with $B\raw -B$, finding the same absurdity. Hence $\bt=0$,
so that $\om(A)$ is real when $A=A^*$. Consequently, by (\ref{GT}) the
function $\hat{A}$ is real-valued, and (\ref{phvstar}) follows as
announced.

We now prove that the Gel'fand transform, and therefore the morphism
$\phv$ in (\ref{phvGT}), is isometric. When $A=A^*$, the axiom
(\ref{cstax2}) reads $\n A^2\n=\n A\n^2$. This implies that $\n
A^{2^m}\n=\n A\n^{2^m}$ for all $m\in\Bbb N$.  Taking the limit in
(\ref{rAfor}) along the subsequence $n=2^m$ then yields \be r(A)=\n
A\n.  \ee In view of (\ref{defrA}) and (\ref{sgAGT}), this implies \be
\n\hat{A}\n_{\infty}=\n A\n. \ll{GTisom} \ee For general $A\in\A$ we
note that $A^*A$ is self-adjoint, so that we may use the previous
result and (\ref{cstax2}) to compute
$$
\n A\n^2=\n A^*A\n=\n \widehat{A^*A}\n_{\infty} =\n
\hat{A}^*\hat{A}\n_{\infty} =\n\hat{A}\n_{\infty}^2.
$$
In the third equality we used $\widehat{A^*}=\hat{A}^*$, which we just
proved, and in the fourth we exploited the fact that $C(X)$ is a \ca,
so that (\ref{cstax2}) is satisfied in it. Hence (\ref{GTisom}) holds
for all $A\in\A$.

It follows that $\phv$ in (\ref{phvGT}) is injective, because if
$\phv(A)=0$ for some $A\neq 0$, then $\phv$ would fail to be an
isometry. (A commutative \BA\ for which the Gel`fand transform is
injective is called {semi-simple}. Thus commutative \ca a are
semi-simple.)

We finally prove that the morphism $\phv$ is surjective. We know from
(\ref{GTisom}) that the image $\phv(\A)=\hat{\A}$ is closed in
$C(\DA)$, because $\CA$ is closed (being a \ca, hence a Banach space).
In addition, we know from \ref{GTGT}.2 that $\phv(\A)$ separates
points on $\DA$. Thirdly, since the Gel`fand transform was just shown
to preserve the adjoint, $\phv(\A)$ is closed under complex
conjugation by (\ref{stcx}). Finally, since $\hat{I}=1_X$ by
(\ref{phvI}) and (\ref{GT}), the image $\phv(\A)$ contains $1_X$.  The
surjectivity of $\phv$ now follows from the following {\bf
Stone-Weierstrass theorem}, which we state without proof.
\begin{Lemma}\ll{SW} 
Let $X$ be a compact Hausdorff space, and regard $C(X)$ as a
commutative \ca\ as explained above.  A $C^*$-subalgebra of $C(X)$
which separates points on $X$ and contains $1_X$ coincides with
$C(X)$. \end{Lemma}

Being injective and surjective, the morphism $\phv$ is bijective, and
is therefore an isomorphism. The uniqueness of $X$ is the a
consequence of the following result.
\begin{Proposition}\ll{uniqX}
Let $X$ be a compact Hausdorff space, and regard $C(X)$ as a
commutative \ca\ as explained above. Then $\Dl(C(X))$ (equipped with
the Gel`fand topology) is homeomorphic to $X$. \end{Proposition}

Each $x\in X$ defines a linear map $\om_x:C(X)\raw \C$ by
$\om_x(f):=f(x)$, which is clearly multiplicative and nonzero.  Hence
$x\raw\om_x$ defines a map $E$ (for Evaluation) from $X$ to
$\Dl(C(X))$, given by \be E(x): f\raw f(x). \ll{ev} \ee Since a
compact Hausdorff space is normal, Urysohn's lemma says that $C(X)$
separates points on $X$ (i.e., for all $x\neq y$ there is an $f\in
C(X)$ for which $f(x)\neq f(y)$). This shows that $E$ is injective.

We now use the compactness of $X$ and Theorem \ref{G1} to prove that
$E$ is surjective. The maximal ideal $\GI_x:=\GI_{\om_x}$ in $C(X)$
which corresponds to $\om_x\in \Dl(C(X))$ is obviously \be
\GI_x=\{f\in C(X)|\, f(x)=0\}.  \ee Therefore, when $E$ is not
surjective there exists a maximal ideal $\GI\subset C(X)$ which for
each $x\in X$ contains at a function $f_x$ for which $f_x(x)\neq 0$
(if not, $\GI$ would contain an ideal $\GI_x$ which thereby would not
be maximal). For each $x$, the set $\CO_x$ where $f_x$ is nonzero is
open, because $f$ is continuous. This gives a covering
$\{\CO_x\}_{x\in X}$ of $X$. By compactness, there exists a finite
subcovering $\{\CO_{x_i}\}_{i=1,\ldots, N}$. Then form the function
$g:=\sum_{i=1}^N |f_{x_i}|^2$. This function is strictly positive by
construction, so that it is invertible (note that $f\in C(X)$ is
invertible iff $f(x)\neq 0$ for all $x\in X$, in which case
$f\inv(x)=1/f(x)$).  But $\GI$ is an ideal, so that, with all
$f_{x_i}\in\GI$ (since all $f_x\in\GI$) also $g\in\GI$. But an ideal
containing an invertible element must coincide with $\A$ (see the
comment after \ref{defideal}), contradicting the assumption that $\GI$
is a maximal ideal.

Hence $E$ is surjective; since we already found it is injective, $E$
must be a bijection. It remains to be shown that $E$ is a
homeomorphism.  Let $X_o$ denote $X$ with its originally given
topology, and write $X_G$ for $X$ with the topology induced by
$E\inv$.  Since $\hat{f}\circ E=f$ by (\ref{ev}) and (\ref{GT}), and
the Gel'fand topology on $\Dl(C(X))$ is the weakest topology for which
all functions $\hat{f}$ are continuous, we infer that $X_G$ is weaker
than $X_o$ (since $f$, lying in $C(X_o)$, is continuous).  Here a
topology $\CT_1$ is called weaker than a topology $\CT_2$ on the same
set if any open set of $\CT_1$ contains an open set of $\CT_2$.  This
includes the possibility $\CT_1=\CT_2$.

Without proof we now state a result from topology.
\begin{Lemma}
Let a set $X$ be Hausdorff in some topology $\CT_1$ and compact in a
topology $\CT_2$. If $\CT_1$ is weaker than $\CT_2$ then
$\CT_1=\CT_2$. \end{Lemma}

Since $X_o$ and $X_G$ are both compact and Hausdorff (the former by
assumption, and the latter by Proposition \ref{DAcpt}), we conclude
from this lemma that $X_0=X_G$; in other words, $E$ is a
homeomorphism. This concludes the proof of \ref{uniqX}.\enp

Proposition \ref{uniqX} shows that $X$ as a topological space may be
extracted from the Banach-algebraic structure of $C(X)$, up to
homeomorphism. Hence if $C(X)\simeq C(Y)$ as a \ca, where $Y$ is a
second compact Hausdorff space, then $X\simeq Y$ as topological
spaces. Given the isomorphism $\A\simeq C(X)$ constructed above, a
second isomorphism $\A\simeq C(Y)$ is therefore only possible if
$X\simeq Y$. This proves the final claim of Theorem \ref{CCA}.\enp

The condition that a compact topological space be Hausdorff is
sufficient, but not necessary for the completeness of $C(X)$ in the
sup-norm. However, when $X$ is not Hausdorff yet $C(X)$ is complete,
the map $E$ may fail to be injective since in that case $C(X)$ may
fail to separate points on $X$.

On the other hand, suppose $X$ is locally compact but not compact, and
consider $\A=C_b(X)$; this is the space of all continuous bounded
functions on $X$. Equipped with the operations (\ref{supnorm}),
(\ref{baco}), and (\ref{stcx}) this is a commutative \ca. The map
$E:X\raw \Dl(C_b(X))$ is now injective, but fails to be surjective
(this is suggested by the invalidity of the proof we gave for
$C(X)$). Indeed, it can be shown that $\Dl(C_b(X))$ is homeomorphic to
the Ceh-Stone compactification of $X$.

Let us now consider what happens to Theorem \ref{CCA} when $\A$ has no
unit.  Following the strategy we used in proving Theorem \ref{GTGT0},
we would like to add a unit to $\A$.  As in the case of a general \BA\
(cf.\ section \ref{Bab}), we form $\AI$ by (\ref{defAI}), define
multiplication by (\ref{muAI}), and use the natural involution \be
(A+\lm\I)^*:=A^*+\ovl{\lm}\I. \ll{invAI} \ee However, the
straightforward norm (\ref{norminAI}) cannot be used, since it is not
a $C^*$-norm in that axiom (\ref{cstax2}) is not satisfied.  Recall
Definition \ref{defbobs}.
\begin{Lemma}\ll{cstunit}
Let $\A$ be a \ca.
\begin{enumerate}
\item
The map $\rh:\A\raw\B(\A)$ given by \be \rh(A)B:=AB \ll{rhABAB} \ee
establishes an isomorphism between $\A$ and $\rh(\A)\subset\B(\A)$.
\item
When $\A$ has no unit, define a norm on $\AI$ by \be \n
A+\lm\I\n:=\n\rh(A)+\lm\I\n, \ll{cninai} \ee where the norm on the
right-hand side is the operator norm (\ref{banalnorm}) in $\B(\A)$,
and $\I$ on the right-hand side is the unit operator in $\B(\A)$.
With the operations (\ref{muAI}) and (\ref{invAI}), the norm
(\ref{cninai}) turns $\AI$ into a \ca\ with unit.
\end{enumerate}
\end{Lemma}

By (\ref{cstax1}) we have $\n\rh(A)B\n=\n AB\n\,\leq\,\n A\n\:\n B\n$
for all $B$, so that $\n\rh(A)\n\,\leq\,\n A\n$ by (\ref{banalnorm}).
On the other hand, using (\ref{cstax2}) and (\ref{astisa}) we can
write
$$
\n A\n =\n AA^*\n/\n A\n=\n\rh(A)\frac{A^*}{\n
A\n}\n\,\leq\,\,\rh(A)\n;
$$
in the last step we used (\ref{opprop}) and $\n (A^*/\n A\n)\n=1$.
Hence \be \n\rh(A)\n=\n A\n. \ll{rhisom} \ee Being isometric, the map
$\rh$ must be injective; it is clearly a homomorphism, so that we have
proved \ref{cstunit}.1.

It is clear from (\ref{muAI}) and (\ref{invAI}) that the map
$A+\lm\I\raw\rh(A)+\lm\I$ (where the symbol $\I$ on the left-hand side
is defined below (\ref{muAI}), and the $\I$ on the right-hand side is
the unit in $\B(\A)$) is a morphism.  Hence the norm (\ref{cninai})
satisfies (\ref{cstax1}), because (\ref{Cstax1}) is satisfied in
$\B(\A)$.  Moreover, in order to prove that the norm (\ref{cninai})
satisfies (\ref{cstax2}), by Lemma \ref{ineqcs} it suffices to prove
that \be \n\rh(A)+\lm\I\n^2\,\leq\,\n (\rh(A)+\lm\I)^*(\rh(A)+\lm\I)\n
\ll{moet} \ee for all $A\in\A$ and $\lm\in\C$. To do so, we use a
trick similar to the one involving (\ref{idine}), but with inf
replaced by sup.  Namely, in view of (\ref{banalnorm}), for given
$A\in\B(\CB)$ and $\ep>0$ there exists a $v\in\CV$, with $\n v\n=1$,
such that $\n A\n^2-\ep \leq\,\n Av\n^2$. Applying this with
$\CB\raw\A$ and $A\raw \rh(A)+\lm\I$, we infer that for every $\ep>0$
there exists a $B\in\A$ with norm 1 such that
$$
\n\rh(A)+\lm\I\n^2-\ep \leq\,\n (\rh(A)+\lm\I)B\n^2= \n AB+\lm
B\n^2=\n (AB+\lm B)^*(AB+\lm B)\n.
$$
Here we used (\ref{cstax2}) in $\A$. Using (\ref{rhABAB}), the
right-hand side may be rearranged as
$$
\n \rh(B^*) \rh(A^*+\ovl{\lm}\I)\rh(A+\lm\I)B\n\,\leq\,\n
\rh(B^*)\n\:\n (\rh(A)+\lm\I)^*(\rh(A)+\lm\I)\n\:\n B\n.
$$
Since $\n\rh(B^*)\n=\n B^*\n=\n B\n=1$ by (\ref{rhisom}) and
(\ref{astisa}), and $\n B\n=1$ also in the last term, the inequality
(\ref{moet}) follows by letting $\ep\raw 0$. \enp

Hence the \ca ic version of Theorem \ref{BAun} is
\begin{Proposition}\ll{caun}
For every \ca\ without unit there exists a unique unital \ca\ $\AI$
and an isometric (hence injective) morphism $\A\raw\AI$, such that
$\AI/\A\simeq\C$. \end{Proposition}

The uniqueness of $\AI$ follows from Corollary \ref{normunique} below.
On the other hand, in view of the fact that both (\ref{norminAI}) and
(\ref{cninai}) define a norm on $\AI$ satisfying the claims of
Proposition \ref{BAun}, we conclude that the unital \BA\ $\AI$ called
for in that proposition is not, in general, unique.

In any case, having established the existence of the unitization of an
arbitrary non-unital \ca, we see that, in particular, a commutative
non-unital \ca\ has a unitization. The passage from Theorem \ref{GTGT}
to Theorem \ref{GTGT0} may then be repeated in the \ca ic setting; the
only nontrivial point compared to the situation for \BA s is the
generalization of Lemma \ref{SW}. This now reads
\begin{Lemma}\ll{SW0} 
Let $X$ be a locally compact Hausdorff space, and regard $C_0(X)$ as a
commutative \ca\ as explained below Definition \ref{defc0}.

A $C^*$-subalgebra $\A$ of $C_0(X)$ which separates points on $X$, and
is such that for each $x\in X$ there is an $f\in\A$ such that
$f(x)\neq 0$, coincides with $C_0(X)$. \end{Lemma}
 
At the end of the day we then find
\begin{Theorem}\ll{CCA0} Let $\A$ be a commutative \ca\ without
unit. There is a locally compact Hausdorff space $X$ such that $\A$ is
(isometrically) isomorphic to $C_0(X)$. This space is unique up to
homeomorphism. \end{Theorem} \su{Spectrum and functional
calculus\ll{sfc}} We return to the general case in which a \ca\ $\A$
is not necessarily commutative (but assumed unital), but analyze
properties of $\A$ by studying certain commutative subalgebras.  This
will lead to important results.

For each element $A\in\A$ there is a smallest $C^*$-subalgebra
$C^*(A,\I)$ of $\A$ which contains $A$ and $\I$, namely the closure of
the linear span of $\I$ and all operators of the type $A_1\ldots A_n$,
where $A_i$ is $A$ or $A^*$.  Following the terminology for operators
on a \Hs, an element $A\in\A$ is called {\bf normal} when
$[A,A^*]=0$. The crucial property of a normal operator is that
$C^*(A,\I)$ is commutative. In particular, when $A$ is self-adjoint,
$C^*(A,\I)$ is simply the closure of the space of all polynomials in
$A$.  It is sufficient for our purposes to restrict ourselves to this
case.
\begin{Theorem}\ll{tsp}
Let $A=A^*$ be a self-adjoint element of a unital \ca.
\begin{enumerate}
\item
The spectrum $\sg_{\A}(A)$ of $A$ in $\A$ coincides with the spectrum
$\sg_{C^*(A,\I)}(A)$ of $A$ in $C^*(A,\I)$ (so that we may
unambiguously speak of the spectrum $\sg(A)$).
\item
The spectrum $\sg(A)$ is a subset of $\R$.
\item
The structure space $\Dl(C^*(A,\I))$ is homeomorphic with $\sg(A)$, so
that $C^*(A,\I)$ is isomorphic to $C(\sg(A))$. Under this isomorphism
the Gel'fand transform $\hat{A}:\sg(A)\raw\R$ is the identity function
${\rm id}_{\sg(A)}:t\raw t$.
\end{enumerate} \end{Theorem}

Recall (\ref{GA}). Let $A\in G(\A)$ be normal in $\A$, and consider
the \ca\ $C^*(A,A\inv,\I)$ generated by $A$, $A\inv$, and $\I$. One
has $(A\inv)^*=(A^*)\inv$, and $A$, $A^*$, $A\inv$, $(A^*)\inv$ and
$\I$ all commute with each other. Hence $C^*(A,A\inv,\I)$ is
commutative; it is the closure of the space of all polynomials in $A$,
$A^*$, $A\inv$, $(A^*)\inv$, and $\I$. By Theorem \ref{CCA} we have
$C^*(A,A\inv,\I)\simeq C(X)$ for some compact Hausdorff space $X$.
Since $A$ is invertible and the Gel'fand transform (\ref{GT}) is an
isomorphism, $\hat{A}$ is invertible in $C(X)$ (i.e., $\hat{A}(x)\neq
0x$ for all $x\in X$).  However, for any $f\in C(X)$ that is nonzero
throughout $X$ we have $0< \n f\n_{\infty}^{-2}ff^*\leq 1$ pointwise,
so that $0\leq 1_X- \n f\n_{\infty}^{-2}ff* < 1$ pointwise, hence
$$
\n 1_X- ff*/\n f\n_{\infty}^2\n_{\infty}\,<1.
$$
Here $f^*$ is given by (\ref{stcx}).  Using Lemma \ref{l1}, in terms
of $\I=1_X$ we may therefore write \be \frac{1}{f}=\frac{f^*}{\n
f\n_{\infty}^2}\sum_{k=0}^{\infty} \left(\I-\frac{ff^*}{\n
f\n_{\infty}^2}\right)^k.  \ee Hence $\hat{A}\inv$ is a
norm-convergent limit of a sequence of polynomials in $\hat{A}$ and
$\hat{A}^*$. Gel'fand transforming this result back to
$C^*(A,A\inv,\I)$, we infer that $A\inv$ is a norm-convergent limit of
a sequence of polynomials in $A$ and $A^*$. Hence $A\inv$ lies in
$C^*(A,\I)$, and $C^*(A,A\inv,\I)=C^*(A,\I)$.

Now replace $A$ by $A-z$, where $z\in\C$. When $A$ is normal $A-z$ is
normal. So if we assume that $A-z\in G(\A)$ the argument above
applies, leading to the conclusion that the resolvent $\rh_{\A}(A)$ in
$\A$ coincides with the resolvent $\rh_{C^*(A,\I)}(A)$ in
$C^*(A,\I)$. By Definition \ref{defspectrum} we then conclude that
$\sg_{\A}(A)=\sg_{C^*(A,\I)}(A)$.

According to Theorem \ref{CCA}, the function $\hat{A}$ is real-valued
when $A=A^*$. Hence by \ref{GTGT}.3 the spectrum $\sg_{C^*(A,\I)}(A)$
is real, so that by the previous result $\sg(A)$ is real.

Finally, given the isomorphism $C^*(A,\I)\simeq C(X)$ of Theorem
\ref{CCA} (where $X=\Dl(C^*(A,\I))$), according to \ref{GTGT}.3 the
function $\hat{A}$ is a surjective map from $X$ to $\sg(A)$.  We now
prove injectivity.  When $\om_1,\om_2\in X$ and $\om_1(A)=\om_2(A)$,
then, for all $n\in\N$, we have $$
\om_1(A^n)=\om_1(A)^n=\om_2(A)^n=\om_2(A^n)
$$
  by iterating (\ref{pABAB}) with $B=A$. Since also
$\om_1(\I)=\om_2(\I)=1$ by (\ref{phvI}), we conclude by linearity that
$\om_1=\om_2$ on all polynomials in $A$.  By continuity (cf.\
\ref{elDl}.2) this implies that $\om_1=\om_2$ on $C^*(A,\I)$, since
the linear span of all polynomials is dense in $C^*(A,\I)$.  Using
(\ref{GT}), we have proved that $\hat{A}(\om_1)=\hat{A}(\om_2)$
implies $\om_1=\om_2$.

Since $\hat{A}\in C(X)$ by \ref{GTGT}.1, $\hat{A}$ is continuous. To
prove continuity of the inverse, one checks that for $z\in\sg(A)$ the
functional $\hat{A}\inv(z)\in \Dl(C^*(A,\I))$ maps $A$ to $z$ (and
hence $A^n$ to $z^n$, etc.). Looking at (\ref{Gtop}), one then sees
that $\hat{A}\inv$ is continuous. In conclusion, $\hat{A}$ is a
homeomorphism. The final claim in \ref{tsp}.3 is then obvious.\enp

An immediate consequence of this theorem is the {\bf continuous
functional calculus}.
\begin{Corollary}\ll{cor1}
For each self-adjoint element $A\in\A$ and each $f\in C(\sg(A))$ there
is an operator $f(A)\in\A$, which is the obvious expression when $f$
is a polynomial (and in general is given via the uniform approximation
of $f$ by polynomials), such that \bea \sg(f(A)) & = & f(\sg(A));
\ll{spmath} \\ \n f(A)\n & = & \n f\n_{\infty}. \ll{normfA} \eea In
particular, the norm of $f(A)$ in $C^*(A,\I)$ coincides with its norm
in $\A$.
\end{Corollary}

 Theorem \ref{tsp}.3 yields an isomorphism $C(\sg(A))\raw C^*(A,\I)$,
which is precisely the map $f\raw f(A)$ of the continuous functional
calculus. The fact that this isomorphism is isometric (see \ref{CCA})
yields (\ref{normfA}).  Since $f(\sg(A))$ is the set of values of $f$
on $\sg(A)$, (\ref{spmath}) follows from (\ref{sgAGT}), with $A\raw
f(A)$.

The last claim follows by combining \ref{tsp}.1 with (\ref{spmath})
and (\ref{normfA}).\enp
\begin{Corollary}\ll{normunique}
The norm in a \ca\ is unique (that is, given a \ca\ $\A$ there is no
other norm in which $\A$ is a \ca). \end{Corollary}

First assume $A=A^*$, and apply (\ref{normfA}) with $f={\rm
id}_{\sg(A)}$.  By definition (cf.\ (\ref{rAb})), the sup-norm of
${\rm id}_{\sg(A)}$ is $r(A)$, so that \be \n A\n=r(A)\:\:\:\:
(A=A^*).\ll{nAnrA} \ee Since $A^*A$ is self-adjoint for any $A$, for
general $A\in\A$ we have, using (\ref{cstax2}), \be \n
A\n=\sqrt{r(A^*A)}. \ll{normun} \ee Since the spectrum is determined
by the algebraic structure alone, (\ref{normun}) shows that the norm
is determined by the algebraic structure as well.\enp

Note that Corollary \ref{normunique} does not imply that a given \sta\
can be normed only in one way so as to be completed into a \ca\ (we
will, in fact, encounter an example of the opposite situation). In
\ref{normunique} the completeness of $\A$ is assumed from the outset.
\begin{Corollary}\ll{cor3}
A morphism $\phv:\A\raw\B$ between two \ca s satisfies \be \n
\phv(A)\n\,\leq\,\n A\n, \ll{morbound} \ee and is therefore
automatically continuous. \end{Corollary}

When $z\in\rh(A)$, so that $(A-z)\inv$ exists in $\A$, then
$\phv(A-z)$ is certainly invertible in $\B$, for (\ref{phvmult})
implies that $(\phv(A-z))\inv=\phv((A-z)\inv)$. Hence
$\rh(A)\subseteq\rh(\phv(A))$, so that \be
\sg(\phv(A))\subseteq\sg(A). \ll{sgsubset} \ee Hence $r(\phv(A))\leq
r(A)$, so that (\ref{morbound}) follows from (\ref{normun}).\enp

For later use we note
\begin{Lemma}\ll{phvflem}
When $\phv:\A\raw\B$ is a morphism and $A=A^*$ then \be
f(\phv(A))=\phv(f(A)) \ee for all $f\in C(\sg(A))$ (here $f(A)$ is
defined by the continuous functional calculus, and so is $f(\phv(A))$
in view of (\ref{sgsubset})).\end{Lemma}

 The property is true for polynomials by (\ref{phvmult}), since for
those $f$ has its naive meaning.  For general $f$ the result then
follows by continuity.\enp \su{Positivity in \ca s\ll{posca}} A
bounded operator $A\in\BH$ on a \Hs\ $\H$ is called positive when
$(\Ps,A\Ps)\geq 0$ for all $\Ps\in\H$; this property is equivalent to
$A^*=A$ and $\sg(A)\subseteq\R^+$, and clearly also applies to closed
subalgebras of $\BH$. In quantum mechanics this means that the
expectation value of the observable $A$ is always positive.

 Classically, a function $f$ on some space $X$ is positive simply when
$f(x)\geq 0$ for all $x\in X$. This applies, in particular, to
elements of the commutative \ca\ $C_0(X)$ (where $X$ is a locally
compact Hausdorff space). Hence we have a notion of positivity for
certain concrete \ca s, which we would like to generalize to arbitrary
abstract \ca s. Positivity is one of the most important features in a
\ca; it will, for example, play a central role in the proof of the
Gel'fand Neumark theorem. In particular, one is interested in finding
a number of equivalent characterizations of positivity.
\begin{Definition}\ll{defpos}
An element $A$ of a \ca\ $\A$ is called {\bf positive} when $A=A^*$
and its spectrum is positive; i.e., $\sg(A)\subset\R^+$. We write
$A\geq 0$ or $A\in\A^+$, where \be \A^+:=\{A\in\Ar|\,
\sg(A)\subset\R^+\}. \ll{defposcone} \ee \end{Definition}

It is immediate from Theorems \ref{GTGT}.3 and \ref{tsp}.3 that
$A\in\Ar$ is positive iff its Gel'fand transform $\hat{A}$ is
pointwise positive in $C(\sg(A))$.
\begin{Proposition}\ll{coco}
The set $\A^+$ of all positive elements of a \ca\ $\A$ is a {\bf
convex cone}; that is,
\begin{enumerate}
\item
when $A\in\A^+$ and $t\in\R^+$ then $tA\in\A^+$;
\item
when $A,B\in\A^+$ then $A+B\in\A^+$;
\item
$\A^+\cap -\A^+=0$.
\end{enumerate}\end{Proposition}

The first property follows from $\sg(tA)=t\sg(A)$, which is a special
case of (\ref{spmath}).

Since $\sg(A)\subseteq [0,r(A)]$, we have $|c-t|\,\leq c$ for all
$t\in \sg(A)$ and all $c\geq r(A)$. Hence $\sup_{t\in
\sg(A)}|c1_{\sg(A)}-\hat{A}|\,\leq c$ by \ref{GTGT}.3 and \ref{tsp}.3,
so that $\n c1_{\sg(A)}-\hat{A}\n_{\infty}\leq c$.  Gel'fand
transforming back to $C^*(A,\I)$, this implies $\n c\I-A\n\,\leq c$
for all $c\geq \n A\n$ by \ref{cor1}. Inverting this argument, one
sees that if $\n c\I-A\n\,\leq c$ for some $c\geq \n A\n$, then
$\sg(A)\subset\R^+$.

Use this with $A\raw A+B$ and $c=\n A\n+\n B\n$; clearly $c\geq \,\n
A+B\n$ by \ref{defnorm}.4. Then
$$
\n c\I-(A+B)\n\,\leq\,\n(\n A\n-A)\n+\n(\n B\n-B)\n\,\leq c,
$$ where in the last step we used the previous paragraph for $A$ and
for $B$ separately.  As we have seen, this inequality implies
$A+B\in\A^+$.

Finally, when $A\in\A^+$ and $A\in -\A^+$ it must be that $\sg(A)=0$,
hence $A=0$ by (\ref{nAnrA}) and (\ref{defrA}).  \enp

This is important, because a convex cone in a real vector space is
equivalent to a {\bf linear partial ordering}, i.e., a partial
ordering $\leq$ in which $A\leq B$ implies $A+C\leq B+C$ for all $C$
and $\lm A\leq\lm B$ for all $\lm\in\R^+$. The real vector space in
question is the space $\Ar$ of all self-adjoint elements of $\A$. The
equivalence between these two structures is as follows: given
$\Ar^+:=\A^+$ one defines $A\leq B$ if $B-A\in\Ar^+$, and given $\leq$
one puts $\Ar^+=\{A\in\Ar\, |\, 0\leq A\}$.

For example, when $A=A^*$ one checks the validity of \be -\n
A\n\,\I\leq A\leq\,\n A\n\,\I \ll{orderbound} \ee by taking the
Gel'fand transform of $C^*(A,\I)$. The implication \be -B\leq A\leq
B\:\:\Longrightarrow\: \n A\n\,\leq\,\n B\n \ll{AleqB} \ee then
follows, because $-B\leq A\leq B$ and (\ref{orderbound}) for $A\raw B$
yield $-\n B\n\,\I\leq A\leq\,\n B\n\,\I$, so that $\sg(A)\subseteq
[-\n B\n, \n B\n]$, hence $\n A\n\,\leq\,\n B\n$ by (\ref{nAnrA}) and
(\ref{defrA}). For later use we also record
\begin{Lemma}\ll{aul}
When $A,B\in\A^+$ and $\n A+B\n\,\leq k$ then $\n A\n\,\leq
k$. \end{Lemma}

By (\ref{orderbound}) we have $A+B\leq k\I$, hence $0\leq A\leq k\I-B$
by the linearity of the partial ordering, which also implies that
$k\I-B\leq k\I$, as $0\leq B$. Hence, using $-k\I\leq 0$ (since $k\geq
0$) we obtain $-k\I\leq A\leq k\I$, from which the lemma follows by
(\ref{AleqB}).\enp

We now come to the central result in the theory of positivity in \ca
s, which generalizes the cases $\A=\BH$ and $\A=C_0(X)$.
\begin{Theorem}\ll{pl1}
One has \bea \A^+ & = & \{A^2|\, A\in\Ar\} \ll{poscone3} \\ & = &
\{B^*B|\, B\in\A\}. \ll{poscone2} \eea \end{Theorem}

When $\sg(A)\subset\R^+$ and $A=A^*$ then $\sqrt{A}\in\Ar$ is defined
by the continuous functional calculus for $f=\sqrt{\cdot}$, and
satisfies $\sqrt{A}^2=A$. Hence $\A^+\subseteq\{A^2|\, A\in\Ar\}$. The
opposite inclusion follows from (\ref{spmath}) and \ref{tsp}.2. This
proves (\ref{poscone3}).

The inclusion $\A^+\subseteq \{B^*B|\, B\in\A\}$ is is trivial from
(\ref{poscone3}).
\begin{Lemma}\ll{posdec}
Every self-adjoint element $A$ has a decomposition $A=A_+ - A_-$,
 where $A_+,A_-\in\A^+$ and $A_+ A_-=0$. Moreover, $\n
 A_{\pm}\n\,\leq\,\n A\n$. \end{Lemma}

 Apply the continuous functional calculus with $f={\rm
id}_{\sg(A)}=f_+-f_-$, where ${\rm id}_{\sg(A)}(t)$,
$f_+(t)=\max\{t,0\}$, and $f_-(t)= \max\{-t,0\}$.  Since $\n
f_{\pm}\n_{\infty}\leq r(A)=\n A\n$ (where we used (\ref{nAnrA})), the
bound follows from (\ref{normfA}) with $A\raw A_{\pm}$.  \enp

We use this lemma to prove that $\{B^*B|\, B\in\A\}\subseteq\A^+$.
Apply the lemma to $A=B^*B$ (noting that $A=A^*)$. Then
$$
(A_-)^3=-A_-(A_+-A_-)A_-=-A_-AA_-=-A_-B^*BA_-=-(BA_-)^* BA_-.
$$
Since $\sg(A_-)\subset\R^+$ because $A_-$ is positive, we see from
(\ref{spmath}) with $f(t)=t^3$ that $(A_-)^3\geq 0$. Hence $-(BA_-)^*
BA_-\geq 0$.
\begin{Lemma}\ll{tll}
If $-C^*C\in\A^+$ for some $C\in\A$ then $C=0$. \end{Lemma}

By (\ref{aaa}) we can write $C=D+iE$, $D,E\in\Ar$, so that \be
C^*C=2D^2+2E^2 -CC^*. \ll{bloed} \ee Now for any $A,B\in\A$ one has
\be \sg(AB)\cup \{0\} = \sg(BA)\cup \{0\} .\ll{sgABsgBA} \ee This is
because for $z\neq 0$ the invertibility of $AB-z$ implies the
invertibility of $BA-z$. Namely, one computes that
$(BA-z)\inv=B(AB-z)\inv A-z\inv\I$.  Applying this with $A\raw C$ and
$B\raw C^*$ we see that the assumption $\sg(C^*C)\subset\R^-$ implies
$\sg(CC^*)\subset\R^-$, hence $\sg(-CC^*)\subset\R^+$. By
(\ref{bloed}), (\ref{poscone3}), and \ref{coco}.2 we see that
$C^*C\geq 0$, i.e., $\sg(C^*C)\subset\R^+$, so that the assumption
$-C^*C\in\A^+$ now yields $\sg(C^*C)=0$.  Hence $C=0$ by
\ref{coco}.3.\enp

The last claim before the lemma therefore implies $BA_-=0$. As
$(A_-)^3= -(BA_-)^* BA_-=0$ we see that $(A_-)^3=0$, and finally
$A_-=0$ by the continuous functional calculus with $f(t)=t^{1/3}$.
Hence $B^*B=A_+$, which lies in $\A^+$.\enp

An important consequence of (\ref{poscone2}) is the fact that
inequalities of the type $A_1\leq A_2$ for $A_1,A_2\in\Ar$ are stable
under conjugation by arbitrary elements $B\in\A$, so that $A_1\leq
A_2$ implies $B^*A_1B\leq B^*A_2B$. This is because $A_1\leq A_2$ is
the same as $A_2-A_1\geq 0$; by (\ref{poscone2}) there is an
$A_3\in\A$ such that $A_2-A_1=A_3^*A_3$. But clearly $(A_3B)^*A_3B\geq
0$, and this is nothing but $B^*AB\leq B^*A_2B$. For example, replace
$A$ in (\ref{orderbound}) by $A^*A$, and use (\ref{cstax2}), yielding
$A^*A\leq \n A\n^2\I$. Applying the above principle gives \be
B^*A^*AB\leq\,\n A\n^2 B^*B \ll{BstABA} \ee for all $A,B\in\A$.
\su{Ideals in \ca s} An ideal $\GI$ in a \ca\ $\A$ is defined by
\ref{defideal}. As we have seen, a proper ideal cannot contain $\I$;
in order to prove properties of ideals we need a suitable replacement
of a unit.
\begin{Definition}\ll{defaprun}
An {\bf approximate unit} in a non-unital \ca\ $\A$ is a family
$\{\I_{\lm}\}_{\lm\in\Lm}$, where $\Lm$ is some directed set (i.e., a
set with a partial order and a sense in which $\lm\raw\infty$), with
the following properties:
\begin{enumerate}
\item
\be \I_{\lm}^*=\I_{\lm} \ll{ilsa} \ee and $\sg(\I_{\lm})\subset
[0,1]$, so that \be \n \I_{\lm}\n\,\leq 1; \ll{estau} \ee
\item
\be \lim_{\lm\raw\infty}\n \I_{\lm}A-A\n=\lim_{\lm\raw\infty}\n
A\I_{\lm}-A\n=0 \ll{cpai} \ee for all $A\in\A$.
\end{enumerate}
\end{Definition}

For example, the \ca\ $C_0(\R)$ has no unit (the unit would be
$1_{\R}$, which does not vanish at infinity because it is constant),
but an approximate unit may be constructed as follows: take $\Lm=\N$,
and take $\I_n$ to be a continuous function which is 1 on $[-n,n]$ and
vanishes for $|x|> n+1$.  One checks the axioms, and notes that one
certainly does not have $\I_n\raw 1_{\R}$ in the sup-norm.
\begin{Proposition}\ll{exau}
Every non-unital \ca\ $\A$ has an approximate unit. When $\A$ is
separable (in containing a countable dense subset) then $\Lm$ may be
taken to be countable. \end{Proposition}

One takes $\Lm$ to be the set of all finite subsets of $\A$, partially
ordered by inclusion.  Hence $\lm\in\Lm$ is of the form
$\lm=\{A_1,\ldots,A_n\}$, from which we build the element
$B_{\lm}:=\sum_i A_i^* A_i$. Clearly $B_{\lm}$ is self-adjoint, and
according to \ref{pl1} and \ref{coco}.2 one has $\sg(B)\subset\R^+$,
so that $n\inv\I+B_{\lm}$ is invertible in $\AI$. Hence we may form
\be \I_{\lm}:=B_{\lm}(n\inv\I+B_{\lm})\inv.  \ee Since $B_{\lm}$ is
self-adjoint and $B_{\lm}$ commutes with functions of itself (such as
$(n\inv\I+B_{\lm})\inv$), one has $\I_{\lm}^*=\I_{\lm}$. Although
$(n\inv\I+B_{\lm})\inv$ is computed in $\AI$, so that it is of the
form $C+\mu\I$ for some $C\in\A$ and $\mu\in\C$, one has
$I_{\lm}=B_{\lm}C+\mu B_{\lm}$, which lies in $\A$.  Using the
continuous functional calculus on $B$, with $f(t)=t/(n+t)$, one sees
from (\ref{spmath}) and the positivity of $B_{\lm}$ that
$\sg(\I_{\lm})\subset [0,1]$.

Putting $C_i:=\I_{\lm}A_i-A_i$, a simple computation shows that \be
\sum_i C_iC_i^*=n^{-2}B_{\lm}(n\inv\I+B_{\lm})^{-2}. \ll{CiCi} \ee We
now apply (\ref{normfA}) with $A\raw B_{\lm}$ and $f(t)=n^{-2}t(n\inv
+t)^{-2}$.  Since $f\geq 0$ and $f$ assumes its maximum at $t=1/n$,
one has $\sup_{t\in\R^+} |f(t)|= 1/4n$. As $\sg(B)\subset\R^+$, it
follows that $\n f\n_{\infty}\leq 1/4n$, hence $\n
n^{-2}B_{\lm}(n\inv\I+B_{\lm})^{-2}\n \,\leq 1/4n$ by (\ref{normfA}),
so that $\n \sum_i C_iC_i^*\n\,\leq 1/4n$ by (\ref{CiCi}). Lemma
\ref{aul} then shows that $\n C_iC_i^*\n\,\leq 1/4n$ for each
$i=1,\ldots, n$.  Since any $A\in\A$ sits in some directed subset of
$\Lm$ with $n\raw\infty$, it follows from (\ref{cstax2}) that
$$
\lim_{\lm\raw\infty}\n \I_{\lm}A-A\n^2=\lim_{\lm\raw\infty}\n(
\I_{\lm}A-A)^* \I_{\lm}A-A\n =\lim_{\lm\raw\infty}\n C^*_iC_i\n=0.
$$
 The other equality in (\ref{cpai}) follows analogously.
 
Finally, when $\A$ is separable one may draw all $A_i$ occurring as
elements of $\lm\in\Lm$ from a countable dense subset, so that $\Lm$
is countable.\enp

The main properties of ideals in \ca s are as follows.
\begin{Theorem}\ll{ideals}
Let $\GI$ be an ideal in a \ca\ $\A$.
\begin{enumerate}
\item
If $A\in\GI$ then $A^*\in\GI$; in other words, every ideal in a \ca\
is self-adjoint.
\item
The quotient $\A/\GI$ is a \ca\ in the norm (\ref{normq}), the
multiplication (\ref{mid}), and the involution \be
\ta(A)^*:=\ta(A^*). \ll{starid} \ee
\end{enumerate}\end{Theorem}

Note that (\ref{starid}) is well defined because of \ref{ideals}.1.

Put $\GI^*:=\{A^*|\,A\in\GI\}$. Note that $J\in\GI$ implies
$J^*J\in\GI\cap\GI^*$: it lies in $\GI$ because $\GI$ is an ideal,
hence a left-ideal, and it lies in $\GI^*$ because $\GI^*$ is an
ideal, hence a right-ideal.  Since $\GI$ is an ideal, $\GI\cap\GI^*$
is a $C^*$-subalgebra of $\A$. Hence by \ref{exau} it has an
approximate unit $\{\I_{\lm}\}$.  Take $J\in\GI$.  Using
(\ref{cstax2}) and (\ref{ilsa}), we estimate
$$
 \n J^*-J^*\I_{\lm}\n^2 = \n(J- \I_{\lm}J)(J^*-J^*\I_{\lm})\n=\n
(JJ^*-JJ^*\I_{\lm})- \I_{\lm}(JJ^*-JJ^*\I_{\lm})\n$$
$$
\leq \n(JJ^*-JJ^*\I_{\lm})\n+\n\I_{\lm}(JJ^*-JJ^*\I_{\lm})\n\,\leq\,
\n(J^*J-J^*J\I_{\lm})\n+\n\I_{\lm}\n\:\n (JJ^*-JJ^*\I_{\lm})\n.
$$
As we have seen, $J^*J\in\GI\cap\GI^*$, so that, also using
 (\ref{estau}), both terms vanish for $\lm\raw\infty$. Hence
 $\lim_{\lm\raw\infty} \n J^*-J^*\I_{\lm}\n=0$. But $\I_{\lm}$ lies in
 $\GI\cap\GI^*$, so certainly $\I_{\lm}\in\GI$, and since $\GI$ is an
 ideal it must be that $J^*\I_{\lm}\in\GI$ for all $\lm$. Hence $J^*$
 is a norm-limit of elements in $\GI$; since $\GI$ is closed, it
 follows that $J^*\in\GI$. This proves \ref{ideals}.1.

In view of \ref{idBA}, all we need to prove to establish
\ref{ideals}.2 is the property (\ref{cstax2}). This uses
\begin{Lemma}\ll{normqcs}
Let $\{\I_{\lm}\}$ be an approximate unit in $\GI$, and let
$A\in\A$. Then \be \n\ta(A)\n=\lim_{\lm\raw\infty} \n
A-A\I_{\lm}\n. \ll{ntaAnlim} \ee \end{Lemma}

It is obvious from (\ref{normq}) that \be \n
A-A\I_{\lm}\n\,\geq\,\n\ta(A)\n. \ll{AminAI} \ee To derive the
opposite inequality, add a unit $\I$ to $\A$ if necessary, pick any
$J\in\GI$, and write
$$
\n A-A\I_{\lm}\n=\n (A+J)(\I-\I_{\lm})+J(\I_{\lm}-\I)\n\,\leq\,\n
A+J\n\:\n \I-\I_{\lm}\n+ \n J\I_{\lm}-J\n.$$ Note that \be \n
\I-\I_{\lm}\n\,\leq 1 \ll{nIlnlI} \ee by \ref{defaprun}.1 and the
proof of \ref{coco}.  The second term on the right-hand side goes to
zero for $\lm\raw\infty$, since $J\in\GI$.  Hence \be
\lim_{\lm\raw\infty} \n A-A\I_{\lm}\n\,\leq\,\n A+J\n. \ll{nAJn} \ee
For each $\ep>0$ we can choose $J\in\GI$ so that (\ref{idine})
holds. For this specific $J$ we combine (\ref{AminAI}), (\ref{nAJn}),
and (\ref{idine}) to find
$$
\lim_{\lm\raw\infty} \n A-A\I_{\lm}\n-\ep \,\leq\,\n
\ta(A)\n\,\leq\,\n A-A\I_{\lm}\n.
$$
Letting $\ep\raw 0$ proves (\ref{ntaAnlim}).\enp

We now prove (\ref{cstax2}) in $\A/\GI$.  Successively using
(\ref{ntaAnlim}), (\ref{cstax2}) in $\AI$, (\ref{nIlnlI}),
(\ref{ntaAnlim}), (\ref{mid}), and (\ref{starid}), we find
$$
\n\ta(A)\n^2= \lim_{\lm\raw\infty} \n
A-A\I_{\lm}\n^2=\lim_{\lm\raw\infty} \n
(A-A\I_{\lm})^*(A-A\I_{\lm})\n$$ $$ =\lim_{\lm\raw\infty}\n
(\I-\I_{\lm})A^*A (\I-\I_{\lm})\n\,\leq\, \lim_{\lm\raw\infty}
\n\I-\I_{\lm}\n\: \n A^*A (\I-\I_{\lm})\n\,\leq\,
\lim_{\lm\raw\infty}\n A^*A (\I-\I_{\lm})\n $$ $$ =
\n\ta(A^*A)\n=\n\ta(A)\ta(A^*)\n=\n\ta(A)\ta(A)^*\n. $$ Lemma
\ref{ineqcs} then implies (\ref{cstax2}).\enp

This seemingly technical result is very important.
\begin{Corollary}\ll{idmor}
The kernel of a morphism between two \ca s is an ideal. Conversely,
every ideal in a \ca\ is the kernel of some morphism. Hence every
morphism has norm 1. \end{Corollary}

The first claim is almost trivial, since $\phv(A)=0$ implies
$\phv(AB)=\phv(BA)=0$ for all $B$ by (\ref{phvmult}). Also, since
$\phv$ is continuous (see \ref{cor3}) its kernel is closed.

The converse follows from Theorem \ref{ideals}, since $\GI$ is the
kernel of the canonical projection $\ta:\A\raw\A/\GI$, where $\A/\GI$
is a \ca, and $\ta$ is a morphism by (\ref{mid}), and (\ref{starid}).

The final claim follows from the preceding one, since $\n\ta\n=1$.\enp

For the next consequence of \ref{ideals} we need a
\begin{Lemma}\ll{injmor}
An injective morphism between \ca s is isometric. In particular, its
range is closed.
\end{Lemma}

Assume there is an $B\in\A$ for which $\n\phv(B)\n\neq\n B\n$. By
(\ref{cstax2}), (\ref{phvmult}), and (\ref{phvstar}) this implies
$\n\phv(B^*B)\n\neq\n B^*B\n$.  Put $A:=B^*B$, noting that $A^*=A$.
By (\ref{normun}) and (\ref{defrA}) we must have
$\sg(A)\neq\sg(\phv(A))$.  Then (\ref{sgsubset}) implies
$\sg(\phv(A))\subset \sg(A)$.  By Urysohn's lemma there is a nonzero
$f\in C(\sg(A))$ which vanishes on $\sg(\phv(A))$, so that
$f(\phv(A))=0$. By Lemma \ref{phvflem} we have $\phv(f(A))=0$,
contradicting the injectivity of $\phv$.\enp
\begin{Corollary}\ll{idmor2}
The image of a morphism $\phv:\A\raw\B$ between two \ca s is
closed. In particular, $\phv(\A)$ is a $C^*$-subalgebra of
$\B$. \end{Corollary}

Define $\ps:\A/\ker(\phv)\raw\B$ by $\ps([A])=\phv(A)$, where $[A]$ is
the equivalence class in $\A/\ker(\phv)$ of $A\in\A$.  By the theory
of vector spaces, $\ps$ is a vector space isomorphism between
$\A/\ker(\phv)$ and $\phv(\A)$, and $\phv=\ps\circ\ta$. In particular,
$\ps$ is injective.  According to \ref{idmor} and \ref{ideals}.2, the
space $\A/\ker(\phv)$ is a \ca.  Since $\phv$ and $\ta$ are morphisms,
$\ps$ is a \ca\ morphism. Hence $\ps(\A/\ker(\phv))$ has closed range
in $\B$ by \ref{injmor}.  But $\ps(\A/\ker(\phv))=\phv(\A)$, so that
$\phv$ has closed range in $\B$.  Since $\phv$ is a morphism, its
image is a \sta\ in $\B$, which by the preceding sentence is closed in
the norm of $\B$. Hence $\phv(\A)$, inheriting all operations in $\B$,
is a \ca.\enp \su{States\ll{states}} The notion of a state in the
sense defined below comes from \qm, but states play a central role in
the general theory of abstract \ca s also.
\begin{Definition}\ll{defstate}
A {\bf state} on a unital \ca\ $\A$ is a linear map $\om:\A\raw\C$
which is {\bf positive} in that $\om(A)\geq 0$ for all $A\in\A^+$, and
{\bf normalized} in that \be \om(\I)=1. \ll{norst} \ee The {\bf state
space} $\SA$ of $\A$ consists of all states on $\A$. \end{Definition}

For example, when $\A\subseteq\BH$ then every $\Ps\in\H$ with norm 1
defines a state $\ps$ by \be \ps(A):=(\Ps,A\Ps). \ll{vectorstate} \ee
Positivity follows from Theorem \ref{pl1}, since $\ps(B^*B)=\n
B\Ps\n^2\geq 0$, and normalization is obvious from
$\ps(\I)=(\Ps,\Ps)=1$.
\begin{Theorem}\ll{riesz}
The state space of $\A=C(X)$ consists of all probability measures on
$X$. \end{Theorem}

By the Riesz theorem of measure theory, each positive linear map
$\om:C(X)\raw\C$ is given by a regular positive measure $\mu_{\om}$ on
$X$. The normalization of $\om$ implies that
$\om(1_X)=\mu_{\om}(X)=1$, so that $\mu_{\om}$ is a probability
measure. \enp

The positivity of $\om$ with \ref{pl1} implies that
$(A,B)_{\om}:=\om(A^*B)$ defines a pre-inner product on $\A$. Hence
from (\ref{csb}) we obtain \be |\om(A^*B)|^2\leq\om(A^*A)\om(B^*B),
\ll{css} \ee which will often be used. Moreover, for all $A\in\A$ one
has \be \om(A^*)=\ovl{\om(A)}, \ll{ovlom} \ee as
$\om(A^*)=\om(A^*\I)=(A,\I)_{\om}=\ovl{(\I,A)_{\om}}=\om(A)$.

Partly in order to extend the definition of a state to non-unital \ca
s, we have
\begin{Proposition}\ll{stext}
A linear map $\om:\A\raw\C$ on a unital \ca\ is positive iff $\om$ is
bounded and \be \n\om\n=\om(\I).\ll{nomom1} \ee In particular:
\begin{enumerate}
\item
A state on a unital \ca\ is bounded, with norm 1.
\item
An element $\om\in\A^*$ for which $\n\om\n=\om(\I)=1$ is a state on
$\A$.
\end{enumerate}
\end{Proposition}

When $\om$ is positive and $A=A^*$ we have, using (\ref{orderbound}),
the bound $|\om(A)|\,\leq \om(\I) \n A\n$.  For general $A$ we use
(\ref{css}) with $A=\I$, (\ref{cstax2}), and the bound just derived to
find
$$
|\om(B)|^2\leq\om(B^*B)\om(\I)\leq \om(\I)^2 \n B^*B\n= \om(\I)^2 \n
B\n^2.
$$
Hence $\n\om\n\,\leq \om(\I)$. Since the upper bound is reached by
$B=\I$, we have (\ref{nomom1}).

To prove the converse claim, we first note that the argument around
(\ref{btb}) may be copied, showing that $\om$ is real on $\Ar$. Next,
we show that $A\geq 0$ implies $\om(A)\geq 0$. Choose $s>0$ small
enough, so that $\n \I-sA\n\,\leq 1$. Then (assuming $\om\neq 0$)
$$
1\geq\,\n \I-sA\n=\frac{\n\om\n}{\om(\I)}\n \I-sA\n
\,\geq\,\frac{|\om(\I-sA)|}{\om(\I)}.$$

Hence $|\om(\I)-s\om(A)|\,\leq \om(\I)$, which is only possible when
$\om(A)\geq 0$.\enp

We now pass to states on \ca s without unit. Firstly, we look at a
state in a more general context.
\begin{Definition}\ll{defposmap}
A {\bf positive map} $\CQ:\A\raw\B$ between two \ca s is a linear map
with the property that $A\geq 0$ implies $\CQ(A)\geq 0$ in
$\B$. \end{Definition}
\begin{Proposition}\ll{poscon}
A positive map between two \ca s is bounded (continuous).
 \end{Proposition}

Let us first show that boundedness on $\A^+$ implies boundedness on
$\A$.  Using (\ref{aaa}) and \ref{posdec}, we can write \be
A=A'_+-A'_- +iA''_+-iA''_-, \ll{posdec2} \ee where $A'_+$ etc.\ are
positive.  Since $\n A'\n\,\leq\,\n A\n$ and $\n A''\n\,\leq\,\n A\n$
by (\ref{aaa}), we have $\n B\n\,\leq\,\n A\n$ for
$B=A'_+,A'_-,A''_+$, or $A''_-$ by \ref{posdec}.  Hence if
$\n\CQ(B)\n\,\leq\,C \n B\n$ for all $B\in\A^+$ and some $C>0$, then
$$
\n\CQ(A)\n \,\leq\,\n\CQ(A'_+)\n+
\n\CQ(A'_-)\n+\n\CQ(A''_+)\n+\n\CQ(A''_-)\n\,\leq 4C\n A\n.
$$

Now assume that $\CQ$ is not bounded; by the previous argument it is
not bounded on $\A^+$, so that for each $n\in\N$ there is an
$A_n\in\A_1^+$ so that $\n\CQ(A_n)\n\,\geq n^3$ (here $\A^+_1$
consists of all $A\in\A^+$ with $\n A\n\,\leq 1$). The series
$\sum_{n=0}^{\infty} n^{-2}A_n$ obviously converges to some
$A\in\A^+$. Since $\CQ$ is positive, we have $\CQ(A)\geq n^{-2}
\CQ(A_n)\geq 0$ for each $n$. Hence by (\ref{AleqB})
$$
\n\CQ(A)\n\,\geq\, n^{-2} \n\CQ(A_n)\n\,\geq n
$$
for all $n\in\N$, which is impossible since $\n\CQ(A)\n$ is some
finite number. Thus $\CQ$ is bounded on $\A^+$, and therefore on $\A$
by the previous paragraph.\enp

Choosing $\B=\C$, we see that a state on a unital \ca\ is a special
case of a positive map between \ca s; Proposition \ref{poscon} then
provides an alternative proof of \ref{stext}.1.  Hence in the
non-unital case we may replace the normalization condition in
\ref{defstate} as follows.
\begin{Definition}\ll{defstate2}
A {\bf state} on a \ca\ $\A$ is a linear map $\om:\A\raw\C$ which is
positive and has norm 1. \end{Definition}

This definition is possible by \ref{poscon}, and is consistent with
\ref{defstate} because of (\ref{nomom1}). The following result is very
useful; cf.\ \ref{caun}.
\begin{Proposition}\ll{extstate}
A state $\om$ on a \ca\ without unit has a unique extension to a state
$\om_{\I}$ on the unitization $\AI$. \end{Proposition}

The extension in question is defined by \be
\om_{\I}(A+\lm\I):=\om(A)+\lm. \ll{statex} \ee This obviously
satisfies (\ref{norst}); it remains to prove positivity.

Since a state $\om$ on $\A$ is bounded by \ref{poscon}, we have
$|\om(A-A\I_{\lm})|\raw 0$ for any approximate unit in $\A$.  The
derivation of (\ref{css}) and (\ref{ovlom}) may then be copied from
the unital case; in particular, one still has
$|\om(A)|^2\leq\om(A^*A)$.  Combining this with (\ref{ovlom}), we
obtain from (\ref{statex}) that
$$
\om_{\I}((A+\lm\I)^*(A+\lm\I))\geq\,|\om(A)+\ovl{\lm}|^2\,\geq 0.
$$
Hence $\om$ is positive by (\ref{poscone2}).\enp

There are lots of states:
\begin{Lemma}\ll{lots}
For every $A\in\A$ and $a\in\sg(A)$ there is a state $\om_a$ on $\A$
for which $\om(A)=a$. When $A=A^*$ there exists a state $\om$ such
that $|\om(A)|=\n A\n$.
\end{Lemma}

If necessary we add a unit to $\A$ (this is justified by
\ref{extstate}).  Define a linear map $\til{\om}_a:\C A+\C\I\raw\C$ by
$\til{\om}_a(\lm A+\mu\I):=\lm a+\mu$. Since $a\in\sg(A)$ one has $\lm
a+\mu\in\sg(\lm A+\mu\I)$; this easily follows from the definition of
$\sg$. Hence (\ref{nq2}) with $A\raw \lm A+\mu\I$ implies
$|\til{\om}_a(\lm A+\mu\I)|\,\leq \n (\lm A+\mu\I)\n$. Since
$\til{\om}_a(\I)=1$, it follows that $\n\til{\om}\n=1$. By the
Hahn-Banach Theorem \ref{HB}, there exists an extension $\om_a$ of
$\til{\om}$ to $\A$ of norm 1.  By \ref{stext}.2 $\om_a$ is a state,
which clearly satisfies $\om_a(A)= \til{\om}_a(A)=a$.

Since $\sg(A)$ is closed by \ref{thmsp}.2, there is an $a\in\sg(A)$
for which $r(A)=|a|$. For this $a$ one has $|\om(A)|=|a|=r(A)=\n A\n$
by (\ref{nAnrA}).  \enp

An important feature of a state space $\SA$ is that it is a {\bf
convex set}.  A convex set $C$ in a vector space ${\cal C}$ is a
subset of $\CV$ such that the convex sum $\lm v +(1-\lm)w$ belongs to
$\cal C$ whenever $v,w\in {\cal C}$ and $\lm\in [0,1]$. Repeating this
process, it follows that $\sum_i p_i v_i$ belongs to $\cal C$ when all
$p_i\geq 0$ and $\sum_i p_i=1$, and all $v_i\in{\cal C}$.  In the
unital case it is clear that $\SA$ is convex, since both positivity
and normalization are clearly preserved under convex sums. In the
non-unital case one arrives at this conclusion most simply via
\ref{extstate}.

We return to the unital case.  Let $\SA$ be the state space of a
unital \ca\ $\A$. We saw in \ref{stext} that each element $\om$ of
$\SA$ is continuous, so that $\SA\subset\A^*$.  Since $w^*$-limits
obviously preserve positivity and normalization, we see that $\SA$ is
closed in $\A^*$ if the latter is equipped with the
$w^*$-topology. Moreover, $\SA$ is a closed subset of the unit ball of
$\A^*$ by \ref{stext}.1, so that $\SA$ is compact in the (relative)
$w^*$-topology by the Banach-Alaoglu theorem.

It follows that the state space of a unital \ca\ is a {\bf compact
convex set}.  The very simplest example is $\A=\C$, in which case
$\SA$ is a point.

The next case is $\A=\C\oplus\C=\C^2$. The dual is $\C^2$ as well, so
that each element of $(\C^2)^*$ is of the form $\om(\lm\dot{+}\mu)=
c_1\lm_1+c_2\lm_2$. Positive elements of $\C\oplus\C$ are fo the form
$\lm\dot{+}\mu$ with $\lm\geq 0$ and $\mu\geq 0$, so that a positive
functional must have $c_1\geq 0$ and $c_2\geq 0$. Finally, since
$\I=1\dot{+}1$, normalization yields $c_1+c_2=1$. We conclude that
$\CS(\C\oplus\C)$ may be identified with the interval $[0,1]$.

Now consider $\A=\M^2(\C)$. We identify $\M^2(\C)$ with its dual
through the pairing $\om(A)=\Tr \om A$. It follows that $\SA$ consists
of all positive $2\x 2$ matrices $\rh$ with $\Tr\rh=1$; these are the
density matrices of \qm. To identify $\SA$ with a familiar compact
convex set, we parametrize \be \rh=\half \left(
\begin{array}{cc}
1 + x & y+iz \\ y-iz & 1 -x
\end{array}
\right) , \ll{parrho} \ee where $x,y,z\in\R$. The positivity of this
matrix then corresponds to the constraint $x^2 + y^2 + z^2\leq
1$. Hence $\CS(\M^2(\C))$ is the unit ball in $\R^3$.
\su{Representations and the GNS-construction\ll{repgns}} The material
of this section explains how the usual \Hs\ framework of \qm\ emerges
from the $C^*$-algebraic setting.  \begin{Definition}\ll{repofca} A
{\bf representation}\index{representation!of a $C^*$-algebra} of $\A$
on a Hilbert space $\H$ is a (complex) linear map $\pi:\A\rightarrow
\B(\H)$ satisfying \bea & & \pi(A\cdot B)=\pi(A) \pi(B); \nn \\ & &
\pi(A^*)=\pi(A)^* \ll{defrep} \eea for all $A,B\in\A$.
\end{Definition}
A \rep\ $\pi$ is automatically continuous, satisfying the bound \be
\n\pi(A)\n\,\leq\,\n A\n. \ll{repbounded} \ee This is because $\pi$ is
a morphism; cf.\ (\ref{morbound}).  In particular, $\n\pi(A)\n =\n
A\n$ when $\pi$ is faithful by Lemma \ref{injmor}.

There is a natural equivalence relation in the set of all \rep s of
$\A$: two \rep s $\pi_1,\pi_2$ on Hilbert spaces $\H_1,\H_2$,
respectively, are called {\bf equivalent} if there exists a unitary
isomorphism $U:\H_1\rightarrow\H_2$ such that $U\pi_1(A)U^*=\pi_2(A)$
for all $A\in\A$.

The map $\pi(A)=0$ for all $A\in\A$ is a \rep; more generally, such
trivial $\pi$ may occur as a summand. To exclude this possibility, one
says that a \rep\ is {\bf non-degenerate} if 0 is the only vector
annihilated by all representatives of $\A$.

 A \rep\ $\pi$ is called {\bf cyclic} if its carrier space $\H$
 contains a {\bf cyclic vector} $\Om$ for $\pi$; this means that the
 closure of $\pi(\A)\Om$ (which in any case is a closed subspace of
 $\H$) coincides with $\H$.
\begin{Proposition}\ll{nondegcyclic}
Any non-degenerate \rep\ $\pi$ is a direct sum of cyclic \rep s.
\end{Proposition}
The proof uses a lemma which appears in many other proofs as well.
\begin{Lemma}\ll{DCTlemma}
Let $\M$ be a \sta\ in $\BH$, take a nonzero vector $\Ps\in\H$, and
let $p$ be the projection onto the closure of $\M\Ps$. Then $p\in\M'$
(that is, $[p,A]=0$ for all $A\in\M$).
\end{Lemma}
If $A\in\M$ then $Ap\H\subseteq p\H$ by definition of $p$. Hence
$p^{\perp}Ap=0$ with $p^{\perp}=\I-p$; this reads $Ap=pAp$.  When
$A=A^*$ then $$ (Ap)^*=pA=(pAp)^*=pAp=Ap,
$$ so that $[A,p]=0$. By (\ref{aaa}) this is true for all $A\in\M$.
\enp

Apply this lemma with $\M=\pi(\A)$; the assumption of non-degeneracy
guarantees that $p$ is nonzero, and the conclusion implies that $A\raw
p\pi(A)$ defines a sub\rep\ of $\A$ on $p\H$. This sub\rep\ is clearly
cyclic, with cyclic vector $\Ps$.  This process may be repeated on
$p^{\perp}\H$, etc.  \enp

If $\pi$ is a non-degenerate \rep\ of $\A$ on $\H$, then any unit
vector $\Ps\in\H$ defines a state $\ps\in \SA$, referred to as a {\bf
vector state} relative to $\pi$, by means of (\ref{vectorstate}).
Conversely, from any state $\om\in\SA$ one can construct a cyclic
\rep\ $\pi_{\om}$ on a Hilbert space $\H_{\om}$ with cyclic vector
$\Om_{\om}$ in the following way.  We restrict ourselves to the unital
case; the general case follows by adding a unit to $\A$ and using
\ref{extstate}.
\begin{Construction}\ll{GNSconstruction}
\begin{enumerate}
\item
   Given $\om\in\SA$, define the sesquilinear form $(\, ,\,)_0$ on
$\A$ by \be (A,B)_0:=\om(A^* B).\ll{AB0omBA} \ee Since $\om$ is a
state, hence a positive functional, this form is positive
semi-definite (this means that $(A,A)_0\geq 0$ for all $A$). Its null
space \be {\cal N}_{\om}=\{A\in\A\, |\, \om(A^*A)=0\} \ll{maxleftid}
\ee is a closed left-ideal in $\A$.
\item
 The form $(\, ,\,)_0$ projects to an inner product $(\, ,\,)_{\om}$
on the quotient $\A/{\cal N}_{\om}$. If $V:\A\rightarrow \A/{\cal
N}_{\om}$ is the canonical projection, then by definition \be
(VA,VB)_{\om}:=(A,B)_0. \ll{VAVB} \ee The \Hs\ $\H_{\om}$ is the
closure of $\A/{\cal N}_{\om}$ in this inner product.
\item
 The \rep\ $\pi_{\om}(\A)$ is firstly defined on $\A/{\cal
N}_{\om}\subset \H_{\om}$ by \be \pi_{\om}(A) VB:= VA B; \ll{gnsrepeq}
\ee it follows that $\pi_{\om}$ is continuous. Hence $\pi_{\om}(A)$
may be defined on all of $\H_{\om}$ by continuous extension of
(\ref{gnsrepeq}), where it satisfies (\ref{defrep}).
\item
The cyclic vector is defined by $\Om_{\om}=V\I$, so that \be
(\Om_{\om},\pi_{\om}(A)\Om_{\om})=\om(A) \: \: \:\:\forall A\in\A.
\ll{gnscruceq} \ee
\end{enumerate}
\end{Construction}

We now prove the various claims made here.  First note that the null
space ${\cal N}_{\om}$ of $(\, ,\,)_0$ can be defined in two
equivalent ways; \be {\cal N}_{\om}:=\{A\in\A\, |\,
(A,A)_0=0\}=\{A\in\A \, |\, (A,B)_0=0\: \forall \,
B\in\A\}.\ll{twoformsno} \ee The equivalence follows from the
Cauchy-Schwarz inequality (\ref{css}).  The equality
(\ref{twoformsno}) implies that $\CN_{\om}$ is a left-ideal, which is
closed because of the continuity of $\om$. This is important, because
it implies that the map $\rh(A):\A\raw\A$ defined in (\ref{rhABAB})
quotients well to a map from $\A/\CN_{\om}$ to $\A/\CN_{\om}$; the
latter map is $\pi_{\om}$ defined in (\ref{gnsrepeq}). Since $\rh$ is
a morphism, it is easily checked that $\pi_{\om}$ is a morphism as
well, satisfying (\ref{defrep}) on the dense subspace $\A/\CN_{\om}$
of $\H_{\om}$.

To prove that $\pi_{\om}$ is continuous on $\A/\CN_{\om}$, we compute
  $\n\pi_{\om}(A)\Ps\n^2$ for $\Ps=VB$, where $A,B\in\A$.  By
  (\ref{VAVB}) and step 2 above, one has
  $\n\pi_{\om}(A)\Ps\n^2=\om(B^*AA*B)$. By (\ref{BstABA}) and the
  positivity of $\om$ one has $\om(B^*AA*B)\leq \n A\n^2
  \om(B^*B)$. But $\om(B^*B)=\n\Ps\n^2$, so that
  $\n\pi_{\om}(A)\Ps\n\,\leq \n A\n\:\n\Ps\n$, upon which \be \n
  \pi_{\om}(A)\n\,\leq\, \n A\n \ee follows from (\ref{BHnorm}).

For later use we mention that the GNS-construction yields \be
(\pi_{\om}(A)\Om_{\om},\pi_{\om}(B)\Om_{\om})=\om(A^*B). \ll{gnsip}
\ee Putting $B=A$ yields \be \n\pi_{\om}(A)\Om_{\om}\n^2=\om(A^*A),
\ll{gnscor} \ee which may alternatively be derived from
(\ref{gnscruceq}) and the fact that $\pi_{\om}$ is a \rep.
\begin{Proposition}\ll{spc}
  If $(\pi(\A),\H)$ is cyclic then the GNS-represen\-ta\-tion
 $(\pi_{\om}(\A),\H_{\om})$ defined by any vector state $\Om$
 (corresponding to a cyclic unit vector $\Om\in\H$) is unitarily
 equivalent to $(\pi(\A),\H)$.
\end{Proposition}

This is very simple to prove: the operator $U:\H_{\om}\rightarrow \H$
implementing the equivalence is initially defined on the dense
subspace $\pi_{\om}(\A)\Om_{\om}$ by
$U\pi_{\om}(A)\Om_{\om}=\pi(A)\Om$; this operator is well-defined, for
$\pi_{\om}(A)\Om_{\om}=0$ implies $\pi(A)\Om=0$ by the
GNS-construction.  It follows from (\ref{gnscruceq}) that $U$ is
unitary as a map from $\H_{\om}$ to $U\H_{\om}$, but since $\Om$ is
cyclic for $\pi$ the image of $U$ is $\H$. Hence $U$ is unitary.  It
is trivial to verify that $U$ intertwines $\pi_{\om}$ and $\pi$.  \enp
\begin{Corollary}\ll{spc5}
If the \Hs s $\H_1$, $\H_2$ of two cyclic \rep s $\pi_1,\pi_2$ each
contain a cyclic vector $\Om_1\in\H_1$, $\Om_2\in\H_2$, and
$$
\om_1(A):=(\Om_1,\pi_1(A)\Om_1)=(\Om_2,\pi_2(A)\Om_2)=:\om_2(A)
$$ for all $A\in\A$, then $\pi_1(\A)$ and $\pi_2(\A)$ are equivalent.
\end{Corollary}

By \ref{spc} the \rep\ $\pi_1$ is equivalent to the GNS-\rep\
$\pi_{\om_1}$, and $\pi_2$ is equivalent to $\pi_{\om_2}$. On the
other hand, $\pi_{\om_1}$ and $\pi_{\om_2}$ are induced by the same
state, so they must coincide.\enp \su{The Gel'fand-Neumark
theorem\ll{GeNa}} One of the main results in the theory of \ca s is
\begin{Theorem}\ll{GNT}
A \ca\ is isomorphic to a subalgebra of $\B(\H)$, for some Hilbert
space $\H$.
\end{Theorem}

The GNS-construction leads to a simple proof this theorem, which uses
the following notion.
\begin{Definition}\ll{defunivrep} 
The {\bf universal representation} $\pi_{\mbox{\tiny u}}$ of a \ca\
$\A$ is the direct sum of all its GNS-representations $\pi_{\om}$,
$\om\in \SA$; hence it is defined on the \Hs\ $\H_{\mbox{\tiny
u}}=\oplus_{\om\in\SA}\H_{\om}$.
\end{Definition}

Theorem \ref{GNT} then follows by taking $\H=\H_{\mbox{\tiny u}}$; the
desired isomorphism is $\pi_{\mbox{\tiny u}}$.  To prove that
$\pi_{\mbox{\tiny u}}$ is injective, suppose that $\pi_{\mbox{\tiny
u}}(A)=0$ for some $A\in\A$. By definition of a direct sum, this
implies $\pi_{\om}(A)=0$ for all states $\om$.  Hence
$\pi_{\om}(A)\Om_{\om}=0$, hence $\n\pi_{\om}(A)\Om_{\om}\n^2=0$; by
(\ref{gnscor}) this means $\om(A^*A)=0$ for all states $\om$, which
implies $\n A^*A\n=0$ by Lemma \ref{lots}, so that $\n A\n=0$ by
(\ref{cstax2}), and finally $A=0$ by the definition of a norm.

Being injective, the morphism $\pi_{\mbox{\tiny u}}$ is isometric by
Lemma \ref{injmor}.\enp

While the universal \rep\ leads to a nice proof of \ref{GNT}, the \Hs\
 $\H_{\mbox{\tiny u}}$ is absurdly large; in practical examples a
 better way of obtaining a faithful \rep\ always exists. For example,
 the best faithful representation of $\BH$ is simply its defining one.

Another consequence of the GNS-construction, or rather of
 \ref{defunivrep}, is
\begin{Corollary}\ll{posposrep}
An operator $A\in\A$ is positive (that is, $A\in\Ar^+$) iff
$\pi(A)\geq 0$ for all cyclic \rep s $\pi$.
\end{Corollary}
\su{Complete positivity\ll{CPmaps}} We have seen that a positive map
$\CQ$ (cf.\ Definition \ref{defposmap} generalizes the notion of a
state, in that the $\C$ in $\om:\A\raw\C$ is replaced by a general
\ca\ $\B$ in $\CQ:A\raw\B$.  We would like to see if one can
generalize the GNS-construction.  It turns out that for this purpose
one needs to impose a further condition on $\CQ$.

We first introduce the \ca\ $\M^n(\A)$ for a given \ca\ $\A$ and
$n\in\N$.  The elements of $\M^n(\A)$ are $n\x n$ matrices with
entries in $\A$; multiplication is done in the usual way, i.e,
$(MN)_{ij}:=\sum_k M_{ik}N_{kj}$, with the difference that one now
multiplies elements of $\A$ rather than complex numbers. In
particular, the order has to be taken into account.  The involution in
$\M^n(\A)$ is, of course, given by $(M^*)_{ij}=M_{ji}^*$, in which the
involution in $\A$ replaces the usual complex conjugation in $\C$.
One may identify $\M^n(\A)$ with $\A\ot\M^n(\C)$ in the obvious way.

When $\pi$ is a faithful \rep\ of $\A$ (which exists by Theorem
\ref{GNT}), one obtains a faithful realization $\pi_n$ of $\M^n(\A)$
on $\H\ot\C^n$, defined by linear extension of
$\pi_n(M)v_i:=\pi(M_{ij})v_j$; we here look at elements of $\H\ot\C^n$
as $n$-tuples $(v_1,\ldots,v_n)$, where each $v_i\in\H$.  The norm $\n
M\n$ of $M\in\M^n(\A)$ is then simply defined to be the norm of
$\pi_n(M)$.  Since $\pi_n(\M^n(\A))$ is a closed \sta\ in
$\B(\H\ot\C^n)$ (because $n<\infty$), it is obvious that $\M^n(\A)$ is
a \ca\ in this norm. The norm is unique by Corollary \ref{normunique},
so that this procedure does not depend on the choice of $\pi$.
\begin{Definition}\ll{defCP}
 A linear map $\CQ:\A\raw \B$ between \ca s is called {\bf completely
positive} if for all $n\in \Bbb N$ the map $\CQ_n: \M^n(\A)\raw
\M^n(\B)$, defined by $(\CQ_n(M))_{ij}:= \CQ(M_{ij})$, is positive.
\end{Definition}

For example, a morphism $\phv$ is a completely positive map, since
when $\Bbb A=\Bbb B^*\Bbb B$ in $\M^n(\A)$, then $\phv(\Bbb
A)=\phv(\Bbb B)^*\phv(\Bbb B)$, which is positive in $\M^n(\B)$.  In
particular, any \rep\ of $\A$ on $\H$ is a completely positive map
from $\A$ to $\BH$.

 If we also assume that $\A$ and $\B$ are unital, and that $\CQ$ is
normalized, we get an interesting generalization of the
GNS-construction, which is of central importance for quantization
theory. This generalization will appear as the proof of the following
{\bf Stinespring theorem}.
\begin{Theorem}\ll{Stinespring}
Let $\CQ:\A\raw \B$ be a completely positive map between \ca s with
unit, such that $\CQ(\I)=\I$.  By Theorem \ref{GNT}, we may assume
that $\B$ is faithfully represented as a subalgebra $\B\simeq
\plg(\B)\subseteq \B(\Hlg)$, for some \Hs\ $\Hlg$.

There exists a \Hs\ $\H^{\chi}$, a \rep\ $\pi^{\chi}$ of $\A$ on
$\Hug$, and a partial isometry $W:\Hlg \raw \Hug$ (with $W^*W=\I$),
such that \be \plg(\CQ(A))=W^* \pug(A)W \:\:\: \forall
A\in\A. \ll{stineeq} \ee Equivalently, with $p:=WW^*$ (the target
projection of $W$ on $\Hug$), $\til{\H}_{\ch}:=p\Hug\subset \Hug$, and
$U:\Hlg\raw \til{\H}_{\ch}$ defined as $W$, seen as map not from
$\Hlg$ to $\Hug$ but as a map from $\Hlg$ to $\til{\H}_{\ch}$, so that
$U$ is unitary, one has \be U\plg(\CQ(A))U\inv=p\pug(A)p.\ll{fundefbt}
\ee
\end{Theorem}

 The proof consists of a modification of the GNS-construction.  It
uses the notion of a {\bf partial isometry}. This is a linear map
$W:\H_1\raw\H_2$ between two \Hs s, with the property that $\H_1$
contains a closed subspace $\CK_1$ such that $(W\Ps,W\Ph)_2=
(\Ps,\Ph)_1$ for all $\Ps,\Ph\in\CK_1$, and $W=0$ on $\CK_1^{\perp}$.
Hence $W$ is unitary from $\CK_1$ to $W\CK_1$. It follows that
$WW^*=[\CK_2]$ and $W^*W=[\CK_1]$ are projections onto the image and
the kernel of $W$, respectively.

 We denote elements of $\Hlg$ by $v,w$, with inner product
$(v,w)_{\ch}$.
\begin{Construction}\ll{stinespringcon}
\begin{enumerate}
\item
 Define the sesquilinear form $(\, ,\,)_0^{\ch}$ on $\A\ot\Hlg$
(algebraic tensor product) by (sesqui-)linear extension of \be (A\ot
v,B\ot w)_0^{\ch}:= (v,\plg(\CQ(A^*B))w)_{\ch}. \ll{woody1} \ee Since
$\CQ$ is completely positive, this form is positive semi-definite;
denote its null space by $\CN_{\ch}$.
\item
 The form $(\, ,\,)_0^{\ch}$ projects to an inner product $(\,
,\,)^{\ch}$ on $\A\ot \Hlg/\CN_{\ch}$. If $V_{\ch}:\A\ot\Hlg\raw
\A\ot\Hlg /\CN_{\chi}$ is the canonical projection, then by definition
\be (V_{\ch}(A\ot v),V_{\ch}(B\ot w))^{\ch}:=(A\ot v,B\ot w)_0^{\ch}.
\ee The \Hs\ $\Hug$ is the closure of $\A\ot \Hlg/\CN_{\ch}$ in this
inner product.
\item
 The \rep\ $\pug(\A)$ is initially defined on $\A\ot \Hlg/\CN_{\ch}$
by linear extension of \be \pi^{\ch}(A) V_{\ch}(B\ot w):= V_{\ch}(A
B\ot w); \ll{gnsrepeq2} \ee this is well-defined, because
$\pi^{\ch}(A)\CN_{\ch}\subseteq \CN_{\ch}$.  One has the bound \be \n
\pi^{\ch}(A)\n\,\leq\, \n A\n, \ll{stinebound} \ee so that
$\pi^{\ch}(A)$ may be defined on all of $\H^{\ch}$ by continuous
extension of (\ref{gnsrepeq2}).  This extension satisfies
$\pi^{\ch}(A^*)=\pi^{\ch}(A)^*$.  \item The map $W:\Hlg \raw \Hug$,
defined by \be Wv:=V_{\ch} \I\ot v \ll{WPsVI} \ee is a partial
isometry. Its adjoint $W^*:\Hug\raw\Hlg$ is given by (continuous
extension of) \be W^*V_{\ch}A\ot v=\plg(\CQ(A))v, \ll{wocomp} \ee from
which the properties $W^*W=\I$ and (\ref{stineeq}) follow.
\end{enumerate}
\end{Construction}

To show that the form defined by (\ref{woody1}) is positive, we write
\be \sum_{i,j} (A_i\ot v_i,A_j\ot v_j)_0^{\ch}=\sum_{i,j}
(v_i,\plg(\CQ(A_i^*A_j))v_j)_{\ch}. \ll{rhsof} \ee Now consider the
element $\Bbb A$ of $\M^n(\A)$ with matrix elements ${\Bbb
A}_{ij}=A^*_iA_j$. Looking in a faithful \rep\ $\pi_n$ as explained
above, one sees that $$ (z,{\Bbb
A}z)=\sum_{i,j}(z_i,\pi(A^*_iA_j)z_j)=\sum_{i,j}(\pi(A_i)z_i,\pi(A_j)z_j)=
\n Az\n^2\,\geq 0 $$ where $Az=\sum_i A_i z_i$. Hence ${\Bbb A}\geq
0$.  Since $\CQ$ is completely positive, it must be that $\Bbb B$,
defined by its matrix elements ${\Bbb B}_{ij}:=\CQ(A_i^*A_j)$, is
positive in $\M^n(\B)$. Repeating the above argument with $\Bbb A$ and
$\pi$ replaced by $\Bbb B$ and $\plg$, respectively, one concludes
that the right-hand side of (\ref{rhsof}) is positive.
 
To prove (\ref{stinebound}) one uses (\ref{BstABA}) in
$\M_n(\A)$. Namely, for arbitrary $A,B_1,\ldots,B_n\in\A$ we conjugate
the inequality $0\leq A^*A\I_n\leq\,\n A\n^2\I_n$ with the matrix
$\Bbb B$, whose first row is $(B_1,\ldots,B_n)$, and which has zeros
everywhere else; the adjoint ${\Bbb B}^*$ is then the matrix whose
first column is $(B^*_1,\ldots,B^*_n)^T$, and all other entries
zero. This leads to $0\leq {\Bbb B}^*A^*A\Bbb B\leq\,\n A\n^2{\Bbb
B}^*\Bbb B$.  Since $\CQ$ is completely positive, one has $\CQ_n({\Bbb
B}^*A^*A\Bbb B)\leq\,\n A\n^2\CQ_n({\Bbb B}^*\Bbb B)$.  Hence in any
\rep\ $\plg(\B)$ and any vector $(v_1,\ldots v_n)\in\Hlg\ot\C^n$ one
has \be \sum_{i,j}(v_i,\plg(\CQ( B^*_i A^*AB_j))v_j)\leq \n A\n^2
\sum_{i,j}(v_i,\plg(\CQ(B^*_iB_j))v_j) .  \ll{Cpineq} \ee With
$\Ps=\sum_i V_{\ch}B_i\ot v_i$, from (\ref{woody1}),
(\ref{gnsrepeq2}), and (\ref{Cpineq}) one then has
$$
\n\pug(A)\Ps\n^2=\sum_{i,j}(AB_i\ot v_i,AB_j\ot v_j)_0^{\ch}=
\sum_{i,j}(v_i,\plg(\CQ(B^*_iA^*AB_j))v_j)_{\ch}$$
$$
\leq \,\n A\n^2\sum_{i,j}(v_i,\plg(\CQ(B^*_iB_j))v_j)_{\ch}= \n
 A\n^2\sum_{i,j}(B_i\ot v_i,B_j\ot v_j)^{\ch}_0 $$ $$ = \n
 A\n^2(V_{\ch}\sum_{i}B_i\ot v_i,V_{\ch}\sum_j B_j\ot v_j)^{\ch} =\n
 A\n^2\;\n\Ps\n^2.
$$

To show that $W$ is a partial isometry, use the definition to compute
$$
(Wv,Ww)^{\ch}=(V_{\ch}\I\ot v,V_{\ch}\I\ot w)^{\ch}=(\I\ot v,\I\ot
w)^{\ch}_0 =(v,w)_{\ch},$$ where we used (\ref{woody1}) and
$\CQ(\I)=\I$.

To check (\ref{wocomp}), one merely uses the definition of the
adjoint, viz.\ $(w, W^*\Ps)_{\ch}=(Ww,\Ps)^{\ch}$ for all $w\in\Hlg$
and $\Ps\in\Hug$. This trivially verified.

To verify (\ref{stineeq}), we use (\ref{WPsVI}) and (\ref{wocomp}) to
compute
$$
W^*\pug(A)Wv=W^*\pug(A)V_{\ch}(\I\ot v)=W^*V_{\ch}(A\ot
v)=\plg(\CQ(A))v.
$$

Being a partial isometry, one has $p=WW^*$ for the projection $p$ onto
the image of $W$, and, in this case, $W^*W=\I$ for the projection onto
the subspace of $\Hlg$ on which $W$ is isometric; this subspace is
$\Hlg$ itself Hence (\ref{fundefbt}) follows from (\ref{stineeq}),
since
$$
U\plg(\CQ(A))U\inv=W\plg(\CQ(A))W^*=WW^*\pug(A)WW^*=p\pug(A)p.
$$\enp

When $\CQ$ fails to preserve the unit, the above construction still
applies, but $W$ is no longer a partial isometry; one rather has $\n
W\n^2=\n \CQ(\I)\n$. Thus it is no longer possible to regard $\Hlg$ as
a subspace of $\Hug$.

If $\A$ and perhaps $\B$ are non-unital the theorem holds if $\CQ$ can
be extended (as a positive map) to the unitization of $\A$, such that
the extension preserves the unit $\I$ (perhaps relative to the
unitization of $\B$).  When the extension exists but does not preserve
the unit, one is in the situation of the previous paragraph.

The relevance of Stinespring's theorem for \qm\ stems from the
following result.
\begin{Proposition}\ll{pacp}
Let $\A$ be a commutative unital \ca. Then any positive map
$\CQ:\A\raw$ is completely positive.
\end{Proposition}

By Theorem \ref{CCA} we may assume that $\A=C(X)$ for some locally
compact Hausdorff space $X$. We may then identify $\M^n(C(X))$ with
$C(X,\M^n(\C))$.  The proof then proceeds in the following steps:
\begin{enumerate}
\item
Elements of the form $F$, where $F(x)=\sum_i f_i (x)M_i$ for $f_i\in
C(X)$ and $M_i\in \M^n(\C)$, and the sum is finite, are dense in
$C(X,\M^n(\C))$.
\item
Such $F$ is positive iff all $f_i$ and $M_i$ are positive.
\item 
Positive elements $G$ of $C(X,\M^n(\C))$ can be norm-approximated by
positive $F$'s, i.e., when $G\geq 0$ there is a sequence $F_k\geq 0$
such that $\lim_k F_k=G$.
\item
$\CQ_n(F)$ is positive when $F$ is positive.
\item
$\CQ_n$ is continuous.
\item
If $F_k\raw G\geq 0$ in $C(X,\M^n(\C))$ then $\CQ(G)=\lim_k \CQ(F_k)$
is a norm-limit of positive elements, hence is positive.
\end{enumerate}

We now prove each of these claims.
\begin{enumerate}
\item
Take $G\in C(X,\M^n(\C))$ and pick $\ep>0$. Since $G$ is continuous,
the set $$\CO^{\ep}_x:=\{y\in X, \n G(x)-G(y)\n\,<\ep\}$$ is open for
each $x\in X$. This gives an open cover of $X$, which by the
compactness of $X$ has a finite subcover $\{\CO^{\ep}_{x_1},\ldots
\CO^{\ep}_{x_l}\}$.  A {\bf partition of unity} subordinate to the
given cover is a collection of continuous positive functions
$\phi_i\in C(X)$, where $i=1,\ldots,l$, such that the support of
$\phv_i$ lies in $\CO^{\ep}_{x_i}$ and $\sum_{i=1}^l\phv_i(x)=1$ for
all $x\in X$.  Such a partition of unity exists.

Now define $F_l\in C(X,\M^n(\C))$ by \be F_l(x):=\sum_{i=1}^l
\phv_i(x)G(x_i).  \ee Since $\n G(x_i)-G(x)\n\,<\ep$ for all $x\in
\CO^{\ep}_{x_i}$, one has
$$
\n F_l(x)-G(x)\n=\n \sum_{i=1}^l\phv_i(x)( G(x_i)-G(x))\n \,\leq
\sum_{i=1}^l \phv_i(x)\n G(x_i)-G(x)\n\,<\sum_{i=1}^l
\phv_i(x)\ep=\ep.$$ Here the norm is the matrix norm in $\M^n(\C)$.
Hence $$\n F_l-G\n=\sup_{x\in X} \n F_l(x)-G(x)\n\,<\ep.$$
\item
An element $F\in C(X,\M^n(\C))$ is positive iff $F(x)$ is positive in
$\M^n(\C)$ for each $x\in X$. In particular, when $F(x)=f(x)M$ for
some $f\in C(X)$ and $M\in \M^n(\C)$ then $F$ is positive iff $f$ is
positive in $C(X)$ and $M$ is positive in $\M^n(\C)$. By \ref{coco}.2
we infer that $F$ defined by $F(x)=\sum_i f_i (x)M_i$ is positive when
all $f_i$ and $M_i$ are positive.
\item
When $G$ in item 1 is positive then each $G(x_i)$ is positive, as we
have just seen.
\item
On $F$ as specified in \ref{pacp}.1 one has $\CQ_n(F)=\sum_i
\CQ(f_i)\ot M_i$. Now each operator $B_i\ot M$ is positive in
$\M^n(\B)$ when $B_i$ and $M$ are positive (as can be checked in a
faithful \rep).  Since $\CQ$ is positive, it follows that $\CQ_n$ maps
each positive element of the form $F=\sum_i f_iM_i$ into a positive
member of $\M^n(\B)$.
\item
We know from \ref{poscon} that $\CQ$ is continuous; the continuity of
$\CQ_n$ follows because $n<\infty$.
\item
A norm-limit $A=\lim_n A_n$ of positive elements in a \ca\ is
positive, because by (\ref{poscone2}) we have $A_n=B_n^*B_n$, and
$\lim B_n=B$ exist because of (\ref{cstax2}). Finally, $A=B^*B$ by
continuity of multiplication, i.e., by (\ref{cstax1}).
\end{enumerate}\enp
\su{Pure states and irreducible \rep s} We return to the discussion at
the end of \ref{states}. One sees that the compact convex sets in the
examples have a natural boundary.  The intrinsic definition of this
boundary is as follows.
\begin{Definition}\ll{defpurestate}   
An {\bf extreme point} in a convex set $K$ (in some vector space) is a
member $\om$ of $K$ which can only be decomposed as \be
\om=\lm\om_1+(1-\lm)\om_2, \ll{decom} \ee $\lm\in (0,1)$, if
$\om_1=\om_2=\om$. The collection $\eb$ of extreme points in $K$ is
called the {\bf extreme boundary} of $K$.  An extreme point in the
state space $K=\SA$ of a \ca\ $\A$ is called a {\bf pure state}.  A
state that is not pure is called a {\bf mixed state}.

When $K=\SA$ is a state space of a \ca\ we write $\PA$, or simply
 $\P$, for $\eb$, referred to as the {\bf pure state space} of $\A$.
\end{Definition}

Hence the pure states on $\A=\C\oplus\C$ are the points $0$ and 1 in
$[0,1]$, where 0 is identified with the functional mapping
$\lm\dot{+}\mu$ to $\lm$, whereas 1 maps it to $\mu$.  The pure states
on $\A=\M^2(\C)$ are the matrices $\rh$ in (\ref{parrho}) for which
$x^2+y^2+z^2=1$; these are the projections onto one-dimensional
subspaces of $\C^2$.

More generally, we will prove in \ref{1irrepkh} that the state space
of $\M^n(\C)$ consists of all positive matrices $\rh$ with unit trace;
the pure state space of $\M^n(\C)$ then consists of all
one-dimensional projections.  This precisely reproduces the notion of
a pure state in \qm.  The first part of Definition \ref{defpurestate}
is due to Minkowski; it was von Neumann who recognized that this
definition is applicable to \qm.

We may now ask what happens to the GNS-construction when the state
$\om$ one constructs the \rep\ $\pi_{\om}$ from is pure. In
preparation:
\begin{Definition}\ll{defirrepca}
 A \rep\ $\pi$ of a \ca\ $\A$ on a Hilbert space $\H$ is called {\bf
irreducible} if a closed subspace of $\H$ which is stable under
$\pi(\A)$ is either $\H$ or 0.
\end{Definition}

This definition should be familiar from the theory of group \rep s.
It is a deep fact of \ca s that the qualifier `closed' may be omitted
from this definition, but we will not prove this.  Clearly, the
defining \rep\ $\pi_d$ of the matrix algebra $\M^N$ on $\C^N$ is
irreducible. In the infinite-dimensional case, the defining \rep s
$\pi_d$ of $\B(\H)$ on $\H$ is irreducible as well.
\begin{Proposition}\ll{eqdefsofirrca}
  Each of the following conditions is equivalent to the irreducibility
of $\pi(\A)$ on $\H$:
\begin{enumerate}
 \item The commutant of $\pi(\A)$ in $\B(\H)$ is $\{\lm\I\, |\,
\lm\in\C\}$; in other words, $\pi(\A)''=\B(\H)$ ({\bf Schur's lemma});
\item
 Every vector $\Om$ in $\H$ is cyclic for $\pi(\A)$ (recall that this
means that $\pi(\A)\Om$ is dense in $\H$).
\end{enumerate} 
\end{Proposition}

The commutant $\pi(\A)'$ is a \sta\ in $\BH$, so when it is nontrivial
it must contain a self-adjoint element $A$ which is not a multiple of
$\I$. Using Theorem \ref{DCT} below and the spectral theorem, it can
be shown that the projections in the spectral resolution of $A$ lie in
$\pi(\A)'$ if $A$ does. Hence when $\pi(\A)'$ is nontrivial it
contains a nontrivial projection $p$. But then $p\H$ is stable under
$\pi(\A)$, contradicting irreducibility.  Hence ``$\pi$ irreducible
$\Raw$ $\pi(\A)'=\C\I$''.

Conversely, when $\pi(\A)'=\C\I$ and $\pi$ is reducible one finds a
contradiction because the projection onto the alleged nontrivial
stable subspace of $\H$ commutes with $\pi(\A)$. Hence
``$\pi(\A)'=\C\I$ $\Raw$ $\pi$ irreducible''.

When there exists a vector $\Ps\in\H$ for which $\pi(\A)\Ps$ is not
dense in $\H$, we can form the projection onto the closure of
$\pi(\A)\Ps$. By Lemma \ref{DCTlemma}, with $\M=\pi(\A)$, this
projection lies in $\pi(\A)'$, so that $\pi$ cannot be irreducible by
Schur's lemma.  Hence ``$\pi$ irreducible $\Raw$ every vector
cyclic''.  The converse is trivial.\enp

We are now in a position to answer the question posed before
\ref{defirrepca}.
\begin{Theorem}\ll{pureirr}
The GNS-\rep\ $\pi_{\om}(\A)$ of a state $\om\in\SA$ is irreducible
iff $\om$ is pure.
\end{Theorem}

When $\om$ is pure yet $\pi_{\om}(\A)$ reducible, there is a
nontrivial projection $p\in\pi_{\om}(\A)'$ by Schur's lemma. Let
$\Om_{\om}$ be the cyclic vector for $\pi_{\om}$. If $p\Om_{\om}=0$
then $Ap\Om_{\om}=pA\Om_{\om}=0$ for all $A\in\A$, so that $p=0$ as
$\pi_{\om}$ is cyclic.  Similarly, $p^{\perp}\Om_{\om}=0$ is
impossible. We may then decompose $\om=\lm\ps +(1-\lm)\ps^{\perp}$,
where $\ps$ and $\ps^{\perp}$ are states defined as in
(\ref{vectorstate}), with $\Ps:=p\Om_{\om}/\n p\Om_{\om}\n$,
$\Ps^{\perp}:=p^{\perp}\Om_{\om}/\n p^{\perp}\Om_{\om}\n$, and $\lm=\n
p^{\perp}\Om_{\om}\n^2$.  Hence $\om$ cannot be pure. This proves
``pure $\Raw$ irreducible''.

In the opposite direction, suppose $\pi_{\om}$ is irreducible, with
(\ref{decom}) for $\om_1,\om_2\in\SA$ and $\lm\in [0,1]$.  Then
$\lm\om_1-\om =(1-\lm)\om_2$, which is positive; hence
$\lm\om_1(A^*A)\leq\om(A^*A)$ for all $A\in\A$. By (\ref{css}) this
yields \be
|\lm\om_1(A^*B)|^2\leq\lm^2\om_1(A^*A)\om_1(B^*B)\leq\om(A^*A)\om(B^*B)
\ll{lmom1} \ee for all $A,B$. This allows us to define a quadratic
form (i.e., a sesquilinear map) $\hat{Q}$ on $\pi_{\om}(\A)\Om_{\om}$
by \be
\hat{Q}(\pi_{\om}(A)\Om_{\om},\pi_{\om}(B)\Om_{\om}):=\lm\om_1(A^*B). \ll{hatQ}
\ee This is well defined: when
$\pi_{\om}(A_1)\Om_{\om}=\pi_{\om}(A_2)\Om_{\om}$ then
$\om((A_1-A_2)^*(A_1-A_2))=0$ by (\ref{gnscor}), so that
$$
|\hat{Q}(\pi_{\om}(A_1)\Om_{\om},\pi_{\om}(B)\Om_{\om})-
\hat{Q}(\pi_{\om}(A_2)\Om_{\om},\pi_{\om}(B)\Om_{\om})|^2=
|\lm\om_1((A_1-A_2)^*B)|^2 \leq 0 $$ by (\ref{lmom1}); in other words,
$\hat{Q}(\pi_{\om}(A_1)\Om_{\om},\pi_{\om}(B)\Om_{\om})=
\hat{Q}(\pi_{\om}(A_2)\Om_{\om},\pi_{\om}(B)\Om_{\om})$. Similarly for
$B$.  Furthermore, (\ref{lmom1}) and (\ref{gnscor}) imply that
$\hat{Q}$ is bounded in that \be | \hat{Q}(\Ps,\Ph)|\,\leq C
\n\Ps\n\:\n\Ph\n, \ll{Qbounded} \ee for all $\Ps,\Ph\in
\pi_{\om}(\A)\Om_{\om}$, with $C=1$. It follows that $\hat{Q}$ can be
extended to all of $\H_{\om}$ by continuity. Moreover, one has \be
\hat{Q}(\Ph,\Ps)=\ovl{\hat{Q}(\Ps,\Ph)} \ll{Qherm} \ee by
(\ref{ovlom}) with $A\raw A^*B$ and $\om\raw\om_1$.
\begin{Lemma}\ll{Qform}
Let a quadratic form $\hat{Q}$ on a \Hs\ $\H$ be bounded, in that
(\ref{Qbounded}) holds for all $\Ps,\Ph\in\H$, and some constant
$C\geq 0$.  There is a bounded operator $Q$ on $\H$ such that
$\hat{Q}(\Ps,\Ph)=(\Ps,Q\Ph)$ for all $\Ps,\Ph\in\H$, and $\n
Q\n\,\leq C$. When (\ref{Qherm}) is satisfied $Q$ is self-adjoint.
\end{Lemma}

Hold $\Ps$ fixed. The map $\Ph\raw \hat{Q}(\Ps,\Ph)$ is then bounded
by (\ref{Qbounded}), so that by the Riesz-Fischer theorem there exists
a unique vector $\Om$ such that $\hat{Q}(\Ps,\Ph)=(\Om,\Ph)$. Define
$Q$ by $Q\Ps=\Om$.  The self-adjointness of $Q$ in case that
(\ref{Qherm}) holds is obvious.

Now use (\ref{Qbounded}) to estimate
$$
\n Q\Ps\n^2=(Q\Ps,Q\Ps)=\hat{Q}(Q\Ps,\Ps)\leq C\n Q\n\:\n\Ps\n^2;
$$ 
taking the supremum over all $\Ps$ in the unit ball yields $\n
Q\n^2\,\leq C\n Q\n^2$, whence $\n Q\n\,\leq C$.  \enp

Continuing with the proof of \ref{pureirr}, we see that there is a
self-adjoint operator $Q$ on $\H_{\om}$ such that \be
(\pi_{\om}(A)\Om_{\om},Q\pi_{\om}(B)\Om_{\om})=\lm\om_1(A^*B). \ll{Qpom}
\ee It is the immediate from (\ref{defrep}) that $[Q,\pi_{\om}(C)]=0$
for all $C\in\A$.  Hence $Q\in\pi_{\om}(\A)'$; since $\pi_{\om}$ is
irreducible one must have $Q=t\I$ for some $t\in\R$; hence
(\ref{Qpom}), (\ref{hatQ}), and (\ref{gnsip}) show that $\om_1$ is
proportional to $\om$, and therefore equal to $\om$ by normalization,
so that $\om$ is pure.\enp

From \ref{spc} we have the
\begin{Corollary}\ll{spc2}
  If $(\pi(\A),\H)$ is irreducible then the GNS-represen\-ta\-tion
$(\pi_{\om}(\A),\H_{\om})$ defined by any vector state $\ps$
(corresponding to a unit vector $\Ps\in\H$) is unitarily equivalent to
$(\pi(\A),\H)$.
\end{Corollary}

Combining this with \ref{pureirr} yields
\begin{Corollary}\ll{spc3}
Every \irrep\ of a \ca\ comes from a pure state via the
GNS-construction.
\end{Corollary}

A useful reformulation of the notion of a pure state is as follows.
\begin{Proposition}\ll{puredom}
A state is pure iff $0\leq\rh\leq\om$ for a positive functional $\rh$
implies $\rh=t\om$ for some $t\in\R^+$.
\end{Proposition}

We assume that $\A$ is unital; if not, use \ref{caun} and
\ref{extstate}. For $\rh=0$ or $\rh=\om$ the claim is obvious.  When
$\om$ is pure and $0\leq\rh\leq\om$, with $0\neq\rh\neq\om$, then
$0<\rh(\I)<1$, since $\om-\rh$ is positive, hence $\n
\om-\rh\n=\om(\I)-\rh(I)=1-\rh(\I)$. Hence $\rh(\I)$ would imply
$\om=\rh$, whereas $\rh(\I)=0$ implies $\rh=0$, contrary to
assumption.  Hence $(\om-\rh)/(1-\rh(\I))$ and $\rh/\rh(\I)$ are
states, and
$$
\om=\lm\frac{\om-\rh}{1-\rh(\I)} +(1-\lm)\frac{\rh}{\rh(\I)}
$$
with $\lm=1-\rh(\I)$. Since $\om$ is pure, by \ref{defpurestate} we
have $\rh=\rh(\I)\om$.

Conversely, if (\ref{decom}) holds then $0\leq\lm\om_1\leq\om$ (cf.\
the proof of \ref{pureirr}), so that $\lm\om_1=t\om$ by assumption;
normalization gives $t=\lm$, hence $\om_1=\om=\om_2$, and $\om$ is
pure.\enp

The simplest application of this proposition is
\begin{Theorem}\ll{pureX}
The pure state space of the commutative \ca\ $C_0(X)$ (equipped with
the relative $w^*$-topology) is homeomorphic to $X$.
\end{Theorem}

In view of Proposition \ref{uniqX} and Theorems \ref{CCA} and
\ref{CCA0}, we merely need to establish a bijective correspondence
between the pure states and the multiplicative functionals on
$C_0(X)$.  The case that $X$ is not compact may be reduced to the
compact case by passing from $\A=C_0(X)$ to $\AI=C(\til{X})$; cf.\
\ref{caun} and \ref{GTGT0} etc. This is possible because the unique
extension of a pure state on $C_0(X)$ to a state on $C(\til{X})$
guaranteed by \ref{extstate} remains pure. Moreover, the extension of
a multiplicative functional defined in (\ref{omtil}) coincides with
the extension $\om_{\I}$ of a state defined in (\ref{statex}), and the
functional $\infty$ in (\ref{infy}) clearly defines a pure state.
 
Thus we put $\A=C(X)$. Let $\om_x\in\Dl(C(X))$ (cf.\ the proof of
\ref{uniqX}), and suppose a functional $\rh$ satisfies
$0\leq\rh\leq\om_x$. Then $\ker(\om_x)\subseteq\ker(\rh)$, and
$\ker(\rh)$ is an ideal. But $\ker(\om_x)$ is a maximal ideal, so when
$\rh\neq 0$ it must be that $\ker(\om_x)=\ker(\rh)$. Since two
functionals on any vector space are proportional when they have the
same kernel, it follows from \ref{puredom} that $\om_x$ is pure.

Conversely, let $\om$ be a pure state, and pick a $g\in C(X)$ with
$0\leq g\leq 1_X$.  Define a functional $\om_g$ on $C(X)$ by
$\om_g(f):=\om(fg)$. Since $\om(f)-\om_g(f)= \om(f(1-g))$, and $0\leq
1-g\leq 1_X$, one has $0\leq\om_g\leq\om$.  Hence $\om_g =t\om$ for
some $t\in\R^+$ by \ref{puredom}. In particular,
$\ker(\om_g)=\ker(\om)$.  It follows that when $f\in\ker(\om)$, then
$fg\in\ker(\om)$ for all $g\in C(X)$, since any function is a linear
combination of functions for which $0\leq g\leq 1_X$. Hence
$\ker(\om)$ is an ideal, which is maximal because the kernel of a
functional on any vector space has codimension 1.  Hence $\om$ is
multiplicative by Theorem \ref{G1}.\enp

It could be that no pure states exist in $\SA$; think of an open
convex cone. It would follow that such a \ca\ has no \irrep s.
Fortunately, this possibility is excluded by the {\bf Krein-Milman
theorem} in functional analysis, which we state without proof.  The
{\bf convex hull} ${\rm co}(V)$ of a subset $V$ of a vector space is
defined by \be {\rm co}(V):=\{\lm v+(1-\lm) w\, |\, v,w\in V, \lm\in
[0,1]\}. \ll{cohull} \ee \begin{Theorem}\ll{kreinmilman} A compact
convex set $K$ embedded in a locally convex vector space is the
closure of the convex hull of its extreme points. In other words,
$K=\ovl{\mbox{\rm co}}(\eb)$.
\end{Theorem}

It follows that arbitrary states on a \ca\ may be approximated by
finite convex sums of pure states.  This is a spectacular result: for
example, applied to $C(X)$ it shows that arbitrary probability
measures on $X$ may be approximated by finite convex sums of point
(Dirac) measures.  In general, it guarantees that a \ca\ has lots of
pure states. For example, we may now refine Lemma \ref{lots} as
follows
\begin{Theorem}\ll{lotsp}
For every $A\in\Ar$ and $a\in\sg(A)$ there is a pure state $\om_a$ on
$\A$ for which $\om_a(A)=a$. There exists a pure state $\om$ such that
$|\om(A)|=\n A\n$.
\end{Theorem}

We extend the state in the proof of \ref{lots} to $C^*(A,\I)$ by
multiplicativity and continuity, that is, we put
$\til{\om}_a(A^n)=a^n$ etc. It follows from \ref{pureX} that this
extension is pure. One easily checks that the set of all extensions of
$\til{\om}_a$ to $\A$ (which extensions we know to be states; see the
proof of \ref{lots}) is a closed convex subset $K_a$ of $\SA$; hence
it is a compact convex set. By the Krein-Milman theorem
\ref{kreinmilman} it has at least one extreme point $\om_a$.  If
$\om_a$ were not an extreme point in $\SA$, it would be decomposable
as in (\ref{decom}).  But it is clear that, in that case, $\om_1$ and
$\om_2$ would coincide on $C^*(A,\I)$, so that $\om_a$ cannot be an
extreme point of $K_a$.\enp

We may now replace the use of \ref{lots} by \ref{lotsp} in the proof
of the Gel'fand-Neumark Theorem \ref{GNT}, concluding that the
universal \rep\ $\pi_{\mbox{\tiny u}}$ may be replaced by
$\pi_{\mbox{\tiny r}}:=\oplus_{\om\in\PA}\pi_{\om}$.  We may further
restrict this direct sum by defining two states to be {\bf equivalent}
if the corresponding GNS-\rep s are equivalent, and taking only one
pure state in each equivalence class. Let us refer to the ensuing set
of pure states as $[\PA]$.  We then have \be \A\simeq\pi_{\mbox{\tiny
r}}(\A):=\oplus_{\om\in [\PA]} \pi_{\om}(\A).  \ll{refdec} \ee It is
obvious that the proof of \ref{GNT} still goes through.

The simplest application of this refinement is
\begin{Proposition}\ll{strfdca}
Every finite-dimensional \ca\ is a direct sum of matrix algebras.
\end{Proposition}

For any morphism $\phv$, hence certainly for any \rep\ $\phv=\pi$, one
has the isomorphism $\phv(\A)\simeq \A/\ker(\phv)$. Since
$\A/\ker(\pi)$ is finite-dimensional, it must be that $\pi(\A)$ is
isomorphic to an algebra acting on a finite-dimensional vector space.
Furthermore, it follows from Theorem \ref{DCT} below that
$\pi(\A)''=\pi(\A)$ in every finite-dimensional \rep\ of $\A$, upon
which \ref{eqdefsofirrca}.1 implies that $\pi(\A)$ must be a matrix
algebra (as $\BH$ is the algebra of $n\x n$ matrices for
$\H=\C^n$). Then apply the isomorphism \ref{refdec}.  \enp \su{The
\ca\ of compact operators} It would appear that the appropriate
generalization of the $C^*$-algebra $\M^n(\C)$ of $n\x n$ matrices to
infinite-dimensional \Hs s $\H$ is the \ca\ $\BH$ of all bounded
operators on $\H$. This is not the case. For one thing, unlike
$\M^n(\C)$ (which, as will follow from this section, has only one
\irrep\ up to equivalence), $\BH$ has a huge number of inequivalent
\rep s; even when $\H$ is separable, most of these are realized on
non-separable \Hs s.

 For example, it follows from \ref{spc2} that any vector state $\ps$
on $\BH$ defines an \irrep\ of $\BH$ which is equivalent to the
defining \rep. On the other hand, we know from \ref{lotsp} and the
existence of bounded self-adjoint operators with continuous spectrum
(such as any multiplication operator on $L^2(X)$, where $X$ is
connected), that there are many other pure states whose GNS-\rep\ is
not equivalent to the defining \rep\ $\pi$.  Namely, when $A\in\BH$
and $a\in\sg(A)$, but $a$ is not in the discrete spectrum of $A$ as an
operator on $\H$ (i.e., there is no eigenvector $\Ps_a\in\H$ for which
$A\Ps_a=a\Ps_a$), then $\pi_{\om_a}$ cannot be equivalent to
$\pi$. For it is easy to show from (\ref{gnscruceq}) that
$\Om_{\om_a}\in\H_{\om_a}$ is an eigenvector of $\pi_{\om_a}(A)$ with
eigenvalue $a$. In other words, $a$ is in the continuous spectrum of
$A=\pi(A)$ but in the discrete spectrum of $\pi_{\om_a}(A)$, which
excludes the possibility that $\pi(\A)$ and $\pi_{\om_a}(\A)$ are
equivalent (as the spectrum is invariant under unitary
transformations).

Another argument against $\BH$ is that it is non-separable in the
nom-topology even when $\H$ is separable. The appropriate
generalization of $\M^n(\C)$ to an infinite-dimensional \Hs\ $\H$
turns out to be the \ca\ $\B_0(\H)$ of compact operator on $\H$. In
non-commutative geometry elements of this \ca\ play the role of
infinitesimals; in general, $\B_0(\H)$ is a basic building block in
the theory of \ca s.  This section is devoted to an exhaustive study
of this \ca.
\begin{Definition}\ll{defb0h}
Let $\H$ be a \Hs.  The \sta\ $\B_f(\H)$ of {\bf finite-rank operators
on $\H$} is the (finite) linear span of all finite-dimensional
projections on $\H$. In other words, an operator $A\in\BH$ lies in
$\B_f(\H)$ when $A\H:=\{A\Ps|\, \Ps\in\H\}$ is finite-dimensional.

The {\bf \ca\ $\B_0(\H)$ of compact operators} on $\H$ is the
norm-closure of $\B_f(\H)$ in $\BH$; in other words, it is the
smallest \ca\ of $\BH$ containing $\B_f(\H)$. In particular, the norm
in $\KH$ is the operator norm (\ref{BHnorm}).  An operator $A\in\BH$
lies in $\B_0(\H)$ when it can be approximated in norm by finite-rank
operators.
\end{Definition}
It is clear that $\B_f(\H)$ is a \sta, since $p^*=p$ for any
projection $p$.  The third item in the next proposition explains the
use of the word `compact' in the present context.
\begin{Proposition}\ll{pco}
\begin{enumerate}
\item
The unit operator $\I$ lies in $\B_0(\H)$ iff $\H$ is
finite-dimensional.
\item
The \ca\ $\KH$ is an ideal in $\BH$.
\item
If $A\in\KH$ then $A{\cal B}_1$ is compact in $\H$ (with the
norm-topology).  Here ${\cal B}_1$ is the unit ball in $\H$, i.e., the
set of all $\Ps\in\H$ with $\n\Ps\n\,\leq 1$.
\end{enumerate}
\end{Proposition}

Firstly, for any sequence (or net) $A_n\in\B_f(\H)$ we may choose a
unit vector $\Ps_n\in(A_n\H)^{\perp}$. Then $(A_n-\I)\Ps=-\Ps$, so
that $\n (A_n-\I)\Ps\n=1$. Hence $\sup_{\n\Ps\n=1}\n
(A_n-\I)\Ps\n\,\geq 1$, hence $\n A_n-\I\n\raw 0$ is impossible by
definition of the norm (\ref{BHnorm}) in $\BH$ (hence in $\KH$).

Secondly, when $A\in\B_f(\H)$ and $B\in\BH$ then $AB\in\B_f(\H)$,
since $AB\H=A\H$.  But since $BA=(A^*B^*)^*$, and $\B_f(\H)$ is a
\sta, one has $A^*B^*\in\B_f(\H)$ and hence $BA\in\B_f(\H)$. Hence
$\B_f(\H)$ is an ideal in $\BH$, save for the fact that it is not
norm-closed (unless $\H$ has finite dimension).  Now if $A_n\raw A$
then $A_nB\raw AB$ and $BA_n\raw BA$ by continuity of multiplication
in $\BH$.  Hence $\KH$ is an ideal by virtue of its definition.

Thirdly, note that the weak topology on $\H$ (in which $\Ps_n\raw\Ps$
iff $(\Ph,\Ps_n)\raw (\Ph,\Ps)$ for all $\Ph\in\H$) is actually the
$w^*$-topology under the duality of $\H$ with itself given by the
Riesz-Fischer theorem. Hence the unit ball ${\cal B}_1$ is compact in
the weak topology by the Banach-Alaoglu theorem. So if we can show
that $A\in\KH$ maps weakly convergent sequences to norm-convergent
sequences, then $A$ is continuous from $\H$ with the weak topology to
$\H$ with the norm-topology; since compactness is preserved under
continuous maps, it follows that $A{\cal B}_1$ is compact.
 
Indeed, let $\Ps_n\raw\Ps$ in the weak topology, with $\n\Ps_n\n=1$
for all $n$.  Since $$ \n\Ps\n^2=(\Ps,\Ps)=\lim_n (\Ps,\Ps_n)\leq
\n\Ps\n\:\n\Ps_n\n=\Ps,
$$ 
one has $\n\Ps\n\,\leq 1$.  Given $\ep>0$, choose $A_f\in\B_f(\H)$
 such that $\n A-A_f\n\,<\ep/3$, and put $p:=[A_f\H]$, the
 finite-dimensional projection onto the image of $A_f$. Then
$$
\n A\Ps_n-A\Ps\n=\n (A-A_f)\Ps_n + (A-A_f)\Ps+ A_f(\Ps_n-\Ps)\n\,\leq
\third\ep+\third\ep+\n A_f\n\: \n p(\Ps_n-\Ps)\n.
$$
Since the weak and the norm topology on a finite-dimensional \Hs\
coincide, they coincide on $p\H$, so that we can find $N$ such that
$\n p(\Ps_n-\Ps)\n\,<\ep/3$ for all $n>N$.  Hence $\n
A\Ps_n-A\Ps\n\,<\ep$.\enp
\begin{Corollary}\ll{copev}
A self-adjoint operator $A\in\KH$ has an eigenvector $\Ps_a$ with
eigenvalue $a$ such that $|a|=\n A\n$.
\end{Corollary}

Define $f_A:\CB_1\raw\R$ by $f_A(\Ps):=\n A\Ps\n^2$. When
$\Ps_n\raw\Ps$ weakly with $\n\Ps_n\n=1$, then
$$
|f_A(\Ps_n)-f_A(\Ps)|=| (\Ps_n,A^*A(\Ps_n-\Ps))-(\Ps-\Ps_n,A^*A\Ps)|
\,\leq \n A^*A (\Ps_n-\Ps)\n+ |(\Ps-\Ps_n,A^*A\Ps)|.
$$
The first term goes to zero by the proof of \ref{pco}.3 (noting that
 $A^*A\in\KH$), and the second goes to zero by definition of weak
 convergence. Hence $f_A$ is continuous.  Since $\CB_1$ is weakly
 compact, $f_A$ assumes its maximum at some $\Ps_a$. This maximum is
 $\n A\n^2$ by (\ref{BHnorm}). Now the Cauchy-Schwarz inequality with
 $\Ps=1$ gives $\n A\Ps\n^2=(\Ps, A^*A\Ps)\leq \n A^*A\Ps\n$, with
 equality iff $A^*A\Ps$ is proportional to $\Ps$. Hence when $A^*=A$
 the property $\n A\n^2=\n A\Ps_a\n^2$ with $\n\Ps_a\n=1$ implies
 $A^2\Ps_a=a^2\Ps_a$, where $a^2=\n A\n^2$.  The spectral theorem or
 the continuous functional calculus with $f(A^2)=\sqrt{A^2}=A$ implies
 $A\Ps_a=a\Ps_a$. Clearly $|a|=\n A\n$.  \enp
\begin{Theorem}\ll{cptev}
A self-adjoint operator $A\in\BH$ is compact iff $A=\sum_i a_i
[\Ps_i]$ (norm-convergent sum), where each eigenvalue $a_i$ has finite
multiplicity. Ordering the eigenvalues so that $a_i\leq a_j$ when
$i>j$, one has $\lim_{i\raw\infty} |a_i|=0$. In other words, the set
of eigenvalues is discrete, and can only have 0 as a possible
accumulation point.
\end{Theorem}

This ordering is possible because by \ref{copev} there is a largest
eigenvalue.

Let $A\in\KH$ be self-adjoint, and let $p$ be the projection onto the
closure of the linear span of all eigenvectors of $A$. As in Lemma
\ref{DCTlemma} one sees that $[A,p]=0$, so that $(pA)^*=pA$. Hence
$p^{\perp}A=(\I-p)A$ is self-adjoint, and compact by \ref{pco}.2.  By
\ref{copev} the compact self-adjoint operator $p^{\perp}A$ has an
eigenvector, which must lie in $p^{\perp}\H$, and must therefore be an
eigenvector of $A$ in $p^{\perp}\H$.  By assumption this eigenvector
can only be zero. Hence $\n p^{\perp}A\n=0$ \ref{copev}, which implies
that $A$ restricted to $p^{\perp}\H$ is zero, which implies that all
vectors in $p^{\perp}\H$ are eigenvectors with eigenvalue zero. This
contradicts the definition of $p^{\perp}\H$ unless
$p^{\perp}\H=0$. This proves ``$A$ compact and self-adjoint $\Raw$ $A$
diagonalizable''.

Let $A$ be compact and self-adjoint, hence diagonalizable. Normalize
the eigenvectors $\Ps_i:=\Ps_{a_i}$ to unit length. Then
$\lim_{i\raw\infty} (\Ps,\Ps_i)=0$ for all $\Ps\in\H$, since the
$\Ps_i$ form a basis, so that \be (\Ps,\Ps)=\sum_i |(\Ps,\Ps_i)|^2,
\ll{PsPs} \ee which clearly converges. Hence $\Ps_i\raw 0$ weakly, so
$\n A\Ps_i\n=|a_i|\raw 0$ by (the proof of) \ref{pco}.3. Hence
$\lim_{i\raw\infty} |a_i|=0$.  This proves ``$A$ compact and
self-adjoint $\Raw$ $A$ diagonalizable with $\lim_{i\raw\infty}
|a_i|=0$''.

Let now $A$ be self-adjoint and diagonalizable, with
$\lim_{i\raw\infty} |a_i|=0$.  For $N<\infty$ and $\Ps\in\H$ one then
has
$$
\n (A-\sum_{i=1}^N a_i [\Ps_i] ) \Ps\n^2 =\n \sum_{i=N+1}^{\infty} a_i
(\Ps_i,\Ps)\Ps_i\n^2\, \leq\, \sum_{i=N+1}^{\infty} |a_i|^2 \:
|(\Ps,\Ps_i)|^2\,\leq |a_N|^2 \sum_{i=N+1}^{\infty}|(\Ps,\Ps_i)|^2.
$$
Using (\ref{PsPs}), this is $\leq |a_N|^2 (\Ps,\Ps)$, so that
$\lim_{N\raw\infty} \n A-\sum_{i=1}^N a_i[\Ps_i]\n=0$, because
$\lim_{N\raw\infty} |a_N|=0$.  Since the operator $\sum_{i=1}^N
a_i[\Ps_i]$ is clearly of finite rank, this proves that $A$ is
compact.  Hence ``$A$ self-adjoint and diagonalizable with
$\lim_{i\raw\infty} |a_i|=0$ $\Raw$ $A$ compact''.

Finally, when $A$ is compact its restriction to any closed subspace of
$\H$ is compact, which by \ref{pco}.1 proves the claim about the
multiplicity of the eigenvalues.  \enp

We now wish to compute the state space of $\KH$. This involves the
study of a number of subspaces of $\BH$ which are not \ca s, but which
are ideals of $\BH$, except for the fact that they are not closed.
\begin{Definition}\ll{pnorms}
The {\bf Hilbert-Schmidt norm} $\n A\n_2$ of $A\in\BH$ is defined by
\be \n A\n^2_2:=\sum_i \n A{\bf e}_i\n^2, \ll{hsnorm} \ee where
$\{{\bf e}_i\}_i$ is an arbitrary basis of $\H$; the right-hand side
is independent of the choice of the basis. The {\bf Hilbert-Schmidt
class} $\B_2(\H)$ consists of all $A\in\BH$ for which $\n
A\n_2<\infty$.

The {\bf trace norm} $\n A\n_1$ of $A\in\BH$ is defined by \be \n
A\n_1:= \n (A^*A)^{\quar}\n_2^2, \ll{tracenorm} \ee where
$(A^*A)^{\quar}$ is defined by the continuous functional calculus.
The {\bf trace class} $\B_1(\H)$ consists of all $A\in\BH$ for which
$\n A\n_2<\infty$.
\end{Definition}

To show that (\ref{hsnorm}) is independent of the basis, we take a
second basis $\{{\bf u}_i\}_i$, with corresponding resolution of the
identity $\I=\sum_i [{\bf u}_i]$ (weakly). Aince $\I=\sum_i [{\bf
e}_i]$ we then have
$$
\n A\n^2_2:=\sum_{i,j} ({\bf e}_j,{\bf u}_i)({\bf u}_i,A^*A{\bf e}_j)
=\sum_{i,j}(A^*A {\bf u}_i,{\bf e}_j)({\bf e}_j,{\bf u}_i)= \sum_i \n
A{\bf u}_i\n^2.
$$ 

If $A\in\B_1(\H)$ then \be \Tr A:= \sum_i ({\bf e}_i,A{\bf e}_i)
\ll{deftrace} \ee is finite and independent of the basis (when
$A\notin \B_1(\H)$, it may happen that $\Tr A$ depends on the basis;
it may even be finite in one basis and infinite in another).
Conversely, it can be shown that $A\in\B_1(\H)$ when $\Tr_+ A<\infty$,
where $\Tr_+$ is defined in terms of the decomposition (\ref{posdec2})
by $\Tr_+ A:= \Tr A'_+-\Tr A'_- +i\Tr A''_+-i\Tr A''_-$.  For
$A\in\B_1(\H)$ one has $\Tr_+ A=\Tr A$.  One always has the equalities
\bea \n A\n_1 & = & \Tr |A|; \ll{tnntn}\\ \n A\n_2 & = & \Tr |A|^2=
\Tr A^*A, \ll{hsntr} \eea where \be |A|:=\sqrt{A^*A}. \ll{Aabs} \ee In
particular, when $A\geq 0$ one simply has $\n A\n_1=\Tr A$, which does
not depend on the basis, whether or not $A\in\B_1(\H)$. The properties
\be \Tr A^*A=\Tr AA^* \ll{trcycl} \ee for all $A\in\BH$, and \be \Tr
UAU^*=\Tr A \ee for all positive $A\in\BH$ and all unitaries $U$,
follow from (\ref{deftrace}) by manipulations similar to those
establishing the basis-independence of (\ref{hsnorm}).  Also, the
linearity property \be \Tr (A+B)=\Tr A+\Tr B \ee for all
$A,B\in\B_1(\H)$ is immediate from (\ref{deftrace}).

It is easy to see that the Hilbert-Schmidt norm is indeed a norm, and
that $\B_2(\H)$ is complete in this norm. The corresponding properties
for the trace norm are nontrivial (but true), and will not be needed.
In any case, for all $A\in\BH$ one has \bea \n A\n & \leq & \n A\n_1;
\ll{tnb}\\ \n A\n & \leq & \n A\n_2.  \ll{hsnb} \eea To prove this, we
use our old trick: although $\n B\n\,\geq \n B\Ps\n$ for all unit
vectors $\Ps$, for every $\ep>0$ there is a $\Ps_{\ep}\in\H$ of norm 1
such that $\n B\n^2\leq \,\n B\Ps_{\ep}\n^2 +\ep$.  Put
$B=(A^*A)^{\quar}$, and note that $\n (A^*A)^{\quar}\n^2=\n A\n$ by
(\ref{cstax2}). Completing $\Ps_{\ep}$ to a basis $\{{\bf e}_i\}_i$,
we have
$$
\n A\n=\n (A^*A)^{\quar}\n^2\,\leq\, \n
(A^*A)^{\quar}\Ps_{\ep}\n^2+\ep \,\leq \sum_i \n (A^*A)^{\quar}{\bf
e}_i\n^2+\ep=\n A\n_1+\ep.
$$
Letting $\ep\raw 0$ then proves (\ref{tnb}). The same trick with $ \n
A\n\leq\, \n A\Ps_{\ep}\n +\ep$ establishes (\ref{hsnb}).

The following decomposition will often be used.
\begin{Lemma}\ll{poldec}
Every operator $A\in\BH$ has a {\bf polar decomposition} \be A=U|A|,
\ll{pdeq} \ee where $|A|=\sqrt{A^*A}$ (cf.\ (\ref{Aabs})) and $U$ is a
partial isometry with the same kernel as $A$.
\end{Lemma}

First define $U$ on the range of $|A|$ by $U|A|\Ps:=A\Ps$. Then
compute
$$
(U|A|\Ps,U|A|\Ph)=(A\Ps,A\Ph)=(\Ps,A^*A\Ph)=(\Ps,|A|^2\Ph)=
(|A|\Ps,|A|\Ph).
$$
Hence $U$ is an isometry on $\ran(|A|)$. In particular, $U$ is well
defined, for this property implies that if $|A|\Ps_1=|A|\Ps_2$ then
$U|A|\Ps_1=U|A|\Ps_2$. Then extend $U$ to the closure of $\ran(|A|)$
by continuity, and put $U=0$ on $\ran(|A|)^{\perp}$.  One easily
verifies that \be |A|=U^*A, \ll{pd2} \ee and that $U^*U$ is the
projection onto the closure of $\ran(|A|)$, whereas $UU^*$ is the
projection onto the closure of $\ran(A)$.  \enp
\begin{Proposition}
One has the inclusions \be \B_f(\H)\subseteq
\B_1(\H)\subseteq\B_2(\H)\subseteq\B_0(\H)\subseteq \BH, \ll{incl} \ee
with equalities iff $\H$ is finite-dimensional.
\end{Proposition}

We first show that $\B_1(\H)\subseteq\B_0(\H)$. Let $A\in \B_1(\H)$.
Since $\sum_i ({\bf e}_i,|A|{\bf e}_i)<\infty$, for every $\ep>0$ we
can find $N(\ep)$ such that $\sum_{i>N(\ep)} ({\bf e}_i,|A|{\bf
e}_i)<\ep$.  Let $p_{N(\ep)}$ be the projection onto the linear span
of all ${\bf e}_i$, $i>N(\ep)$. Using (\ref{cstax2}) and (\ref{tnb}),
we have
$$
\n\: |A|^{\half}p_{N(\ep)}\n^2=\n p_{N(\ep)}|A|p_{N(\ep)}\n\,\leq\, \n
 p_{N(\ep)}|A|p_{N(\ep)}\n_1\,<\ep,
$$ 
so that $|A|^{\half} p^{\perp}_{N(\ep)}\raw |A|^{\half}$ in the
operator-norm topology. Since the star is norm-continuous by
(\ref{astisa}), this implies $p^{\perp}_{N(\ep)}|A|^{\half}\raw
|A|^{\half}$.  Now $p^{\perp}_{N(\ep)}|A|^{\half}$ obviously has
finite rank for every $\ep>0$, so that $|A|^{\half}$ is compact by
Definition \ref{defb0h}. Since $A=U |A|^{\half}|A|^{\half}$ by
(\ref{pdeq}), Proposition \ref{pco}.2 implies that $A\in\KH$.

The proof that $\B_2(\H)\subseteq\B_0(\H)$ is similar: this time we
have
$$
\n\: |A|p_{N(\ep)}\n^2 =\n p_{N(\ep)}|A|^2 p_{N(\ep)}\n\,\leq\, \n
 p_{N(\ep)}|A|^2 p_{N(\ep)}\n_2\,<\ep,
$$ 
so that $|A| p^{\perp}_{N(\ep)}\raw |A|$, with the same conclusion.

Finally, we use Theorem \ref{cptev} to rewrite (\ref{tnntn}) and
(\ref{hsnorm}) as \bea \n A\n_1 & = & \sum_i a_i; \ll{aitr}\nn \\ \n
A\n_2 & = & \sum_i a^2_i, \ll{hsetc} \eea where the $a_i$ are the
eigenvalues of $|A|$. This immediately gives \be \n A\n_2\,\leq\,\n
A\n_1, \ee implying $\B_1(\H)\subseteq\B_2(\H)$.

Finally, the claim about proper inclusions is trivially established by
producing examples on the basis of \ref{cptev} and (\ref{hsetc}).
\enp

The chain of inclusions (\ref{incl}) is sometimes seen as the
non-commutative analogue of
$$
\ell_c(X)\subseteq \ell^1(X)\subseteq \ell^2(X)\subseteq\ell_0(X)
\subseteq\ell^{\infty}(X),
$$
where $X$ is an infinite discrete set. Since $\ell_1(X)=\ell_0(X)^*$
and $\ell^{\infty}(X)=\ell_1(X)^*=\ell_0(X)^{**}$, this analogy is
strengthened by the following result.
\begin{Theorem}\ll{dualities}
One has $\KH^*=\B_1(\H)$ and $\B_1(\H)^*=\KH^{**}=\BH$ under the
pairing \be \hat{\rh}(A)=\Tr \rh A=\hat{A}(\rh).  \ee Here
$\hat{\rh}\in \KH^*$ is identified with $\rh\in \B_1(\H)$, and
$\hat{A}\in \B_1(\H)^*$ is identified with $A\in\BH$.
\end{Theorem}

The basic ingredient in the proof is the following lemma, whose proof
is based on the fact that $\B_2(\H)$ is a \Hs\ in the inner product
\be (A,B):=\Tr A^*B. \ll{hsip} \ee To show that this is well defined,
use (\ref{csb}) and (\ref{hsntr}).
\begin{Lemma}\ll{cruclem}
For $\rh\in\B_1(\H)$ and $A\in\BH$ one has \be |\Tr A\rh|\,\leq \n
A\n\: \n \rh\n_1. \ll{cleq} \ee
\end{Lemma}

Using (\ref{pdeq}) for $\rh$ and (\ref{csb}) for the inner product
(\ref{hsip}), as well as (\ref{trcycl}) and (\ref{tracenorm}), we
estimate
$$
|\Tr A\rh|^2 =|\Tr AU |\rh|^{\half}|\rh|^{\half}|= |(( AU
|\rh|^{\half})^*,|\rh|^{\half})|
$$ $$ \leq\, \n\:|\rh|^{\half}\n^2_2\;\n ( AU |\rh|^{\half})^*\n^2_2
=\n \rh\n_1\Tr ( |\rh|^{\half} U^*A^*AU |\rh|^{\half}).
$$
Now observe that if $0\leq A_1\leq A_2$ then $\Tr A_2\leq\Tr A_2$ for
all $A_1,A_2\in\KH$, since on account of \ref{cptev} one has $A_1\leq
A_2$ iff all eigenvalues of $A_1$ are $\leq$ all eigenvalues of
$A_2$. Then use (\ref{aitr}). From (\ref{BstABA}) we have
$|\rh|^{\half} U^*A^*AU |\rh|^{\half}\leq \n AU\n^2 \rh$, so from the
above insight we arrive at
$$
\Tr ( |\rh|^{\half} U^*A^*AU |\rh|^{\half})\leq \n \rh\n_1\,\n
AU\n^2\,\leq\, \n A\n^2,
$$ 
since $U$ is a partial isometry. Hence we have (\ref{cleq}).  \enp

We now prove $\KH^*=\B_1(\H)$. It is clear from \ref{cruclem} that \be
\B_1(\H)\subseteq \KH^*, \ll{B1Hincl} \ee with \be \n\hat{\rh}\n\,\leq
\n\rh\n_1. \ll{hatrhleq} \ee To prove that $\KH^*\subseteq\B_1(\H)$,
we use (\ref{hsnb}). For $\hat{\rh}\in\KH^*$ and
$A\in\B_2(\H)\subseteq\KH$ we therefore have
$$
|\hat{\rh}(A)|\,\leq \,\n\hat{\rh}\n\:\n A\n\,\leq \,\n\hat{\rh}\n\:\n
A\n_2.
$$
Hence $\hat{\rh}\in\B_2(\H)^*$; since $\B_2(\H)$ is a \Hs, by
Riesz-Fischer there is an operator $\rh\in\B_2(\H)$ such that
$\hat{\rh}(A)=\Tr \rh A$ for all $A\in\B_2(\H)$.  In view of
(\ref{incl}), we need to sharpen $\rh\in\B_2(\H)$ to
$\rh\in\B_1(\H)$. To do so, choose a finite-dimensional projection
$p$, and note that $p|\rh|\in\B_f(\H)\subseteq \B_1(\H)$; the presence
of $p$ even causes the sum in (\ref{deftrace}) to be finite in a
suitable basis.  Now use the polar decomposition $\rh=U|\rh|$ with
(\ref{pd2}) to write
$$
\Tr p|\rh|=\Tr pU^*\rh=\Tr\rh pU^*=\hat{\rh}(pU^*);
$$
changing the order inside the trace is justified by naive arguments,
since the sum in (\ref{deftrace}) is finite.  Using the original
assumption $\hat{\rh}\in\KH^*$, we have \be |\Tr
p|\rh|\:|\,\leq\,\n\hat{\rh}\n\: \n pU^*\n\,\leq\, \n\hat{\rh}\n\: \n
p\n=\n\hat{\rh}\n\, \ll{Trprh} \ee since $U$ is a partial isometry,
whereas $\n p\n=1$ in view of (\ref{cstax2}) and $p=p^2=p^*$. Now
choose a basis of $\H$, and take $p$ to be the projection onto the
subspace spanned by the first $N$ elements; from (\ref{deftrace}) and
(\ref{Trprh}) we then have
$$
|\Tr p|\rh|\:|=|\sum_{i=1}^N ({\bf e}_i,|\rh|{\bf e}_i)|\,\leq\,
\n\hat{\rh}\n.
$$
It follows that the sequence $s_N:=|\sum_{i=1}^N ({\bf e}_i,|\rh|{\bf
e}_i)|$ is bounded, and since it is positive it must have a limit. By
(\ref{Trprh}) and (\ref{tnntn}) this means that $\n \rh\n_1\,\leq\,
\n\hat{\rh}\n$, so that $\rh\in\B_1(\H)$, hence
$\KH^*\subseteq\B_1(\H)$.  Combining this with (\ref{B1Hincl}) and
(\ref{hatrhleq}), we conclude that $\KH^*=\B_1(\H)$ and $\n \rh\n_1=
\n\hat{\rh}\n$.

We turn to the proof of $\B_1(\H)^*=\BH$. It is clear from
\ref{cruclem} that $\BH\subseteq\B_1(\H)^*$, with \be
\n\hat{A}\n\,\leq\,\n A\n. \ll{nhatAn} \ee To establish the converse,
pick $\hat{A}\in\B_1(\H)^*$ and $\Ps,\Ph\in\H$, and define a quadratic
form $Q_A$ on $\H$ by \be Q_A(\Ps,\Ph):=\hat{A}(|\Ph><\Ps|). \ll{bh3}
\ee Here the operator $|\Ph><\Ps|$ is defined by $|\Ph><\Ps|\Om:=
(\Ps,\Om)\Ph$. For example, when $\Ps$ has unit length, $|\Ps><\Ps|$
is the projection $[\Ps]$, and in general $|\Ps><\Ps|=\n\Ps\n^2[\Ps]$.
Note that $(|\Ph><\Ps|)^*=|\Ps><\Ph|$, so that
$$
|\:|\Ph><\Ps|\:|=\sqrt{(|\Ph><\Ps|)^*|\Ph><\Ps|}=\sqrt{(\Ph,\Ph)|\Ps><\Ps|}=
\n\Ph\n\: \n\Ps\n [\Ps].
$$
Since, for any projection $p$, the number $\Tr p$ is the dimension of
$p\H$ (take a basis whose elements lie either in $p\H$ or in
$p^{\perp}\H$), we have $\Tr [\Ps]=1$. Hence from (\ref{tnntn}) we
obtain \be \n |\Ph><\Ps|\n_1=\n\Ph\n\: \n\Ps\n .  \ll{bh2} \ee Since
$\hat{A}\in\B_1(\H)^*$ by assumption, one has \be
|\hat{A}(|\Ph><\Ps|)|\,\leq \n\hat{A}\n\: \n |\Ph><\Ps|\n_1. \ll{bh1}
\ee Combining (\ref{bh1}), (\ref{bh2}), and (\ref{bh3}), we have \be
|Q_A(\Ps,\Ph)|\,\leq \n\hat{A}\n \:\n\Ph\n\: \n\Ps\n .  \ee Hence by
Lemma \ref{Qform} and (\ref{bh3}) there is an operator $A$, with \be
\n A\n\,\leq\,\n\hat{A}\n, \ll{htaA2} \ee such that
$\hat{A}(|\Ph><\Ps|)=(\Ps,A\Ph)$.  Now note that $(\Ps,A\Ph)=\Tr
|\Ph><\Ps| A$; this follows by evaluating (\ref{tnntn}) over a basis
containing $\n\Ph\n\inv|\Ph>$. Hence $\hat{A}(|\Ph><\Ps|)=\Tr
|\Ph><\Ps| A$.  Extending this equation by linearity to the span
$\B_f(\H)$ of all $|\Ph><\Ps|$, and subsequently by continuity to
$\B_1(\H)$, we obtain $\hat{A}\rh=\Tr\rh A$. Hence
$\B_1(\H)^*\subseteq\BH$, so that, with (\ref{B1Hincl}), we obtain
$\B_1(\H)^*=\BH$.  Combining (\ref{nhatAn}) and (\ref{htaA2}), we find
$\n A\n=\n\hat{A}\n$, so that the identification of $\B_1(\H)^*$ with
$\BH$ is isometric.  \enp
\begin{Corollary}\ll{1irrepkh}
\begin{enumerate}
\item
The state space of the \ca\ $\KH$ of all compact operators on some
\Hs\ $\H$ consists of all density matrices, where a {\bf density
matrix} is an element $\rh\in\B_1(\H)$ which is positive ($\rh\geq 0$)
and has unit trace ($\Tr\rh=1$).
\item
The pure state space of $\KH$ consists of all one-dimensional
projections.
\item
The $C^*$-algebra $\B_0(\H)$ possesses only one \irrep, up to unitary
equivalence, namely the defining one.
\end{enumerate}
\end{Corollary}

Diagonalize $\rh=\sum_i p_i [\Ps_i]$; cf.\ \ref{cptev} and \ref{incl}.
Using $A=[\Ps_i]$, which is positive, the condition $\hat{\rh}(A)\geq
0$ yields $p_i\geq 0$. Conversely, when all $p_i\geq 0$ the operator
$\rh$ is positive. The normalization condition
$\n\hat{\rh}\n=\n\rh\n_1= \sum p_i=1$ (see \ref{defstate2}) and
(\ref{aitr})) yields \ref{1irrepkh}.1.

The next item \ref{1irrepkh}.2 is then obvious from
\ref{defpurestate}.

Finally, \ref{1irrepkh}.3 follows from \ref{1irrepkh}.2 and
Corollaries \ref{spc3} and \ref{spc2}.  \enp

Corollary \ref{1irrepkh}.3 is one of the most important results in the
theory of \ca s. Applied to the finite-dimensional case, it shows that
the \ca\ $\M^n(\C)$ of $n\x n$ matrices has only one \irrep.

The opposite extreme to a pure state on $\KH$ is a {\bf faithful
 state} $\hat{\rh}$, for which by definition the left-ideal
 $\CN_{\rh}$ defined in (\ref{AB0omBA}) is zero. In other words, one
 has $\Tr A^*A>0$ for all $A\neq 0$.
\begin{Proposition}
The GNS-\rep\ $\pi_{\rh}$ corresponding to a faithful state
$\hat{\rh}$ on $\KH$ is unitarily equivalent to the \rep\
$\hat{\pi}_{\rh}(\KH)$ on the Hilbert space $\B_2(\H)$ of
Hilbert-Schmidt operators given by left-multiplication, i.e., \be
\hat{\pi}_{\rh}(A)B:=AB. \ll{hsrep} \ee
\end{Proposition}

It is obvious from (\ref{hsnorm}) that for $A\in\BH$ and
$B\in\B_2(\H)$ one has \be \n AB\n_2\,\leq\, \n A\n\: \n B\n_2, \ee so
that the \rep\ (\ref{hsrep}) is well-defined (even for $A\in\BH$
rather than merely $A\in\KH$). Moreover, when $A,B\in\B_2(\H)$ one has
\be \Tr AB=\Tr BA.\ll{trab} \ee This follows from (\ref{trcycl}) and
the identity \be AB=\quar\sum_{n=0}^3 i^n(B+i^nA^*)^*(B+i^n A^*).  \ee
When $\rh\in\B_1(\H)$ and $\rh\geq 0$ then $\rh^{1/2}\in\B_2(\H)$; see
(\ref{tnntn}) and (\ref{hsntr}).  It is easily seen that $\rh^{1/2}$
is cyclic for $\hat{\pi}_{\rh}(\KH)$ when $\hat{\rh}$ is faithful.
Using (\ref{hsip}) and (\ref{trab}) we compute
$$
(\rh^{1/2},\hat{\pi}_{\rh}(A)\rh^{1/2})=\Tr
\rh^{1/2}\hat{\pi}_{\rh}(A)\rh^{1/2}=\Tr \rh A=\hat{\rh}(A).
$$
The equivalence between $\pi_{\rh}$ and $\hat{\pi}_{\rh}$ now follows
from \ref{spc5} or \ref{spc}.\enp

For an alternative proof, use the GNS construction itself.  The map
$A\rightarrow A\rh^{1/2}$, with $\rh\in \B_1(\H)$, maps $\B_0(\H)$
into $\B_2(\H)$, and if $\hat{\rh}$ is faithful the closure (in norm
derived from the inner product (\ref{hsip})) of the image of this map
is $\B_2(\H)$.  \su{The double commutant theorem} The so-called double
commutant theorem was proved by von Neumann in 1929, and remains a
central result in operator algebra theory. For example, although it is
a statement about von Neumann algebras, it controls the
(ir)reducibility of \rep s.  Recall that the commutant $\M'$ of a
collection $\M$ of bounded operators consists of all bounded operators
which commute with all elements of $\M$; the bicommutant $\M''$ is
$(\M')'$.

We first give the finite-dimensional version of the theorem; this is
already nontrivial, and its proof contains the main idea of the proof
of the infinite-dimensional case as well.
\begin{Proposition}\ll{DCT1}
Let $\M$ be a \sta\ (and hence a \ca) in $\M^n(\C)$ containing $\I$
(here $n<\infty$). Then $\M''=\M$.
\end{Proposition}

The idea of the proof is to take $n$ arbitrary vectors
$\Ps_1,\ldots,\Ps_n$ in $\C^n$, and, given $A\in\M''$, construct a
matrix $A_0\in\M$ such that $A\Ps_i=A_0\Ps_i$ for all
$i=1,\ldots,n$. Hence $A=A_0\in\M$. We will write $\H$ for $\C^n$.

Choose some $\Ps=\Ps_1\in\H$, and form the linear subspace $\M\Ps$ of
$\H$.  Since $\H$ is finite-dimensional, this subspace is closed, and
we may consider the projection $p=[\M\Ps]$ onto this subspace. By
Lemma \ref{DCTlemma} one has $p\in\M'$. Hence $A\in\M''$ commutes with
$p$. Since $\I\in\M$, we therefore have $\Ps=\I\Ps\in\M\Ps$, so
$\Ps=p\Ps$, and $A\Ps=Ap\Ps= pA\Ps\in\M\Ps$. Hence $A\Ps=A_0\Ps$ for
some $A_0\in\M$.

Now choose $\Ps_1,\ldots,\Ps_n\in\H$, and regard
$\Ps_1\dot{+}\ldots\dot{+}\Ps_n$ as an element of
$\H^n:=\oplus^n\H\simeq\H\ot\C^n$ (the direct sum of $n$ copies of
$\H$), where $\Ps_i$ lies in the $i$'th copy. Furthermore, embed $\M$
in $\B(\H^n)\simeq \M^n(\BH)$ by $A\raw\dl(A):=A\I_n^{\ot}$ (where
$\I_n^{\ot}$ is the unit in $\M^n(\BH)$); this is the diagonal matrix
in $\M^n(\BH)$ in which all diagonal entries are $A$.

Now use the first part of the proof, with the substitutions $\H\raw
\H^n$, $\M\raw\dl(\M)$, $A\raw\Bbb A:=\dl(A)$, and $\Ps\raw
\Ps_1\dot{+}\ldots\dot{+}\Ps_n$.  Hence given
$\Ps_1\dot{+}\ldots\dot{+}\Ps_n$ and $\dl(A)\in\dl(\M)$ there exists
${\Bbb A}_0\in \dl(\M)''$ such that \be
\dl(A)(\Ps_1\dot{+}\ldots\dot{+}\Ps_n)={\Bbb
A}_0(\Ps_1\dot{+}\ldots\dot{+}\Ps_n). \ll{vnue} \ee For arbitrary
$\Bbb B\in\M^n(\BH)$, compute $([\Bbb
B,\dl(A)])_{ij}=[B_{ij},A]$. Hence $\dl(\M)'=\M^n(\M')$. It is easy to
see that $\M^n(\M')'=\M_n(\M'')$, so that \be \dl(\M)''=\dl(\M'').
\ee Therefore, ${\Bbb A}_0=\dl(A)_0$ for some $A_0\in\M$. Hence
(\ref{vnue}) reads $A\Ps_i=A_0\Ps_i$ for all $i=1,\ldots,n$.\enp

As it stands, Proposition \ref{DCT1} is not valid when $\M^n(\C)$ is
replaced by $\BH$, where $\dim(\H)=\infty$. To describe the
appropriate refinement, we define two topologies on $\BH$ which are
weaker than the norm-topology we have used so far (and whose
definition we repeat for convenience).
\begin{Definition}\ll{topBH}
\begin{itemize}
\item 
The {\bf norm-topology} on $\BH$ is defined by the criterion for
 convergence $A_{\lm}\rightarrow A$ iff $\n A_{\lm}-A\n\raw 0$. A
 basis for the norm-topology is given by all sets of the form \be
 \CO^n_{\ep}(A):=\{ B\in\BH|\, \n B-A\n\,<\ep\}, \ee where $A\in\BH$
 and $\ep>0$.
\item
The {\bf strong topology} on $\BH$ is defined by the convergence
 $A_{\lm}\rightarrow A$ iff $\n (A_{\lm}-A)\Ps\n\raw 0$ for all
 $\Ps\in\H$. A basis for the strong topology is given by all sets of
 the form \be \CO^s_{\ep}(A,\Ps_1,\ldots,\Ps_n):=\{ B\in\BH|\, \n
 (B-A)\Ps_i\n\,<\ep\;\forall i=1,\ldots,n\}, \ll{strongnbd} \ee where
 $A\in\BH$, $\Ps_1,\ldots,\Ps_n\in\H$, and $\ep>0$.
\item
The {\bf weak topology} on $\BH$ is defined by the convergence
 $A_{\lm}\rightarrow A$ iff $|(\Ps, (A_{\lm}-A)\Ps)| \raw 0$ for all
 $\Ps\in\H$. A basis for the weak topology is given by all sets of the
 form \be \CO^w_{\ep}(A,\Ps_1,\ldots,\Ps_n,\Ph_1,\ldots,\Ph_n):=\{
 B\in\BH|\, |(\Ph_i, (A_{\lm}-A)\Ps_i)| \,<\ep\;\forall
 i=1,\ldots,n\}, \ee where $A\in\BH$,
 $\Ps_1,\ldots,\Ps_n,\Ph_1,\ldots,\Ph_n\in\H$, and $\ep>0$.
\end{itemize}
\end{Definition}

These topologies should all be seen in the light of the general theory
of locally convex topological vector spaces. These are vector spaces
whose topology is defined by a family $\{p_{\al}\}$ of semi-norms;
recall that a semi-norm on a vector space $\CV$ is a function
$p:\CV\raw\R$ satisfying \ref{defnorm}.1, 3, and 4.  A net
$\{v_{\lm}\}$ in $\CV$ converges to $v$ in the topology generated by a
given iff $p_{\al}(v_{\lm}-v)\raw 0$ for all $\al$.

The norm-topology is defined by a single semi-norm, namely the
 operator norm, which is even a norm.  Its open sets are generated by
 $\ep$-balls in the operator norm, whereas the strong and the weak
 topologies are generated by finite intersections of $\ep$-balls
 defined by semi-norms of the form $p^s_{\Ps}(A):=\n A\Ps\n$ and
 $p^w_{\Ps,\Ph}(A):=|(\Ph,A\Ps)|$, respectively.  The equivalence
 between the definitions of convergence stated in \ref{topBH} and the
 topologies defined by the open sets in question is given in theory of
 locally convex topological vector spaces.

The estimate (\ref{opprop}) shows that norm-convergence implies strong
convergence.  Using the Cauch-Schwarz inequality (\ref{csb}) one sees
that strong convergence implies weak convergence.  In other words, the
norm topology is stronger than the strong topology, which in turn is
stronger than the weak topology.
\begin{Theorem}\ll{DCT}
Let $\M$ be a \sta\ in $\BH$, containing $\I$. The following are
equivalent:
\begin{enumerate}
\item
$\M''=\M$;
\item
$\M$ is closed in the weak operator topology;
\item
$\M$ is closed in the strong operator topology.
\end{enumerate}
\end{Theorem}

It is easily verified from the definition of weak convergence that the
commutant $\GN'$ of a \sta\ $\GN$ is always weakly closed: for if
$A_{\lm}\raw A$ weakly with all $A_{\al}\in\GN$, and $B\in\GN$, then
$$
(\Ph,[A,B]\Ps)=(\Ph,AB\Ps)-(B^*\Ph,A\Ps)= \lim_{\al}
(\Ph,A_{\lm}B\Ps)-(B^*\Ph,A_{\lm}\Ps)=\lim_{\al}(\Ph,[A_{\lm},B]\Ps)=0.
$$

If $\M''=\M$ then $\M={\frak N}'$ for ${\frak N}=\M'$, so that $\M$ is weakly
closed. Hence ``$1\Raw 2$''.

Since the weak topology is weaker than the strong topology, ``$2\Raw
3$'' is trivial.

To prove ``$3\Raw 1$'', we adapt the proof of \ref{DCT1} to the
infinite-dimensional situation.  Instead of $\M\Ps$, which may not be
closed, we consider its closure $\ovl{\M\Ps}$, so that
$p=[\ovl{\M\Ps}]$. Hence $A\in\M''$ implies $A\in\ovl{\M\Ps}$; in
other words, for every $\ep>0$ there is an $A_{\ep}\in\M$ such that
$\n (A-A_{\ep})\Ps\n\,<\ep$.  For $\H^n$ this means that
$$
\n\dl(A-A_{\ep})(\Ps_1\dot{+}\ldots\dot{+}\Ps_n)\n^2=\sum_{i=1}^n \n
(A-A_{\ep})\Ps_i\n^2\,<\ep^2.
$$
Noting the inclusion
$$
\{\sum_{i=1}^n \n (A-B)\Ps_i\n^2\,<\ep^2\}\subseteq
\CO^s_{\ep}(A,\Ps_1,\ldots,\Ps_n)
$$
(cf.\ (\ref{strongnbd})), it follows that $A_{\ep}\raw A$ for $\ep\raw
0$. Since all $A_{\ep}\in\M$ and $\M$ is strongly closed, this implies
that $A\in\M$, so that $\M''\subseteq\M$.  With the trivial inclusion
$\M\subseteq\M''$, this proves that $\M''=\M$.  \enp
\section{Hilbert $C^*$-modules and induced \rep s}\setc{equation}{0}
\su{Vector bundles}\ll{Vb}
 This chapter is concerned with the
 `non-commutative analogue' of a vector bundle. Let us first recall
 the notion of an ordinary vector bundle; this is a special case of
 the following \begin{Definition}\ll{defbundle} A {\bf
 bundle}\index{bundle} $\SB(X,F,\ta)$ consists of topological spaces
 $\SB$ (the {\bf total space}),\index{total space!of a bundle} $X$
 (the {\bf base}),\index{base!of a bundle} $F$ (the {\bf typical
 fiber}\index{typical fiber}), and a continuous surjection
 $\ta:\SP\raw X$ with the following property: each $x\in X$ has a
 neighbourhood $\CN_{\al}$ such that there is a homeomorphism
 $\ps_{\al}:\ta\inv(\CN_{\al})\raw \CN_{\al}\x F\subset X\x F$ for
 which $\ta=\ta_X\circ \ps_{\al}$ (where $\ta_X:X\x F\raw X$ is the
 projection onto the first factor).
\end{Definition}

The maps $\ps_{\al}$ are called {\bf local trivialization}s. We
factorize $\ps_{\al}=(\ta,\ps_{\al}^F)$, so that $\ps_{\al}^F$
restricted to $\ta\inv(x)$ provides a homeomorphism between the latter
and the typical fiber $F$. Each subset $\ta\inv(x)$ is called a {\bf
fiber}\index{fiber} of $\SB$. One may think of $\SB$ as $X$ with a
copy of $F$ attached at each point.

The simplest example of a bundle over a base $X$ with typical fiber
$F$ is the {\bf trivial bundle} $\SB=X\x F$, with $\ta(x,f):=x$.
According to the definition, any bundle is locally trivial in the
specified sense.
\begin{Definition}\ll{defvb}
A {\bf vector bundle}\index{vector!bundle} is a bundle in which
\begin{enumerate}
\item
each fiber is a finite-dimensional vector space, such that the
 relative topology of each fiber coincides with its topology as a
 vector space;
\item
each local trivialization $\ps_{\al}^F:\ta\inv(x)\raw F$ (where
 $x\in\CN_{\al}$) is linear.
\end{enumerate}
A {\bf complex vector bundle} is a vector bundle with typical fiber
$\C^m$, for some $m\in\N$.
\end{Definition}

 We will generically denote vector bundles by the letter $\SV$, with
typical fiber $F=V$.  The simplest vector bundle over $X$ with fiber
$V=\C^n$ is the trivial bundle $\SV=X\x \C^n$.  This bundle leads to
possibly nontrivial sub-bundles, as follows.  Recall the definition of
$\M^n(\A)$ in \ref{CPmaps}, specialized to $\A=C(X)$ in the proof of
\ref{pacp}. If $X$ is a compact Hausdorff space, then
$\M^n(C(X))\simeq C(X,\M^n(\C))$ is a \ca. Let $X$ in addition be
connected. One should verify that a matrix-valued function $p\in
C(X,\M^n(\C))$ is an idempotent (that is, $p^2=p$) iff each $p(x)$ is
an idempotent in $\M^n(\C)$. Such an idempotent $p$ defines a vector
bundle $\SV_p$, whose fiber above $x$ is $\ta\inv(x):=p(x)\C^n$. The
space $\SV_p$ inherits a topology and a projection $\ta$ (onto the
first co-ordinate) from $X\x \C^n$, relative to which all axioms for a
vector bundle are satisfied. Note that the dimension of $p(x)$ is
independent of $x$, because $p$ is continuous and $X$ is connected.

The converse is also true.
\begin{Proposition}\ll{Swan}
Let $\SV$ be a complex vector bundle over a connected compact
Hausdorff space $X$, with typical fiber $\C^m$.  There is an integer
$n\geq m$ and an idempotent $p\in C(X,\M^n(\C))$ such that
$\SV\subseteq X\x\C^n$, with $\ta\inv(x)=p(x)\C^n$.
\end{Proposition}

The essence of the proof is the construction of a complex vector
bundle $\SV'$ such that $\SV\oplus\SV'$ is trivial (where the direct
sum is defined fiberwise); this is the bundle $X\x\C^n$.

Following the philosophy of non-commutative geometry, we now try to
describe vector bundles in terms of \ca s. The first step is the
notion of a {\bf section} of $\SV$; this is a map $\Ps:X\raw \SV$ for
which $\ta(\Ps(x))=x$ for all $x\in X$. In other words, a section maps
a point in the base space into the fiber above the point. Thus one
defines the space $\Gm(\SV)$ of all continuous sections of $\SV$. This
is a vector space under pointwise addition and scalar multiplication
(recall that each fiber of $\SV$ is a vector space). Moreover, when
$X$ is a connected compact Hausdorff space, $\Gm(\SV)$ is a
right-module for the commutative \ca\ $C(X)$: one obtains a linear
action $\pi_R$ of $C(X)$ on $\Gm(\SV)$ by 
\be
\pi_R(f)\Ps(x):=f(x)\Ps(x). \ll{pirpil}
 \ee Since $C(X)$ is commutative, this is,
of course, a left-action as well.

For example, in the trivial case one has the obvious isomorphisms \be
\Gm(X\x \C^m)\simeq C(X,\C^m)\simeq C(X)\ot\C^m\simeq \oplus^m C(X).
\ee A fancy way of saying this is that $\Gm(X\x \C^m)$ is a {\bf
finitely generated free module} for $C(X)$.  Here a free (right-)
module $\CE$ for an algebra $\A$ is a direct sum $\CE=\oplus^n\A$ of a
number of copies of $\A$ itself, on which $\A$ acts by
right-multiplication, i.e., \be \pi_R(B) A_1\oplus\ldots \oplus
A_n:=A_1B\oplus\ldots \oplus A_n B.  \ee If this number is finite one
says that the free module is finitely generated.

When $\SV$ is non-trivial, one obtains $\Gm(\SV)$ as a certain
modification of a finitely generated free module for $C(X)$. For any
algebra $\A$, and idempotent $p\in \M^n(\A)$, the action of $p$ on
$\oplus^n \A$ commutes with the action by $\A$ given by
right-multiplication on each component.  Hence the vector space $p
\oplus^m \A$ is a right- $\A$-module, called {\bf projective}. When
$m<\infty$, one calls $p \oplus^m \A$ a {\bf finitely generated
projective module} for $\A$.

In particular, when $\SV=X\x\C^n$ and $\SV_p$ is the vector bundle
described prior to \ref{Swan}, we see that \be \Gm(\SV_p)=p \oplus^n
C(X) \ll{sersw} \ee under the obvious (right-) action of $C(X)$.

This lead to the {\bf Serre-Swan theorem}.
\begin{Theorem}\ll{Serre}
Let $X$ be a connected compact Hausdorff space. There is a bijective
correspondence between complex vector bundles $\SV$ over $X$ and finitely
generated projective modules $\CE(\SV)=\Gm(\SV)$ for $C(X)$.
\end{Theorem}

This is an immediate consequence of Proposition \ref{Swan}: any vector
bundle is of the form $\SV_p$, leading to $\Gm(\SV_p)$ as a finitely
generated projective $C(X)$-module by (\ref{sersw}).  Conversely,
given such a module $p \oplus^n C(X)$, one has $p\in C(X,\M^n(\C))$,
and thereby a vector bundle $\SV_p$ as described prior to
\ref{Swan}.\enp

Thus we have achieved our goal of describing vector bundles over $X$
purely in terms of concepts pertinent to the \ca\ $C(X)$.
Let us now add further structure.
\begin{Definition}\ll{hvb}
A {\bf Hermitian vector bundle} is a complex vector bundle $\SV$ with
an inner product $(\, ,\,)_x$ defined on each fiber $\ta\inv(x)$,
which continuously depends on $x$.  More precisely, for all
$\Ps,\Ph\in\Gm(\SV)$ the function $x\raw (\Ps(x),\Ph(x))_x$ lies in
$C(X)$.
\end{Definition}

Using the local triviality of $\SV$ and the existence of a partition
of unity, it is easily shown that any complex vector bundle over a
paracompact space can be equipped with such a Hermitian
structure. Describing the bundle as $\SV_p$, a Hermitian structure is
simply given by restricting the natural inner product on each fiber
$\C^n$ of $X\x \C^n$ to $\SV_p$.  One may then choose the idempotent
$p\in C(X,\C^n)$ so as to be a projection with respect to the usual
involution on $C(X,\C^n)$ (i.e., one has $p^*=p$ in addition to
$p^2=p$).  Any other Hermitian structure on $\SV_p$ may be shown to be
equivalent to this canonical one.

There is no reason to restrict the dimension of the fibers so as to be
finite-dimensional.  A {\bf Hilbert bundle} is defined by replacing
`finite-dimensional vector space' in \ref{defvb}.1 by `Hilbert space',
still requiring that all fibers have the same dimension (which may be
infinite). A Hilbert bundle with finite-dimensional fibers is
evidently the same as a Hermitian vector bundle. The simplest example
of a Hilbert bundle is a \Hs, seen as a bundle over the base space
consisting of a single point.

The following class of Hilbert bundles will play a central role in the
theory of induced group \rep s.
\begin{Proposition}\ll{homvb}
Let $H$ be a closed subgroup of a locally compact group $G$, and take
a unitary \rep\ $\Ulg$ of $H$ on a \Hs\ $\Hlg$. Then $H$ acts on
$G\x\Hlg$ by $h:(x,v)\raw (xh\inv,\Ulg(h)v)$, and the quotient 
\be
\SHG:=G\x_H\Hlg=(G\x\Hlg)/H \ll{SHG}
\ee
 by this action is a Hilbert bundle over $X=G/H$,
with projection 
\be
\ta_{\ch}([x,v]_H):= [x]_H \ll{deftch}
\ee
 and typical fiber $\Hlg$.
\end{Proposition}

Here $[x,v]_H$ is the equivalence class in $G\x_H \Hlg$ of $(x,v)\in
G\x\Hlg$, and $[x]_H=xH$ is the equivalence class in $G/H$ of $x\in
G$. Note that the projection $\ta_{\ch}$ is well defined.
 
The proof relies on the fact that $G$ is a bundle over $G/H$ with
projection 
\be
\ta(x)=[x]_H \ll{deftauGH}
\ee
 and typical fiber $H$.  This fact, whose
proof we omit, implies that every $q\in G/H$ has a neighbourhood
$\CN_{\al}$, so that
$\ps_{\al}=(\ta,\ps_{\al}^H):\ta\inv(\CN_{\al})\raw \CN_{\al}\x H$ is
a diffeomorphism, which satisfies \be
\ps_{\al}^H(xh)=\ps_{\al}^H(x)h. \ll{mmwd} \ee This leads to a map
$\ps_{\al}^{\ch}:\ta_{\ch}\inv(\CN_{\al})\raw \CN_{\al}\x \Hlg$, given
by $\ps_{\al}^{\ch}([x,v]_H):= ([x]_H, \Ulg(\ps_{\al}^H(x))v)$. This map
is well defined because of (\ref{mmwd}), and is a local trivialization
of $G\x_H \Hlg$.  All required properties are easily checked.  \enp
\su{Hilbert $C^*$-modules\ll{HCM}} What follows generalizes the notion
of a Hilbert bundle in such a way that the commutative \ca\ $C(X)$ is
replaced by an arbitrary \ca\ $\B$. This is an example of the strategy
of non-commutative geometry.
\begin{Definition}\ll{hcm}
A {\bf Hilbert $C^*$-module}\index{Hilbert!$C^*$-module} over a \ca\
$\B$ consists of
\begin{itemize}
\item
A complex linear space $\CE$.
\item
A right-action $\pir$ of $\B$ on $\CE$ (i.e., $\pir$ maps $\B$
linearly into the space of all linear operators on $\CE$, and
satisfies $\pir(AB)= \pir(B)\pir(A)$), for which we shall write $\Ps
B:=\pir(B)\Ps$, where $\Ps\in\CE$ and $B\in\B$.
\item
A sesquilinear map $\la \, ,\,\bb:\CE\x\CE\raw\B$, linear in the
second and anti-linear in the first entry, satisfying \bea \la
\Ps,\Ph\bb^* & = & \la\Ph,\Ps\bb; \ll{hcm1}\\ \la\Ps,\Ph B\bb & = &
\la\Ps,\Ph\bb B; \ll{hcm2}\\ \la\Ps,\Ps\bb & \geq & 0; \ll{hcm3}\\
\la\Ps,\Ps\bb & = & 0 \:\: \Leftrightarrow \Ps=0,\ll{hcm4} \eea for
all $\Ps,\Ph\in\CE$ and $B\in\B$.
\end{itemize} 
The space $\CE$ is complete in the norm \be \n\Ps\n:=
\n\la\Ps,\Ps\bb\n^{\half}.\ll{normincm} \ee We say that $\CE$ is a
{\bf Hilbert $\B$-module}\index{Hilbert!$\B$-module}, and write
$\CE\rlh\B$.
\end{Definition}

One checks that (\ref{normincm}) is indeed a norm: $\n\Ps\n^2$ equals
 $\sup \{\om(\la\Ps,\Ps\bb)\}$, where the supremum is taken over all
 states $\om$ on $\B$. Since each map $\Ps\raw
 \sqrt{\om(\la\Ps,\Ps\bb)}$ is a semi-norm (i.e., a norm except for
 positive definiteness) by (\ref{hcm3}), the supremum is a semi-norm,
 which is actually positive definite because of Lemma \ref{lots} and
 (\ref{hcm4}).

The $\B$-action on $\CE$ is automatically non-degenerate: the property
$\Ps B=0$ for all $B\in\B$ implies that $\la\Ps,\Ps\bb B=0$ for all
$B$, hence $\la\Ps,\Ps\bb=0$ (when $\B$ is unital this is follows by
taking $B=\I$; otherwise one uses an approximate unit in
$\B$), so that $\Ps=0$ by (\ref{hcm4}).

When all conditions in \ref{hcm} are met except (\ref{hcm4}), so that
 $\n\cdot\n$ defined by (\ref{normincm}) is only a semi-norm, one
 simply takes the quotient of $\CE$ by its subspace of all null
 vectors and completes, obtaining a \HCM\ in that way.

It is useful to note that (\ref{hcm1}) and (\ref{hcm2}) imply that \be
\la\Ps B,\Ph\bb=B^*\la\Ps,\Ph\bb. \ll{hcmuseful} \ee
\begin{Example}\ll{hcmex}
\begin{enumerate}
\item
Any \ca\ $\A$ is a $\A$-module $\A\rlh\A$ over itself, with $\la
A,B\ra_{\A}:=A^*B$. Note that the norm (\ref{normincm}) coincides with
the $C^*$-norm by (\ref{cstax2}).
\item
 Any Hilbert space $\H$ is a Hilbert $\C$-module $\H\rlh\C$ in its
inner product.
\item   
Let $\sf H$ be a Hilbert bundle $\sf H$ over a compact Hausdorff space
$X$.  The space of continuous sections $\CE=\Gm({\sf H})$ of $\sf H$
is a Hilbert $C^*$-module $\Gm({\sf H})\rlh C(X)$ over $\B=C(X)$; for
$\Ps,\Ph\in \Gm_0({\sf H})$ the function $\la \Ps,\Ph\ra_{C(X)}$ is
defined by \be \la \Ps,\Ph\ra_{C(X)}: x\raw (\Ps(x),\Ph(x))_x,
\ll{HCMfVB} \ee where the inner product is the one in the fiber
$\ta\inv (x)$. The right-action of $C(X)$ on $\Gm({\sf H})$ is defined
by (\ref{pirpil}).
\end{enumerate}
\end{Example}
In the third example the norm in $\Gm({\sf H})$ is $\n\Ps\n\, = \sup_x
\n \Ps(x)\n$, where $\n \Ps(x)\n=(\Ps(x),\Ps(x))_x^{\half}$, so that
it is easily seen that $\CE$ is complete.

Many \HCM s of interest will be constructed in the following way.
Recall that a pre-$C^*$-algebra is a $\mbox{}^*$-algebra satisfying
all properties of a \ca\ except perhaps completeness.  Given a
pre-$C^*$-algebra $\til{\B}$, define a {\bf pre-Hilbert
$\til{\B}$-module}\index{pre-Hilbert $\til{\B}$-module}
$\til{\CE}\rlh\til{\B}$ as in Definition \ref{hcm}, except that the
final completeness condition is omitted.
\begin{Proposition}\ll{hcmineq}
In a pre-Hilbert $\til{\B}$-module (and hence in a Hilbert
$\B$-module) one has the inequalities \bea \n\Ps B\n & \leq &
\n\Ps\n\,\n B\n; \ll{hce1}\\ \la\Ps,\Ph\bb \la\Ph,\Ps\bb & \leq &
\n\Ph\n^2\, \la\Ps,\Ps\bb; \ll{hce2}\\ \n\la\Ps,\Ph\bb\n & \leq &
\n\Ps\n\,\n\Ph\n. \ll{hce3} \eea \end{Proposition}

To prove (\ref{hce1}) one uses (\ref{hcmuseful}), (\ref{BstABA}),
 (\ref{AleqB}), and (\ref{cstax2}).  For (\ref{hce2}) we substitute
 $\Ph\la\Ph,\Ps\bb-\Ps$ for $\Ps$ in the inequality $\la\Ps,\Ps\bb\geq
 0$. Expanding, the first term equals
 $\la\Ps,\Ph\bb\la\Ph,\Ph\bb\la\Ph,\Ps\bb$. Then use (\ref{BstABA}),
 and replace $\Ph$ by $\Ph/\n\Ph\n$.  The inequality (\ref{hce3}) is
 immediate from (\ref{hce2}).  \enp
\begin{Corollary}\ll{HCMcom}
A pre-Hilbert $\til{\B}$-module $\til{\CE}\rlh\til{\B}$ can be
completed to a Hilbert $\B$-module. \end{Corollary}

 One first completes $\til{\CE}$ in the norm (\ref{normincm}),
obtaining $\CE$. Using (\ref{hce1}), the $\til{\B}$-action on
$\til{\CE}$ extends to a $\B$-action on $\CE$. The completeness of
$\B$ and (\ref{hce3}) then allow one to extend the $\til{\B}$-valued
sesquilinear form on $\til{\CE}$ to a $\B$-valued one on $\CE$. It is
easily checked that the required properties hold by continuity.\enp

In Example \ref{hcmex}, it is almost trivial to see that $\A$ and $\H$
  are the closures of $\til{\A}$ (defined over
 $\til{\A}$) and  of a dense subspace $\CD$,  respectively.

A \HCM\ $\CE\rlh\B$ defines a certain \ca\ $C^*(\CE,\B)$, which plays
an important role in the induction theory in \ref{RI}.  A map
$A:\CE\raw\CE$ for which there exists a map $A^*:\CE\raw\CE$ such that
\be \la \Ps,A\Ph\bb=\la A^*\Ps,\Ph\bb \ll{defadjhcm} \ee for all
$\Ps,\Ph\in\CE$ is called {\bf adjointable}\index{adjointable
operator}.
\begin{Theorem}\ll{csthcm}
 An adjointable map is automatically $\C$-linear, $\B$-linear (that
is, $(A\Ps)B=A(\Ps B)$ for all $\Ps\in\CE$ and $B\in\B$), and bounded.
The adjoint of an adjointable map is unique, and the map $A\raw A^*$
defines an involution on the space $C^*(\CE,\B)$ of all adjointable
maps on $\CE$.

Equipped with this involution, and with the norm (\ref{banalnorm}),
defined with respect to the norm (\ref{normincm}) on $\CE$, the space
$C^*(\CE,\B)$ is a \ca.

Each element $A\in C^*(\CE,\B)$ satisfies the bound \be \la
A\Ps,A\Ps\bb \,\leq\,\n A\n^2\la\Ps,\Ps\bb \ll{rieffelbound} \ee for
all $\Ps\in\CE$.  The (defining) action of $C^*(\CE,\B)$ on $\CE$ is
non-degenerate. We write $C^*(\CE,\B)\raw\CE\rlh\B$.
\end{Theorem}

The property of $\C$-linearity is immediate.  To establish
$\B$-linearity one uses (\ref{hcmuseful}); this also shows that
$A^*\in C^*(\CE,\B)$ when $A\in C^*(\CE,\B)$.

To prove boundedness of a given adjointable map $A$, fix $\Ps\in\CE$
and define $T_{\Ps}:\CE\raw\B$ by $T_{\Ps}\Ph:=\la A^*A\Ps,\Ph\bb$.
It is clear from (\ref{hce3}) that $\n T_{\Ps}\n\,\leq\,\n A^*A\Ps\n$,
so that $T_{\Ps}$ is bounded. On the other hand, since $A$ is
adjointable, one has $T_{\Ps}\Ph=\la \Ps,A^*A\Ph\bb$, so that, using
(\ref{hce3}) once again, one has $\n T_{\Ps}\Ph\n\,\leq\, \n
A^*A\Ph\n\,\n\Ps\n$. Hence $\sup \{\n T_{\Ps}\n\, |\, \n\Ps\n
=1\}<\infty$ by the principle of uniform boundedness (here it is
essential that $\CE$ is complete). It then follows from
(\ref{normincm}) that $\n A\n<\infty$.

Uniqueness and involutivity of the adjoint are proved as for \Hs s;
the former follows from (\ref{hcm4}), the latter in addition requires
(\ref{hcm1}).

The space $C^*(\CE,\B)$ is norm-closed, as one easily verifies from
(\ref{defadjhcm}) and (\ref{normincm}) that if $A_n\raw A$ then
$A^*_n$ converges to some element, which is precisely $A^*$. As a
norm-closed space of linear maps on a Banach space, $C^*(\CE,\B)$ is a
Banach algebra, so that its satisfies (\ref{cstax1}).  To check 
(\ref{cstax2}) one infers from (\ref{normincm}) and the definition
(\ref{defadjhcm}) of the adjoint that $\n A\n^2\,\leq\,\n A^*A\n$;
then use Lemma \ref{ineqcs}.

Finally, it follows from (\ref{hcm3}), (\ref{poscone2}), and
 (\ref{defadjhcm}) that for fixed $\Ps\in\CE$ the map $A\raw
 \la\Ps,A\Ps\bb$ from $C^*(\CE,\B)$ to $\B$ is positive.  Replacing
 $A$ by $A^*A$ in (\ref{orderbound}) and using (\ref{cstax2}) and
 (\ref{defadjhcm}) then leads to (\ref{rieffelbound}).

To prove the final claim, we note that, for fixed $\Ps,\Ph\in\CE$, the
map $Z\raw \Ps\la\Ph,Z\bb$ is in $C^*(\CE,\B)$. When the right-hand
side vanishes for all $\Ps,\Ph$ it must be that $\la\Ph,Z\bb=0$ for
all $\Ph$, hence for $\Ph =Z$, so that $Z=0$. Here we used the fact
that $\Ps B=0$ for all $\Ps$ and $B$ in the linear span of
$\la\CE,\CE\bb$ implies $B=0$, for by (\ref{hcm2}) it implies that
$\la\Ps,\Ps\bb B=0$.  \enp

Under a further assumption (which is by no means always met in our
examples) one can completely characterize $C^*(\CE,\B)$. A \HCM\ over
$\B$ is called {\bf self-dual}\index{self-dual \HCM} when every
bounded $\B$-linear map $\phv:\CE\raw\B$ is of the form
$\phv(\Ps)=\la\Ph,\Ps\bb$ for some $\Ph\in\CE$.
\begin{Proposition}\ll{pa2}
In a self-dual Hilbert $C^*$-module $\CE\rlh\B$ the \ca\ $C^*(\CE,\B)$
coincides with the space ${\cal L}(\CE)^{\B}$ of all bounded
$\C$-linear and $\B$-linear maps on $\CE$.\end{Proposition}

In view of Theorem \ref{csthcm} we only need to show that a given map
  $A\in{\cal L}(\CE)^{\B}$ is adjointable.  Indeed, for fixed
  $\Ps\in\CE$ define $\phv_{A,\Ps}:\CE\raw\B$ by
  $\phv_{A,\Ps}(Z):=\la\Ps,AZ\bb$. By self-duality this must equal
  $\la \Ph,Z\bb$ for some $\Ph$, which by definition is $A^*\Ps$.\enp

In the context of Example \ref{hcmex}.1, one may wonder what
$C^*(\A,\A)$ is.  The map $\rh:\A\raw\B(\A)$ given by (\ref{rhABAB})
is easily seen to map $\A$ into $C^*(\A,\A)$. This map is isometric
(hence injective).  Using (\ref{defadjhcm}), one infers that
$A\rh(B)=\rh(AB)$ for all $A,B\in\A$. Hence $\rh(\A)$ is an ideal in
$C^*(\A,\A)$.  When $\A$ has a unit, one therefore has
$C^*(\A,\A)=\rh(\A)\simeq\A$; cf.\ the proof of \ref{cstunit}.

When $\A$ has no unit, $C^*(\A,\A)$ is the so-called {\bf multiplier
algebra}\index{multiplier algebra} of $\A$.  One may compute this
object by taking a faithful non-degenerate \rep\ $\pi: \A\raw\BH$; it
can be shown that $C^*(\A,\A)$ is isomorphic to the idealizer of
$\pi(\A)$ in $\BH$ (this is the set of all $B\in\BH$ for which
$B\pi(A)\in\pi(\A)$ for all $A\in\A$). One thus obtains \bea
C^*(C_0(X),C_0(X)) & = & C_b(X); \ll{mula1} \\ C^*(\B_0(\H),\B_0(\H))&
= & \B(\H). \ll{mula2} \eea Eq.\ (\ref{mula1}) follows by taking
$\pi(C_0(X))$ to be the \rep\ on $L^2(X)$ by multiplication operators
(where $L^2$ is defined by a measure with support $X$), and
(\ref{mula2}) is obtained by taking $\pi(\KH)$ to be the defining
\rep; see the paragraph following \ref{defb0h}.

In Example \ref{hcmex}.2 the \ca\ $C^*(\H,\C)$ coincides with
$\B(\H)$, because every bounded operator has an adjoint. Its
subalgebra $\B_0(\H)$ of compact operators has an analogue in the
general setting of \HCM s as well.  \su{The \ca\ of a \HCM\ll{qdp}} In
preparation for the imprimitivity theorem, and
also as a matter of independent interest, we introduce the analogue
for \HCM s of the \ca\ $\B_0(\H)$ of compact operators on a \Hs.  This
is the \ca\ most canonically associated to a \HCM.
\begin{Definition}\ll{defcoonhcm}
The \ca\ $C^*_0(\CE,\B)$ of ``compact''operators on a Hil\-bert
$C^*$-module $\CE\rlh\B$ is the $C^*$-subalgebra of $C^*(\CE,\B)$
 generated by the adjointable maps
of the type $T^{\B}_{\Ps,\Ph}$, where $\Ps,\Ph\in\CE$, and \be
T^{\B}_{\Ps,\Ph}Z:=\Ps\la \Ph,Z\bb. \ll{defTPsPh} \ee We write
$C^*_0(\CE,\B)\rlh\CE\rlh\B$, and call this a {\bf dual pair}\index{
dual pair}.
\end{Definition}

The word ``compact'' appears between quotation marks because in
general elements of $C_0^*(\CE,\B)$ need not be compact operators.
The significance of the notation introduced at the end of the
definition will emerge from Theorem \ref{HAM} below.  Using the
(trivially proved) properties \bea (T^{\B}_{\Ps,\Ph})^* & = &
T^{\B}_{\Ph,\Ps}; \ll{TstPsPh}\\ AT^{\B}_{\Ps,\Ph}& = &
T^{\B}_{A\Ps,\Ph}; \\ T^{\B}_{\Ps,\Ph}A & = & T^{\B}_{\Ps,A^*\Ph},
\ll{TBcPsPh} \eea where $A\in C^*(\CE,\B)$, one verifies without
difficulty that $C^*_0(\CE,\B)$ is a (closed 2-sided) ideal in
$C^*(\CE,\B)$, so that it is a \ca\ by Theorem \ref{csthcm}.  From
(\ref{hce1}) and (\ref{hce3}) one finds the bound \be \n
T^{\B}_{\Ps,\Ph}\n\,\leq\,\n\Ps\n\,\n\Ph\n. \ll{TBnineq} \ee One sees
from the final part of the proof of Theorem \ref{csthcm} that
$C^*_0(\CE,\B)$ acts non-degenerately on $\CE$.

When $C_0^*(\CE,\B)$ has a unit it must coincide with $C^*(\CE,\B)$.
\begin{Proposition}\ll{cptAA}
\begin{enumerate}
\item
When $\CE=\B=\A$ (see Example \ref{hcmex}.1) one has \be
C^*_0(\A,\A)\simeq \A.\ll{CstBBisB} \ee This leads to the dual pair
$\A\rlh\A\rlh\A$.
\item
For $\CE=\H$ and $\B=\C$ (see Example \ref{hcmex}.2) one obtains \be
 C_0^*(\H,\C)=\B_0(\H), \ll{C0HCB0H} \ee whence the dual pair
 $\KH\rlh\H\rlh\C$.
\end{enumerate}
\end{Proposition}

One has $T^{\A}_{\Ps,\Ph}=\rh(\Ps\Ph^*)$; see (\ref{rhABAB}).  Since
$\rh:\A\raw\B(\A)$ is an isometric morphism, the map $\phv$ from the
linear span of all $T^{\A}_{\Ps,\Ph}$ to $\A$, defined by linear
extension of $\phv(T^{\A}_{\Ps,\Ph})=\Ps\Ph^*$, is an isometric
morphism as well. It is, in particular, injective. When $\A$ has a
unit it is obvious that $\phv$ is surjective; in the non-unital case
the existence of an approximate unit implies that the linear span of
all $\Ps\Ph^*$ is dense in $\A$.  Extending $\phv$ to $C^*_0(\A,\A)$
by continuity, one sees from Corollary \ref{idmor2} that
$\phv(C^*_0(\A,\A))=\A$.

Eq.\ (\ref{C0HCB0H}) follows from Definition \ref{defb0h} and the
fact that the linear span of all $T^{\C}_{\Ps,\Ph}$ is $\B_f(\H)$.
\enp

A \HCM\ $\CE$ over $\B$ is called {\bf full}\index{full \HCM} when the
collection $\{\la\Ps,\Ph\bb\}$, where $\Ps,\Ph$ run over $\CE$, is
dense in $\B$.  A similar definition applies to pre-\HCM s.
 
Given a complex linear space $\CE$, the conjugate space $\ovl{\CE}$ is
equal to $\CE$ as a real vector space, but has the conjugate action of
complex scalars.
\begin{Theorem}\ll{HAM}
Let $\CE$ be a full Hilbert $\B$-module. The expression \be
\la\Ps,\Ph\ra_{C_0^*(\CE,\B)}:=T^{\B}_{\Ps,\Ph} \ll{Defibi} \ee in
combination with the right-action $\pir(A)\Ps:=A^*\Ps$, where $A\in
C_0^*(\CE,\B)$, defines $\ovl{\CE}$ as a full \HCM\ over
$C_0^*(\CE,\B)$. In other words, from $\CE\rlh\B$ one obtains
$\ovl{\CE}\rlh C_0^*(\CE,\B)$.  The left-action $\pil(B)\Ps:=\Ps B^*$
of $\B$ on $\ovl{\CE}$ implements the isomorphism \be C^*_0(\ovl{\CE},
C_0^*(\CE,\B))\simeq\B.\ll{C0C0Bc} \ee
\end{Theorem}

We call $\A:= C_0^*(\CE,\B)$; in the references to (\ref{hcm1}) etc.\
below one should substitute $\A$ for $\B$ when appropriate.  The
properties (\ref{hcm1}), (\ref{hcm2}), and (\ref{hcm3}) follow from
(\ref{TstPsPh}), (\ref{TBcPsPh}), and Lemma \ref{ptir}, respectively.

To prove (\ref{hcm4}), we use (\ref{Defibi}) with $\Ph=\Ps$,
(\ref{defTPsPh}) with $Z=\Ps$, (\ref{hcm2}), (\ref{hcmuseful}), and
(\ref{normincm}) to show that $\la\Ps,\Ps\ra_{\A}=0$ implies
$\n\la\Ps,\Ps\ra_{\B}^3\n=0$. Since $\la\Ps,\Ps\ra_{\B}$ is positive
by (\ref{hcm3}), this implies $\la\Ps,\Ps\ra_{\B}=0$, hence $\Ps=0$ by
(\ref{hcm4}).

It follows from (\ref{hcmuseful}) and (\ref{TBcPsPh}) that each
$\pil(B)$ is adjointable with respect to $\la\, ,\, \ba$. Moreover,
applying (\ref{normincm}), (\ref{Defibi}), (\ref{TBnineq}), and
(\ref{hce1}) one finds that $\pil(B)$ is a bounded operator on
$\ovl{\CE}$ with respect to $\n\cdot\n_{\A}$, whose norm is majorized
by the norm of $B$ in $\B$. The map $\pil$ is injective because $\CE$
is non-degenerate as a right-$\B$-module.

Let $\ovl{\CE}_c$ be the completion of $\ovl{\CE}$ in
$\n\cdot\n_{\A}$; we will shortly prove that $\ovl{\CE}_c=\ovl{\CE}$.
It follows from the previous paragraph that $\pil(B)$ extends to an
operator on $\ovl{\CE}_c$ (denoted by the same symbol), and that
$\pil$ maps $\B$ into $C^*(\ovl{\CE}_c,\A)$.  It is trivial from its
definition that $\pil$ is a morphism.  Now observe that \be
\pil(\la\Ps,\Ph\bb)=T^{\A}_{\Ps,\Ph}, \ll{piLTA} \ee for the
definitions in question imply that \be
T^{\A}_{\Ps,\Ph}Z=\Ps\la\Ph,Z\ba=T^{\B}_{Z,\Ph}\Ps=Z\la
\Ph,\Ps\bb.\ll{TATBZ} \ee The fullness of $\CE\rlh\B$ and the
definition of $C^*_0(\ovl{\CE}_c,\A)$ imply that $\pil:\B\raw
C^*_0(\ovl{\CE}_c,\A)$ is an isomorphism.  In particular, it is
norm-preserving by Lemma \ref{injmor}.

The space $\CE$ is equipped with two norms by applying
(\ref{normincm}) with $\B$ or with $\A$; we write $\n\cdot\n_{\B}$ and
$\n\cdot\n_{\A}$.  From (\ref{Defibi}) and (\ref{TBnineq}) one derives
\be \n\Ps\n_{\A}\,\leq\,\n \Ps\n_{\B}. \ll{nAnBineq} \ee

For $\Ps\in\CE$ we now use (\ref{normincm}), the isometric nature of
$\pil$, and (\ref{piLTA}) to find that 
\be
\n\Ps\n_{\B}= \n T^{\A}_{\Ps,\Ps}\n^{\half}.
\ee 
 From (\ref{TBnineq}) with $\B\raw\A$ one
then derives the converse inequality to (\ref{nAnBineq}), so that
$\n\Ps\n_{\A}=\n \Ps\n_{\B}$.  Hence $\ovl{\CE}_c=\ovl{\CE}$, as $\CE$
is complete in $\n\cdot\n_{\B}$ by assumption.  The completeness of
$\CE$ as a Hilbert $\B$-module is equivalent to the completeness of
$\ovl{\CE}$ as a Hilbert $\A$-module.

We have now proved (\ref{C0C0Bc}). Finally noticing that as a \HCM\
over $\A$ the space $\ovl{\CE}$ is full by definition of $
C_0^*(\CE,\B)$, the proof of Theorem \ref{HAM} is ready.  \enp

For later reference we record the remarkable identity \be
\la\Ps,\Ph\ra_{C^*_0(\CE,\B)} Z= \Ps\la\Ph,Z\bb, \ll{bimpr0} \ee which
is a restatement of (\ref{TATBZ}).  \su{Morita equivalence\ll{qme}}
The imprimitivity theorem establishes an isomorphism between the
respective \rep\ theories of two \ca s that stand in a certain
equivalence relation to each other.
\begin{Definition}\ll{defsmeq}
Two \ca s $\A$ and $\B$ are {\bf
Morita-equivalent}\index{Morita-equivalence! } when there exists a
full \HCM\ $\CE\rlh\B$ under which $\A\simeq C_0^*(\CE,\B)$. We
write $\A\Meq\B$ and $\A\rlh\CE\rlh\B$.\end{Definition}
\begin{Proposition}\ll{qmeiser}
Morita equivalence is an equivalence relation in the class of all \ca
s.
\end{Proposition}

The reflexivity property $\B\Meq\B$ follows from (\ref{CstBBisB}),
which establishes the dual pair $\B\rlh\B\rlh\B$.  Symmetry is implied
by (\ref{C0C0Bc}), proving that $\A\rlh\CE\rlh\B$ implies
$\B\rlh\ovl{\CE}\rlh\A$.

The proof of transitivity is more involved.  When $\A\Meq\B$ and
$\B\Meq\GC$ we have the chain of dual pairs
$$
\A\rlh\CE_1\rlh\B\rlh\CE_2\rlh\GC.
$$
We then form the linear space $\CE_1\ot_{\B}\CE_2$ (which is the
quotient of $\CE_1\ot\CE_2$ by the ideal $\CI_{\B}$ generated by all
vectors of the form $\Ps_1 B\ot\Ps_2-\Ps_1\ot B\Ps_2$), which carries
a right-action $\pir^{\ot}(\GC)$ given by \be
\pir^{\ot}(C)(\Ps_1\ot_{\B}\Ps_2):= \Ps_1\ot_{\B}(\Ps_2 C).\ll{piTor}
\ee Moreover, we can define a sesquilinear map $\la\,
,\,\ra^{\ot}_{\GC}$ on $\CE_1\ot_{\B}\CE_2$ by \be \la
\Ps_1\ot_{\B}\Ps_2,\Ph_1\ot_{\B}\Ph_2\ra^{\ot}_{\GC}:=
\la\Ps_2,\la\Ps_1,\Ph_1\bb\Ph_2\ra_{\GC}. \ll{PsBcPs} \ee With
(\ref{piTor}) this satisfies (\ref{hcm1}) and (\ref{hcm2}); as
explained prior to (\ref{hcmuseful}), one may therefore construct a
\HCM, denoted by $\CE_{\ot}\rlh\GC$. (Remarkably, if one looks at
(\ref{PsBcPs}) as defined on $\CE_1\ot\CE_2$, the null space of
(\ref{normincm}) is easily seen to contain $\CI_{\B}$, but in fact
coincides with it, so that in constructing $\CE_{\ot}$ one only needs
to complete $\CE_1\ot_{\B}\CE_2$.)

Apart from the right-action $\pir^{\ot}(\GC)$, the space $\CE_{\ot}$
carries a left-action $\pi^{\ot}_L(\A)$: the operator \be
\pil^{\ot}(A)(\Ps_1\ot_{\B}\Ps_2):= (A\Ps_1)\ot_{\B}\Ps_2 \ll{piTor2}
\ee is bounded on $\CE_1\ot_{\B}\CE_2$ and extends to $\CE_{\ot}$.  We
now claim that \be C^*_0(\CE_{\ot},\GC)=\pi^{\ot}_L(\A).  \ee Using
(\ref{defTPsPh}), the definition of $\ot_{\B}$, and (\ref{hcm2}), it
is easily shown that \be
\pil^{\ot}(T^{\B}_{\Ps_1\la\Ps_2,\Ph_2\bb,\Ph_1})\Om_1\ot_{\B}\Om_2=
\Ps_1\ot_{\B}\la\Ps_2,\Ph_2\la\Ph_1,\Om_1\bb\bb\Om_2. \ll{difc1} \ee
Now use the assumption $C_0^*(\CE_2,\GC)=\B$; as in (\ref{Defibi}),
with $\B\raw\GC$, and $\CE\raw\CE_2$, this yields
$\la\Ps,\Ph\bb=T^{\GC}_{\Ps,\Ph}$.  Substituting this in the
right-hand side of (\ref{difc1}), and using (\ref{defTPsPh}) with
$\B\raw\GC$, the right-hand side of (\ref{difc1}) becomes
$\Ps_1\ot_{\B}\Ps_2\la\Ph_2\la\Ph_1,\Om_1\bb,\Om_2\ra_{\GC}$.  Using
$\Ps B^*=\pi_L(B)\Ps$ (see \ref{HAM}), (\ref{defadjhcm}) with
$\B\raw\GC$, (\ref{PsBcPs}), and (\ref{defTPsPh}) with $\B\raw\GC$, we
eventually obtain \be T^{\GC}_{\Ps_1\ot_{\B}\Ps_2,\Ph_1\ot_{\B}\Ph_2}
=\pil^{\ot}(T^{\B}_{\Ps_1\la\Ps_2,\Ph_2\bb,\Ph_1}). \ll{eqfrom2in} \ee
This leads to the inclusion
$C^*_0(\CE_{\ot},\GC)\subseteq\pi^{\ot}_L(\A)$.  To prove the opposite
inclusion, one picks a double sequence $\{\Ps_2^i,\Ph_2^i\}$ such that
$\sum_i^N T^{\GC}_{\Ps_2^i,\Ph_2^i}$ is an approximate unit in
$\B=C^*_0(\CE_2,\GC)$.  One has $\lim_N \sum_i^N
\Ps^i_2\la\Ph^i_2,Z\ra_{\GC}=Z$ from (\ref{defTPsPh}), and a short
computation using (\ref{defTPsPh}) with (\ref{PsBcPs}) then yields
$$
\lim_N \sum_i^N T^{\GC}_{\Ps_1\ot_{\B}\Ps^i_2,\Ph_1\ot_{\B}\Ph^i_2}
=\pil^{\ot}(T^{\B}_{\Ps_1,\Ph_1}).
$$
Hence $\pi^{\ot}_L(\A)\subseteq C^*_0(\CE_{\ot},\GC)$, and combining
both inclusions one finds (\ref{eqfrom2in}).

Therefore, one has the dual pair $\A\rlh\CE_{\ot}\rlh\GC$, implying
that $\A\Meq\GC$. This proves transitivity.\enp

Here is a simple example of this concept.
\begin{Proposition}\ll{cptmec}
The \ca\ $\B_0(\H)$ of compact operators is Morita-equivalent to $\C$,
  with dual pair $\B_0(\H)\rlh\H\rlh\C$.  In particular, the matrix
  algebra $\M^n(\C)$ is Morita-equivalent to $\C$.\end{Proposition}

This is immediate from (\ref{C0HCB0H}).  In the finite-dimensional
case one has $\M^n(\C)\rlh\C^n\rlh\C$, where $\M^n(\C)$ and $\C$ act
on $\C^n$ in the usual way. The double \HCM\ structure is completed by
specifying \bea \la z,w\ra_{\C} & = & \ovl{z}^i w^i; \nn \\ (\la
z,w\ra_{\M^n(\C)})_{ij} & = & z^i\ovl{w}^j, \ll{CKHfd} \eea from which
one easily verifies (\ref{bimpr0}). \enp

Since $\M^n(\C)\Meq \C$ and $\C\Meq\M^m(\C)$, one has
$\M^n(\C)\Meq\M^m(\C)$.  This equivalence is implemented by the dual
pair $\M^n(\C)\rlh \M^{n\x m}(\C)\rlh\M^m(\C)$, where $\M^{n\x m}(\C)$
is the space of complex matrices with $n$ rows and $m$ columns. We
leave the details as an exercise.

In practice the following way to construct dual pairs, and therefore
Morita equivalences, is useful.
\begin{Proposition}\ll{riefbimod}
Suppose one has
\begin{itemize}
\item
two pre-\ca s $\til{\A}$ and $\til{\B}$;
\item
 a full pre-Hilbert $\til{\B}$-module $\til{\CE}$;
\item
a left-action of $\til{\A}$ on $\til{\CE}$, such that
 $\ovl{\til{\CE}}$ can be made into a full pre-Hilbert
 $\til{\A}$-module with respect to the right-action
 $\pir(A)\Ps:=A^*\Ps$;
\item
the identity \be \la\Ps,\Ph\ra_{\til{\A}} Z=\Ps\la\Ph,Z\ra_{\til{\B}}
\ll{bimpr} \ee (for all $\Ps,\Ph,Z\in\til{\CE}$) relating the two
\HCM\ structures;
\item
the bounds \bea \la \Ps B,\Ps B \ra_{\til{\A}} & \leq & \n B\n^2\la
\Ps,\Ps\ra_{\til{\A}}; \ll{rieffelbound22}\\ \la A\Ps,A\Ps
\ra_{\til{\B}} & \leq & \n A\n^2\la \Ps,\Ps\ra_{\til{\B}}
\ll{rieffelbound3} \eea for all $A\in\til{\A}$ and $B\in\til{\B}$.
\end{itemize}

Then $\A\Meq\B$, with dual pair $\A\rlh\CE\rlh\B$, where $\CE$ is the
 completion of $\til{\CE}$ as a Hilbert $\B$-module.
\end{Proposition}

Using Corollary \ref{HCMcom} we first complete $\til{\CE}$ to a
Hilbert $\B$-module $\CE$.  By (\ref{rieffelbound3}), which implies
$\n A\Ps\n\,\leq\,\n A\n\,\n\Ps\n$ for all $A\in\til{\A}$ and
$\Ps\in\til{\CE}$, the action of $\til{\A}$ on $\til{\CE}$ extends to
an action of $\A$ on $\CE$.  Similarly, we complete $\ovl{\til{\CE}}$
to a Hilbert $\A$-module $\ovl{\CE}_c$; by (\ref{rieffelbound22}) the
left-action $\pil(B)\Ps:=\Ps B^*$ extends to an action of $\B$ on
$\ovl{\CE}_c$.  As in the proof of Theorem \ref{HAM}, one derives
(\ref{nAnBineq}) and its converse for $\Ps\in\til{\CE}$, so that the
$\B$-completion $\CE$ of $\til{\CE}$ coincides with the
$\A$-completion $\ovl{\CE}_c$ of $\ovl{\til{\CE}}$; that is,
$\ovl{\CE}_c=\ovl{\CE}$.

Since $\ovl{\CE}$ is a full pre-Hilbert $\til{\A}$-module, the
$\A$-action on $\CE$ is injective, hence faithful. It follows from
(\ref{bimpr}), Theorem \ref{HAM}, and (once again) the fullness of
$\ovl{\CE}$, that $\A\simeq C^*_0(\CE,\B)$.  In particular, each
$A\in\A$ automatically satisfies (\ref{defadjhcm}).\enp

Clearly, (\ref{bimpr}) is inspired by (\ref{bimpr0}), into which it is
turned after use of this proposition. We will repeatedly use
Proposition \ref{riefbimod} in what follows; see \ref{isoCstGG}
and \ref{macglimm}.
\su{Rieffel induction\ll{RI}}
To formulate and prove the imprimitivity theorem we need a basic
technique, which is of interest also in a more general context.
 Given a Hilbert $\B$-module $\CE$, the
goal of the {\bf Rieffel induction} procedure described in this section is to
construct a \rep\ $\pug$ of $C^*(\CE,\B)$ from a \rep\ $\plg$ of $\B$.
In order to explicate that the induction procedure is a generalization
of the GNS-construction \ref{GNSconstruction}, we first induce from
a state $\om_{\ch}$ on $\B$, rather than from a \rep\ $\plg$.
\begin{Construction}\ll{RIC}
Suppose one has a \HCM\ $\CE\rlh\B$.
\begin{enumerate}
\item
   Given a state $\om_{\ch}$ on $\B$, define the sesquilinear form
$\wt{(\, ,\,)}_0^{\ch}$ on $\CE$ by \be
\wt{(\Ps,\Ph)}_0^{\ch}:=\om_{\ch}(\la\Ps,\Ph\bb).\ll{PsPh0} \ee Since
$\om_{\ch}$ and $\la\, ,\bb$ are positive (cf.\ (\ref{hcm3})), this
form is positive semi-definite.  Its null space is \be
\til{\CN}_{\ch}=\{\Ps\in\CE\,|\,
\wt{(\Ps,\Ps)}_0^{\ch}=0\}. \ll{rief0} \ee
\item
 The form $\wt{(\, ,\,)}_0^{\ch}$ projects to an inner product
$\wt{(\, ,\,)}^{\ch}$ on the quotient $\CE/\til{\CN}_{\ch}$. If
$\til{V}_{\ch}:\CE\rightarrow \CE/\til{\CN}_{\ch}$ is the canonical
projection, then by definition \be
\wt{(\til{V}_{\ch}\Ps,\til{V}_{\ch}\Ph)}^{\ch}:=\wt{(\Ps,\Ph)}_0^{\ch}. \ll{rief02}
\ee The \Hs\ $\til{\H}^{\ch}$ is the closure of $\CE/\til{\CN}_{\ch}$
in this inner product.
\item
 The \rep\ $\til{\pi}^{\ch}(C^*(\CE,\B))$ is firstly defined on $\CE
/\til{\CN}_{\ch}\subset \til{\H}^{\ch}$ by \be \pi^{\ch}(A)
\til{V}_{\ch}\Ps:= \til{V}_{\ch}A \Ps; \ll{rindeq} \ee it follows that
$\til{\pi}^{\ch}$ is continuous.  Since $\CE/\til{\CN}_{\ch}$ is dense
in $\til{\H}^{\ch}$, the operator $\til{\pi}^{\ch}(A)$ may be defined
on all of $\til{\H}^{\ch}$ by continuous extension of (\ref{rindeq}),
where it satisfies (\ref{repofca}).
\end{enumerate} 
\end{Construction}

 The GNS-construction \ref{GNSconstruction} is a special case of
\ref{RIC}, obtained by choosing $\CE=\B=\A$, as explained in Example
\ref{hcmex}.1.

The analogue of (\ref{twoformsno}) of course applies here.  The
continuity of $\til{\pi}^{\ch}$ follows from (\ref{rindeq}) and
(\ref{rief02}), which imply that
$\n\til{\pi}^{\ch}(A)\til{V}_{\ch}\Ps\n^2=\wt{(A\Ps,A\Ps)}^{\ch}_0$. Using
(\ref{PsPh0}), (\ref{rieffelbound}), and (\ref{hce3}) in succession,
one finds that \be \n\til{\pi}^{\ch}(A)\n\,\leq\,\n A\n. \ll{npugAn0}
\ee On the other hand, from the proof of Theorem \ref{GNT} one sees
that \be \n A\n^2 = \sup \{|\om(A^*A)|\: \; |\, \om\in\SA\}. \ll{nAomA} \ee
  Applying
(\ref{nAomA}) to $\B$, used with the definition of $\n A\n$ for $A\in
C^*(\CE,\B)$, implies that \be \n A\n =\sup \{ \n
\til{\pi}^{\ch}(A)\n,\, \om_{\ch}\in {\cal S}(\B)\}.  \ll{npugAn} \ee
A similar argument combined with Corollary \ref{normunique} shows that
$\til{\pi}^{\ch}$ is faithful (hence norm-preserving) when $\om_{\ch}$
is.  As a corollary, one infers a useful property, which will be used,
e.g., in the proof of Theorem \ref{HAM}.
\begin{Lemma}\ll{ptir}
Let $A\in C^*(\CE,\B)$ satisfy $\la\Ps,A\Ps\bb\geq 0$ for all
$\Ps\in\CE$.  Then $A\geq 0$.\end{Lemma}

Take a faithful state $\om_{\ch}$ on $\B$; the condition implies that
$\til{\pi}^{\ch}(A)\geq 0$.\enp

When one starts from a \rep\ $\plg(\B)$ rather than from a state, the
general construction looks as follows.
\begin{Construction}\ll{rieind2}
Start from a \HCM\ $\CE\rlh\B$.
\begin{enumerate}
\item
Given a \rep\ $\plg(\B)$ on a \Hs\ $\H_{\ch}$, with inner product $(\,
,\,)_{\ch}$, the sesquilinear form $(\, ,\,)_0^{\ch}$ is defined on
$\CE\ot\Hlg$ (algebraic tensor product) by sesquilinear extension of
\be (\Ps\ot v,\Ph\ot w)_0^{\ch}:= (v,\plg(\la\Ps,\Ph\bb)
w)_{\ch},\ll{Psotv} \ee where $v,w\in\Hlg$.  This form is positive
semi-definite, because $(\, ,\,)_{\ch}$ and $\la\, ,\bb$ are.  The
null space is \be \CN_{\ch}=\{\til{\Ps}\in\CE\ot\Hlg|\,
(\til{\Ps},\til{\Ps})_0^{\ch}=0\}.  \ll{rief0b} \ee As in
(\ref{twoformsno}), we may equally well write \be
\CN_{\ch}=\{\til{\Ps}\in\CE\ot\Hlg|\,
(\til{\Ps},\til{\Ph})_0^{\ch}=0\: \forall\, \til{\Ph}\in\CE\ot\Hlg\}.
\ll{rief0c} \ee
\item
 The form $(\, ,\,)_0^{\ch}$ projects to an inner product $(\,
 ,\,)^{\ch}$ on the quotient $\CE\ot\Hlg/\CN_{\ch}$, defined by \be
 (V_{\ch}\til{\Ps},V_{\ch}\til{\Ph})^{\ch}
 :=(\til{\Ps},\til{\Ph})_0^{\ch},\ll{rief02b} \ee where
 $V_{\ch}:\CE\ot\Hlg\rightarrow \CE\ot\Hlg/\CN_{\ch}$ is the canonical
 projection.  The \Hs\ $\Hug$ is the closure of $\CE\ot\Hlg/\CN_{\ch}$
 in this inner product.
\item
 The \rep\ $\pug(C^*(\CE,\B))$ is then defined on $\Hug$ by continuous
extension of \be \pug(A)V_{\ch}\til{\Ps}:=
V_{\ch}(A\ot\I_{\ch}\til{\Ps} ), \ll{deftilpich} \ee where $\I_{\ch}$
is the unit operator on $\Hlg$; the extension in question is possible,
since \be \n\pug(A)\n\,\leq\,\n A\n. \ll{riefinorm} \ee
\end{enumerate}
\end{Construction}

To prove that the form defined in (\ref{Psotv}) is positive
semi-definite, we assume that $\plg(\B)$ is cyclic (if not, the
argument below is repeated for each cyclic summand; see
\ref{nondegcyclic}).  With $\til{\Ps}=\sum_i\Ps_i v_i$ and
$v_i=\plg(B_i)\Om$ (where $\Om$ is a cyclic vector for $\plg(\B)$),
one then uses (\ref{Psotv}), (\ref{hcmuseful}), and (\ref{hcm2}) to
find $(\til{\Ps},\til{\Ps})_0^{\ch}=(v,\plg(\la \Ph,\Ph\bb)v)_{\ch}$
with $\Ph:=\sum_i \Ps_i B_i$. Hence $(\til{\Ps},\til{\Ps})_0^{\ch}\geq
0$ by (\ref{hcm2}) and the positivity of $\plg:\B\raw\B(\Hlg)$.

Similarly, one computes $\n \pug(A)V_{\ch}\til{\Ps}\n^2=(v,\plg(\la
A\Ph,A\Ph\bb)v)_{\ch}$ from (\ref{rief02b}) and (\ref{deftilpich});
according to (\ref{rieffelbound}) and the property
$\n\plg(A)\n\,\leq\,\n A\n$ (cf.\ the text after \ref{repofca}),
this is bounded by $\n A\n^2 (v,\plg(\la \Ph,\Ph\bb)v)_{\ch}$.  Since
the second factor equals $\n V_{\ch}\til{\Ps}\n^2$, this proves
(\ref{riefinorm}).\enp

 Paraphrasing the comment after the first version of the construction,
 $\pug$ is faithful when $\plg$ is.  Also, it is not difficult to
 verify that $\pug$ is non-degenerate when $\plg$ is.

To interrelate the above two formulations, one assumes that
$\pi_{\ch}$ is cyclic, with cyclic vector $\Om_{\ch}$. Then define a
linear map $\til{U}:\CE\raw\CE\ot\Hlg$ by \be \til{U}\Ps:=\Ps\ot
\Om_{\ch}.  \ee According to (\ref{PsPh0}), (\ref{Psotv}), and
(\ref{gnscruceq}), this map has the property \be
(\til{U}\Ps,\til{U}\Ph)_0^{\ch}=\wt{(\til{\Ps},\til{\Ph})}_0^{\ch}. \ll{vvuseful}
\ee By (\ref{rief02}) and (\ref{rief02b}) the map $\til{U}$ therefore
quotients to a unitary isomorphism $U:\til{\H}^{\ch}\raw\Hug$, which
by (\ref{rindeq}) and (\ref{deftilpich}) duly intertwines
$\til{\pi}^{\ch}$ and $\pug$.

Of course, any subspace of $C^*(\CE,\B)$ may be subjected to the
induced \rep\ $\pug$. This particularly applies when one has a given
(pre-) \ca\ $\A$ and a $\mbox{}^*$-homomorphism $\pi:\A\raw
C^*(\CE,\B)$, leading to the \rep\ $\pug(\A)$ on $\Hug$. Further to an
earlier comment, one verifies that $\pug$ is non-degenerate when $\pi$
and $\plg$ are.  With slight abuse of notation we will write $\pug(A)$
for $\pug(\pi(A))$.  The situation is depicted in Figure \ref{figri}.

\begin{figure}
\begin{center}
\begin{picture}(8,3.5)
\put(0,2.5){$\A$} \put(2.5,2.5){$\CE$} \put(5,2.5){$\B$}
\put(7.5,2.5){$\Hlg$} \put(5,0){$\Hug$}
\put(5.1,2){\vector(0,-1){1.5}} \put(0.8,2.6){\vector(1,0){1.3}}
\put(1,2){\vector(2,-1){3}} \put(3.5,2.6){$\rlh$}
\put(5.8,2.6){\vector(1,0){1.3}} \put(5.3,1.25){induction}
\put(1.25,3){$\pi$} \put(6.25,3){$\plg$} \put(1.8,.8){$\pug$}
\end{picture}
\end{center}
\caption{Rieffel induction} \ll{figri}
\end{figure}
\su{The imprimitivity theorem}\ll{impthmr} After this preparation, we
pass to the {\bf imprimitivity theorem}.
\begin{Theorem}\ll{riimp} There is a bijective correspondence
between the non-dege\-ne\-ra\-te \rep s of Morita-equivalent \ca s
$\A$ and $\B$, preserving direct sums and irreducibility. This
correspondence is as follows.

Let the pertinent dual pair be $\A\rlh\CE\rlh\B$.  When
$\pi_{\sg}(\A)$ is a \rep\ on a \Hs\ $\H_{\sg}$ there exists a \rep\
$\plg(\B)$ on a \Hs\ $\Hlg$ such that $\pi_{\sg}$ is equivalent to the
Rieffel-induced \rep\ $\pug$ defined by (\ref{deftilpich}) and the
above dual pair.

In the opposite direction, a given \rep\ $\plg(\B)$ is equivalent to
   the Rieffel-induced \rep\ $\pi^{\sg}$, defined with respect to some
   \rep\ $\pi_{\sg}(\A)$ and the dual pair $\B\rlh\ovl{\CE}\rlh\A$.

Taking $\pi_{\sg}(\A)=\pi^{\ch}(\A)$ as just defined, one has
$\pi^{\sg}(\B)\simeq \pi_{\ch}(\B)$. Conversely, taking
$\pi_{\ch}(\B)=\pi^{\sg}(\B)$, one has $\pi^{\ch}(\A)\simeq
\pi_{\sg}(\A)$.\end{Theorem}

 See Figure \ref{rifig}.
\begin{figure}
\begin{center}
\begin{picture}(8,5.5)
\put(0,5){$\A$} \put(2.5,5){$\CE$} \put(5,5){$\B$} \put(7.5,5){$\Hlg$}
\put(0,2.5){$\B$} \put(2.5,2.5){$\ovl{\CE}$} \put(5,2.5){$\A$}
\put(7.5,2.5){$\Hug:=\H_{\sg}$} \put(1.3,5.1){$\rlh$}
\put(3.8,5.1){$\rlh$} \put(5.6,5.1){\vector(1,0){1.5}}
\put(1.3,2.6){$\rlh$} \put(3.8,2.6){$\rlh$}
\put(5.6,2.6){\vector(1,0){1.5}} \put(5.5,4.5){\vector(1,-1){1.5}}
\put(6.25,5.5){$\plg$} \put(6,3){$\pug$} \put(5,0){$\B$}
\put(7.5,0){$\H^{\sg}$} \put(5.5,2){\vector(1,-1){1.5}}
\put(6,.5){$\pi^{\sg}$} \put(5.6,.1){\vector(1,0){1.5}}
\end{picture}
\end{center}
\caption{Quantum imprimitivity theorem: $\H^{\sg}\simeq \H_{\ch}$ and
$\pi^{\sg}\simeq \plg$} \ll{rifig}
\end{figure}
  Starting with $\plg(\B)$, we construct $\pug(\A)$ with Rieffel
induction from the dual pair $\A\rlh\CE\rlh\B$, relabel this \rep\ as
$\pi_{\sg}(\A)$, and move on to construct $\pi^{\sg}(\B)$ from Rieffel
induction with respect to the dual pair $\B\rlh\ovl{\CE}\rlh\A$.  We
then construct a unitary map $U:\H^{\sg}\raw\Hlg$ which intertwines
$\pi^{\sg}$ and $\pi_{\ch}$.

We first define $\til{U}:\ovl{\CE}\ot\CE\ot\Hlg\raw\Hlg$ by linear
extension of \be \til{U}\Ps\ot\Ph\ot
v:=\plg(\la\Ps,\Ph\bb)v. \ll{tilUPs} \ee Note that $\til{U}$ is indeed
$\C$-linear.  Using (\ref{tilUPs}), the properties (\ref{phvstar})
and (\ref{phvmult}) with $\phv\raw\plg$, (\ref{Psotv}), and
(\ref{defTPsPh}), one obtains \be (\til{U}\Ps_1\ot\Ph_1\ot
v_1,\til{U}\Ps_2\ot\Ph_2\ot v_2)_{\ch} = (\Ph_1\ot
v_1,T^{\B}_{\Ps_1,\Ps_2}\Ph_2\ot v_2)^{\ch}_0. \ll{nokmm} \ee Now use
the assumption $\A=C^*_0(\CE,\B)$ to use (\ref{bimpr0}), and
subsequently (\ref{rief02b}) and (\ref{deftilpich}), all read from
right to left. The right-hand side of (\ref{nokmm}) is then seen to be
equal to $(V_{\ch}\Ph_1\ot v_1,\pug(\la\Ps_1,\Ps_2\ra_{\A})
V_{\ch}\Ph_2\ot v_2)^{\ch}$. Now put $\pug=\pi_{\sg}$ and
$\Hug=\H_{\sg}$, and use (\ref{Psotv}) and (\ref{rief02b}) from right
to left, with $\ch\raw\sg$. This shows that \be
(\til{U}\Ps_1\ot\Ph_1\ot v_1,\til{U}\Ps_2\ot\Ph_2\ot v_2)_{\ch}=
(V_{\sg}(\Ps_1\ot V_{\ch}\Ph_1\ot v_1), V_{\sg}(\Ps_2\ot
V_{\ch}\Ph_2\ot v_2))^{\sg}.  \ee In particular, $\til{U}$ annihilates
$\Ps\ot \til{\Ph}$, where $\til{\Ph}\in \CE\ot\Hlg$, whenever
$\til{\Ph}\in\CN_{\ch}$ or $\Ps\ot
V_{\ch}\til{\Ph}\in\CN_{\sg}$. Hence we see from the construction
firstly of $\Hug=\H_{\sg}$ from $\CE\ot\Hlg$, and secondly of
$\H^{\sg}$ from $\ovl{\CE}\ot\H_{\sg}$ (cf.\ \ref{rieind2}), that
$\til{U}$ descends to an isometry $U:\H^{\sg}\raw\Hlg$, defined by
linear extension of \be UV_{\sg}(\Ps\ot V_{\ch}\Ph\ot
v):=\til{U}\Ps\ot\Ph\ot v=\plg(\la\Ps,\Ph\bb)v.  \ll{defUUtil} \ee

Using the assumptions that the \HCM\ $\CE\rlh\B$ is full and that the
\rep\ $\plg(\B)$ is non-degenerate, we see that the range of $\til{U}$
and hence of $U$ is dense in $\Hlg$, so that $U$ is unitary.

To verify that $U$ intertwines $\pi^{\sg}$ and $\pi_{\ch}$, we use
(\ref{defUUtil}) and (\ref{deftilpich}), with $\ch\raw\sg$, to compute
\be U \pi^{\sg}(B)V_{\sg}(\Ps\ot V_{\ch}\Ph\ot v)=\plg(\la
\pi_L(B)\Ps,\Ph\bb)v, \ll{nokmm3} \ee where the left-action of
$B\in\B$ on $\Ps\in\ovl{\CE}$ is as defined in \ref{HAM}. Thus writing
$\pi_L(B)\Ps=\Ps B^*$, using (\ref{hcmuseful}), (\ref{phvmult}) with
$\phv\raw\plg$, and (\ref{defUUtil}) from right to left, the
right-hand side of (\ref{nokmm3}) is seen to be
$\plg(B)UV_{\sg}(\Ps\ot V_{\ch}\Ph\ot v)$. Hence $U \pi^{\sg}(B)=
\plg(B)U$ for all $B\in\B$.

Using the proof that the Morita equivalence relation is symmetric (see
\ref{qmeiser}), one immediately sees that the construction works in
the opposite direction as well.

It is easy to verify that $\plg=\pi_{\ch^1}\oplus \pi_{\ch^2}$ leads
to $\pug=\pi^{\ch^1}\oplus \pi^{\ch^2}$. This also proves that the
bijective correspondence $\plg(\B)\lraw\pug(\A)$ preserves
irreducibility: when $\plg$ is irreducible and $\pug$ isn't, one puts
$\pug=\pi_{\sg}$ as above, decomposes $\pi_{\sg}= \pi_{\sg^1}\oplus
\pi_{\sg^2}$, then decomposes the induced \rep\ $\pi^{\sg}(\B)$ as
$\pi^{\sg}=\pi^{\sg^1}\oplus \pi^{\sg^2}$, and thus arrives at a
contradiction, since $\pi^{\sg}\simeq \plg$.  \enp
 
Combined with Proposition \ref{cptmec}, this theorem leads to a new
proof of Corollary \ref{1irrepkh}.
\su{Group $C^*$-algebras\ll{groupcas}}
In many interesting applications, and also in the theory of induced
\rep\ as originally formulated for groups by Frobenius and Mackey, the
\ca\ $\B$ featuring in the definition of a \HCM\ and in Rieffel induction
is a so-called group \ca. 

We start with the definition of the group algebra $C^*(G)$ of a finite group
with $n(G)$ elements;
one then usually writes $\C(G)$ instead of $C^*(G)$.
As a vector space, $\C(G)$ consist of all complex-valued functions on $G$,
so that $\C(G)=\C^{n(G)}$. 
This is made into a \sta\ by the {\bf convolution}
\be
f*g(x):=\sum_{y,z\in G|yz=x} f(y)g(z) \ll{cfg}
\ee
and the involution
\be
f^*(x):=\ovl{f(x\inv)}. \ll{definvga}
\ee
It is easy to check that the multiplication $*$ is associative as a consequence
of the associativity of the product in $G$. In similar vein, the operation
defined by (\ref{definvga}) is an involution because of the properties
$(x\inv)\inv=x$ and $(xy)\inv=y\inv x\inv$ at the group level.

 A \rep\ $\pi$ of $\C(G)$  on a \Hs\
$\H$ is defined  as a morphism $\pi:\C(G)\raw \BH$. 
\begin{Proposition}\ll{Upithmf}
There is a bijective correspondence between non-degene\-ra\-te \rep s
$\pi$ of the $\mbox{}^*$-algebra $\C(G)$ 
 and unitary \rep s $U$ of $G$, which preserves unitary equivalence and
direct sums (and therefore preserves irreducibility).  This
correspondence is given in one direction by 
\be
\pi(f):=\sum_{x\in G} f(x)U(x) , \ll{Utopif}
\ee 
and in the other by
\be
U(x):=\pi(\dl_x), \ll{pitoUf}
\ee
where $\dl_x(y):=\dl(xy)$.
\end{Proposition}

It is elementary to verify that $\pi$ is indeed a \rep\ of $\C(G)$ when
$U$ is a unitary \rep\ of $G$, and {\em vice versa}.
Putting $x=e$ in (\ref{pitoUf}) yields $\pi(\dl_e)=\I$, so that $\pi$ 
cannot be degenerate. 

When $U_1(x)=VU_2(x)V^*$ for all $x\in G$ then evidently 
$\pi_1(f)=V\pi_2(f)V^*$ for all $f\in\C(G)$. The converse follows by choosing
$f=\dl_x$. Similarly, $\pi(f)=\pi_1(f)\oplus \pi_2(f)$ for all $f$ iff
$U(x)=U_1(x)\oplus U_2(x)$ for all $x$.
\enp

We can define a $C^*$-norm on $\C(G)$ by taking any faithful \rep\ $\pi$,
and putting $\n f\n:=\n\pi(f)\n$. Since $\C(G)$ is a finite-dimensional
vector space it is complete in this norm, which therefore 
is independent of the choice of $\pi$ by Corollary \ref{normunique}.

Let now $G$ be an arbitrary locally compact group (such as a
finite-dimensional Lie group). We also assume that $G$ is {\bf
unimodular}\index{unimodular}; that is, each left Haar measure is also
right-invariant. This assumption is not necessary, but simplifies most
of the formulae.  We denote Haar measure by $dx$; it is unique up to
normalization. Unimodularity implies that the Haar measure is invariant
under inversion $x\raw x\inv$.
When $G$ is compact we choose the normalization so that
$\int_G dx=1$.  The Banach space $L^1(G)$ and the Hilbert space
$L^2(G)$ are defined with respect to the Haar measure.

The {\bf  convolution product} is defined, initially on $C_c(G)$, by 
\be
f*g(x):=\int_G dy\, f(xy\inv)g(y); \ll{defconv}
\ee
it is evident that for a finite group this expression  specializes to 
(\ref{cfg}). 
The involution is given by (\ref{definvga}).
As in the finite case, one checks that these operations make $\C_c(G)$ a
\sta; this time one needs the invariance of the Haar measure at various
steps of the proof.
\begin{Proposition}\ll{conl1}
The operations (\ref{defconv}) and  (\ref{definvga}) are continuous in
the $L^1$-norm; one has
\bea
\n f*g\n_1 & \leq &  \n f\n_1\:\n g\n_1; \ll{concl1} \\
\n f^*\n_1 & = &  \n f\n_1.\ll{concl2}
\eea
Hence $L^1(G)$ is a Banach \sta\ under the continuous extensions of 
(\ref{defconv}) and  (\ref{definvga}) from $C_c(G)$ to $L^1(G)$.
\end{Proposition}

Recall the definition of a Banach \sta\ below \ref{defcstaralgebra}.

It is obvious from invariance of  the Haar measure under $x\raw x\inv$
that $\n f^*\n_1=\n f\n_1$, so that the involution is certainly continuous.
The proof of (\ref{concl1}) is a straightforward generalization of the
case $G=\R$; cf.\ (\ref{convR}). This time we have 
$$
\n f*g\n_1=\int_{G}dx\,|\int_{G}dy\, f(xy\inv)g(y)|\leq \int_{G}dy\,
|g(y)| \int_{G}dx\,|f(xy\inv)|$$
$$
=\int_{G}dy\, |g(y)| \int_{G}dx\,|f(x)|=\n f\n_1\,\n g\n_1,$$
which is (\ref{concl1}).\enp

In order to equip $L^1(G)$ with a $C^*$-norm, we construct a faithful \rep\ on a \Hs.
\begin{Proposition}\ll{lrepl1}
For $f\in L^1(G)$ the operator $\pi_L(f)$ on $L^2(G)$, defined by 
\be
\pi_L(f)\Ps:=f*\Ps .\ll{defpil1}
\ee
is bounded, satisfying $\n \pi_L(f)\n\, \leq\, \n f\n_1$.
The linear map $\pi_L:L^1(G)\raw\B(L^2(G))$ is a faithful \rep\ of $L^1(G)$, seen as
a Banach \sta\ as in \ref{conl1}.
\end{Proposition}

 Introducing the {\bf left-regular representation}
$U_L$ of $G$ on $L^2(G)$
by 
\be U_L(y)\Ps(x):=\Ps(y\inv x), \ll{deflrep}
\ee
it follows that
\be
\pi_L(f)=\int_G dx\, f(x)U_L(x). \ll{deflrepbis}
\ee
The boundedness of $\pi_L(f)$ then follows from  Lemma \ref{Upibounded} below.
One easily verifies that $\pi_L(f*g)=\pi_L(f)\pi_L(g)$ and
$\pi_L(f^*)=\pi_L(f)^*$.

To prove that $\pi_L$ is faithful, we first show that $L^1(G)$ possesses  the analogue of an
approximate unit (see \ref{defaprun} for \ca s). When $G$ is finite, the delta-function
$\dl_e$ is a unit in $\C(G)$. For general locally compact groups one would like to take the Dirac
$\dl$-`function' as a unit, but this distribution is not in $L^1(G)$.
\begin{Lemma}\ll{apul1}
The Banach \sta\ $L^1(G)$ has an approximate unit $\I_{\lm}$ in the sense that 
(\ref{ilsa}) - (\ref{cpai}) hold for all $A\in L^1(G)$, and $\n\cdot\n=\n\cdot\n_1$.
\end{Lemma}

 Pick a basis of neighbourhoods ${\cal N}_{\lm}$ of
$e$, so that each ${\cal N}_{\lm}$ is invariant under $x\raw x\inv$; this
basis is partially ordered by
inclusion. Take
$\I_{\lm}=N_{\lm}\ch_{{\cal N}_{\lm}}$, which is the characteristic
function of ${\cal N}_{\lm}$ times a normalization factor ensuring
that $\n \I_{\lm}\n_1=1$. Eq.\ (\ref{ilsa}) then holds by virtue of (\ref{definvga}) and the
invariance of ${\cal N}_{\lm}$ under inversion.  By construction, the inequality (\ref{estau}) 
holds as an equality.
One has $\I_{\lm}*f(x)=N_{\lm}\int_{\CN_{\lm}}dy \,f(y\inv x)$ and 
$f*\I_{\lm}(x)=N_{\lm}\int_{\CN_{\lm}}dy\, f(xy\inv$. 
For $f\in C_c(G)$ one therefore has $\lim_{\lm} \I_{\lm}*f=f$ and
$\lim_{\lm} f*\I_{\lm}=f$ pointwise (i.e., for fixed $x$). 
The Lebesgue dominated convergence theorem then leads to  (\ref{cpai}) for all
$A\in C_c(G)$, and therefore for all $A\in L^1(G)$, since $C_c(G)$ is dense in $L^1(G)$.
\enp

To finish the proof of \ref{lrepl1}, we now note from 
(\ref{defpil1}) that $\pi_L(f)=0$ implies $f*\Ps=0$ for all $\Ps\in
L^2(G)$, and hence certainly for $\Ps=\I_{\lm}$.  Hence $\n f\n_1=0$
by Lemma \ref{apul1}, so that $f=0$ and $\pi_L$ is injective. 
 \enp
\begin{Definition}\ll{defcstrG} 
The {\bf  reduced group $C^*$-algebra}
 $C^*_{r}(G)$ is  the smallest \ca\ in $\B(L^2(G))$
containing $\pi_L(C_c(G))$. In other words, $C^*_{r}(G)$ is the
 closure of the latter in the norm 
\be
\n f\n_r:=\n\pi_L(f)\n.
\ee 
\end{Definition}

Perhaps the simplest example of a reduced group algebra
is obtained by taking $G=\R$.
\begin{Proposition}
One has the isomorphism
\be
C^*_r(\R)\simeq C_0(\R).\ll{cstrn1}
\ee
\end{Proposition}

 It follows from the discussion preceding \ref{defc0} that 
the Fourier transform (\ref{FT1}) maps $L^1(G)$ into a subspace of $C_0(\R)$ which separates
points on $\R$. It is clear that for every $p\in\R$ there is an $f\in L^1(\R)$ for which
$\hat{f}(p)\neq 0$. In order to apply Lemma \ref{SW0}, we need to verify that
\be
\n f\n_r=\n\hat{f}\n_{\infty}.
\ll{2norms}
\ee
Since the Fourier  transform
turns convolution
into pointwise multiplication, the left-regular \rep\ $\pi_L$ on $L^2(\R)$ is Fourier-transformed
into the action on $L^2(\R)$ by multiplication operators.
Hence (\ref{cstrn1}) follows from Lemma \ref{SW0}. \enp

This example generalizes to arbitrary locally compact abelian
groups. Let $\hat{G}$ be the set of all irreducible unitary \rep s
$U_{\gm}$ of $G$; such \rep s are necessarily one-dimensional, so that
$\hat{G}$ is nothing but the set of characters on $G$.  The
generalized Fourier transform $\hat{f}$ of $f\in L^1(G)$ is a function
on $\hat{G}$, defined as \be \hat{f}(\gm):=\int_G dx\,
f(x)U_{\gm}(x). \ll{abpltr} \ee By the same arguments as for $G=\R$,
one obtains \be C^*_r(G)\simeq C_0(\hat{G}). \ll{CrGC0hatG} \ee

We return to the general case, where $G$ is not necessarily abelian.
We have now found a \ca\ which may play the role of $\C(G)$ for locally compact groups.
Unfortunately, the analogue of Proposition \ref{Upithmf} only holds for a limited class
of groups. Hence we need a different construction. Let us agree that here and in what follows,
a unitary \rep\  of a topological group is always meant to be continuous.
\begin{Lemma}\ll{Upibounded} 
Let $U$ be an arbitrary unitary
  repre\-sen\-ta\-tion of $G$ on a \Hs\ $\H$.
Then $\pi(f)$, defined by
\be
\pi(f):=\int_G dx\, f(x)U(x) \ll{Utopi}
\ee
 is bounded, with 
\be
\n \pi(f)\n\, \leq\, \n f\n_1 \ll{npifnnfn1}.
\ee 
\end{Lemma}

The integral (\ref{Utopi}) is most simply defined weakly, that is, by its matrix
elements $$
(\Ps,\pi(f)\Ph):=\int_G dx\, f(x)(\Ps,U(x)\Ph).
$$
Since $U$ is unitary, we have $|(\Ps,\pi(f)\Ps)|\,\leq\, (F,F)_{L^2(G)}$
for all $\Ps\in\H$, where $F(x):=\, \n\Ps\n\sqrt{|f(x)|}$.  The Cauchy-Schwarz
inequality then leads to $|(\Ps,\pi(f)\Ps)|\,\leq\,\n f\n_1\n\Ps\n^2$. 
Lemma \ref{Qform} then leads to (\ref{npifnnfn1}).\enp

Alternatively, one may define (\ref{Utopi}) as a {\bf Bochner integral}.
We explain this notion in a more general context. 
\begin{Definition}\ll{bochner}
Let $X$ be a measure
space and let $\CB$ be a  Banach space.
 A function $f:X\raw \CB$
 is Bochner-integrable with respect to a measure $\mu$ on $X$ iff 
\begin{itemize}
\item 
$f$ is weakly measurable (that is, for each functional $\om\in \CB^*$
the function $x\raw \om(f(x))$ is measurable);
\item
 there is
a null set $X_0\subset X$ such that $\{f(x)|x\in X\backslash
X_0\}$ is separable; 
\item
 the function defined by $x\raw \n f(x)\n$
is integrable.  
\end{itemize}
\end{Definition}

It will always be directly clear from this whether a
given operator- or vector-valued integral may be read as a Bochner
integral; if not, it is understood as a weak integral, in a sense
always obvious from the context.  The Bochner integral $\int_X
d\mu(x) f(x)$ can be manipulated as if it were an ordinary
(Lebesgue) integral. For example, one has 
\be
\n \int_X d\mu(x)\, f(x)\n\,\leq \int_X d\mu(x)\,\n  f(x)\n. \ll{bocleb}
\ee
Thus reading (\ref{Utopi}) as a  Bochner integral, (\ref{npifnnfn1}) 
is immediate
from (\ref{bocleb}).

 The following result generalizes the correspondence between $U_L$ in
(\ref{deflrep}) and $\pi_L$ in (\ref{deflrepbis}) to arbitrary \rep s. 
\begin{Theorem}\ll{Upithm}
There is a bijective correspondence between non-degene\-ra\-te \rep s
$\pi$ of the Banach $\mbox{}^*$-algebra $L^1(G)$ which satisfy
(\ref{npifnnfn1}), and unitary \rep s $U$ of $G$.  This
correspondence is given in one direction by (\ref{Utopi}), and in the
other by
\be
U(x)\pi(f)\Om:=\pi(f^x)\Om, \ll{pitoU}
\ee
where $f^x(y):=f(x\inv y)$. 
This bijection preserves direct sums, and therefore
irreducibility. \end{Theorem}

  Recall from \ref{nondegcyclic} that any non-degenerate \rep\ of a \ca\ is a
direct sum of cyclic \rep s; the proof also applies to
$L^1(G)$. Thus $\Om$ in (\ref{pitoU}) stands for a cyclic vector of
a certain cyclic summand of $\H$, and (\ref{Utopi}) defines $U$ on a
dense subspace of this summand; it will be shown that $U$ is unitary,
so that it can be extended to all of $\H$ by continuity.

Given $U$, it follows from easy calculations
that $\pi(f)$ in (\ref{Utopi}) indeed defines a representation. It is
bounded by Lemma \ref{Upibounded}. The proof of non-degeneracy makes
use of Lemma \ref{apul1}.
Since $\pi$ is continuous,
one has $\lim_{\lm} \pi(\I_{\lm})=\I$ strongly, proving that $\pi$ must be non-degenerate.

To go in the opposite direction we use the approximate unit
 once more;
it follows from (\ref{pitoU}) (from which the continuity of
$U$ is obvious) that $U(x)\pi(f)\Om=
\lim_{\lm}\pi(\I_{\lm}^x)\pi(f)\Om$.
 Hence $U(x)= \lim_{\lm}\pi(\I_{\lm}^x)$ strongly on a dense domain.
The property $U(x)U(y)=U(xy)$ then follows from (\ref{pitoU}) and
(\ref{defconv}). The unitarity of each $U(x)$ follows by direct calculation, or from
the following argument. Since $\n
\pi(\I_{\lm}^x)\n\,\leq
\,\n
\I_{\lm}^x\n_1=1$, we infer that $\n U(x)\n\,\leq 1$ for all $x$.
Hence also $\n U(x\inv)\n\,\leq 1$, which is the same as $\n U(x)\inv\n\,\leq 1$
 We see that $U(x)$ and $U(x)\inv$ are both contractions; this is
only possible when $U(x)$ is unitary.

Finally, if $U$ is reducible there is a projection $E$ such that
$[E,U(x)]=0$ for all $x\in G$. It follows
from (\ref{Utopi}) that $[\pi(f),E]=0$ for all $f$, hence $\pi$ is
reducible. Conversely, if $\pi$ is reducible then
$[E,\pi(\I_{\lm}^x)]=0$ for all $x\in G$; by the previous paragraph
this implies $[E,U(x)]=0$ for all $x$. The final claim then follows from Schur's lemma
\ref{eqdefsofirrca}.1.
\eip 

This theorem suggests looking at a different object from
$C^*_r(G)$.  Inspired by \ref{defunivrep} one puts
\begin{Definition}\ll{defcstg}
The {\bf group $C^*$-algebra}  $C^*(G)$ is the
 closure of the Banach \sta\  algebra
$L^1(G)$ in the norm
\be
\n f\n:= \n \pi_u(f)\n, \ll{normcciG}
\ee
where $\pi_u$ is the direct sum of all non-degenerate \rep s $\pi$ of
$L^1(G)$ which are bounded as in (\ref{npifnnfn1}).

Equivalently,  $C^*(G)$ is the closure of  
$L^1(G)$ in the norm
\be
\n f\n: =\sup_{\pi} \n \pi(f)\n, \ll{normcciG2}
\ee
where the sum is over all \rep s $\pi(L^1(G))$ of the form (\ref{Utopi}), in which 
$U$ is an irreducible unitary \rep\ of $G$, and only one representative
of each equivalence class of such \rep s is included.
 \end{Definition}

The equivalence between the two definitions follows from (\ref{refdec}) and
Theorem \ref{Upithm}.
\begin{Theorem}\ll{rificol}
There is a bijective correspondence between  non-dege\-ne\-rate \rep s $\pi$ of the \ca\
$C^*(G)$  and unitary
  represen\-tations $U$ of $G$, given by (continuous extension of)
(\ref{Utopi}) and (\ref{pitoU}). This correspondence preserves irreducibility.
\end{Theorem}

It is obvious from (\ref{repbounded}) and (\ref{normcciG})  that for any \rep\
$\pi(C^*(G))$ and $f\in L^1(G)$ one has
\be
\n\pi(f)\n\,\leq\,\n f\n\,\leq\,\n f\n_1. \ll{3ineq}
\ee
Hence the restriction $\pi(L^1(G))$ satisfies (\ref{npifnnfn1}), and therefore corresponds
to $U(G)$ by Theorem \ref{Upithm}. Conversely, 
given $U(G)$ one finds $\pi(L^1(G))$ satisfying (\ref{npifnnfn1}) by \ref{Upithm};
it then follows from (\ref{3ineq}) that one may extend $\pi$ to a \rep\ of $C^*(G)$ by
continuity.
\enp

In conjunction with (\ref{2norms}), the second definition of $C^*(G)$
stated in \ref{rificol} implies that for abelian groups $C^*(G)$
always coincides with $C^*_r(G)$. The reason is that for
$\gm\in\hat{G}$ one has $\pi_{\gm}(f)=\hat{f}(\gm)\in\C$, so that the
norms (\ref{normcciG2}) and (\ref{2norms}) coincide. 
In particular, one has
\be
C^*(\R^n)\simeq C_0(\R^n).\ll{redisgrR}
\ee

For general locally compact groups, looking at \ref{defcstrG} we see
that 
\be
C^*_{r}(G)=\pi_L(C^*(G))\simeq C^*(G)/\ker(\pi_L). \ll{cstrgcstg}
\ee

A  Lie group group is
said to be {\bf amenable}\index{amenable} when the equality $C^*_r(G)=C^*(G)$
holds; in other words, $\pi_L(C^*(G))$ is faithful iff $G$
is amenable. 
We have just seen that all locally compact abelian groups are amenable. 
It follows from the Peter-Weyl theorem that
 all compact groups are amenable as well. However, non-compact semi-simple Lie groups
are not amenable.
\su{$C^*$-dynamical systems and crossed products}
An {\bf  automorphism} of a \ca\  $\A$ is an isomorphism between $\A$ and $\A$. 
It follows from Definitions \ref{defspectrum} and \ref{defautacc} that
$\sg(\al(A))=\sg(A)$ for any $A\in\A$ any automorphism $\al$; hence
\be
\n\al(A)\n=\n A\n \ll{autiso}
\ee
by (\ref{normun}).

One $\A$ has a unit, one has
\be
\al(\I)=\I \ll{al11}
\ee
by (\ref{phvmult}) and the uniqueness of the unit. When $\A$ has no unit,
one may extend $\al$ to an automorphism $\al^{\I}$ of 
the unitization $\AI$ by
\be
\al^{\I}(A+\lm\I):=\al(A)+\lm\I. \ll{aluI}
\ee
\begin{Definition}\ll{defautacc}
An {\bf automorphic action}  $\al$ of a group
$G$ on a \ca\ $\A$ is a group homomorphism $x\raw\al_x$ such that each $\al_x$
is an automorphism of $\A$. In other words, one has 
\bea
\al_x\circ\al_y(A) & = & \al_{xy}(A); \ll{alxy} \\
\al_x(AB) & = & \al_x(A)\al_x(A);\ll{al2} \\
\al(A^*)  & = & \al(A)^* \ll{al3}
\eea for all $x,y\in G$ and $A,B\in\A$.

A {\bf $C^*$-dynamical system} $(G,\A,\al)$ consists of a locally compact
 group $G$, a \ca\ $\A$, and an automorphic action of $G$ on $\A$ such
 that for each $A\in\A$ the function from $G$ to $\A$, defined by
 $x\raw \n\al_x(A)\n$, is continuous.
\end{Definition}

   The term `dynamical system' comes from
the example $G=\R$ and $\A=C_0(S)$, where $\R$ acts on $S$ by $t:\sg\raw \sg(t)$, and
$\al_t(f):\sg\raw\sg(t)$. Hence a general $C^*$-dynamical system is a non-commutative
analogue of a dynamical system.
\begin{Proposition}  
Let $(G,\A,\al)$ be a $C^*$-dynamical system, and define $L^1(G,\A,\al)$ as the space of all measurable
functions $f:G\raw\A$ for which
\be
\n f\n_1:=\int_G dx\, \n f(x)\n \ll{norml1ga}
\ee
is finite. The operations
\bea
f*g(x) & := & \int_G dy\, f(y)\al_y(g(y\inv x)); \ll{convtga} \\
f^*(x)  & := & \al_x(f(x\inv)^*) \ll{invtga}
\eea
turn $L^1(G,\A,\al)$ into a Banach $\mbox{}^*$-algebra.
\end{Proposition}  

As usual, we have assumed that $G$ is unimodular; with a slight
modification one may extend these formulae to the non-unimodular case.
The integral (\ref{convtga}) is defined as a Bochner integral; the
assumptions in Definition \ref{bochner} are satisfied as a consequence
of the continuity assumption in the definition of a $C^*$-dynamical
system. To verify the properties (\ref{concl1}) and (\ref{concl2}) one
follows the same derivation as for $L^1(G)$, using (\ref{bocleb}) and
(\ref{autiso}). The completeness of $L^1(G,\A,\al)$ is proved as in
the case $\A=\C$, for which $L^1(G,\A,\al)=L^1(G)$.\enp

In order to generalize Theorem \ref{Upithm}, we need 
\begin{Definition}\ll{defcovrep}
A {\bf covariant representation} of a $C^*$-dynamical system $(G,\A,\al)$ consists of
a pair $(U,\til{\pi})$, where $U$ is a
   unitary \rep\ of $G$, and 
$\til{\pi}$ is a non-degenerate \rep\ of $\A$ which 
for all $x\in G$ and $A\in\A$
 satisfies  
\be
U(x)\til{\pi}(A)U(x)^*=\til{\pi}(\al_x(A)). \ll{covcond5}
\ee
\end{Definition}

Here is an elegant and useful method to construct covariant \rep s.
\begin{Proposition}\ll{concovrep}
Let $(G,\A,\al)$ be a $C^*$-dynamical system, and
suppose one has a state $\om$ on $\A$ which is $G$-invariant in the sense that
\be
\om(\al_x(A))=\om(A) \ll{ominv}
\ee
 for all $x\in G$ and $A\in\A$.
Consider the GNS-\rep\ $\pi_{\om}(\A)$ on a \Hs\ $\H_{\om}$ with cyclic vector
$\Om_{\om}$.
 For $x\in G$, define an operator $U(x)$ on the dense
subspace $\pi_{\om}(\A)\Om_{\om}$ of $\H_{\om}$ by
\be
U(x)\pi_{\om}(A)\Om_{\om}:=\pi_{\om}(\al_x(A))\Om_{\om}. \ll{Ucovrep}
\ee
This operator is well defined, and defines a unitary \rep\ of $G$ on
$\H_{\om}$.
\end{Proposition}

If $\pi_{\om}(A)\Om_{\om}=\pi_{\om}(B)\Om_{\om}$
then $\om((A-B)^*(A-B))=0$ by (\ref{gnscor}). 
Hence $\om(\al_x(A-B)^*\al_x(A-B))=0$ by (\ref{ominv}), so that
$\n \pi_{\om}(\al_x(A-B))\Om_{\om}\n^2=0$ by (\ref{gnscor}). Hence
$\pi_{\om}(\al_x(A))\Om_{\om}=\pi_{\om}(\al_x(B))$, so that
$U(x)\pi_{\om}(A)\Om_{\om}=U(x)\pi_{\om}(B)\Om_{\om}$.

Furthermore, (\ref{alxy}) implies that $U(x)U(y)=U(xy)$, whereas
(\ref{Ucovrep}) and (\ref{ominv}) imply that 
$$
(U(x)\pi_{\om}(A)\Om_{\om},U(x)\pi_{\om}(B)\Om_{\om})=
(\pi_{\om}(A)\Om_{\om},\pi_{\om}(B)\Om_{\om}).
$$
This shows firstly that $U(x)$ is bounded on $\pi_{\om}(\A)\Om_{\om}$, so that
it may be extended to $\H_{\om}$ by continuity. Secondly, $U(x)$ is a
partial isometry, which is unitary from $\H_{\om}$ to the closure of
$U(x)\H_{\om}$. Taking $A=\al_{x\inv}(B)$ in (\ref{Ucovrep}), one sees
that $U(x)\H_{\om}=\pi_{\om}(\A)\Om_{\om}$, whose closure is 
$\H_{\om}$ because $\pi_{\om}$ is cyclic. Hence $U(x)$ is unitary.
\enp

Note that (\ref{Ucovrep}) with (\ref{al11}) or (\ref{aluI}) implies that
\be
U(x)\Om_{\om}=\Om_{\om}.\ll{ominv2}
\ee

Proposition \ref{concovrep} describes the way unitary \rep s of the
 Poincar\'{e} group are constructed in algebraic quantum field theory,
 in which $\om$ is then taken to be the vacuum state on the algebra of
 local observables of the system in question.  Note, however, that not
 all covariant \rep s of a $C^*$-dynamical system arise in this way; a
 given unitary \rep\ $U(G)$ may may not contain the trivial \rep\ as a
 sub\rep; cf.\ (\ref{ominv2}). 

In any case, the generalization of Theorem \ref{Upithm} is as follows.
Recall (\ref{norml1ga}).
\begin{Theorem}\ll{Upithmbis}
Let  $(G,\A,\al)$ be a $C^*$-dynamical system.
There is a bijective correspondence between non-degene\-ra\-te \rep s
$\pi$ of the Banach $\mbox{}^*$-algebra $L^1(G,\A,\al)$ which satisfy
(\ref{npifnnfn1}), and covariant \rep s $(U(G),\til{\pi}(\A))$.  This
correspondence is given  in one direction by 
\be
\pi(f)=\int_G dx\,\til{\pi}(f(x))U(x);\ll{tgca1}
\ee
in the other direction one defines $Af:x\raw Af(x)$
and $\til{\al}_x(f):y\raw \al_x(f(x\inv y))$,  and puts
\bea
U(x)\pi(f)\Om & = & \pi(\til{\al}_x(f))\Om;  \ll{tgca2}\\
\til{\pi}(A)\pi(f)\Om & = & \pi(Af)\Om, \ll{tgca3}
\eea
where $\Om$ is a cyclic vector for a cyclic summand
of $\pi(C^*(G,\til{\A}))$.

This bijection preserves direct sums, and therefore
irreducibility. \end{Theorem}

The proof of this theorem is analogous to that of \ref{Upithm}.
The approximate unit in $L^1(G,\A,\al)$ is 
 constructed by taking the tensor product of an
 approximate unit in $L^1(G)$ and an  approximate unit in $\A$.
 The rest of the proof may then essentially be read off from \ref{Upithm}.
 \enp
Generalizing \ref{defcstg}, we put
\begin{Definition}\ll{defcrpr}
Let  $(G,\A,\al)$ be a $C^*$-dynamical system.
The {\bf crossed product}  $C^*(G,\A,\al)$ of $G$ and $\A$ is the
 closure of the Banach \sta\  algebra
$L^1(G,\A,\al)$ in the norm
\be
\n f\n:= \n \pi_u(f)\n, \ll{normcciGcp}
\ee
where $\pi_u$ is the direct sum of all non-degenerate \rep s $\pi$ of
$L^1(G,\A,\al)$ which are bounded as in (\ref{npifnnfn1}).

Equivalently,  $C^*(G,\A,\al)$ is the closure of  
$L^1(G,\A,\al)$ in the norm
\be
\n f\n: =\sup_{\pi} \n \pi(f)\n, \ll{normcciG2cp}
\ee
where the sum is over all \rep s $\pi(L^1(G,\A,\al))$ of the form 
(\ref{tgca1}), in which 
$(U,\til{\pi})$ is an irreducible covariant \rep\ of $(G,\A,\al)$,
 and only one representative
of each equivalence class of such \rep s is included.
 \end{Definition}

Here we simply say that a covariant \rep\ $(U,\til{\pi})$ is irreducible
when the only bounded operator commuting with all $U(x)$ and
$\til{\pi}(A)$ is a multiple of the unit. 
The equivalence between the two definitions follows from (\ref{refdec}) and
Theorem \ref{Upithmbis}.
\begin{Theorem}\ll{rificolcp}
Let  $(G,\A,\al)$ be a $C^*$-dynamical system.
There is a bijective correspondence between  non-dege\-ne\-rate \rep s 
$\pi$ of the crossed product
$C^*(G,\A,\al)$  and  covariant \rep s $(U(G),\til{\pi}(\A))$.
This correspondence is
given by (continuous extension of)
(\ref{tgca1}) and (\ref{tgca2}), (\ref{tgca3}). 
This correspondence preserves direct sums, and
therefore irreducibility.
\end{Theorem}

The proof is identical to that of \ref{rificol}.\enp
\su{Transformation group $C^*$-algebras} 
We now come to an important class of crossed products, in which
$\A=C_0(Q)$, where $Q$ is a locally compact Hausdorff space, and
$\al_x$ is defined as follows. 
\begin{Definition}\ll{defgrac}
A (left-) {\bf action} $L$ of a group $G$ on a
space $Q$ is a map $L:G\x Q\raw Q$, satisfying $L(e,q)=q$ and
$L(x,L(y,q))=L(xy,q)$ for all $q\in Q$ and $x,y\in G$. 
If $G$ and $Q$ are locally compact we assume that $L$ is continuous.
If $G$ is a Lie
group and $Q$ is a manifold 
we assume that $L$ is smooth. We write $L_x(q) =xq:=L(x,q)$.
\end{Definition}

We assume the reader is familiar with this concept, at least at a heuristic
level. The main example we shall consider is the canonical action of
$G$ on the coset space $G/H$ (where $H$ is a closed subgroup of $G$).
This action is given by
\be
x[y]_H:=[xy]_H, \ll{xyH}
\ee
where $[x]_H:=xH$; cf.\ \ref{homvb} etc.
For example, when $G=SO(3)$ and $H=SO(2)$ is the subgroup of rotations
around the $z$-axis, one may identify $G/H$ with the unit two-sphere
$S^2$ in $\R^3$. The $SO(3)$-action (\ref{xyH}) is then simply the usual
action on $\R^3$, restricted to $S^2$.

Assume that $Q$ is a locally compact Hausdorff space, so that one may
form the commutative \ca\ $C_0(Q)$; cf.\ \ref{comCS}.  A $G$-action on
$Q$ leads to an automorphic action of $G$ on $C_0(Q)$, given by \be
\al_x(\til{f}):q\raw \til{f}(x\inv q). \ll{alxtilf} \ee Using the fact
that $G$ is locally compact, so that $e$ has a basis of compact
neighbourhoods, it is easy to prove that the continuity of the
$G$-action on $Q$ implies that \be \lim_{x\raw e}
\n\al_x(\til{f})-\til{f}\n=0 \ee for all $\til{f}\in C_c(Q)$. Since
$C_c(Q)$ is dense in $C_0(Q)$ in the sup-norm, the same is true for
$\til{f}\in C_0(Q)$. Hence the function $x\raw \al_x(\til{f})$ from
$G$ to $C_0(Q)$ is continuous at $e$ (as $\al_e(\til{f})=\til{f}$).  Using
(\ref{alxy}) and (\ref{autiso}), one sees that this function is
continuous on all of $G$. Hence $(G,C_0(Q),\al)$ is a $C^*$-dynamical
system.

It is quite instructive to look at covariant \rep s $(U,\til{\pi})$ of
 $(G,C_0(Q),\al)$ in the special case that $G$ is a Lie group and $Q$
 is a manifold. Firstly, given a unitary \rep\ $U$ of a Lie group $G$
 on a \Hs\ $\H$ one can construct a \rep\ of the Lie algebra $\g$ by
 \be dU(X)\Ps:=\ddt U(\Exp(tX))\Ps_{|t=0}. \ll{defdUeq} \ee When $\H$
 is infinite-dimensional this defines an unbounded operator, which is
 not defined on all of $\H$. Eq.\ (\ref{defdUeq}) makes sense when
 $\Ps$ is a {\bf smooth vector} for a $U$; this is an element
 $\Ps\in\H$ for which the map $x\raw U(x)\Ps$ from $G$ to $\H$ is
 smooth.  It can be shown that the set $\H_U^{\infty}$ of smooth
 vectors for $U$ is a dense linear subspace of $\H$, and that the
 operator $idU(X)$ is essentially self-adjoint on
 $\H_U^{\om}$. Moreover, on $\H_U^{\infty}$ one has \be
 [dU(X),dU(Y)]=dU([X,Y]). \ll{crdU} \ee

Secondly, given a Lie group action one defines a linear map $X\raw \xi_X$ 
from $\g$ to the space of all vector fields on $Q$  by  
\be
\xi_X \til{f}(q):=\ddt \til{f}(\Exp(tX)q)_{|t=0}, \ll{deffunvf}
\ee
where $\Exp:\g\raw G$ is the usual exponential map.

The meaning of the covariance condition (\ref{covcond5})
 on the pair $(U,\til{\pi})$  may now be clarified
by re-expressing it in infinitesimal form. 
For  $X\in\g$, $\til{f}\in\cci(Q)$, and $\hbar\in\R\backslash\{0\}$
we put
\bea
\q^{\pi}(\til{X}) & := & i\hbar dU(X);  \ll{defqpi1}\\
\q^{\pi}(\til{f})& := & \til{\pi}(\til{f}). \ll{defqpi2}
\eea
From the commutativity of $C_0(Q)$, (\ref{crdU}), and
 (\ref{covcond5}), respectively, we then obtain 
\bea
& & 
\frac{i}{\hb}[\q^{\pi}(\til{f}),\q^{\pi}(\til{g})]=0; \ll{tilpicon} \\
& & \frac{i}{\hb} [\q^{\pi}(\til{X}),\q^{\pi}(\til{Y})] =\q^{\pi}(-\wt{[X,Y]}); \ll{infdUcon} \\
& & \frac{i}{\hb} [\q^{\pi}(\til{X}),\q^{\pi}(\til{f})] = \q^{\pi}(\xi_X\til{f}). \ll{infcov}
\eea
These equations  hold on the domain $\H_U^{\infty}$, and may be seen as a
generalization of the canonical commutation relations of \qm. To see this,
 consider the case $G=Q=\R^n$, where the $G$-action is given by
$L(x,q):=q+x$. If $X=T_k$ is the $k$'th generator of $\R^n$ one has
$\xi_k:=\xi_{T_k}=\partial/\partial q^k$. Taking
$f=q^l$, the $l$'th co-ordinate function on $\R^n$, one therefore obtains
$\xi_k q^l=\dl_k^l$. The relations (\ref{tilpicon})  - (\ref{infcov})
then become
\bea
& & 
\frac{i}{\hb}[\q^{\pi}(q^k),\q^{\pi}(q^l)]=0; \ll{tilpiconrr} \\
& & \frac{i}{\hb} [\q^{\pi}(\til{T}_k),\q^{\pi}(\til{T}_l)] =0; \ll{infdUconrr} \\
& & \frac{i}{\hb} [\q^{\pi}(\til{T_k}),\q^{\pi}(q^l)] = \dl_k^l. \ll{infcovrr}
\eea
Hence one may identify $\q^{\pi}(q^k)$ and $\q^{\pi}(\til{T_k})$
with the quantum position and momentum observables, respectively.
(It should be remarked that $\q^{\pi}(q^k)$ is an unbounded operator,
but one may show from the \rep\ theory of the Heisenberg group that 
$\q^{\pi}(q^k)$ and $\q^{\pi}(\til{T_k})$ always possess a common dense
domain on which (\ref{tilpiconrr})  - (\ref{infcovrr}) are valid.)
\begin{Definition}\ll{deftgca}
Let $L$ be a continuous action of a locally compact group on a locally compact
space $Q$.
The {\bf transformation group $C^*$-algebra} $C^*(G,Q)$ 
is the crossed product $C^*(G,C_0(Q),\al)$ defined by
the automorphic action (\ref{alxtilf}).
\end{Definition}

Conventionally, the $G$-action $L$ on $Q$ is not indicated
 in the notation  $C^*(G,Q)$, although
the construction clearly depends on it.

One may identify $L^1(G,C_0(Q))$ with a subspace of the space of all
(measurable) functions from $G\x Q$ to $\C$; an element $f$ of the latter
defines $F\in L^1(G,C_0(Q))$ by $F(x)=f(x,\cdot)$. Clearly,
$L^1(G,C_0(Q))$ is then identified with the space of all such functions
$f$ for which
\be
\n f\n_1=\int_G dx\, \sup_{q\in Q}|f(x,q)| \ll{fn1tga}
\ee
is finite; cf.\ (\ref{norml1ga}). In this realization, the operations
(\ref{convtga}) and (\ref{invtga}) read
\bea
f*g(x,q) & = & \int_G dy\, f(y,q)g(y\inv x,y\inv  q); \ll{convactiongr} \\
f^*(x,q) & = & \ovl{f(x\inv,x\inv q)}. 
\ll{invactiongr}
\eea
As always, $G$ is here assumed to be unimodular.
Here is a simple example.
\begin{Proposition}\ll{isoCstGG}
Let a locally compact group $G$ act on $Q=G$ by  $L(x,y):=xy$.
Then $C^*(G,G)\simeq \B_0(L^2(G))$ as \ca s.
\end{Proposition}

We start from $C_c(G\x G)$, regarded as a dense subalgebra of $C^*(G,G)$.
We define a linear map $\pi:C_c(G\x G)\raw \B(L^2(G))$ by
\be
\pi(f)\Ps(x):=\int_G dy\, f(xy\inv,x)\Ps(y). \ll{piCGG}
\ee
One verifies from (\ref{convactiongr}) and (\ref{invactiongr}) that
$\pi(f)\pi(g)=\pi(f*g)$ and $\pi(f^*)=\pi(f)^*$, so that $\pi$ is
a \rep\ of the \sta\ $C_c(G\x G)$. It is easily verified that
the Hilbert-Schmidt-norm (\ref{hsnorm}) of $\pi(f)$ is
\be
\n\pi(f)\n^2_2=\int_G\int_G dx\, dy\, |f(xy\inv,x)|^2.
\ee
Since this is clearly finite for $f\in C_c(G\x G)$, we conclude from
(\ref{incl}) that $\pi(C_c(G\x G))\subseteq \B_0(L^2(G))$.
Since $\pi(C_c(G\x G))$ is dense in $\B_2(L^2(G))$ in the 
 Hilbert-Schmidt-norm (which is a standard fact of \Hs\ theory), and
$\B_2(L^2(G))$ is dense in $\B_0(L^2(G))$ in the usual operator norm
(since by Definition \ref{defb0h} even $\B_f(L^2(G))$ is dense in 
 $\B_0(L^2(G))$), we conclude that the closure of 
 $\pi(C_c(G\x G))$ in the operator norm coincides with $\B_0(L^2(G))$.

Since $\pi$ is evidently faithful, the equality
$\pi(C^*(G,G))=\B_0(L^2(G))$, and therefore the isomorphism
$C^*(G,G)\simeq \B_0(L^2(G))$, follows from the previous paragraph if
we can show that the norm defined by (\ref{normcciG2cp}) coincides
with the operator norm of $\pi(\cdot)$. This, in turn, is the case
if all irreducible \rep s of the \sta\ $C_c(G\x G)$ are unitarily
equivalent to $\pi$.

To prove this, we proceed as in Proposition \ref{riefbimod}, in
which we take $\til{\A}=C_c(G\x G)$, $\til{\B}=\C$, and
$\til{\CE}=C_c(G)$. The pre-\HCM\ $C_c(G)\rlh\C$ is defined
by the obvious $\C$-action on $C_c(G)$, and the inner product
\be
\la \Ps,\Ph\ra_{\C} : =  (\Ps,\Ph)_{L^2(G)}.
\ee
The left-action of $\til{\A}$ on $\til{\CE}$ is $\pi$ as defined in
(\ref{piCGG}), whereas the $C_c(G\x G)$-valued inner product on
$\til{C_c(G)}$ is given by
\be
 \la \Ps,\Ph\ra_{C_c(G\x G)}:=\Ps(y)\ovl{\Ph}(x\inv y).
\ee
 
It is not necessary to consider the bounds (\ref{rieffelbound22})
and (\ref{rieffelbound3}). Following the proof of Theorem \ref{riimp},
one  shows directly that there is a bijective correspondence between the
\rep s of $C_c(G\x G)$ and of $\C$.
\enp
\su{The abstract transitive imprimitivity theorem}
We specialize to the case where $Q=G/H$, where $H$ is a closed subgroup of $G$,
and the $G$-action on $G/H$ is given by (\ref{xyH}). This leads to the
transformation group \ca\ $C^*(G,G/H)$. 
\begin{Theorem}\ll{macglimm}
The transformation group \ca\ $C^*(G,G/H)$ is Morita-equivalent to $C^*(H)$.
\end{Theorem}

We need to construct a full \HCM\ $\CE\rlh C^*(H)$ for which
$C^*_0(\CE,C^*(H))$ is isomorphic to $C^*(G,G/H)$. 
This will be done  on the basis of Proposition \ref{riefbimod}.
For simplicity we assume that both $G$ and $H$ are unimodular.
In \ref{riefbimod} we take 
\begin{itemize}
\item
$\til{\A}=C_c(G,G/H)$, seen as a dense
subalgebra of $\A=C^*(G,G/H)$ as explained prior to (\ref{fn1tga});
\item
$\til{\B}=C_c(H)$, seen as a dense
subalgebra of $\B=C^*(H)$;
\item
$\til{\CE}=C_c(G)$.
\end{itemize}

We make a pre-Hilbert $C_c(H)$-module $C_c(G)\rlh C_c(H)$ by means of
the right-action
\be
\pir(f)\Ps=\Ps f:x\raw \int_H dh\, \Ps(xh\inv) f(h). \ll{CPrm0}
\ee
Here $f\in C_c(H)$ and $\Ps\in C_c(G)$. The $C_c(H)$-valued
inner product on $C_c(G)$ is defined by
\be 
\la\Ps,\Ph\ra_{C_c(H)}: h\raw \int_G dx\, \ovl{\Ps(x)}\Ph(xh).
\ll{CPrm1}
\ee
Interestingly, both formulae may be written in terms of the right-regular
\rep\ $U_R$ of $H$ on $L^2(G)$, given by
\be
U_R(h)\Ps(x):=\Ps(xh). \ll{defURcap}
\ee
Namely, one has 
\be
\pir(f)=\int_H dh\, f(h)U(h^{-1}), \ll{rmw3.5}
\ee
which should be compared with (\ref{Utopi}), and
\be
\la \Ps,\Ph\ra_{C_c(H)}:\, h\raw (\Ps,U(h)\Ph)_{L^2(G)}. \ll{rmw3.6}
\ee

The properties (\ref{hcm1}) and (\ref{hcm2}) are easily verified from
(\ref{definvga}) and (\ref{defconv}), respectively.  To prove
(\ref{hcm3}), we take a vector state $\om_{\ch}$ on $C^*(H)$, with
corresponding unit vector $\Om_{\ch}\in\Hlg$. Hence for $f\in
C_c(H)\subset L^1(H)$ one has \be
\om_{\ch}(f)=(\Om_{\ch},\pi_{\ch}(f)\Om_{\ch})=\int_H dh\, f(h)
(\Om_{\ch},U_{\ch}(h)\Om_{\ch}), \ll{omchlem} \ee where $U_{\ch}$ is
the unitary \rep\ of $H$ corresponding to $\pi_{\ch}(C^*(H))$; see
Theorem \ref{rificol} (with $G\raw H$).  We note that the Haar measure
on $G$ and the one on $H$ define a unique measure $\nu$ on $G/H$,
satisfying \be \int_{G} dx\, f(x) = \int_{G/H} d\nu(q)\, \int_H dh\,
f(s(q)h) \ll{3.2r} \ee for any $f\in C_c(G)$, and any measurable map
$s:G/H\raw G$ for which $\ta\circ s={\rm id}$ (where $\ta:G\raw G/H$
is the canonical projection $\ta(x):=[x]_H=xH$).  Combining
(\ref{omchlem}), (\ref{CPrm1}), and (\ref{3.2r}), we find \be
\om_{\ch}(\la\Ps,\Ps\ra_{C_c(H)})=\int_{G/H} d\nu(q)\, \n \int_H dh\,
\Ps(s(q)h)U_{\ch}(h)\Om_{\ch}\n^2. \ll{a135} \ee Since this is
positive, this proves that $\pi_{\ch}(\la\Ps,\Ps\ra_{C_c(H)})$ is
positive for all \rep s $\pi_{\ch}$ of $C^*(H)$, so that
$\la\Ps,\Ps\ra_{C_c(H)}$ is positive in $C^*(H)$ by Corollary
\ref{posposrep}. This proves (\ref{hcm3}).  Condition (\ref{hcm4})
easily follows from (\ref{a135}) as well, since
$\la\Ps,\Ps\ra_{C_c(H)}= 0$ implies that the right-hand side of
(\ref{a135}) vanishes for all $\ch$. This implies that the function
$(q,h)\raw \Ps(s(q)h)$ vanishes almost everywhere for arbitrary
sections $s$. Since one may choose $s$ so as to be piecewise
continuous, and $\Ps\in C_c(G)$, this implies that $\Ps=0$.

We now come to the left-action $\pil$ of $\til{\A}=C_c(G,G/H)$ on
$C_c(G)$ and the $C_c(G,G/H)$-valued inner product $\la \, ,\,
\ra_{C_c(G,G/H)}$ on $C_c(G)$.  These are given by \bea \pil(f)\Ps(x)
& = & \int_G dy\, f(xy\inv,[x]_H)\Ps(y); \ll{cghimbim0} \\
\la\Ps,\Ph\ra_{C_c(G,G/H)}& : & (x,[y]_H)\raw \int_H dh\,
\Ps(yh)\ovl{\Ph(x\inv yh)}. \ll{cghimbim} \eea Using
(\ref{convactiongr}) and (\ref{invactiongr}), one may check that
$\pil$ is indeed a left-action, and that $\ovl{C_c(G)}\rlh C_c(G,G/H)$
is a pre-\HCM\ with respect to the right-action of $C_c(G,G/H)$ given
by $\pir(f)\Ps:=\pil(f^*)\Ps$; cf.\ \ref{riefbimod}.  Also, using
(\ref{cghimbim}), (\ref{cghimbim0}), (\ref{CPrm0}), and
(\ref{CPrm1}), it is easy to verify the crucial condition
(\ref{bimpr}).

To complete the proof, one needs to show that the \HCM s $C_c(G)\rlh C_c(H)$  and
$\ovl{C_c(G)}\rlh C_c(G,G/H)$ are full, and that the bounds (\ref{rieffelbound22})
and (\ref{rieffelbound3}) are satisfied. This is indeed the case, but an argument that is
sufficiently elementary for inclusion in these notes does not seem to exist.
Enthusiastic readers may find the proof in M.A. Rieffel, Induced representations of
$C^*$-algebras, {\em Adv.\ Math.} {\bf 13} (1974) 176-257.
\enp
\su{Induced group \rep s}
The theory of induced group \rep s provides a mechanism for constructing a unitary \rep\ of a
locally compact group
$G$ from a unitary \rep\ of some closed subgroup $H$. Theorem \ref{macglimm}  then turns out to be
equivalent to a complete characterization of induced group \rep s, in the sense that it gives a
necessary and sufficient criterion for a unitary \rep\ to be induced.

In order to explain the idea of an induced group \rep\ from a geometric point of view, we return
to Proposition \ref{homvb}. The group $G$ acts on the Hilbert bundle $\SHG$ defined by
(\ref{SHG}) by means of
\be
{\sf U}^{\ch}(x):[y,v]_H\raw [xy,v]_H. \ll{sfU}
\ee
Since the left-action $x:y\raw xy$ of $G$ on itself commutes with the right-action
$h:y\raw yh$ of $H$ on $G$, the action (\ref{sfU}) is clearly well defined.

The $G$-action ${\sf U}^{\ch}$ on the vector bundle $\SHG$ induces a
natural $G$-action $U^{(\ch)}$ on the space of continuous sections
$\Gm(\SHG)$ of $\SHG$, defined on $\Ps^{(\ch)}\in\Gm(\SHG)$ by \be
U^{(\ch)}(x)\Ps^{(\ch)}(q):= {\sf U}^{\ch}(x) \Ps^{(\ch)}(x\inv q)).
\ll{irhermann} \ee One should check that $U^{(\ch)}(x)\Ps^{(\ch)}$ is
again a section, in that $\ta_{\ch}(U^{(\ch)}(x)\Ps^{(\ch)}(q))=q$;
see (\ref{deftch}).  This section is evidently continuous, since the
$G$-action on $G/H$ is continuous.

 There is a natural inner product on the space of sections
$\Gm(\SHG)$, given by \be (\Ps^{(\ch)},\Ph^{(\ch)}):=\int_{G/H} d\nu(q)\,
(\Ps^{(\ch)}(q),\Ph^{(\ch)}(q))_{\ch}, \ll{ipshg} \ee where $\nu$ is
the measure on $G/H$ defined by (\ref{3.2r}), and $(\, ,\, )_{\ch}$ is
the inner product in the fiber $\ta_{\ch}\inv (q)\simeq\Hlg$. Note
that different identifications of the fiber with $\Hg$ lead to the
same inner product.  The \Hs\ $L^2(\SHG)$ is the completion of the
space $\Gm_c(\SHG)$ of continuous sections of $\SHG$ with compact
support (in the norm derived from this inner product).

When the measure $\nu$ is $G$-invariant (which is the case, for
example, when $G$ and $H$ are unimodular), the operator $U^{(\ch)}(x)$
defined by (\ref{irhermann}) satisfies \be
(U^{(\ch)}(x)\Ps^{(\ch)},U^{(\ch)}(x)\Ph^{(\ch)})=(\Ps^{(\ch)},\Ph^{(\ch)}). \ll{Uchun}
\ee When $\nu$ fails to be $G$-invariant, it can be shown that it is
still quasi-invariant in the sense that $\nu(\cdot)$ and
$\nu(x\inv\cdot)$ have the same null sets for all $x\in G$.
Consequently, the Radon-Nikodym derivative $q\raw
d\nu(x^{-1}(q))/d\nu(q)$ exists as a measurable function on $G/H$.
One then modifies (\ref{irhermann}) to \be
U^{(\ch)}(x)\Ps^{(\ch)}(q):= \sqrt{\frac{d\nu(x^{-1}(q))}{d\nu(q)}}\,
{\sf U}^{\ch}(x) \Ps^{(\ch)}(x\inv q)).  \ll{irhermann2} \ee
\begin{Proposition}\ll{indrepher}
Let $G$ be a locally compact group with closed subgroup $H$, and let $U_{\ch}$ be a
   unitary \rep\ of $H$ on a \Hs\ $\Hlg$.
Define the Hilbert space $L^2(\SHG)$ of $L^2$-sections of the Hilbert
bundle $\SHG$ as the completion of $\Gm_c(\SHG)$ in the inner product (\ref{ipshg}),
where the measure $\nu$ on $G/H$ is defined by (\ref{3.2r}).

 The map $x\raw U^{(\ch)}(x)$ given by (\ref{irhermann2}) with
(\ref{sfU}) defines
a unitary \rep\ of $G$ on $L^2(\SHG)$. When $\nu$ is $G$-invariant,
the expression (\ref{irhermann2}) simplifies to (\ref{irhermann}). 
\end{Proposition}

One easily verifies that the square-root precisely compensates for the lack of $G$-invariance
of $\nu$, guaranteeing the property (\ref{Uchun}).  Hence $U^{(\ch)}(x)$
is isometric
on  $\Gm_c(\SHG)$, so that it is bounded, and can be extended to $L^2(\SHG)$ by continuity.
Since $U^{(\ch)}(x)$ is invertible, with inverse  $U^{(\ch)}(x\inv)$, it is therefore
a unitary operator. The property $U^{(\ch)}(x)U^{(\ch)}(y)=U^{(\ch)}(xy)$ is easily checked.
\enp

The \rep\ $U^{(\ch)}(G)$ is said to be {\bf induced} by $U_{\ch}(H)$.
\begin{Proposition}\ll{indrepsi}
In the context of \ref{indrepher}, define a \rep\ $\til{\pi}^{(\ch)}(C_0(G/H))$ on $L^2(\SHG)$
by
\be
\til{\pi}^{(\ch)}(\til{f})\Ps^{(\ch)}(q):= \til{f}(q)\Ps^{(\ch)}(q). \ll{tilpiuchbr}
\ee
The pair  $(U^{(\ch)}(G),\til{\pi}^{(\ch)}(C_0(G/H)))$ is a covariant \rep\ of the $C^*$-dynamical
system
$(G,C_0(G/H),\al)$, where $\al$ is given by (\ref{alxtilf}).
\end{Proposition}

Given \ref{indrepher}, this follows from a simple computation.\enp

Note that the \rep\ (\ref{tilpiuchbr}) is nothing but the right-action (\ref{pirpil}) of
$(C_0(G/H))$  on $L^2(\SHG)$; this right-action is at the same time a left-action, because
$(C_0(G/H))$ is commutative.

We now give a more convenient unitarily equivalent realization of this covariant \rep.
For this purpose we note that a section $\Ps^{(\ch)}:Q\raw \SHG$ of the  bundle $\SHG$   may
alternatively be represented
as a  map $\Ps^{\ch}:G\raw \Hg$ which is $H$-equivariant in that
\be
\Ps^{\ch}(xh\inv)=U_{\ch}(h)\Ps^{\ch}(x). \ll{eqforunitary}
\ee
Such a map defines a section $\Ps^{(\ch)}$ by 
\be
\Ps^{(\ch)}(\ta(x))=[x,\Ps^{\ch}(x)]_H, \ll{reldifse}
\ee
where $\ta:G\raw G/H$ is given by (\ref{deftauGH}). The section $\Ps^{(\ch)}$ thus
defined is independent of the choice of
$x\in\ta\inv(\ta(x))$ because of (\ref{eqforunitary}).

For $\Ps^{(\ch)}$ to lie in $\Gm_c(\SHG)$, 
the projection of the support of $\Ps^{\ch}$ from $G$ to
$G/H$ must be compact. 
In this realization  the inner product on $\Gm_c(\SHG)$ is given by 
\be
(\Ps^{\ch},\Ph^{\ch}):=\int_{G/H} d\nu(\ta(x))\, (\Ps^{\ch}(x),\Ph^{\ch}(x))_{\ch}; \ll{iponHG}
\ee
 the integrand indeed only depends on $x$ through $\ta(x)$ because of (\ref{eqforunitary}). 
\begin{Definition}\ll{defHG}
The \Hs\ $\HG$ is the completion  in the inner product (\ref{iponHG})
of the set of continuous functions
$\Ps^{\ch}:G\raw\Hg$ which satisfy the equivariance condition
(\ref{eqforunitary}), and the projection of whose support to $G/H$ is
compact.
\end{Definition}

Given (\ref{reldifse}), we define the induced $G$-action $U^{\ch}$ on $\Ps^{\ch}$ by
\be
[y,U^{\ch}(x)\Ps^{\ch}(y)]_H:=U^{(\ch)}(x)\Ps^{(\ch)}(\ta(y)).\ll{branobra}
\ee
Using (\ref{irhermann}), (\ref{reldifse}), and (\ref{sfU}), as well as the
definition $x\ta(y)=x[y]_H=[xy]_H=\ta(xy)$ of the $G$-action on $G/H$ (cf.\ 
 (\ref{deftauGH})), we obtain
$$
U^{(\ch)}(x)\Ps^{(\ch)}(\ta(y))={\sf U}^{\ch}(x) \Ps^{(\ch)}(x\inv
\ta(y)))= {\sf U}^{\ch}(x)[x\inv y,\Ps^{\ch}(x\inv y)]_H= [y,\Ps^{\ch}(x\inv y)]_H.
$$
Hence we infer from (\ref{branobra}) that
\be
 U^{\ch}(y)\Ps^{\ch}(x)= 
\Ps^{\ch}(y\inv x). \ll{gblattner1}
\ee
Replacing (\ref{irhermann}) by  (\ref{irhermann2}) in the above derivation yields
\be
 U^{\ch}(y)\Ps^{\ch}(x)=\sqrt{\frac{d\nu(\ta(y\inv x))}{d\nu(\ta(x))}}\,
\Ps^{\ch}(y\inv x). \ll{gblattner2}
\ee
Similarly, in the realization $\Hug$ the \rep\ (\ref{tilpiuchbr}) reads
\be
\til{\pi}^{\ch}(\til{f})\Ps^{\ch}(x):= \til{f}([x]_H)\Ps^{\ch}(x). \ll{tilpiuchbr2}
\ee
Analogous to \ref{indrepsi}, we then have
\begin{Proposition}\ll{indrepsi2}
In the context of \ref{indrepher}, define a \rep\
$\til{\pi}^{\ch}(C_0(G/H))$ on $\Hug$ (cf.\ \ref{defHG}) 
by (\ref{tilpiuchbr2}).  The
pair $(U^{\ch}(G),\til{\pi}^{\ch}(C_0(G/H)))$, where $U^{\ch}$ is
given by (\ref{gblattner2}), is a covariant \rep\ of the
$C^*$-dynamical system $(G,C_0(G/H),\al)$, where $\al$ is given by
(\ref{alxtilf}).

This pair is unitarily equivalent to the pair
$(U^{(\ch)}(G),\til{\pi}^{(\ch)}(C_0(G/H)))$ by the unitary map
$V:\Hug\raw\H^{(\ch)}$ given by \be
V\Ps^{\ch}(\ta(x)):=[x,\Ps^{\ch}(x)]_H, \ll{VlowH} \ee in the sense
that \be V U^{\ch}(y)V\inv= U^{(\ch)}(y) \ll{chp1} \ee for all $y\in
G$, and \be V\til{\pi}^{\ch}(\til{f})V\inv= \til{\pi}^{(\ch)}(\til{f})
\ll{chp2} \ee for all $\til{f}\in C_0(G/H)$.
\end{Proposition}

Comparing (\ref{VlowH}) with (\ref{reldifse}), it should be obvious from the argument
leading from (\ref{branobra}) to (\ref{gblattner2}) that (\ref{chp1}) holds.
 An analogous but simpler calculation shows (\ref{chp2}). \enp
\su{Mackey's transitive imprimitivity theorem}
In the preceding section we have seen that the unitary \rep\ $U^{\ch}(G)$ induced by
a unitary \rep\ $\Ulg$ of a closed subgroup $H\subset G$ can be extended to a covariant
\rep\ $(U^{\ch}(G),\til{\pi}^{\ch}(C_0(G/H))$. The original imprimitivity theorem of Mackey,
which historically preceded Theorems \ref{riimp} and \ref{macglimm}, states that all
covariant pairs $(U(G),\til{\pi}(C_0(G/H))$ arise in this way.
\begin{Theorem}\ll{mackim}
Let $G$ be a locally compact group with closed subgroup $H$, and
consider the $C^*$-dynamical system $(G,C_0(G/H),\al)$, where $\al$ is
given by (\ref{alxtilf}).  Recall (cf.\ \ref{defcovrep}) that a
covariant \rep\ of this system consists of a unitary \rep\ $U(G)$ and
a \rep\ $\til{\pi}(C_0(G/H))$, satisfying the covariance condition \be
U(x)\til{\pi}(\til{f})U(x)\inv=\til{\pi}(\til{f}^x) 
\ll{a154}\ee for all $x\in
G$ and $\til{f}\in C_0(G/H$; here $\til{f}^x(q):=\til{f}(x\inv q)$.

Any   unitary \rep\ $\Ulg(H)$ leads to a covariant \rep\
$(U^{\ch}(G),\til{\pi}^{\ch}(C_0(G/H))$ of $(G,C_0(G/H),\al)$, given by
\ref{defHG}, (\ref{gblattner2}) and (\ref{tilpiuchbr2}).
 Conversely, any
 covariant \rep\ $(U,\til{\pi})$ of $(G,C_0(G/H),\al)$ is unitarily equivalent
to a pair of this form. 

This leads to  a bijective correspondence between the space of equivalence classes
of unitary \rep s of $H$ and the space of equivalence classes
of covariant \rep s $(U,\til{\pi})$ of  the $C^*$-dynamical
system $(G,C_0(G/H),\al)$, which preserves direct sums and therefore irreducibility
(here the equivalence relation is unitary equivalence).
\end{Theorem}

The existence of the bijective correspondence with the stated
properties follows by combining Theorems \ref{macglimm} and
\ref{riimp}, which relate the \rep s of $C^*(H)$ and $C^*(G,G/H)$,
with Theorems \ref{rificol} and \ref{rificolcp}, which allow one to
pass from $\pi(C^*(H))$ to $U(H)$ and from $\pi(C^*(G,G/H))$ to
$(U(G),\til{\pi}(C_0(G/H))$, respectively.

The explicit form of the correspondence remains to be established.
Let us start with a technical point concerning Rieffel induction in general.
Using  (\ref{rief02b}), (\ref{PsPh0}),
and (\ref{normincm}), one shows that $\n\til{V}\Ps\n\,\leq\,\n\Ps\n$,
where the  norm on the left-hand side is in $\til{\H}^{\ch}$,
and the norm on the right-hand side is the one defined in (\ref{normincm}).
It follows that the induced space  $\til{\H}^{\ch}$ obtained
by Rieffel-inducing from a pre-\HCM\ is the same as the induced space
constructed  from its completion. The same comment, of course,
applies to $\Hug$.

We will use a gerenal technique that is often useful in problems involving
Rieffel induction. 
\begin{Lemma}\ll{mtfhug}
Suppose one has  a  \Hs\ $\H_*^{\ch}$ (with inner product denoted
by $(\, , \,)_{*}^{\ch}$) and a linear map
$\til{U}:\CE\ot\Hlg\raw \H_*^{\ch}$ satisfying
\be
(\til{U}\til{\Ps},\til{U}\til{\Ph})_{*}^{\ch}=
(\til{\Ps},\til{\Ph})_0^{\ch}  \ll{whatwtil}
\ee
for all $\til{\Ps},\til{\Ph}\in \CE\ot\Hlg$. 

Then  $\til{U}$ quotients to an
isometric map between $\CE\ot\Hlg/ \CN^{\ch}$ and the image of
$\til{U}$ in $\H^{\ch}_*$. When the image is dense this map
extends to a unitary isomorphism $U:\H^{\ch}\raw
\H_*^{\ch}$. Otherwise, $U$ is unitary between $\H^{\ch}$
and the closure of the image of $\til{U}$.

In any case, the \rep\ $\pug(C^*(\CE,\B))$ is
  equivalent to the representation $\pi_*^{\ch}(C^*(\CE,\B))$, defined by
continuous extension of
\be
\pi^{\ch}_*(A)\til{U}\til{\Ps}:=\til{U} (A\ot I_{\ch}\til{\Ps}).
\ll{pichipre}
\ee
\end{Lemma}

 It is obvious that
$\CN^{\ch}= \ker(\til{U})$, so that,
comparing with (\ref{deftilpich}), one indeed has $U\circ \pug=
\pi^{\ch}_* \circ U$.\enp 

We use this lemma in the following way. 
To avoid notational confusion, we continue to denote the \Hs\ $\Hug$ defined 
in Construction \ref{rieind2}, starting from
the pre-\HCM\ $C_c(G)\rlh C_c(H)$ defined in the proof of \ref{macglimm}, by $\Hug$.
The \Hs\ $\Hug$ defined below (\ref{iponHG}), however, will play the role $\H_*^{\ch}$
in \ref{mtfhug}, and will therefore be denoted by this symbol.

 Consider the map $\til{U}:
C_c(G)\ot\Hlg\raw \H_*^{\ch}$ defined  by linear extension of
\be
\til{U}\Ps\ot v(x):=\int_H dh\, \Ps(xh)U_{\ch}(h)v. \ll{tilUfell}
\ee
Note that the equivariance condition (\ref{eqforunitary}) is indeed
satisfied by the left-hand side, as follows from the invariance of the Haar
measure.
 
Using (\ref{Psotv}), (\ref{CPrm1}), and (\ref{Utopi}), with $G\raw H$,
one obtains \be (\Ps\ot v,\Ph\ot w)_0^{\ch}=\int_H dh\,
(\Ps,U_R(h)\Ph)_{L^2(G)} (v,U_{\ch}(h)w)_{\ch}= \int_H dh\, \int_G
dx\, \ovl{\Ps(x)}\Ph(xh)(v,U_{\ch}(h)w)_{\ch}; \ll{fundrig0} \ee cf.\
(\ref{defURcap}).  On the other hand, from (\ref{tilUfell}) and
(\ref{iponHG}) one has \be (\til{U}\Ps\ot v,\til{U}\Ph\ot
w)_{\H_*^{\ch}}=\int_H dh\, \int_{G/H} d\nu(\ta(x))\, \int_H dk\,
\ovl{\Ps(xk)} \Ph(xh) (U_{\ch}(k)v,U_{\ch}(h)w)_{\ch}.  \ee Shifting
$h\raw kh$, using the invariance of the Haar measure on $H$, and using
(\ref{3.2r}), one verifies (\ref{whatwtil}). It is clear that
$\til{U}(C_c(G)\ot\Hlg)$ is dense in $\Hug$, so by Proposition
\ref{mtfhug} one obtains the desired unitary map
$U:\Hug\raw\H_*^{\ch}$.

Using (\ref{pichipre}) and (\ref{cghimbim0}), one finds that the induced \rep\ of
$C^*(G,G/H)$ on $\H_*^{\ch}$ is given by
\be
\pi^{\ch}(f)\Ps^{\ch}(x)  =  \int_G dy\, f(xy\inv,[x]_H)\Ps^{\ch}(y);  \ll{finalrie}
 \ee
this looks just like (\ref{cghimbim0}), with the difference that $\Ps$ in 
(\ref{cghimbim0}) lies in $C_c(G)$, whereas $\Ps^{\ch}$ in (\ref{finalrie}) lies in
$\H_*^{\ch}$. Indeed, one should check that the function $\pi^{\ch}(f)\Ps^{\ch}$ defined by
(\ref{finalrie})  satisfies the equivariance condition (\ref{eqforunitary}).

Finally, it is a simple exercise the verify that the \rep\ $\pi^{\ch}(C^*(G,G/H))$ defined
by (\ref{finalrie}) corresponds to the covariant \rep\ $(U^{\ch}(G),\til{\pi}^{\ch}(C_0(G/H))$
by the correspondence (\ref{tgca1}) - (\ref{tgca3}) of Theorem \ref{Upithmbis}.
\enp
\section{Applications to quantum mechanics}
\setc{equation}{0} \su{The mathematical structure of classical and
quantum mechanics} In classical mechanics one starts from a phase
space $S$, whose points are interpreted as the pure states of the
system. More generally, mixed states are identified with probability
measures on $S$. The observables of the theory are functions on $S$;
one could consider smooth, continuous, bounded, measurable, or some
other other class of real-vaued functions. Hence the space $\Ar$ of
observables may be taken to be $\cin(S,\R)$, $C_0(S,\R)$, $C_b(S,\R)$,
or $L^{\infty}(S,\R)$, etc.

 There is a pairing $\la\, ,\, \ra:\CS\x\Ar\raw\R\cup\infty$ between the state space $\CS$ of
probability measures
$\mu$ on
$S$ and the space $\Ar$ of observables $f$. This pairing is given by 
\be
\la\mu,f\ra:=\mu(f)=\int_S d\mu(\sg)\, f(\sg).
\ee
The physical interpretation of this pairing is that in a state $\mu$ the observable $f$ has
expectation value $\la\mu,f\ra$. In general, this expectation value will be unsharp, in that
$\la\mu,f\ra^2\neq \la\mu,f^2\ra$. However,
  in a pure state
$\sg$ (seen as the Dirac measure
$\dl_{\sg}$ on $S$) the observable $f$ has sharp expectation value 
\be
\dl_{\sg}(f)=f(\sg).
\ee

In elementary \qm\ the state space consists of all density matrices $\rh$ on some
\Hs\ $\H$; the pure states are identified with unit vectors $\Ps$. The observables are taken to be
either all unbounded self-adjoint operators
$A$ on
$\H$, or all bounded self-adjoint operators, or all compact self-adjoint operators, etc.
This time the pairing between states and observables is given by
\be
\la\rh,A\ra=\Tr\rh A.
\ee
In a pure state $\Ps$ one has
\be
\la\Ps,A\ra=(\Ps,A\Ps).
\ee
A key difference between classical and \qm\ is that even in pure states expectation values are
generally unsharp. The only exception is when an observable $A$ has discrete spectrum, and $\Ps$
is an eigenvector of $A$. 

In these examples, the state space has a convex structure, whereas the set of observables
is a real vector space (barring problems with the addition of unbounded operators on a \Hs).
We may, therefore, say that a physical theory consists of 
\begin{itemize}
\item
a convex set $\CS$, interpreted as
the state space;
\item
a real vector space $\Ar$, consisting of the observables;
\item
a pairing  $\la\, ,\, \ra:\CS\x\Ar\raw\R\cup\infty$, which assigns the expectation value
$\la\om,f\ra$ to a state $\om$ and an observable $f$. 
\end{itemize}
In addition, one should specify the dynamics of the theory, but this is not our concern here.

The situation is quite neat if $\CS$ and $\Ar$ stand in some duality
relation.  For example, in the classical case, if $S$ is a locally
compact Hausdorff space, and we take $\Ar=C_0(S,\R)$, then the space
of all probability measures on $S$ is precisely the state space of
$\A=C_0(S)$ in the sense of Definition \ref{defstate}; see Theorem
\ref{riesz}. In the same sense, in \qm\ the space of all density
matrices on $\H$ is the state space of the \ca\ $\KH$ of all compact
operators on $\H$; see Corollary \ref{1irrepkh}.1.  On the other hand,
with the same choice of the state space, if we take $\Ar$ to be the
space $\BH_{\R}$ of all bounded self-adjoint operators on $\H$, then
the space of observables is the dual of the (linear space spanned by
the) state space, rather then {\em vice versa}; see Theorem
\ref{dualities}.

In the \ca ic approach to \qm, a general quantum system is specified
by some \ca\ $\A$, whose self-adjoint elements in $\Ar$
correspond to the observables of the theory. The state space of $\Ar$
is then given by Definition \ref{defstate}. This general setting
allows for the existence of {\bf superselection rules}. We will not go
into this generalization of elementary \qm\ here, and concentrate on
the choice $\A=\BH$.  \su{Quantization} The physical interpretation of
\qm\ is a delicate matter. Ideally, one needs to specify the physical
meaning of any observable $A\in\Ar$. In practice, a given quantum
system arises from a classical system by `quantization'. This means
that one has a classical phase space $S$ and a linear map
$\CQ:\Ar^0\raw {\cal L}(\H)$, where $\Ar^0$ stands for $\cin(S,\R)$,
or $C_0(S,\R)$, etc, and ${\cal L}(\H)$ denotes some space of
self-adjoint operators on $\H$, such as $\KH_{\R}$ or
$\BH_{\R}$. Given the physical meaning of a classical observable $f$,
one then ascribes the same physical interpretation to the
corresponding quantum observable $\CQ(f)$. This provides the physical
meaning of al least all operators in the image of $\CQ$.  It is
desirable (though not strictly necessary) that $\CQ$ preserves
positivity, as well as the (approximate) unit.

It is quite convenient to assume that $\Ar^0=C_0(S,\R)$, which choice discards what happens
at infinity on $S$. We are thus led to the following
\begin{Definition}\ll{defquantization}
Let $X$ be a locally compact Hausdorff space.  A {\bf quantization} of
$X$ consists of a \Hs\ $\H$ and a positive map $\CQ:C_0(X)\raw\BH$.
When $X$ is compact it is required that $\CQ(1_X)=\I$, and when $X$ is
non-compact one demands that $\CQ$ can be extended to the unitization
$C_0(X)_{\I}$ by a unit-preserving positive map.
\end{Definition}

Here $C_0(X)$ and $\BH$ are, of course, regarded as \ca s, with the intrinsic notion of positivity
given by \ref{defpos}. Also recall Definition \ref{defposmap} of a positive map. 
It follows from \ref{posdec} that a positive map automatically preserves self-adjointness, in that
\be
\CQ(\ovl{f})=\CQ(f)^*
\ee
for all $f\in C_0(X)$; this implies that $f\in C_0(X,\R)$ is mapped into
 a self-adjoint operator.

There is an interesting reformulation of the notion of a quantization in the above sense.
\begin{Definition}\ll{POV}
Let $X$ be a set with a $\sg$-algebra $\Sg$ of subsets of $X$.
A {\bf positive-operator-valued measure}\index{positive-operator-valued 
measure} or {\bf POVM}\index{POVM} on $X$ in a \Hs\ $\H$ is 
 a map $\Delta \raw A(\Delta)$ from $\Sg$  to $\BH^+$ (the set of
positive operators on $\H$), satisfying 
$A(\emptyset)=0$, $A(X)=\I$, and
$A(\cup_i\Delta_i)=\sum_i A(\Delta_i)$ for any countable collection of 
disjoint $\Delta_i\in\Sg$ (where
 the infinite sum is taken in the weak operator
topology).

A {\bf projection-valued measure}\index{projection-valued measure} 
or {\bf PVM}\index{PVM} is a POVM which in addition satisfies
$A(\Dl_1\cap\Dl_2)  =  A(\Dl_1)A(\Dl_2)$ for all $\Dl_1,\Dl_2\in\Sg$.
\end{Definition}

Note that the above conditions force $0\leq A(\Delta)\leq \I$.
A PVM is usually written as $\Dl\raw E(\Dl)$; it follows that each
$E(\Dl)$ is a projection (take $\Dl_1=\Dl_2$ in the definition). 
This notion is familiar from the spectral theorem.
\begin{Proposition}\ll{pmpovm}
Let $X$ be a locally compact Hausdorff space, with  Borel structure
$\Sg$. There is a bijective correspondence between
quantizations $\CQ:C_0(X)\raw\BH$, and POVM's  $\Delta \raw A(\Delta)$ on $S$ in $\H$, given by
\be
\CQ(f)=\int_S dA(x)\, f(x). \ll{CQfAx}
\ee

The map $\CQ$ is a \rep\ of $C_0(X)$ iff $\Delta \raw A(\Delta)$ is a PVM.
\end{Proposition}

The precise meaning of (\ref{CQfAx}) will emerge shortly.
Given the assumptions, in view of \ref{GTGT0} and \ref{caun}
we may as well assume that $X$ is compact.

Given $\CQ$, for arbitrary $\Ps\in \H$ one constructs a 
functional $\hat{\mu}_{\Ps,\Ps}$ on $C(X)$ by
$\hat{\mu}_{\Ps,\Ps}(f):=(\Ps,\CQ(f)\Ps)$
Since $\CQ$ is linear and  positive, this functional has the same properties.
Hence the Riesz \rep\ theorem yields a probability 
measure $\mu_{\Ps,\Ps}$ on $X$.
For $\Dl\in\Sg$ one then puts
$(\Ps,A(\Dl)\Ps):=\mu_{\Ps,\Ps}(\Dl)$,  defining an operator
$A(\Dl)$ by polarization. The ensuing map $\Delta \raw A(\Delta)$
is easily checked to have the properties required of a POVM.

Conversely, for each pair $\Ps,\Ph\in\H$
a POVM $\Dl\raw A(\Dl)$ in $\H$ defines 
a signed measure $\mu_{\Ps,\Ph}$ on $X$  by
means of $\mu_{\Ps,\Ph}(\Dl):=(\Ps,A(\Dl)\Ph)$. This yields 
a positive map $\CQ:C(X)\raw\BH$ by 
$(\Ps,\CQ(f)\Ph):=\int_X d\mu_{\Ps,\Ph}(x)\, f(x)$;  the 
 meaning of (\ref{CQfAx}) is expressed by this equation.

Approximating $f,g\in C(X)$ by step functions, one verifies that
the property $E(\Dl)^2=E(\Dl)$ is equivalent to $\CQ(fg)=\CQ(f)\CQ(g)$.
\enp
\begin{Corollary}\ll{neumark}
Let $\Dl\raw A(\Dl)$ be a POVM on a locally compact Hausdorff space
$X$ in a \Hs\ $\Hlg$. There exist a \Hs\ $\Hug$, a projection $p$
on $\Hug$, a unitary map $U:\Hlg\raw p\Hug$, and a PVM $\Dl\raw E(\Dl)$ 
on $\Hug$ such that $UA(\Dl)U\inv=pE(\Dl)p$ for all $\Dl\in\Sg$.
\end{Corollary}

Combine Theorem \ref{Stinespring} with Proposition \ref{pmpovm}. \enp

When  $X$ is the phase space $S$ of a physical system, the physical interpretation of the map
$\Dl\raw A(\Dl)$ is  contained in the statement that the number
\be
p_{\rh}(\Dl):=\Tr \rh A(\Dl) \ll{pipovm}
\ee
is the probability that, in a state $\rh$,
the system in question is localized in $\Dl\subset S$.
 
When $X$ is a configuration space $Q$, it is usually sufficient to
take the positive map $\CQ$ to be a \rep\ $\pi$ of $C_0(Q)$ on
$\H$. By Proposition \ref{pmpovm}, the situation is therefore
described by a PVM $\Dl\raw E(\Dl)$ on $Q$ in $\H$.  The probability
that, in a state $\rh$, the system is localized in $\Dl\subset Q$ is
\be p_{\rh}(\Dl):=\Tr \rh E(\Dl). \ll{pipvm} \ee \su{Stinespring's
theorem and coherent states} By Proposition \ref{pacp}, a quantization
$\CQ:C_0(X)\raw\BH$ is a completely positive map, and Definition
\ref{defquantization} implies that the conditions for Stinespring's
Theorem \ref{Stinespring} are satisfied. We will now construct a class
of examples of quantization in which one can construct an illuminating
explicit realization of the \Hs\ $\Hug$ and the partial isometry $W$.

Let $S$ be a locally compact Hausdorff space (interpreted as a
classical phase space), and consider an embedding $\sg\raw\Ps^{\sg}$
of $S$ into some \Hs\ $\H$, such that each $\Ps^{\sg}$ has unit norm
(so that a pure classical state is mapped into a pure quantum
state). Moreover, there should be a measure $\mu$ on $S$ such that \be
\int_S d\mu(\sg) (\Ps_1,\Ps^{\sg}) (\Ps^{\sg},\Ps_2)=(\Ps_1,\Ps_2).
\ll{qhnormpol} \ee for all $\Ps_1,\Ps_2\in\H$.  The $\Ps^{\sg}$ are
called {\bf coherent states} for $S$.

Condition (\ref{qhnormpol}) guarantees that we may define a POVM on $S$ in $\H$
by
\be
A(\Dl)= \int_{\Dl}  d\mu(\sg)\,[\Ps^{\sg}], \ll{pov}
\ee
where $[\Ps]$ is the projection onto the one-dimensional subspace spanned
by $\Ps$ (in Dirac's notation one would have $[\Ps]=|\Ps><\Ps|$).

The positive map $\CQ$ corresponding to the POVM $\Dl\raw A(\Dl)$ by Proposition \ref{pmpovm}
is given by
\be
\CQ(f)= \int_{S}  d\mu(\sg)\, f(\sg) [\Ps^{\sg}].
\ll{b2}
\ee
  In particular, one has $\CQ(1_S)=\I$. 

For example, when $S=T^*\R^3=\R^6$, so that $\sg=(p,q)$, one may take 
\be
\Ps^{(p,q)}(x)=(\pi)^{-n/4}e^{-\half
ipq+ipx}e^{-(x-q)^2/2}\ll{pqcohst} 
\ee
in $\H=L^2(\R^3)$.
Eq.\ (\ref{qhnormpol})  then holds with $d\mu(p,q)=d^3pd^3q/(2\pi)^3$. 
 Extending the map $\CQ$ from $C_0(S)$ to $\cin(S)$ in a heuristic way, one finds that 
$\CQ(q_i)$
and $\CQ(p_i)$ are just the usual position- and momentum operators in the Schr\"{o}dinger \rep.

In Theorem \ref{Stinespring} we now put $\A=C_0(S)$, $\B=\BH$, $\plg(A)=A$ for all $A$.
We may then verify the statement of the theorem by taking
\be 
\Hug= L^2(S,d\mu) \ll{defHuhbar}.
\ee 
The map $W:\H\raw\Hug$ is then given by 
\be 
W\Ps(\sg):=(\Ps^{\sg},\Ps).  \ll{mapHlts} 
\ee 
It follows from
(\ref{qhnormpol}) that $W$ is a partial isometry.
The \rep\ $\pi(C_0(S))$ is  given by
\be
\pi(f)\Ph(\sg)=f(\sg)\Ph(\sg).
\ee
 Finally, for (\ref{fundefbt}) one has the simple expression
\be
\til{\CQ}(f)=U\CQ(f)U\inv=pfp.\ll{toeplitz}
\ee

Eqs.\ (\ref{defHuhbar}) and (\ref{toeplitz}) form the core of 
the realization of quantum
mechanics on phase space. One realizes the state space as a closed
subspace of $L^2(S)$ (defined with respect to a suitable 
 measure), and defines the quantization of a
classical observable $f\in C_0(S)$ as multiplication by $f$, 
sandwiched between the projection onto the subspace in question.
This should be contrasted with the usual way of doing \qm\ on $L^2(Q)$, where $Q$ is
the configuration space of the system.
 
In specific cases the projection $p=WW^*$ can be explicitly given as well. For example, 
in the case $S=T^*\R^3$ considered above one may pass to complex variables by putting
$z=(q-ip)/\sqrt{2}$. We then map $L^2(T^*\R^3, d^3pd^3q/(2\pi)^3)$ into
$\CK:=L^2(\C^3, d^3z d^3 \ovl{z}\exp(-z\zb)/(2\pi i)^3)$ 
by the unitary operator
$V$, given by
\be
V\Ph(z,\zb):=e^{\half z\zb}\Ph(p=(\zb-z)/\sqrt{2},q=(\zb+z)/\sqrt{2}).
\ee
One may then verify from (\ref{mapHlts}) and (\ref{pqcohst}) that $VpV\inv$ is the projection
onto the space of entire functions in $\CK$. 
\su{Covariant localization in configuration space}
In elementary \qm\ a particle moving on $\R^3$ with spin $j\in\N$
is described by the \Hs\ 
\be
\H^j_{\mbox{\tiny QM}}=L^2(\R^3)\ot \H_j, \ll{Hupj}
\ee
 where $\H_j=\C^{2j+1}$
carries the \irrep\ $U_j(SO(3))$ (usually called $\CD_j$). 
The basic physical observables are represented by unbounded operators
$Q^S_k$ (position), $P^S_k$ (momentum), and $J^S_k$ (angular momentum), where
$k=1,2,3$. These operators  satisfy the commutation relations
(say, on the domain $\CS(\R^3)\ot\H_j$)
\bea
& & [Q^S_k,Q^S_l]  =  0; \ll{qqcom}\\
& & [P^S_k,Q^S_l] =  -i\hbar\dl_{kl}; \ll{qpjcom1}\\
& &  [J^S_k,Q^S_l]=i\hbar\ep_{klm}Q^S_m; \ll{qpjcom2}\\
& & [P^S_k,P^S_l] =  0; \ll{jpcom1}\\
& & [J^S_k,J^S_l]=i\hbar\ep_{klm}J^S_m;\ll{jpcom2}\\
& & [J^S_k,P^S_l]=i\hbar\ep_{klm}P^S_m, \ll{jpcom3}
\eea
justifying their physical interpretation.

The momentum and angular momentum operators
are most conveniently defined in terms of a unitary
\rep\ $U^j_{\mbox{\tiny QM}}$ of the Euclidean group $E(3)=SO(3)\ltimes\R^3$ on $\H^j_{\mbox{\tiny QM}}$,
given by
\be
U^j_{\mbox{\tiny QM}}(R,a)\Ps(q)=U_j(R)\Ps(R\inv(q-a)). \ll{paulirep}
\ee
In terms of the standard generators $P_k$ and $T_k$ of $\R^3$ and $SO(3)$,
respectively, one then has $P^S_k=i\hbar dU^j_{\mbox{\tiny QM}}(P_k)$ and
$J^S_k=i\hbar dU^j_{\mbox{\tiny QM}}(T_k)$; see (\ref{defdUeq}). The  commutation relations (\ref{jpcom1})
- (\ref{jpcom3})
follow from (\ref{crdU}) and the commutation relations in the Lie algebra of $E(3)$.

Moreover, we define a \rep\ $\til{\pi}^j_{\mbox{\tiny QM}}$ of
$C_0(\R^3)$ on $\H^j_{\mbox{\tiny QM}}$ by 
\be
\til{\pi}^j_{\mbox{\tiny QM}}(\til{f})=\til{f}\ot\I_j,\ll{piupj}
\ee
 where
$\til{f}$ is seen as a multiplication operator on $L^2(\R^3)$.
The associated
PVM $\Dl\raw E(\Dl)$ on $\R^3$ in $\H^j_{\mbox{\tiny QM}}$ (see \ \ref{pmpovm}) is 
$E(\Dl)=\ch_{\Dl}\ot\I_j$, in terms of which
the position operators are given by $Q^S_k=\int_{\R^3} dE(x) x_k$;
cf.\ the spectral theorem for unbounded operators. 
Eq.\ (\ref{qqcom}) then reflects the commutativity
of $C_0(\R^3)$, as well as the fact that $\til{\pi}^j_{\mbox{\tiny QM}}$ is a \rep.

Identifying $Q=\R^3$ with $G/H=E(3)/SO(3)$ in the obvious way, one checks
that the canonical left-action of $E(3)$ on $E(3)/SO(3)$ is identified
with its defining action on $\R^3$.
It is then not hard to verify from (\ref{paulirep}) that
the pair $(U^j_{\mbox{\tiny QM}}(E(3)), \til{\pi}^j_{\mbox{\tiny QM}}(C_0(\R^3)))$ is a 
covariant \rep\ of the $C^*$-dynamical system $(E(3), C_0(\R^3),\al)$, with
$\al$ given by (\ref{alxtilf}). 
The commutation relations (\ref{qpjcom1}), (\ref{qpjcom2})  
are a consequence of the
covariance relation (\ref{a154}).

Rather than using the unbounded operators $Q^S_k$, $P^S_k$, and $J^S_k$,
and their commutation relations, we therefore state the situation
in terms of the pair $(U^j_{\mbox{\tiny QM}}(E(3)),\til{\pi}^j_{\mbox{\tiny QM}}(C_0(\R^3)))$.
Such a pair, or, equivalently, a non-degenerate
\rep\ $\pi^j_{\mbox{\tiny QM}}$ of the transformation group
\ca\ $C^*(E(3),\R^3)$ (cf.\ \ref{rificolcp}, then 
by definition describes
 a quantum system which is localizable in $\R^3$.
and covariant under the defining
action of $E(3)$. 
It is natural to require that $\pi^j_{\mbox{\tiny QM}}$
be irreducible, in which case the quantum system itself is said to be
irreducible.
\begin{Proposition}\ll{mackeyspin}
An irreducible quantum system which is localizable in $\R^3$ and
covariant under $E(3)$ is completely characterized by its spin
$j\in\N$.  The corresponding covariant \rep\
$(U^j(E(3)),\til{\pi}^j(C_0(\R^3)))$, given by \ref{defHG},
(\ref{gblattner2}), and (\ref{tilpiuchbr2}),
 is equivalent to the one described by
(\ref{Hupj}), (\ref{paulirep}), and (\ref{piupj}).
\end{Proposition}

This follows from Theorem \ref{mackim}.
The  \rep\ $U^j_{\mbox{\tiny QM}}(E(3))$ defined in (\ref{paulirep}) is unitarily equivalent
to the induced \rep\ $U^j$. To see this, check that the unitary map
$V:\H^j\raw\H^j_{\mbox{\tiny QM}}$ defined by $V\Ps^j(q):=\Ps^j(e,q)$ intertwines
$U^j$ and $U^j_{\mbox{\tiny QM}}$. In addition, it intertwines the \rep\ (\ref{tilpiuchbr2}) with
$\til{\pi}^j_{\mbox{\tiny QM}}$ as defined in (\ref{piupj}).
\enp

This is a neat explanation of spin in \qm.  

Generalizing this approach to an arbitrary homogeneous configuration
space $Q=G/H$, a non-degenerate \rep\ $\pi$ of $C^*(G,G/H)$ 
on a \Hs\ $\H$ describes
a quantum system which is localizable in $G/H$ and covariant under the
canonical action of $G$ on $G/H$. By  \ref{rificolcp}
this is equivalent to a covariant \rep\  $(U(G),\til{\pi}(C_0(G/H)))$
on $\H$,
and by Proposition \ref{pmpovm} one may instead assume one has a 
PVM $\Dl\raw E(\Dl)$ on $G/H$ in $\H$ and  a unitary \rep\ $U(G)$, 
which satisfy 
\be
U(x)E(\Dl)U(x)\inv=E(x\Dl) \ll{covloc2}
\ee
for all $x\in G$ and $\Dl\in\Sg$; cf.\ (\ref{covloc}). 
The physical interpretation of the PVM is given by (\ref{pipvm});
the operators
defined in (\ref{defqpi1}) play the role of quantized momentum
observables. Generalizing Proposition \ref{mackeyspin}, we have
\begin{Theorem}\ll{mackeygh}
An irreducible quantum system which is localizable in $Q=G/H$
and covariant under the canonical action of $G$ 
is characterized by an irreducible unitary \rep\ 
of $H$.
The system of imprimitivity $(U^{\ch}(G),\til{\pi}^j(C_0(G/H)))$
is equivalent to the one described by (\ref{gblattner2})
and  (\ref{tilpiuchbr2}).
\end{Theorem}

This is immediate from Theorem \ref{mackim}.\enp

For example, writing the two-sphere $S^2$ as $SO(3)/SO(2)$, one infers that
$SO(3)$-covariant quantum particles on $S^2$ are characterized by an integer $n\in\Bbb Z$.
For each unitary \irrep\ $U$ of $SO(2)$ is labeled by such an $n$, and given by
$U_n(\th)=\exp(in\th)$.
\su{Covariant quantization on phase space}
Let us return to quantization theory, and ask
what happens  in the presence of a  symmetry group. The following notion, which generalizes
Definition \ref{defcovrep}, is natural in this context.
\begin{Definition}\ll{defsi5}
A {\bf generalized covariant representation} of a $C^*$-dynamical system 
\\ $(G,C_0(X),\al)$,
where $\al$ arises from a continuous $G$-action on $X$ by means of
(\ref{alxtilf}), consists
of a pair $(U,\CQ)$, where $U$ is a
   unitary \rep\ of $G$ on a \Hs\ $\H$, and 
$\CQ:C_0(X)\raw\BH$ is a quantization of $C_0(X)$ (in the sense of Definition
\ref{defquantization}), which for all $x\in G$ and $\til{f}\in C_0(X)$
 satisfies  
 the covariance condition 
\be
U(x)\CQ(\til{f})U(x)^*=\CQ(\al_x(\til{f})). \ll{covcond2}
\ee  
\end{Definition}

This condition may be equivalently stated
in terms of the POVM $\Dl\raw A(\Dl)$ associated to $\CQ$
(cf.\ \ref{pmpovm}) by
\be
U(x)A(\Dl)U(x)\inv=A(x\Dl). \ll{covloc}
\ee

Every (ordinary) covariant representation is evidently a generalized
one as well, since a \rep\ is a particular example of a quantization.
A class of examples of truly generalized  covariant representations
arises as follows.  Let $(U(G),\til{\pi}(C_0(G/H))$ be a
covariant representation on a Hilbert space $\CK$, and suppose that
$U(G)$ is reducible. Pick a projection $p$ in the commutant of $U(G)$;
then $(pU(G),p\til{\pi}p)$ is a generalized covariant \rep\  on
$\H=p\CK$.  Of course, $(U,\til{\pi})$ is  described by
Theorem \ref{mackim}, and must be of the form
$(U^{\ch},\til{\pi}^{\ch})$. 
This class actually turns out to exhaust all possibilities.
What follows generalizes Theorem \ref{mackim}   to the case where
the \rep\ $\til{\pi}$ is replaced by a quantization $\CQ$. 
\begin{Theorem}\ll{posmackimp}
Let $(U(G),\CQ(C_0(G/H)))$ be a generalized 
covariant \rep\ of the $C^*$-dynamical system $(G,C_0(G/H),\al)$, defined
with respect to the canonical $G$-action on $G/H$.

There exists a unitary \rep\ $U_{\ch}(H)$, with corresponding
covariant \rep\ $(U^{\ch},\til{\pi}^{\ch})$ of $(G,C_0(G/H),\al)$ on
the \Hs\ $\Hug$, as described by \ref{defHG}, (\ref{gblattner2}), and
(\ref{tilpiuchbr2}), and a projection $p$ on $\Hug$ in the commutant
of $U^{\ch}(G)$, such that $(pU^{\ch}(G),p\til{\pi}^{\ch}p)$ and
$(U(G),\CQ(C_0(G/H)))$ are equivalent.
\end{Theorem}

We apply Theorem \ref{Stinespring}. To avoid confusion, we denote
the \Hs\ $\Hug$ and the \rep\ $\pug$ in Construction
\ref{stinespringcon} by $\til{\H}^{\ch}$ and $\til{\pi}^{\ch}$,
respectively; the space defined in \ref{defHG} and the induced
\rep\ (\ref{gblattner2}) will still be called $\Hug$ and $\pug$,
as in the formulation of the theorem above. Indeed, our goal is to
show that $(\til{\pi}^{\ch},\til{\H}^{\ch})$ may be identified with
$(\pug,\Hug)$.  We identify $\B$ in \ref{Stinespring} and
\ref{stinespringcon} with $\BH$, where $\H$ is specified in
\ref{posmackimp};  we therefore omit
the \rep\ $\plg$ occurring in \ref{Stinespring} etc., putting $\Hlg=\H$.

For $x\in G$ we define a linear map $\til{U}(x)$ on $C_0(G/H)\ot \H$
by linear extension of
\be
\til{U}(x) f\ot\Ps:= \al_x(f)\ot U(x)\Ps.
\ee
Since $\al_x\circ\al_y=\al_{xy}$, and $U$ is a \rep, 
$\til{U}$ is clearly a $G$-action. Using the covariance
condition (\ref{covcond2}) and the unitarity of $U(x)$, one verifies that
\be
(\til{U}(x) f\ot\Ps,\til{U}(x) g\ot\Ph)^{\ch}_0 =(f\ot\Ps, g\ot\Ph)^{\ch}_0,
\ee
where $(\, , \,)^{\ch}_0$ is defined in (\ref{woody1}). Hence $\til{U}(G)$ quotients to a  
\rep\ $\til{U}^{\ch}(G)$ on  $\til{\H}^{\ch}$.
Computing on $C_0(G/H)\ot \H$ and then passing to the quotient,
 one checks that
$(\til{U}^{\ch},\til{\pi}^{\ch})$ is a covariant \rep\  on 
 $\til{\H}^{\ch}$.
By  Theorem \ref{mackim}, this system must be of the form $(U^{\ch},\til{\pi}^{\ch})$
(up to unitary equivalence).

Finally, the projection $p$ defined in \ref{Stinespring} commutes
with all $\til{U}^{\ch}(x)$.  This is verified from (\ref{WPsVI}),
(\ref{wocomp}), and (\ref{covcond2}).  The claim follows.\enp 
 \section*{Literature}
\addcontentsline{toc}{section}{Literature}
\thispagestyle{myheadings}
\markright{LITERATURE}
\begingroup
\def\item{\vskip2.75pt plus1.375pt minus.6875pt\noindent\hangindent1em}
\hbadness2500 \tolerance 2500
\item 
 Bratteli, O. and D.W. Robinson [1987] {\em Operator
Algebras and Quantum Statistical Mechanics, Vol.\ I: $C^*$- and $W^*$-Algebras, Symmetry Groups,
Decomposition of States}, 2nd ed. Springer, Berlin.\item 
  Bratteli, O. and D.W. Robinson [1981] {\em Operator Algebras and Quantum Statistical Mechanics,
Vol.\ II: Equilibrium States, Models in Statistical Mechanics}. Springer, Berlin.\item 
Connes, A. [1994] {\em Noncommutative Geometry}. Academic Press, San
Diego.\item Davidson, K.R. [1996] {\em $C^*$-Algebras by
Example}. {\em Fields Institute Monographs} {\bf 6}.  Amer.\ Math.\
Soc., Providence (RI).\item Dixmier, J. [1977] {\em
$C^*$-Algebras}. North-Holland, Amsterdam.\item Fell, J.M.G. [1978]
{\em Induced Representations and Banach $\mbox{}^*$-algebra
Bundles. Lecture Notes in Mathematics} {\bf 582}. Springer,
Berlin.\item Fell, J.M.G.  and R.S. Doran [1988] {\em Representations
of $\mbox{}^*$-Algebras, Locally Compact Groups and Banach
$\mbox{}^*$-Algebraic Bundles, Vol.\ 2}. Academic Press, Boston.\item
Kadison, R.V. [1982] Operator algebras - the first forty years.  In:
Kadison, R.V. (ed.) {\em Operator Algebras and Applications}, {\em
Proc.\ Symp.\ Pure Math.} {\bf 38(1)}, pp.\ 1-18.  American
Mathematical Society, Providence.\item Kadison, R.V. [1994] Notes on
the Gelfand-Neumark theorem. In: Doran, R.S. (ed.) {\em
$C^*$-algebras: 1943-1993}. {\em Cont.\ Math.} {\bf 167}, pp.\
21-53. Amer.\ Math.\ Soc., Providence (RI).\item Kadison, R.V. and
J.R. Ringrose [1983] {\em Fundamentals of the Theory of Operator
Algebras I}.  Academic Press, New York.\item Kadison, R.V. and
J.R. Ringrose [1986] {\em Fundamentals of the Theory of Operator
Algebras II}.  Academic Press, New York.\item Lance, E.C. [1995] {\em
Hilbert $C^*$-Modules. A Toolkit for Operator Algebraists. LMS Lecture
Notes} {\bf 210}.  Cambridge University Press, Cambridge.\item Mackey,
G.W. [1963] {\em The Mathematical Foundations of Quantum
Mechanics}. New York, Benjamin.
\item 
Mackey, G.W. [1968] {\em Induced Representations}. Benjamin, New York.\item 
Mackey, G.W. [1978]  {\em Unitary Group Representations in Physics Probability and
Number Theory}. Benjamin, New York.\item 
Pedersen, G.K. [1979] {\em $C^*$-Algebras and their Automorphism Groups}. Academic Press,
London.\item 
Pedersen, G.K. [1989] {\em Analysis Now}. Springer, New York.\item
Takesaki, M. [1979] {\em Theory of Operator Algebras I}. Springer, Heidelberg.
\item  Wegge-Olsen, N.E. [1993] {\em  K-theory and $C^*$-algebras}.
 Oxford University Press, Oxford.
\endgroup

\end{document}